# Project X

## Accelerator Reference Design, Physics Opportunities, Broader Impacts

**June 2013**

# Preface

Particle physics has made enormous progress in understanding the nature of matter and forces at a fundamental level and has unlocked many mysteries of our world. The development of the Standard Model of particle physics has been a magnificent achievement of the field. Many deep and important questions have been answered and yet many mysteries remain. The discovery of neutrino oscillations, discrepancies in some precision measurements of Standard-Model processes, observation of matter-antimatter asymmetry, the evidence for the existence of dark matter and dark energy, all point to new physics beyond the Standard Model. The pivotal developments of our field, including the latest discovery of the Higgs Boson, have progressed within three interlocking frontiers of research–the Energy, Intensity and Cosmic frontiers–where discoveries and insights in one frontier powerfully advance the other frontiers as well.

Project X is an opportunity to continue this legacy through developing a world-leading Intensity Frontier facility at Fermilab that can support a very diverse and rich physics program in this quest. The Project X proton accelerator is ready to construct, and will leverage the substantial investment the U.S. research enterprise has made broadly in developing superconducting radio frequency (SCRF) accelerator technology. This accelerator technology is the most efficient mechanism to prepare and accelerate beams of charged particles and has excellent scaling properties for delivering very high power beams. Laboratories worldwide have now embraced SCRF as core technology for leading facilities in material science, energy research, nuclear physics and particle physics. In the domain of particle physics, SCRF is the enabling technology for both the highest energy beams that drive the Energy Frontier and the highest intensity beams that drive the Intensity Frontier. A key metric for the Intensity Frontier, including a variety of existing and planned proton-beam future facilities, is beam power, and the worldwide power landscape is illustrated in the figure opposite. It is clear from this figure that megawatt-class beam facilities are the emerging standard, and that Project X (green stars) establishes a world leading facility enabled by SCRF accelerator technology.

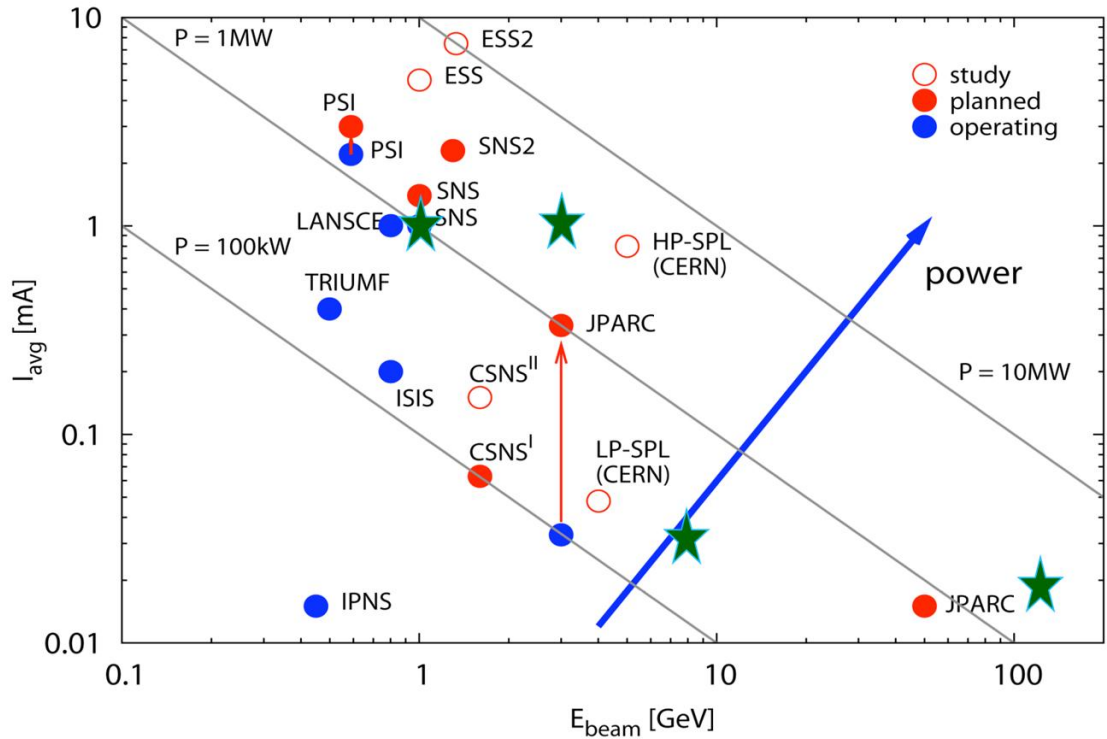

Currently operating, planned, and under-study proton-beam facilities. Beam energy and average current are on the horizontal and vertical axes; the diagonal lines are contours of constant beam power. The Project X footprint is indicated by the green stars. Figure courtesy M. Seidel, PSI.

The Intensity Frontier research program enabled by Project X can address some of the most compelling and fundamental questions in particle physics today including:

*Are there new forces in nature?* Experiments have established flavor-changing processes in quarks and neutrinos, so it seems conceivable that charged leptons change flavor too. With Project X, one can search for these phenomena via muon-to-electron conversion and related processes. Many of the theoretical ideas unifying forces and flavor violation anticipate baryon-number violation, and Project X can extend the limits on neutron-antineutron oscillations by orders of magnitude. These same ideas posit measurable flavor-changing neutral currents that may amplify rare decays such as (charged and neutral) $K \to \pi \nu \bar{\nu}$.

*Are there new properties of matter?* According to the Sakharov conditions, the baryon asymmetry of the universe requires CP-violating interactions, but their strength in the Standard Model is insufficient to account for the observed excess. It is not known whether the missing CP violation takes place in the neutrino sector or the quark sector. Project X will aid both searches, by increasing the reach of neutrino oscillation experiments and by enabling a new suite of searches for nonzero electric dipole moments (EDMs). The latter program is broad, looking for an EDM of the neutron, proton, muon directly, and the electron through isotopes produced by Project X.

*Are there new dimensions?* Many popular extensions of the Standard Model introduce extra dimensions: in the case of supersymmetry, the dimensions are fermionic. The space of non-standard interactions opens up possibilities for the interactions mentioned above: quark and neutrino CP violation and quark-flavor-changing neutral currents with supersymmetry, and flavor-changing neutral currents from a warped fifth spatial dimension. Rare kaon decays, EDMs, and neutron-antineutron oscillations are closely tied to these possibilities.

In addition to probing these fundamental questions the Project X research program includes experiments that can test and enrich our understanding of quantum chromodynamics and the electroweak theory.

The Project X accelerator architecture presents opportunities to stage construction of the complete facility if necessary, with a robust research program at the first stage and each successive stage growing both the versatility of beams available and the intensity of those beams. A key and unique characteristic of the Project X complex is that the multiplicity of beams required by the research program can largely be provided *simultaneously,* enabling multiple experiments to proceed in parallel. Inspired by the research opportunities of Project X and the maturity of SCRF technology, the High Energy Physics Advisory Panel (HEPAP) in a recent review of facilities[1] characterized Project X as being "absolutely central" to the future of particle physics and "ready to initiate construction" in terms of technical maturity.

Project X also provides opportunities for research beyond the domain of particle physics. These potential Broader Impacts include an opportunity to provide irradiation resources for the study of materials important to energy applications and provision of low energy muons for muon spin rotation experiments which can incisively probe the magnetic properties of materials.

The three parts of this book form a comprehensive discussion of Project X. Part 1 is the Reference Design Report (RDR) for the Project X accelerator complex. Part 2 describes the Project X Particle Physics Research Program opportunities. Part 3 describes the Broader Impacts opportunities realized in irradiation resources from a spallation target and low energy muons to support muon-spin-rotation experiments (as the only such facility in the US) for materials science research.

---

[1] HEPAP Facilities Subpanel, "Major High Energy Physics Facilities, 2014-2024, Input to the prioritization of proposed scientific user facilities for the Office of Science", March 2013
http://science.energy.gov/~/media/hep/hepap/pdf/Reports/HEPAP_facilities_letter_report.pdf

# Project X Reference Design Report

## June 2013

### The Project X Collaboration:

| | |
|---|---|
| Argonne National Laboratory | Oak Ridge National Laboratory/SNS |
| Brookhaven National Laboratory | Pacific Northwest National Laboratory |
| Cornell University | SLAC National Accelerator Laboratory |
| Fermi National Accelerator Laboratory | University of Tennessee |
| Lawrence Berkeley National Laboratory | Thomas Jefferson National Accelerator Facility |
| Michigan State University | ILC/America's Regional Team |
| North Carolina State University | |
| | |
| Bhabha Atomic Research Center, Mumbai | Raja Ramanna Center for Advanced Technology, Indore |
| Inter-University Accelerator Center, Delhi | Variable Energy Cyclotron Center, Kolkota |

### Edited by:

Stephen D. Holmes/Fermi National Accelerator Laboratory

# Table of Contents

# I   Introduction

Project X is a high-intensity proton facility being developed to support a world-leading program of Intensity Frontier physics over the next two decades at Fermilab. Project X is an integral part of the U.S. Intensity Frontier Roadmap as described in the P5 report of May 2008 [1] and within the Fermilab Strategic Plan of November 2011 [2].

This document represents Part I of the "Project X Book" describing the Project X accelerator facility and the broad range of physics research opportunities enabled by Project X. Parts II and III provide in-depth descriptions of the physics research program, both within and beyond particle physics [3]. The primary elements of the U.S. program to be supported by Project X include:

*Neutrino Experiments*: Experimental studies of neutrino oscillations and neutrino interaction physics with ultra-intense neutrino beams provided by a high-power proton source with energies up to 120 GeV, utilizing near detectors at the Fermilab site and massive detectors at distant underground laboratories.

*Goal*: At least 2 MW of proton beam power at any energy between 60 to 120 GeV; several hundred kW of proton beam power on target at 8 GeV.

*Kaon, Muon, Nucleon, and Neutron Precision Experiments*: World-leading experiments studying ultra-rare kaon decays, searching for muon-to-electron conversion and nuclear electron dipole moments (EDMs), and exploring neutron properties at very high precision.

*Goal*: MW-class proton beams supporting multiple experiments at 1 and 3 GeV, with flexible capability for providing distinct beam formats to concurrent users while allowing simultaneous operations with the neutrino program.

*Material Science and Nuclear Energy Applications*: High-intensity accelerator, spallation, target and transmutation technology demonstrations will provide critical input into the design of future energy systems, including next generation fission reactors, nuclear waste transmutation systems and future thorium fuel-cycle power systems. Possible applications of muon spin rotation techniques provide sensitive probes of the magnetic structure of materials.

Goal: Provide MW-class proton beams at 1 GeV, coupled with novel targets required to support a broad range of materials science and energy applications.

*Platform for Evolution to Future Frontier Facilities*: A high-intensity proton source will strengthen and modernize the Fermilab injector complex, providing a robust platform upon which to build future frontier facilities. The Neutrino Factory and Muon Collider are examples that would provide world-leading capabilities at the Intensity and Energy Frontiers for many decades to come.

> *Goal*: Provide a straightforward upgrade path for a 4 MW, low-duty-factor source of protons at energies between 5 and 15 GeV.

These four elements are expected to form the basis of the Mission Need statement required for the Department of Energy (DOE) Critical Decision 0 (CD-0), and represent the fundamental design criteria for Project X.

The following chapters present the Reference Design for the Project X accelerator facility. The Reference Design is based on a continuous wave (CW) superconducting (SC) linac providing up to 1 and 3 MW of beam power at 1 and 3 GeV respectively. A pulsed linac provides acceleration of roughly 4% of the beam delivered from the CW linac to the 8 GeV injection energy of the existing Recycler/Main Injector complex. Upgrades to the Recycler and Main Injector support a factor of three increase, beyond current capabilities, in proton beam power at 60 to 120 GeV.

The Reference Design represents a facility that will be unique in the world with unmatched capabilities for the delivery of very high beam power, with flexible beam formats, to multiple users at multiple energies. The utilization of linacs within the Project X facility enables capabilities beyond what is achievable with circular accelerators. It is anticipated that the final configuration and operating parameters of the complex will be further refined through the R&D program in advance of CD-2.

## I.1 Evolution of the Reference Design

The Reference Design represents an evolution of the Project X design concept over a number of years, incorporating continuous input on physics research goals and advances in the underlying technology development programs. The initial Project X goals and associated design concept [4] were primarily driven by technology synergies with the International Linear Collider (ILC) and the 2-MW operational goal for the Main Injector in support of the long baseline neutrino program. This concept was based on an 8-GeV superconducting pulsed linac, paired with the existing Recycler and Main Injector rings. Operations in support of programs beyond long-baseline neutrinos were not considered in this first concept. While some enhancements were introduced in the Project X Initial Configuration Document-1 (ICD-1 [5]) it followed the same path as the initial Project X concept but with an increased beam current. The accelerator complex defined in ICD-1 could drive the long-baseline neutrino program, and provided enhanced capabilities in the muon-to-electron conversion experiment (Mu2e); It did not, however, provide a flexible platform for pursuit of a broader research program in rare muon, kaon, nucleon, and neutron processes based on high duty-factor beams.

ICD-2 [6] addressed the lack of flexibility in ICD-1 by introducing a 2-GeV CW linac and a 2-8 GeV rapid cycling synchrotron (RCS), still paired with the existing Recycler and Main Injector. The intent of ICD-2 was to strengthen the broader physics research program while preserving the 2-MW neutrino beam program from the Main Injector. However, it was soon recognized that the 2-GeV CW linac energy was inadequate for supporting a compelling set of kaon experiments, and that utilization of a rapid-cycling synchrotron would ultimately limit the ability to support future facilities of interest to the U.S. particle physics community, such as the Neutrino Factory and Muon Collider.

These considerations have led to the development of the Reference Design. The Reference Design continues to be based on a superconducting CW linac; however, the energy has been raised to 3 GeV to provide a more robust kaon program, and the selection of RF frequencies has been changed to provide more efficient acceleration. As opportunities for research in materials science and energy applications were identified, the capability for delivering high power beams at 1 GeV was incorporated. In order to retain greater flexibility for future programs the RCS was replaced by a pulsed linac. The configurations of the Main Injector and Recycler remain as developed in the ICD-1 and ICD-2.

## I.2  Assumptions

The following assumptions are made in determining the Project X facility configuration that meets the mission elements defined above:

- Project X will be constructed on the Fermilab site and will utilize the upgraded Main Injector and Recycler for the generation of a long-baseline neutrino beam.
- The long-baseline neutrino program will require a capability of greater than 2 MW proton beam power on a neutrino production target at any energy within the range 60-120 GeV, and will eventually require additional beam power upgrades.

- The precision experiments program requires MW-class proton beams, high duty factors, and flexible bunch patterns. This is best accomplished with a CW linac based on superconducting radio frequency (SRF) technologies.

- The production yield curve of kaons vs. incident proton energy sets the minimum CW linac energy at ~ 3 GeV

- Precision experiments based on nuclei and neutrons, as well as materials science and energy applications programs require MW-class proton beams at 1 GeV.

- A proton electric dipole moment program would require a very low duty factor beam at 236 MeV.

- All experimental programs must operate simultaneously.

- The Recycler operates at a fixed energy of 8 GeV; the upgraded Main Injector nominally accepts beam at 8 GeV, but potentially can be capable of operating at injection energies as low as 6 GeV.

- The NOvA Project will have upgraded the Main Injector (MI) to support a 1.333-second cycle time at 120 GeV, and will have converted the Recycler Ring (RR) to serve as a proton accumulator ring, resulting in 700 kW beam power. Additional upgrades to the MI/RR will be needed for 2-MW operations.

- A neutrino beamline directed towards Lead, South Dakota as part of the Long Baseline Neutrino Experiment (LBNE) will be operating with beam power on target of 700 kW, with shielding and infrastructure designed to accommodate at least 2.3 MW.

- Secondary beam requirements and facilities associated with the experimental program will be established prior to CD-1;

- Project X construction is likely to be staged; however, the reference design represents the capabilities required to meet the fundamental design criteria at the end of the final stage.

## I.3 Facility Scope and Staging

The scope of Project X has been developed to support the mission needs and assumptions outlined above. The primary elements of the accelerator facility are:

- An $H^-$ source consisting of a 30-keV DC ion source, Low Energy Beam Transport (LEBT), 2.1 MeV RFQ, and Medium Energy Beam Transport (MEBT) augmented with a wideband bunch-by-bunch chopper capable of generating bunch trains of arbitrary patterns at to 162.5 MHz;

- A 3-GeV superconducting linac operating in CW mode, and capable of accelerating an average (averaged over >1 μsec) beam current of 2 mA to 1 GeV and 1 mA to 3 GeV, with a peak beam current (averaged over <1 μsec) of 5mA;

- A 3 to 8 GeV pulsed superconducting linac capable of accelerating an average current of 43 μA with a 4.3% duty cycle;

- A pulsed dipole that can switch the 3-GeV beam between injection into the pulsed linac and the rare processes program;

- An RF beam splitter that can extract 1 mA of beam at 1 GeV;

- An RF beam splitter that can deliver the 3 GeV beam to multiple (three or more) experimental areas;

- Upgrades to the 8-GeV Booster to support injection at 1 GeV (Stages 1 and 2);

- Upgrades to the Main Injector/Recycler complex to support a factor of three increase in beam intensity (Stage 3);

- Target facilities required to produce secondary particle beams needed by the experimental program.

Anticipated financial constraints have led to consideration of a staged approach to Project X, based on application of the following principles:

- Each stage must present compelling physics opportunities;

- Each stage should utilize existing elements of the Fermilab complex to the extent possible;

- At the completion of the final stage the full vision of a world leading intensity frontier program at Fermilab should be realized.

A three stage approach to the Reference Design consistent with the above principles has been developed and discussed with the Department of Energy. The stages are envisaged as follows:

*Stage 1*: Construction of a 1-GeV CW linac operating with an average current of 1 mA, providing beams to the existing 8-GeV Booster, to the muon campus currently under construction, and to a new 1-GeV experimental facility.

*Stage 2*: Addition of a 1-3-GeV linac operating with an average current of 1 mA providing beam to a new 3 GeV experimental facility, and accompanied by upgrades of the 1-GeV linac to 2 mA average current and the Booster to 20 Hz capability.

*Stage 3*: Addition of the 3-8-GeV pulsed linac, accompanied by upgrades to the Recycler and Main Injector.

This report describes the Reference Design for all three stages of Project X. The staging strategy, including associated performance goals, is discussed in greater detail in Appendix 1. However, the siting and configuration associated with the Reference Design is strongly influenced by the staging plan, and a particular siting has been selected for this report consistent with the three stage plan. The siting is shown in Figure I-1. As displayed in the figure the 1-GeV, 1-3-GeV, and 3-8-GeV linacs are physically distinct, and are connected by two isochronous arcs, each with bending angle of 180 deg. Further details on the siting and associated conventional construction requirements are discussed in Chapter V.

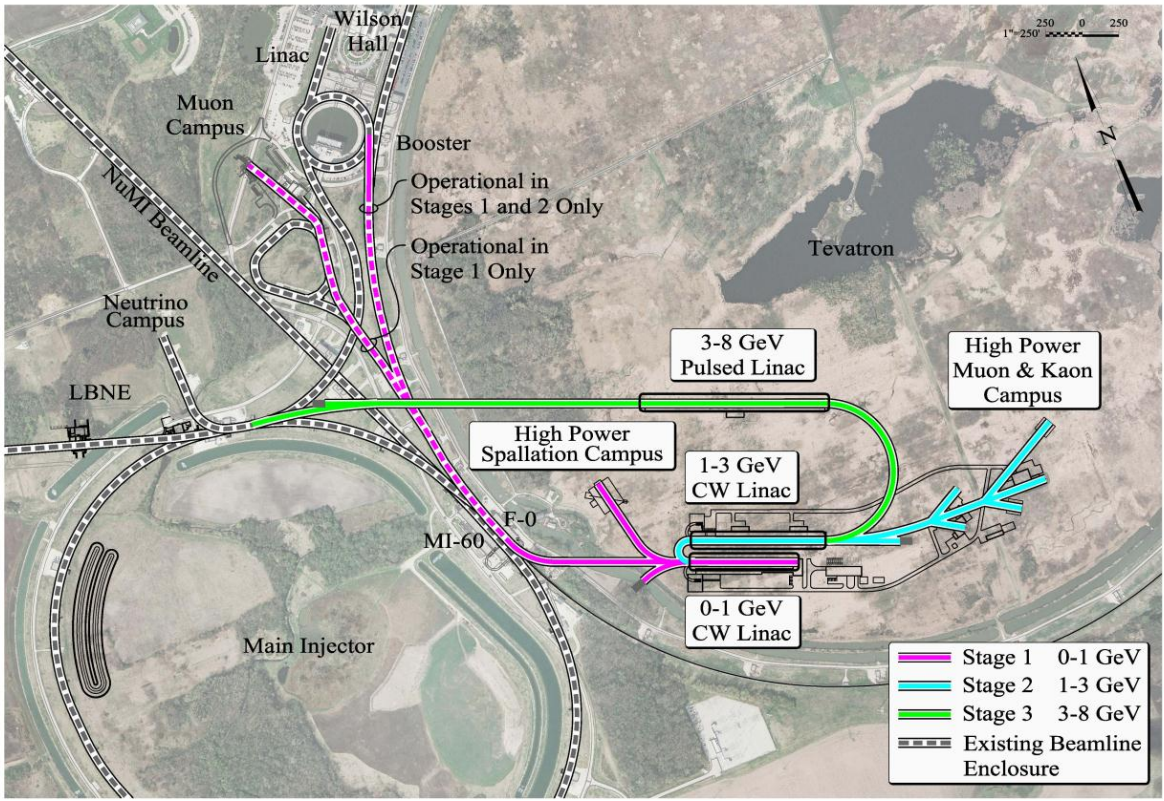

**Figure I-1**: Layout of Project X on the Fermilab Site

## II  Performance Goals

The overall goals of the Reference Design are to provide a context for the Project X R&D program and to define a scope that will provide a basis for supporting the mission needs identified in Section I and for the cost estimate necessary as part of the Critical Decision 0 (CD-0) evaluation – the first step in the critical decision process mandated by DOE order 413.3b. The cost range required for CD-0 will be established via variations of a number of assumptions inherent in the Reference Design, including:

- Options for acceleration from 3 to 8 GeV;
- Options for direct injection from the pulsed linac into the Main Injector accompanied by elimination of the Recycler from the scope of Project X;
- Alternative siting options;
- Alternative staging scenarios.

Beyond CD-0 the Reference Design will provide the starting point for development of the Conceptual Design Report required for CD-1.

### II.1  Technical Goals

The performance goals for Project X are defined through a Functional Requirements Specification (FRS [7]). It is anticipated that the FRS will be approved by the Director of the DOE Office of High Energy Physics (OHEP) and the Fermilab Director as part of the CD-0 process. High level performance goals associated with the Project X Reference Design are listed in Table II-1.

| Performance Parameter | Requirement | Unit |
|---|---|---|
| Linac Beam Power @ 1 GeV | 1 | MW |
| Linac Beam Power @ 3 GeV | 3 | MW |
| Main Injector Beam Power @60-120 GeV | >2 | MW |
| 1 GeV Linac Average Beam Current | 2 | mA |
| 3 GeV Linac Average Beam Current | 1 | mA |
| Bunch Patterns at 1 and 3 GeV | Programmable | |
| Linac Beam Power at 8 GeV | 350 | kW |
| Upgrade Potential at 8 GeV | 4 | MW |
| Availability at 1 and 3 GeV | 90 | % |
| Availability at 60-120 GeV | 85 | % |

**Table II-1**:  Performance Goals for the Project X Accelerator Facility

The proposed experimental program determines the CW linac energies: 1 GeV has been chosen as the optimum for the materials and energy programs, and 3 GeV has been determined to be sufficient to meet the requirements of both the muon and kaon programs. The 8 GeV pulsed linac energy is aligned with the injection energy of the Main Injector/Recycler complex.

Project X will provide beams to the 1 and 3 GeV programs, with variable bunch patterns, concurrently. Table II-2 presents the experimental research requirements associated with the CW linac. Sophisticated beam chopping and RF separation techniques, capable of supporting simultaneous beam delivery four different target halls[1], are being developed to support for these operations.

|  | Beam Energy (GeV) | Train Frequency (MHz) | Pulse Width (nanoseconds) | Inter-Pulse Extinction |
|---|---|---|---|---|
| Materials/Energy | 1 | 10-20 | 0.1-0.2 |  |
| Kaon Decays | 3 | 20-30 | 0.1-0.2 | $<10^{-3}$ |
| Muon Conversion | 1-3 | 1-2 | <100 | $<10^{-9}$ |
| Nuclei/Neutrons | 1-3 | 10-20 | 0.1-0.2 |  |

**Table II-2**: Bunch train requirements for the CW linac-based programs

## II.2 Operational Scenarios and Beam Formatting

Project X will have unique capabilities for delivering high power proton beams simultaneously to a variety of experimental programs with different bunch structure and intensity requirements. The CW linac will provide, simultaneously, high duty factor proton beams to the 1 GeV and 3 GeV experimental programs, with 4.3 % of the 3 GeV beam delivered to the pulsed linac for acceleration to 8 GeV and injection into the Recycler/Main Injector complex. The Main Injector will be capable of delivering 2 MW of beam power at any energy between 60 and 120 GeV for neutrino programs. We briefly outline the sub-systems and techniques to be used in implementing these capabilities through the staged construction and development of the Project X program.

---

[1] In this document we consider one end user at 1 GeV and three at 3 GeV. However the number of users can be increased, as required, by the serial installation of additional RF splitters.

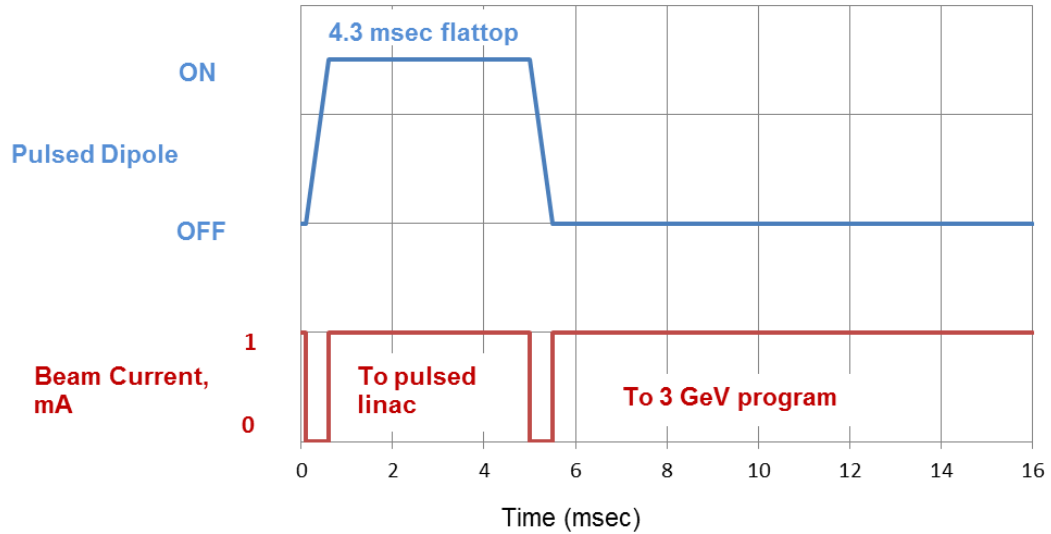

**Figure II-1**: A schematic timeline for linac beam current (first 16 ms of the 100 ms cycle). The pulsed magnet's rise and fall time is 0.5 ms.

Figure II-1 shows the CW linac timeline schematically. There are two timeline intervals: (1) the 4.3 ms long time interval associated with a 10 Hz injection rate into the pulsed linac, and (2) the ~95 ms long time interval associated with the 3 GeV rare-processes experiments (discussed below). Six 4.3 ms pulses, spaced by 100 ms, are provided for acceleration in the pulsed linac and transfer to the Recycler/Main Injector in support of the long baseline neutrino program. This beam is directed into the pulsed linac by a pulsed dipole with a rise time of 0.5 ms. In order to create these gaps for the pulsed dipole, the beam is interrupted by the chopper located within the Low Energy Beam Transport line. The linac average current during this pulse is 1 mA. For the duration of the ~95 ms between pulsed linac pulses the beam is directed toward the rare processes program, again with an average current of 1 mA. In Stages 1, and 2 the timeline for the 1 GeV linac current will be similar to that in Figure II-1 with beam directed towards the Booster rather than the pulsed linac. The flattop for the Booster injection will be 1 ms at a rate of 15 Hz.

Several front end subsystems are employed to provide beam users with variable bunch patterns concurrently. These subsystems are described in detail in the following sections; here we briefly describe the main considerations. For optimal linac operation the power of the RF system should be matched to the required beam power. If the average beam intensity stays constant but the peak intensity varies with time so that the beam power (temporarily) exceeds the power available from the RF system, the beam energy begins to droop. Fortunately, superconducting cavities have a comparatively large stored energy that strongly suppresses the energy variations if the beam intensity variations are sufficiently fast. For the accelerating gradients chosen within the CW linac the stored energy is sufficient to limit the energy gain variation to less than 0.1% over a 1 μs period at 5 mA beam current. This means that currents of up to 5 mA can be sustained for less than 1 μs as long as the average current is maintained at 2 mA (1 mA beyond 1 GeV) over timescales of 1 μs and greater. All presently suggested experiments require significantly faster beam intensity variations (or bunching patterns) than 1 μs, leading to these variations being "invisible" to the accelerating structures.

The Reference Design utilizes a DC ion source capable of delivering 0-5 mA of H⁻ ions. After acceleration and bunching in a 162.5 MHz RFQ the beam is chopped by a wideband chopper (at 2.1 MeV) that allows populating the linac rf buckets in arbitrary patterns at 162.5 MHz. Utilization of RF separation at the exit of the linac allows each experiment to receive a desired bunch pattern. The only limitations are that: (1) the beam current, when averaged over time spans in excess of ~1 μs, should not exceed 2 mA (1 mA beyond 1 GeV), (2) all bunches should have the same intensity,[2] and (3) the peak current should not exceed 5 mA.

Figure II-2 presents a possible beam structure to support simultaneous operations of a muon conversion experiment and a nucleon based experiment at 1 GeV as part of Stage 1 of Project X (see Appendix I for discussion of Stages). The bunch pattern shown repeats itself every 1 μs, with the bunch train generated by the front end (black, red) shown at the top of the figure. An RF splitter operating at $(n + 1/2) \times 162.5$ MHz is deployed at the end of the 1 GeV linac, applying kicks in opposite directions to the black bunches (arriving at a splitter RF phase of 270°) and the red bunches (arriving at 90°). The result after the splitter is a stream of (black) bunches at a frequency of 40 MHz and a pulse of (red) bunches with a width of ~40 ns repeating at 1 MHz. The latter pattern would be appropriate for an upgrade of the Mu2e experiment. For a peak current of 3.64 mA ($1.4 \times 10^8$ protons per bunch) the average current accelerated to 1 GeV is 1 mA, with 910 kW of beam power contained in the black beam and 90 kW in the red beam.

---

[2] A possibility of partial bunch scraping and, consequently, bunch current regulation will be experimentally studied at Project X Injector Experiment (PXIE) and outside the scope of this RDR.

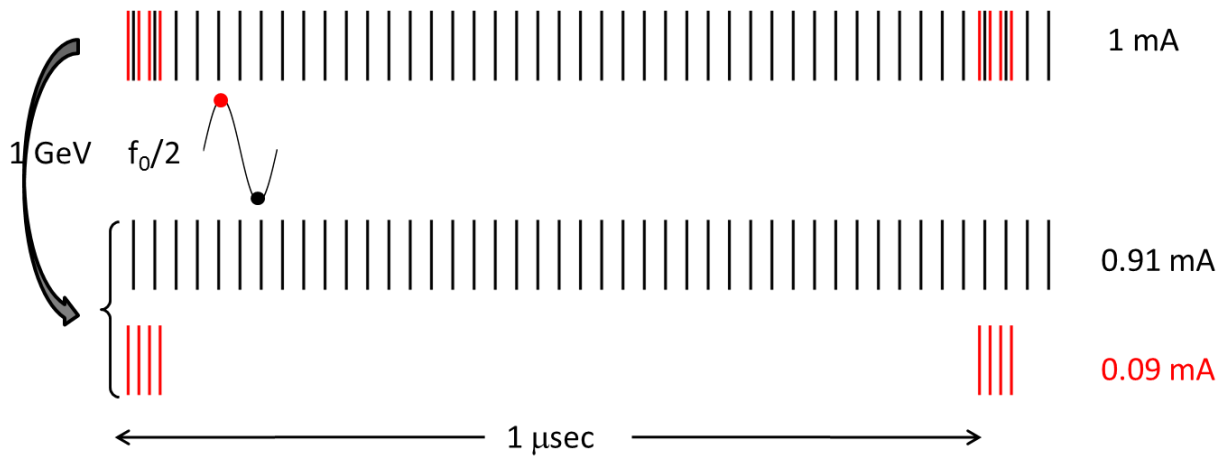

**Figure II-2**: Bunch structure of the 1 GeV CW linac beam at Stage 1. The pattern shown repeats with a 1 μs period. The linac bunch structure (top) is deconvolved into a beam supporting the materials/energy/nuclei program (black bunches) and a beam supporting a next generation muon-to-electron conversion experiment (red). The individual bunch intensity is $14\times10^7$ with a 50 ps (FW) bunch length. The total beam power delivered is 910 kW (black) and 90 kW (red).

Operations become somewhat more complex following completion of the 3 GeV linac at Stage 2. It is envisioned to support three experiments at 3 GeV: a muon conversion experiment, a rare kaon experiment, and a nuclear physics experiment(s). In parallel the 1 GeV program established in Stage 1 is continued. Two RF splitters are employed to support joint 1 and 3 GeV operations – the splitter at 1 GeV operating at $(n+1/2) \times 162.5$ MHz, and at 3 GeV a splitter operating at $(n\pm1/8) \times 162.5$ MHz. Figure II-3 shows a possible arrangement – the bunch patterns are reproduced every 0.5 μs, with the bunch train generated by the front end (black, red, blue, and green bunches) shown at the top of the figure. This bunch train is separated at 1 GeV into two trains with the black bunches receiving a kick from the splitter corresponding to arrival at 90° phase and serving as the beam for the 1 GeV program while the red, blue, and green bunches arrive at 270° phase and are sent to the 1-3 GeV linac. At 3 GeV this train is separated into three trains, again depending upon the arrival phase at the splitter. For a peak current of 5 mA ($1.9\times10^8$ protons per bunch) the average current is 2 mA to 1 GeV and 1 mA from 1 to 3 GeV, with 1 MW delivered to the 1 GeV program, and 3 MW delivered to the 3 GeV program and allocated between the red, blue, and green programs as 750/1500/750 kW. The red bunch train is appropriate to a muon to electron conversion experiment – a burst of four bunches (~75ns) separated by 0.5 μs. The blue and green trains provide continuous beam at approximately 20 MHz and 10 MHz respectively, structures appropriate for rare kaon decays and a variety of nuclear physics measurements.

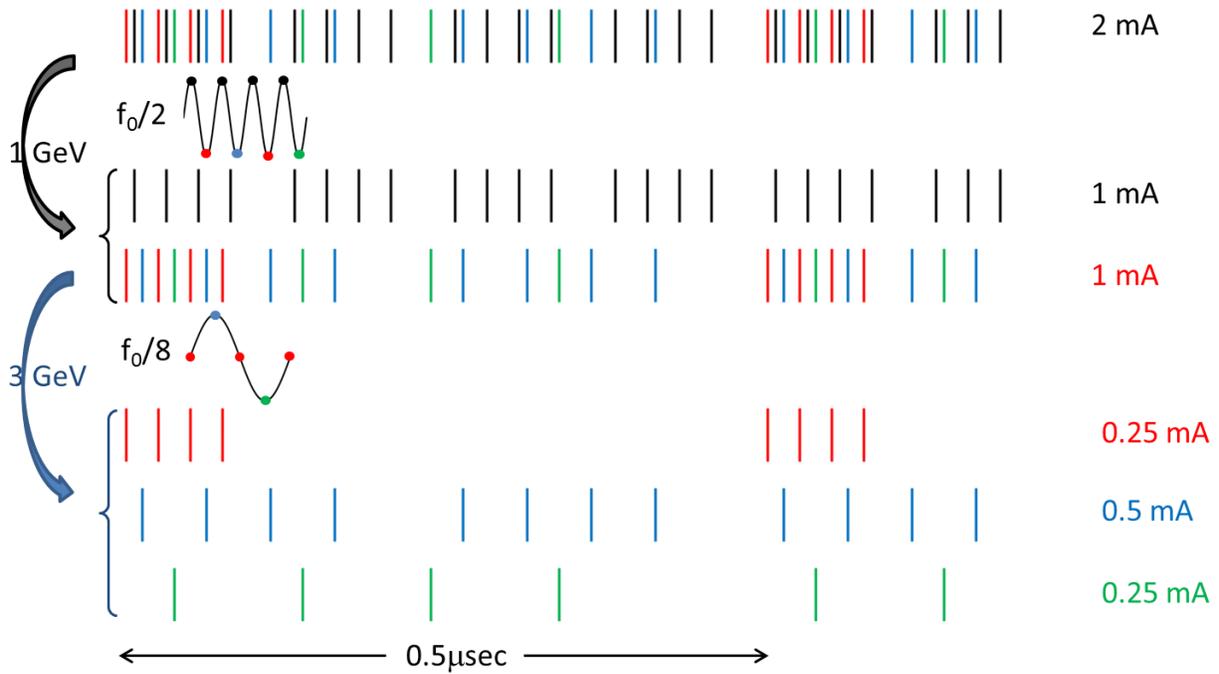

**Figure II-3**: Bunch structure of the 1 GeV CW linac beam at Stage 2. The pattern shown repeats with a 1 µs period. The top diagram shows the bunch pattern in the linac up to 1 GeV; this pattern is initially deconvolved into a beam supporting the materials/energy program at 1 GeV (black bunches) and a beam continuing on to the 1-3 GeV linac (red, blue, green). At 3 GeV the beam is further deconvolved to provide different bunch patterns to the three experiments with red pulses delivered to a muon conversion experiment, blue to a rare kaon decay experiment, and green to a third (presently unspecified) experiment. The individual bunch intensity is $19 \times 10^7$ with a 40 ps (FW) bunch length.

The patterns shown in Figure II-2 and Figure II-3 are created in the MEBT by a wideband chopper that disposes up to 80% of the beam delivered from the RFQ in a pre-determined, repetitive pattern, creating a continuous load of up to 8 kW on the MEBT absorber. The functioning of the front-end system, including the wideband chopper that provide great flexibility in developing variable bunch patterns, is described in Chapters III and IV.

# III  Accelerator Facility Design

Figure III-1 shows the configuration of all linear acceleration elements within Project X. A room temperature (RT) section accelerates beam to 2.1 MeV and creates the desired bunch structure for injection into the superconducting linac. The CW linac accelerates 2 mA (1 mA in Stage 1) of beam current to 1 GeV and 1 mA to 3 GeV. The pulsed linac accepts beam at 3 GeV and accelerates it to 8 GeV with a 4.3% duty factor. The pulsed linac provides beam for injection into the Recycler via an 8 GeV transport line. After accumulation in the Recycler beam is transferred to the Main Injector and accelerated to anywhere between 60-120 GeV. Beam is available for experiments at any of the following energies simultaneously: 1, 3, 8 and 60-120 GeV. The operational parameters for the CW linac are given in Table III-1.

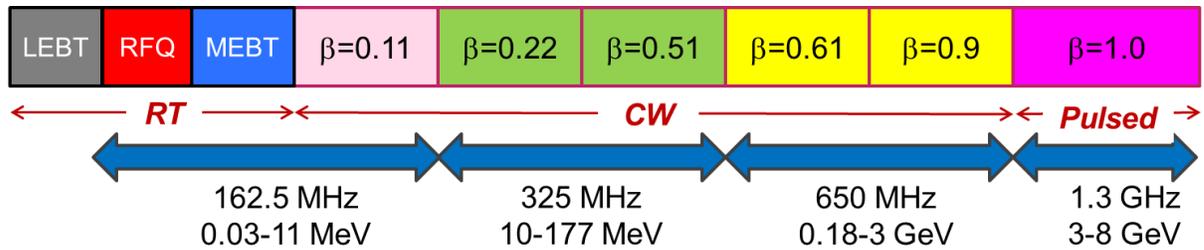

**Figure III-1**: The Project X Linacs

| CW Linac | Requirement | |
|---|---|---|
| Particle species | H⁻ | |
| Input beam energy (Kinetic) | 2.1 | MeV |
| Output beam energy (Kinetic) | 3.0 | GeV |
| Pulse repetition rate | 162.5 | MHz |
| RF pulse length | CW | |
| Beam pulse length | Programmable | |
| Average beam current to 1 GeV | 2 | mA |
| Average beam current to 3 GeV | 1 | mA |
| 3 GeV rms norm. transverse emittance, $\varepsilon_x/\varepsilon_y$ | 0.6/0.4 | mm-mrad |
| 3 GeV rms norm. longitudinal emittance | 0.65/1.9 | mm-mrad /keV-ns |
| 3 GeV rms bunch length | 3 | ps |

**Table III-1**: CW Linac Parameters

## III.1 CW Linac

The CW linac comprises the sections labeled "RT" and "CW" as shown in Figure III-1. The CW linac accelerates H⁻ ions from an ion source to 3 GeV, at which point they are directed either towards the 3 GeV experimental areas or towards the 3-8 GeV pulsed linac. The average beam current is 2 mA to 1 GeV and 1 mA to 3 GeV, with peak currents up to 5 mA for periods of less than 1 μs. The entire CW linac includes the following major elements:

1) The warm front-end;
2) One accelerating SC cryomodule based 162.5 MHz Half-Wave Resonators (HWR);
3) Two accelerating SC sections based on 325 MHz Single-Spoke Resonators (SSR1 & SSR2);
4) Two accelerating SC sections of 650 MHz elliptical cavities with $\beta$ = 0.61 and 0.9 (LB650 and HB650)

### III.1.1 Warm Front End

The front-end of the Project X linac provides H⁻ beam to the first superconducting module within the CW linac. The front-end must provide beam formatted correctly for utilization in the 1 and 3 GeV experimental programs, and for injection into the Recycler.

The front end consists of an ion source, Low Energy Beam Transport (LEBT), Radio Frequency Quadrupole accelerator (RFQ), and Medium Energy Beam Transport (MEBT). The nominal beam current from the ion source through the RFQ is 5 mA. The ion source delivers 5 mA of H⁻ at 30 keV. The 162.5 MHz RFQ accepts and accelerates this beam to 2.1 MeV. A wideband chopper situated within the MEBT removes ~ 80% of the bunches emanating from the RFQ in order to form appropriate bunch patterns for acceleration in the CW linac as described in Chapter II.2.

The RFQ energy of 2.1 MeV is chosen because it is below the neutron production threshold for most materials. At the same time this energy is sufficiently large to mitigate the space charge effects in the MEBT at currents as high as 10 mA (the maximum capability of the source). The choice of a comparatively low energy for the LEBT (30 keV) allows reducing the length of RFQ adiabatic buncher, and, consequently, achieving sufficiently small longitudinal emittance so that at the exit of the RFQ the beam phase space will be close to emittance equipartitioning. To mitigate the space charge effects in the LEBT compensation of beam space charge by residual gas ions can be applied either for the full LEBT length or over its initial part only.

### Low Energy Beam Transport (LEBT)

The layout of the ion source and LEBT is shown in Figure III-2. Two ion sources are installed to improve the beam availability. Each source can be removed for repairs, reinstalled, and conditioned without interrupting the operation of the other source. The LEBT transports the beam from the ion source to the RFQ entrance. It also matches the beam optical functions to the RFQ. In addition, the LEBT forms a low-duty factor beam as required for commissioning and tuning of the downstream beam line and interrupts the beam as part of the machine protection system (MPS).

The LEBT functional requirement specifications are listed in [8]. The LEBT includes 3 solenoids (for each leg), a slow switching dipole magnet, a chopper assembly (a kicker followed by a beam absorber), and diagnostics to characterize and to tune the beam. The length of the beam line, ~3 m, insures that the gas migration from the ion source to the RFQ is tolerable. Fast machine protection and pulsed beam operation are achieved via the chopper assembly, which is comprised of a kicker followed by an absorber. In some scenarios, it can be used also as a pre-chopper to assist the MEBT chopping system. Note that the primary machine protection mechanism is to disable the beam from the ion source by turning off its extraction and bias voltages, with the LEBT chopper serving as a fast beam switch during the ion source turn off time.

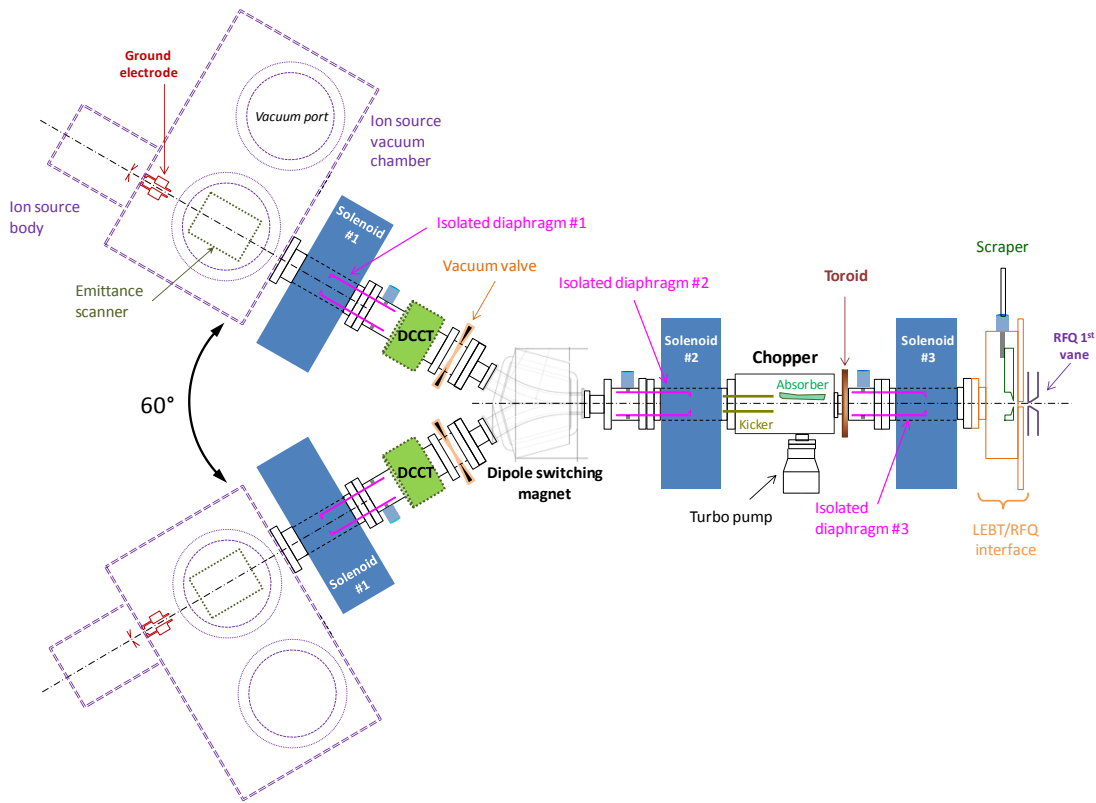

**Figure III-2**: Schematic of the LEBT with two ion sources.

The LEBT optics design (Figure III-3) incorporates two regions. The beam is nearly fully neutralized from the exit of the ion source to immediately upstream of the (chopper) kicker. Depending on the operational mode the beam can be either neutralized or un-neutralized downstream of the kicker to the RFQ entrance. The edge focusing of the switching dipole is adjusted to minimize asymmetry between horizontal and vertical focusing. In the un-neutralized mode, the secondary ions created in the downstream region are removed by a constant electric field on the kicker plates, and the upstream ions are trapped by a positive voltage on the insulated diaphragm #2. In the neutralized mode, the kicker plates as well as the insulated diaphragm #2 have normally ground potential, while the insulated diaphragm #3 or the scraper is biased positively to prevent the ion escape longitudinally.

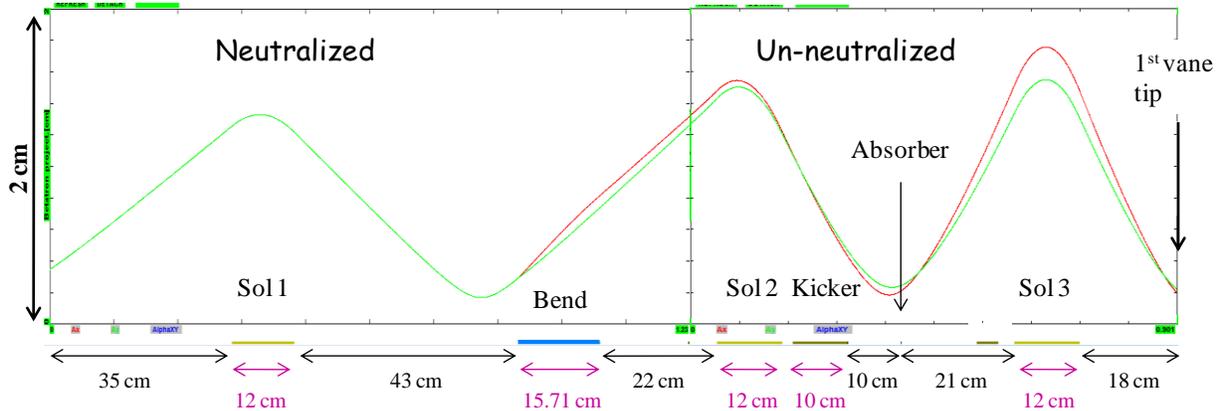

**Figure III-3**: Beam horizontal (red) and vertical (green) envelopes (rms) of the partially un-neutralized LEBT (with OptiM [9])

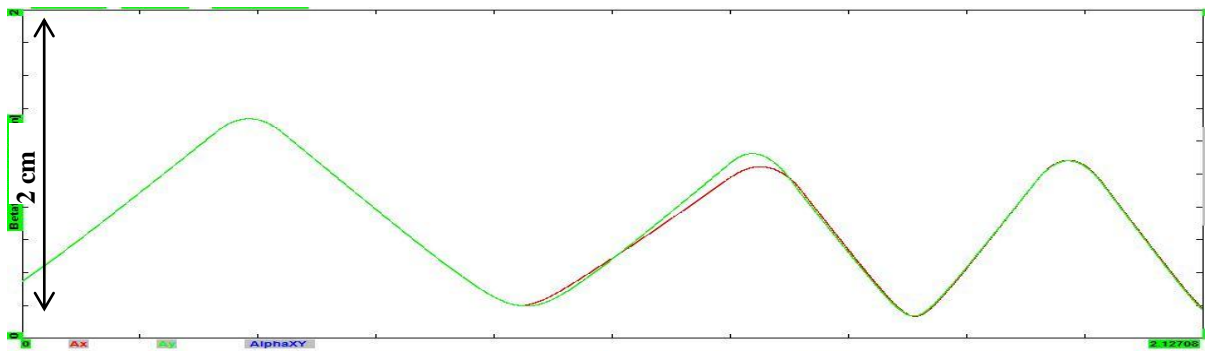

**Figure III-4**: Beam horizontal (red) and vertical (green) envelopes (rms) in the LEBT with neutralized transport only. All optical elements are identical to those in Figure III-3; only the solenoid currents were adjusted.

The LEBT scheme is flexible enough to accommodate both versions by adjusting potentials and solenoid currents (see Figure III-4 for simulations of a fully neutralized transport case). The transport with the un-neutralized downstream section is beneficial for minimizing the difference between short pulse and CW operations, as well as for minimizing transient effects during beam switching on; but un-neutralized beam transport does result in emittance growth as can be seen in Figure III-5 and Figure III-6 . The relative benefits of each scenario will be clarified via experiments undertaken at PXIE (Project X Injector Experiment [10]).

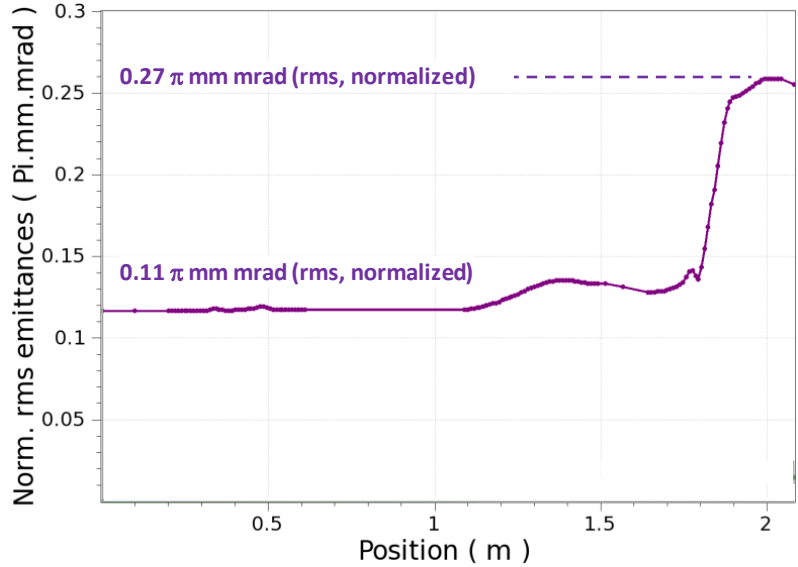

**Figure III-5**: Emittance evolution along the LEBT for a 5 mA, 30 keV (with TraceWin) for partially neutralized transport. The start of the emittance growth shows where the transition between the neutralized and un-neutralized beam transport occurs.

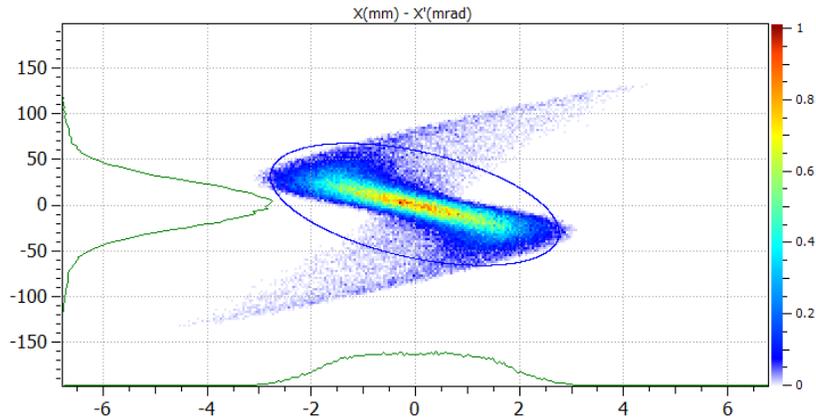

**Figure III-6**: Phase space distribution (x-x') at the entrance of the RFQ (with Tracewin) for partially neutralized transport. The initial distribution was Gaussian.

*Radio Frequency Quadrupole*

The 162.5 MHz RFQ accelerates H⁻ ions with beam currents up to 10 mA from 30 keV to 2.1 MeV. The beam dynamics design of the RFQ is optimized using the measured beam distribution from the D-Pace H- ion source [11]. The design has over 96% transmission for beam current from 1 to 15 mA. At 5 mA (nominal current) 99.8% beam capture is achieved with transverse and longitudinal emittances (rms, norm) of 0.15 mm-mrad and 0.64 keV-ns (0.204 mm-mrad), respectively. The beam dynamics design was conducted using PARMTEQM and TraceWin; Figure III-7 shows the simulated $3\sigma$ beam envelopes at 5 mA, from the ion source through the exit of the RFQ. Error analyses indicate that the RFQ design has adequate tolerance for mechanical and field errors, as well as errors in TWISS parameters.

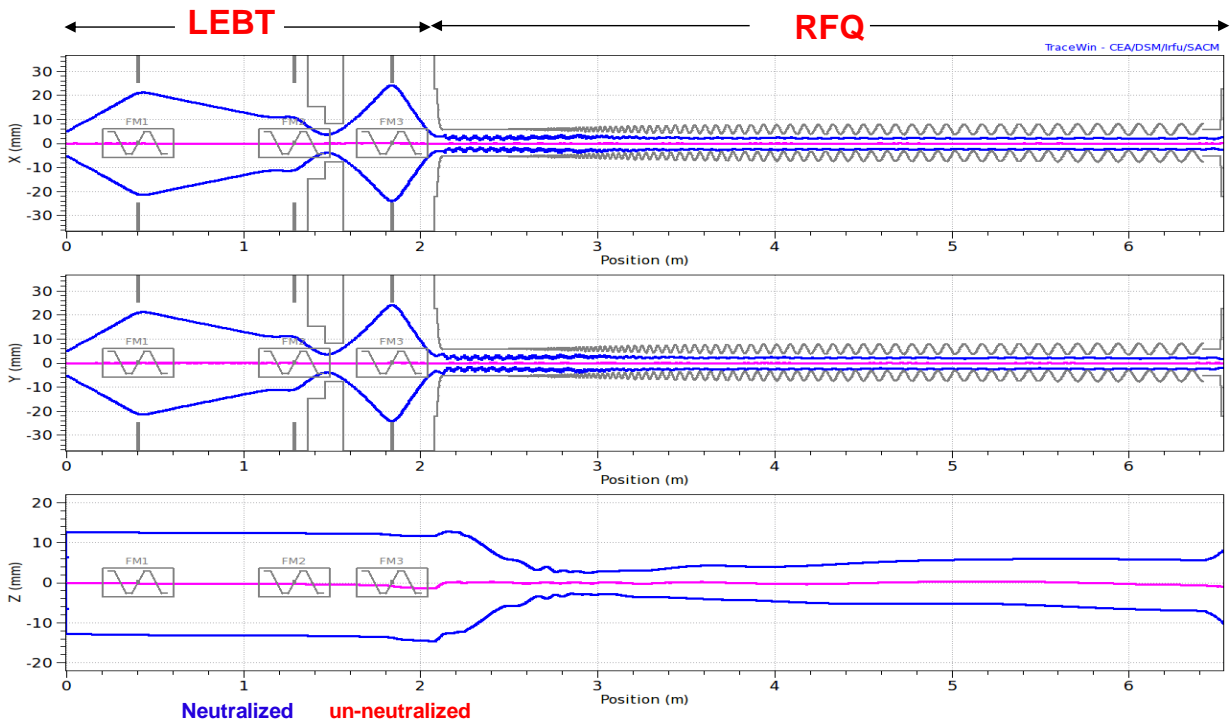

**Figure III-7**: TraceWin simulations of $3\sigma$ beam envelopes (horizontal, vertical, and longitudinal) through the LEBT and RFQ for a 5-mA beam current.

*Medium Energy Beam Transport (MEBT)*

The required bunch structure for Project X operations will be formed in the MEBT, the ~10m section between RFQ and HWR cryomodule. The heart of the MEBT is a wideband chopping system that directs unwanted bunches to an absorber according to a pre-selected pattern and passes the bunches chosen to be accelerated into the SRF linac with minimum distortions. Beam chopping in the MEBT is used in other facilities (e.g. SNS [12]), but the concept of a bunch-by-bunch selection results in significantly more demanding requirements to the chopping system. In addition, the MEBT provides proper optical matching between the RFQ and the SRF section, includes instrumentation to measure the properties of the beam coming out of the RFQ and transported to the SRF cavities, and has means of protecting the SRF section from excessive beam loss.

The nominal H⁻ beam current in the MEBT input is 5 mA. However the design of MEBT absorber supports damping up to 10 mA. Consequently, the operation with beam current twice larger than nominal is expected to be possible. Such an operational regime is also supported by beam transport simulations with 10 mA current. Experimental studies to be carried out in PXIE should demonstrate a possibility of 10 mA operation. To minimize emittance growth ($\Delta\varepsilon/\varepsilon < 10\%$) the MEBT transverse focusing is made sufficiently smooth. It is provided by equidistantly placed quadrupole triplets with the only exception of two doublets located immediately after RFQ (Figure III-8) which match the RFQ beam envelopes to the MEBT periodic focusing structure. Each triplet or doublet is followed by a pair of dipole correctors. The complete list of functional requirements is in Ref. [13]. The specifications for the quadrupoles and correctors are listed in Ref. [14]. The spaces between neighboring triplets or doublets are referred to as MEBT sections. The period in the regular part is 1140 mm. This leaves a 650-mm long (flange-to-flange) space for various equipment (350 mm in the section between doublets labeled #0 in Figure III-8).

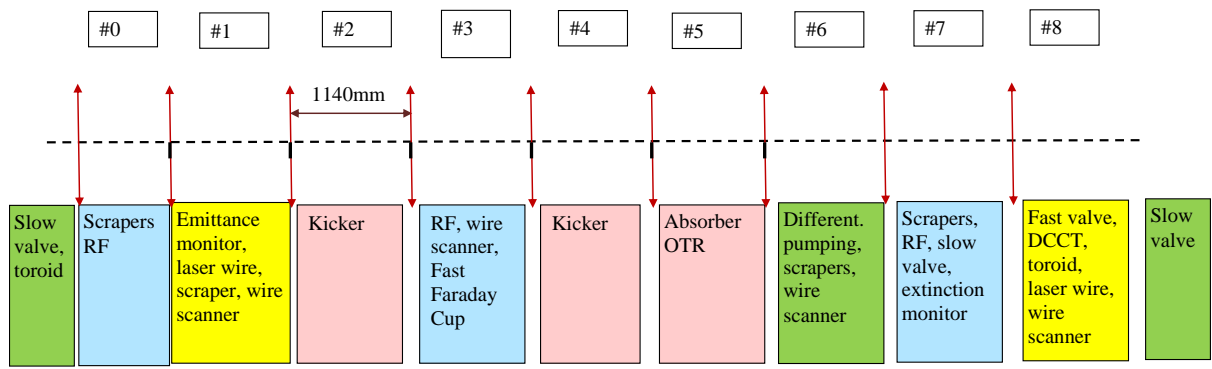

**Figure III-8**: The MEBT structure. Sections are colored according to their main functions. The red vertical arrows schematically show the transverse focusing (doublets or triplets) elements.

The undesired beam bunches will be removed in the MEBT by a chopping system, represented in Figure III-8 by the red boxes. The chopper consists of two identical 50 cm long kickers separated by a 180º transverse phase advance and an absorber (90º from the last kicker). In the broadband, travelling-wave kicker, the transverse electric field propagates with the phase velocity equal to the speed of H⁻ ions (~20 mm/ns, β = 0.0668). Such a two-kicker scheme allows creating a vertical kick sufficiently large for separation of passed and removed bunches with moderate voltage on the kicker plates. Detailed specifications for the kicker can be found in Ref. [15].

Figure III-9 presents the simulated transverse beam envelopes in the MEBT for both passing and chopped bunches. The chopped bunches are directed to the absorber which is displaced vertically from the beam trajectory. Presently two versions of the kicker, which differ by the structure's impedance, are being investigated [16]. To keep the beam properly bunched and matched to the first SC section, the MEBT includes 3 identical bunching cavities [17].

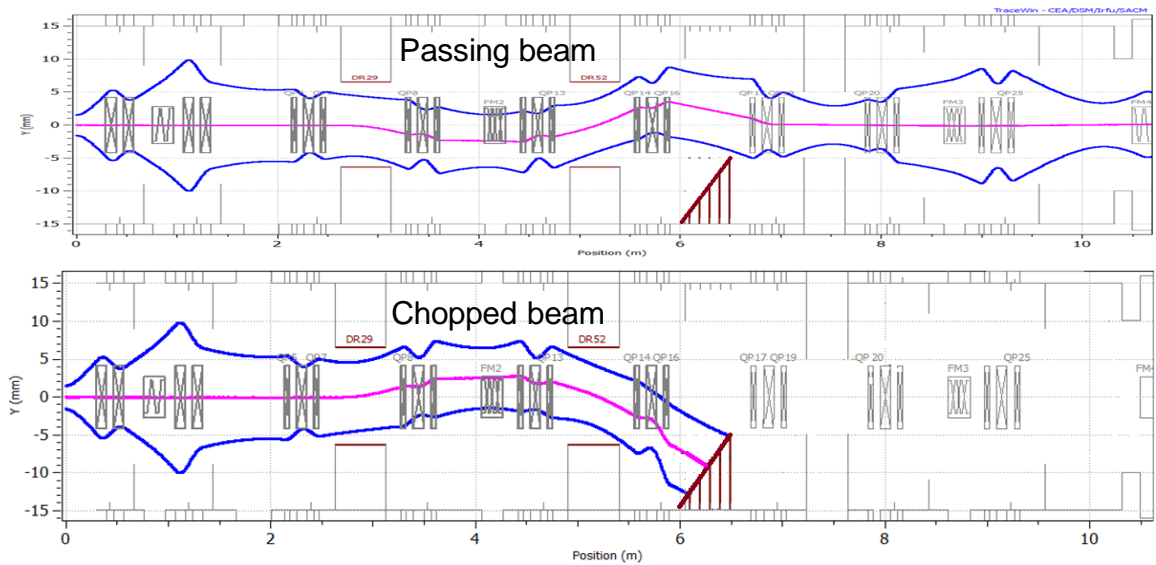

**Figure III-9**: The simulated transverse 3σ beam envelopes in the MEBT.

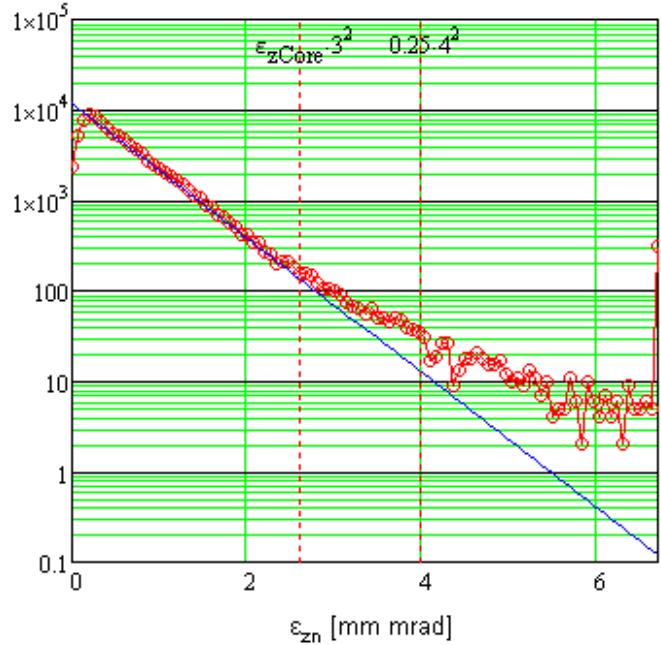

**Figure III-10**: Beam longitudinal distribution at the end of RFQ simulated for 5 mA beam current. Vertical lines mark the $1\sigma$ and $4\sigma$ boundaries for 0.25 mm-mrad (rms) normalized emittance.

Some of the Project X experiments are expected to require extremely good extinction for removed bunches. The target value is smaller than $10^{-9}$, i.e. less than one particle per bunch. A finite population in the longitudinal RFQ tails (see Figure III-10) can be the main limitation to achieve the required beam extinction. Although there is very good rejection of the tails in the first SC cryomodule the weak longitudinal focusing in the MEBT and its long length allow momentum tails of allowed bunches to move to the nearby rejected bunches and be accepted in their RF bucket. Figure III-11 presents the phase space at the location of bunch chopper for particles that are accepted for further acceleration. Particles in the lanes above and below the main bunch will be accepted in the nearby buckets. As one can see to be accepted to these buckets the tails have to extend beyond $\sim 10\sigma$. None of the $10^6$ particles used in the simulation approached this limit. The final result for the extinction of removed bunches has to be determined by experiment. The suppression of longitudinal RFQ tails is expected to be the key in achieving the extinction goal.

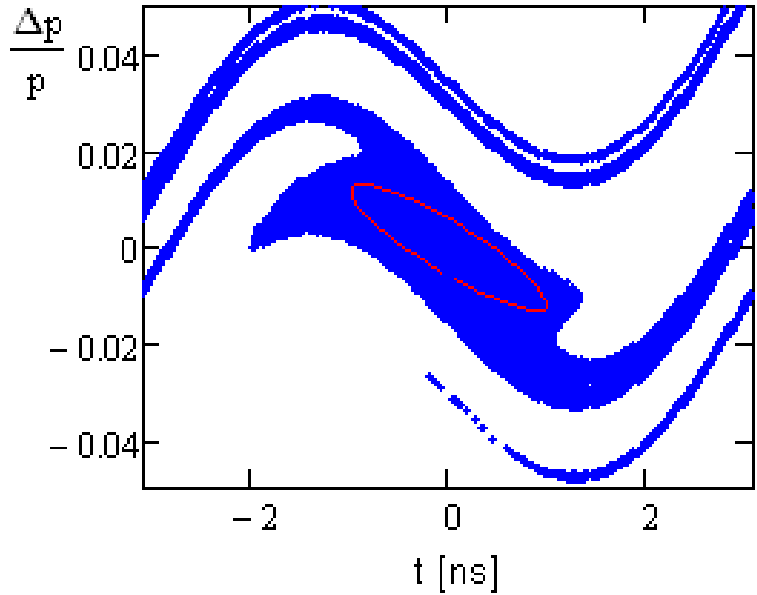

**Figure III-11**: Longitudinal phase space at the beam chopper location for particles accepted for further acceleration (blue dots). The red line designates the 4σ bunch boundary.

### III.1.2 CW Linac Physics Design

The CW linac starts immediately downstream of the MEBT. It accelerates the beam from 2.1 MeV to 3 GeV and consists of two SC linacs connected by an arc bending the beam by 180 deg. The first linac belongs to the Project X Stage I. It accelerates the beam to 1 GeV. In Stage I the beam can be RF separated between a few 1 GeV experiments or sent to one of them. The beam can be also delivered to the Fermilab Booster and muon campus. Later, when Stage II is implemented, the beam will be RF separated between the 1 GeV experimental area and the second linac accelerating the beam to 3 GeV. Two linacs are connected by the isochronous bending arc. This arc belongs to Stage II. The arc introduction addresses two problems. First, it minimizes civil construction in Stage I and, consequently, results in a cost reduction. Second, it allows one to proceed with the construction of Stage II without an interruption of Stage I operation.

*Accelerating cavities*

Five types of superconducting cavities are used in the CW part of Project X to cover the entire velocity range required for acceleration of H⁻ (or protons) from 2.1 MeV to 3 GeV. The selection of cavity frequencies and cell configuration are chosen to maximize acceleration efficiency for each accelerating structure, minimize cost of the accelerator and its operation, and to address other factors helping to minimize the beam loss.

The primary efficiency factor is the transit time factor, which has a dependence on the beam velocity, β, as shown in Figure III-12 for different number of cells in a cavity. One can see that the range of betas over which the beam can be efficiently accelerated increases with decreasing number of cells. On the other hand, too small number of cells reduces effective gradient and increases costs due to end effects. As one can see from Figure III-12 the maximum of acceleration is achieved for the velocities slightly larger than $\beta_G$, which is called the geometric-beta. The velocity where the maximum is achieved, $\beta_{opt}$, is called the optimal beta. For a periodic structure with harmonic distribution of electric field along the axis, $E \propto \sin(\omega z / \beta c)\exp(i\omega t)$, the transit time factor can be expressed by the following formula:

$$T(\beta) = \frac{2\beta}{\pi n}\left( \frac{\sin(\pi n(\beta-\beta_G)/(2\beta))}{\beta-\beta_G} - (-1)^n \frac{\sin(\pi n(\beta+\beta_G)/(2\beta))}{\beta+\beta_G} \right), \qquad (1)$$

where $n$ is the number of cells in the cavity operating at $\pi$-mode. The transit time factor is normalized such that $T(\beta_G) = 1$. The above expression approximates very well the transient factors obtained by numerical integration of actual time dependent electric field for the Project X cavities. The $\beta_G$ presented below were obtained by fitting Eq. (1) to the numerical integration results. For multi-cell elliptic resonators this definition yields slightly larger values of $\beta_G$ than the more commonly used definition of $\beta_G$ as the ratio of cavity period to half-wavelength.

Recent developments of the 1.3-GHz ILC cavity technology at Fermilab [18] and elsewhere make it a preferred choice for the acceleration of ultra-relativistic beams and, in particular, the preferred choice for the Project X pulsed linac, accelerating protons from 3 to 8 GeV. This selection determined the base frequency of the project to be 1300 MHz. The choice of other frequencies was set by a requirement that all other frequencies have to be harmonically related to the base, which yielded 162.5, 325 and 650 MHz. Such a choice results in a comparatively smooth frequency increase during beam acceleration, accommodating bunch shortening due to adiabatic damping.

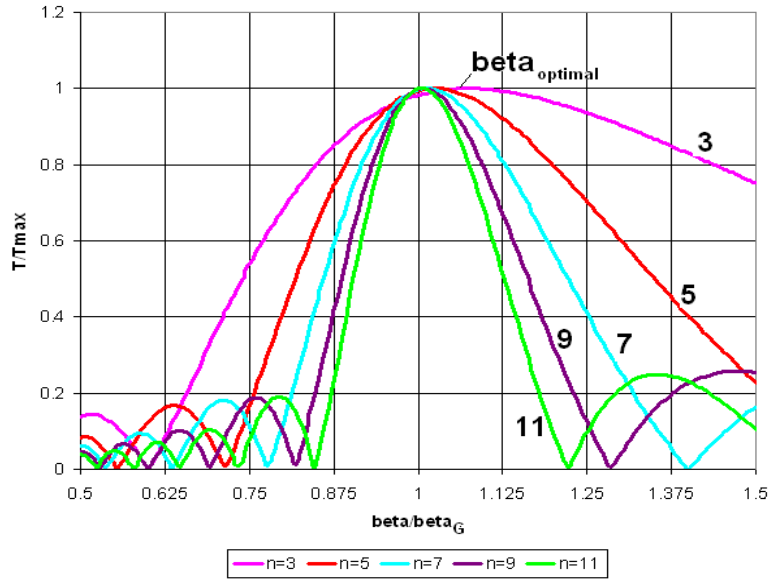

**Figure III-12**: Transit time factor versus the ratio of the beta to the geometric beta, $\beta/\beta_G$, for different number of cells in a cavity, *n*.

| Name | $\beta_G$ | $\beta_{opt}$ | Freq (MHz) | Type of cavity | Energy range (MeV) |
|---|---|---|---|---|---|
| HWR | 0.094 | 0.112 | 162.5 | Half wave resonator | 2.1-11 |
| SSR1 | 0.186 | 0.222 | 325 | Single-spoke resonator | 11-38 |
| SSR2 | 0.431 | 0.515 | 325 | Single-spoke resonator | 38-177 |
| LB650 | 0.61* | 0.647 | 650 | Elliptic 5-cell | 177 - 480 |
| HB650 | 0.9* | 0.95 | 650 | Elliptic 5-cell | 480 - 3000 |

* To be consistent with previously written documents $\beta_G$ for the elliptic cavities is defined as the ratio of regular cell length to half-wavelength.

**Table III-2**: Accelerating cavity types for the CW linac.

Table III-2 and Figure III-13 presents the cavity types utilized, and their deployment, through the CW linac. Acceleration starts with the half-wave resonators (HWR) operating at 162.5 MHz. It is followed by two types of single spoke resonators (SSR1 and SSR2) operating at 325 MHz, and finally by two types of elliptical 5-cell cavities, low beta (LB650) and high beta (HB650), at the 650 MHz. Figure III-14 presents the transient factors for the CW linac. The voltage gain in each next cavity type is significantly larger than in the previous one. That determines that the transition happens earlier than the transit time factors for two types are equal.

The choice of the RFQ frequency was determined by a possibility of bunch-by-bunch chopping which would be beyond the present "state-of-the art" implementation for 325 MHz but is feasible for 162.5 MHz. The use of the same frequency for the first superconducting cryomodule results in reduced transverse defocusing from cavity fields and reduced longitudinal focusing. This yields the possibility of increasing the cavity voltage for the first SC cavities, otherwise limited by beam dynamics. As a consequence the number of cavities, and linac length, required to get the beam accelerated to 11 MeV is reduced by more than a factor of 2. Note that even this frequency choice does not allow using all available voltage for first few cavities. In particular, the first cavity uses about half of nominal voltage.

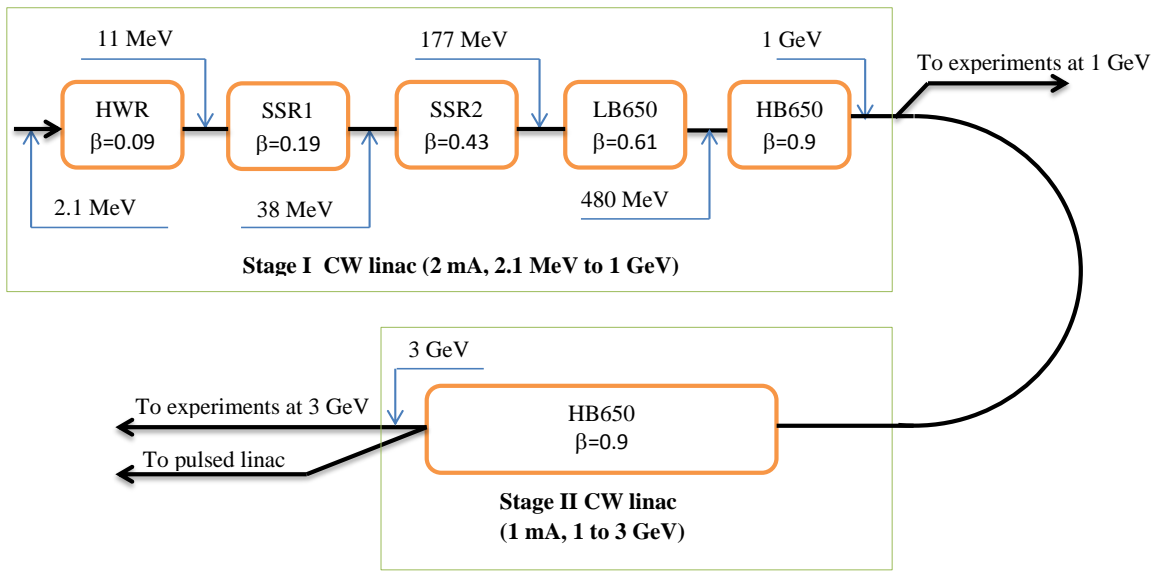

**Figure III-13**: SRF technology map for CW portion of Project X. β refers to the cavity optimum beta for HWR, SSR1, and SSR2, and the cavity geometrical beta for LB650 and HB650.

In summary, the primary benefits of the cavity arrangement described above are:
- The longitudinal beam dynamics at each transition is simplified by limiting the frequency jump to no more than a factor of two from one cavity type to another.
- The lower frequency sections provide sufficient aperture for minimizing uncontrolled beam loss to tolerable levels (0.1 - 0.2 W/m). In particular, it allows for operation at reduced focusing strength which minimizes the intra-beam stripping.

- The choice of cavities with a low frequency at the beginning of the linac reduces effects of the acceleration cavities' focusing/defocusing to an acceptable level, thus, allowing for operation at maximum possible accelerating gradients.

However, there are also some disadvantages related to the preference of 650 MHz to 1300 MHz frequency for the major part of CW linac:

- Microphonics presents a more serious issue at lower frequencies.
- Cavities for 650 MHz are more expensive (more niobium) than at 1300 MHz, but the cost increase is compensated (with the accuracy we can presently account) by a smaller number of the cavities and RF sources.

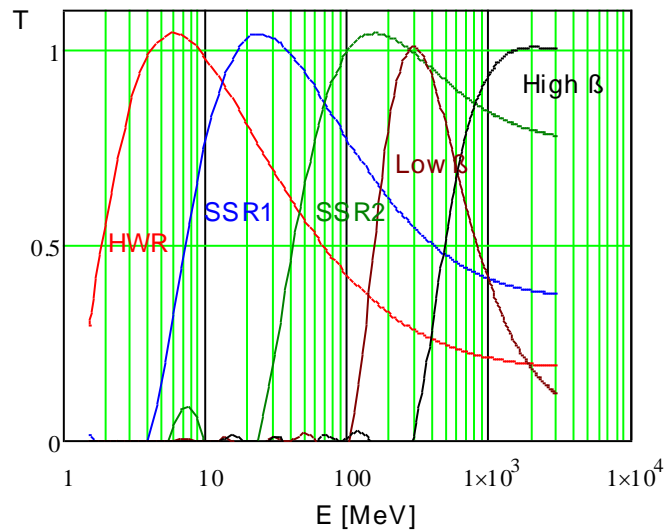

**Figure III-14**: Transient factors for CW SC Project X cavities.

The operating gradient is chosen to provide a peak surface magnetic field that allows operation below high-field Q-slope; see Figure III-15 taken from [19]. For the frequency of 162.5 MHz we adopted the maximum magnetic field of about 40 mT; while for the frequencies of 325 and 625 MHz the peak magnetic field was chosen below about 70 mT. For all frequencies the peak surface electric field is lower than 40 MV/m [20] in order to avoid the risk of strong field emission (see details in Section IV).

The transition from the 325 MHz single-spoke cavities to the 650 MHz section based on elliptical cavities is chosen at the energy of about 170 MeV, because for lower energies elliptical cavities lose efficiency. It is inefficient to accelerate H⁻ ions from 170 MeV to 3 GeV using only one cavity type and, thus, two families of 650 MHz cavities have been chosen.

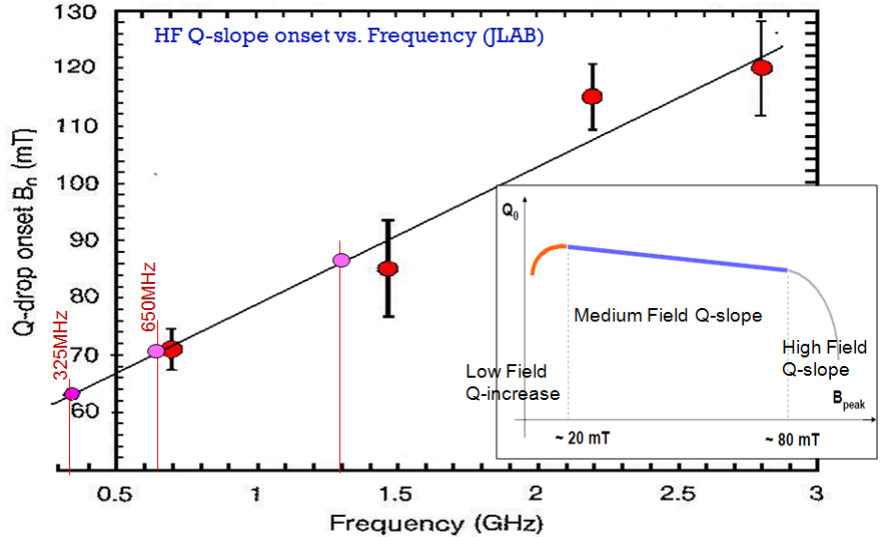

**Figure III-15**: High field Q-slope versus frequency.

Figure III-16 demonstrates the optimization of the transition energy between the two families of 650 MHz cavities and their geometric betas assuming a linear dependence of the field enhancement factors versus beta [21]. The upper figure shows the number of cavities required as a function of the betas of the two sections. The lower figure shows the energy gain per cavity vs. beam energy for $\beta_G$=0.61 cavities (red curve) and $\beta_G$=0.9 cavities (blue curve). The fully optimized geometric betas for the two 650 MHz sections are 0.64 and 0.9 respectively (upper figure), and the optimal transition energy is 466 MeV (lower figure). The initial synchronous phase is -30°, and it increases as the square root of the energy. More exact simulations taking into account realistic enhancement factors show optimal choice of geometric betas of 0.61 and 0.9.

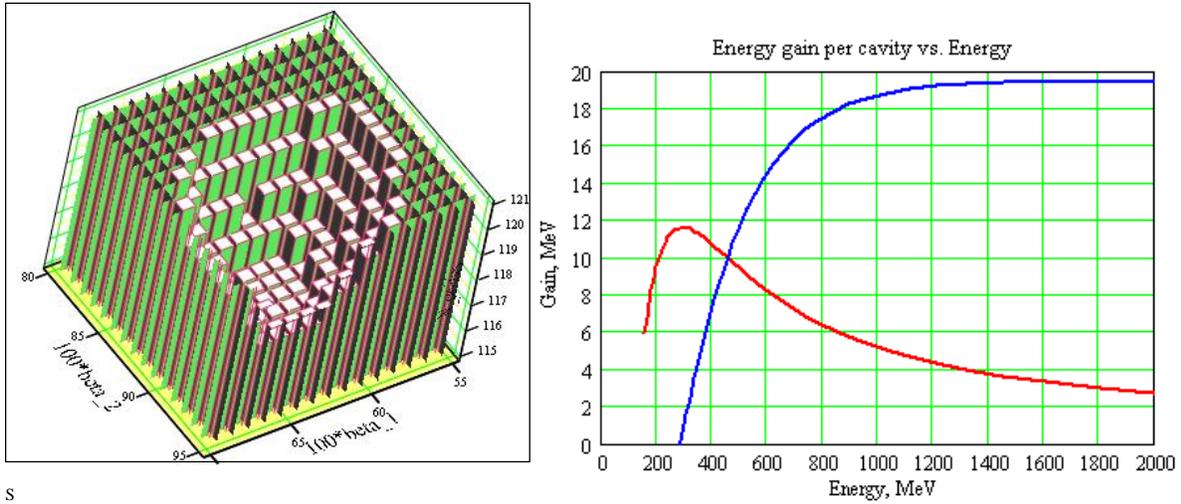

**Figure III-16**: Left: Number of cavities required versus cavity beta in the two 650 MHz sections.  Right: Energy gain per cavity versus particle energy for beta=0.61 (red curve) and beta=0.9 (blue curve) cavities.

The resultant characteristics of the six cryomodule types identified for the Project X CW linac are summarized in Table III-3. The cavity $Q_0$'s are based on an assumed operating temperature of 2K and a residual surface resistance of 10 n$\Omega$.  The energy gain per cavity throughout the CW linac is shown in Figure III-17. The gap in the energy gain for HB cavities is related to the longitudinal matching around 1 GeV energy.

| Sec-tion | Freq. (MHz) | Energy (MeV) | Cav/mag/CM | Gradient (MV/m) | Energy Gain (MeV) | $Q_0$@2K ($10^{10}$) | CM Config. | CM length (m) |
|---|---|---|---|---|---|---|---|---|
| HWR | 162.5 | 2.1-11 | 8 /8/1 | 8.2 | 1.7 | 0.5 | 8 x (sc) | 5.8 |
| SSR1 | 325 | 11-38 | 16 /8/ 2 | 10 | 2.05 | 0.2 | 4 x (csc) | 5.2 |
| SSR2 | 325 | 38-177 | 35 /21/ 7 | 11.2 | 5.32 | 1.2 | sccsccsc | 6.5 |
| LB650 | 650 | 177-480 | 30 /20*/ 5 | 16.5 | 11.6 | 1.5 | ccc-fd-ccc | 7.1 |
| HB650 | 650 | 480-1000 | 42 / 16†/ 7 | 17 | 17.6 | 2.0 | cccccc | 9.5 |
| HB650 | 650 | 1000-3000 | 120 / 30†/ 15 | 17 | 17.6 | 2.0 | cccccccc | 11.2 |

*5 superconducting and 5 warm (external to the cryomodule) doublets

†All doublets are warm, i.e. external to the cryomodule

**Table III-3**: Accelerating cryomodule requirements for the CW linac. Within the CM configuration column "c" refers to an individual accelerating cavity, "s" refers to a solenoid magnet, "f" refers to a single quadrupole magnet, and "fd" refers to a quadrupole doublet.

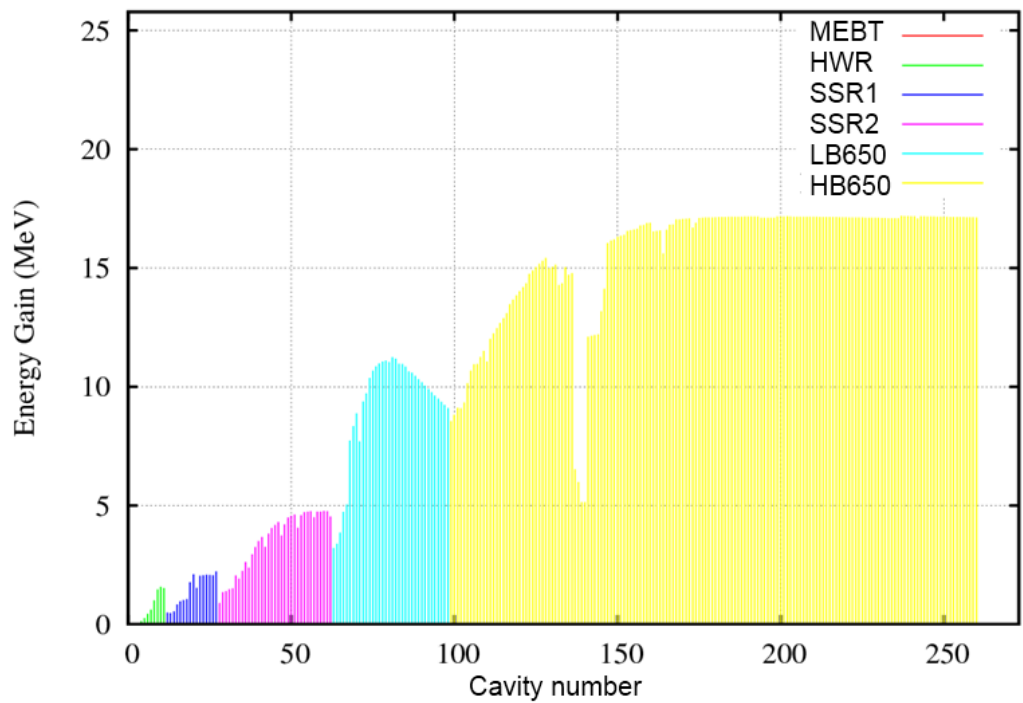

**Figure III-17**: The energy gain per cavity in the CW linac.

### III.1.3 CW Linac – Cryomodule Requirements

Cavities and focusing elements, as necessary, are grouped within cryomodules. In the 162.5 and 325 MHz sections the transverse focusing is provided by superconducting solenoids. In the 650 MHz sections quadrupole doublets (FD lattice) are used. Normal conducting doublets are located between cryomodules for both low-beta (LB650) and high-beta (HB650) cryomodules. Additional superconducting doublets are located in the center of low-beta cryomodules. The arrangement of focusing periods by cryomodule type is shown in Table III-3.

For beam diagnostics purposes, and for beam steering, each magnet package includes vertical and horizontal built-in correctors and a 3-coordinates beam position monitor (BPM). All cryomodules are separated by warm sections. These warm sections are used for additional diagnostics (bunch transverse and longitudinal profile monitors, beam loss monitors, *etc*.) and for collimation required to avoid uncontrolled beam loss in SC cryomodules. The makeup for each of the warm insertions will be determined by requirements of safe and reliable operations, diagnostics, collimation, and cryogenic segmentation constraints.

Integrated estimates of RF power consumption and cryogenic losses for the CW linac are shown in Table III-4, and plotted by cavity in Figure III-18 and Figure III-19.

| Section | R/Q ($\Omega$) | $Q_0$@2K ($10^{10}$) | $E_{peak}$ (MV/m) | $B_{peak}$ (mT) | Max. RF power per cavity* (kW) | Max. RF power per CM (kW) | Cryo-losses per CM (W) |
|---|---|---|---|---|---|---|---|
| HWR | 272 | 0.5 | 38 | 41 | 4.9 | 39 | 24 |
| SSR1 | 242 | 0.5 | 38 | 58 | 5.5 | 44 | 27 |
| SSR2 | 275 | 1.2 | 40 | 70 | 17 | 85 | 43** |
| LB 650 | 378 | 1.5 | 37.5 | 70 | 34 | 204 | 145** |
| HB 650 | 638 | 2.0 | 35.2 | 64 | 50 | 300 | 147** |
| HB 650 | 638 | 2.0 | 35.2 | 64 | 31 | 248 | 196** |

\* Power is computed for 2 mA to 1 GeV linac and 1 mA from 1-3 GeV. Allowances for transmission loss and microphonics suppression are included.

\*\* To be updated at the technical design stage. Only dynamic losses are taken into account.

**Table III-4**: Integrated RF power consumption and cryogenic losses in the CW linac.

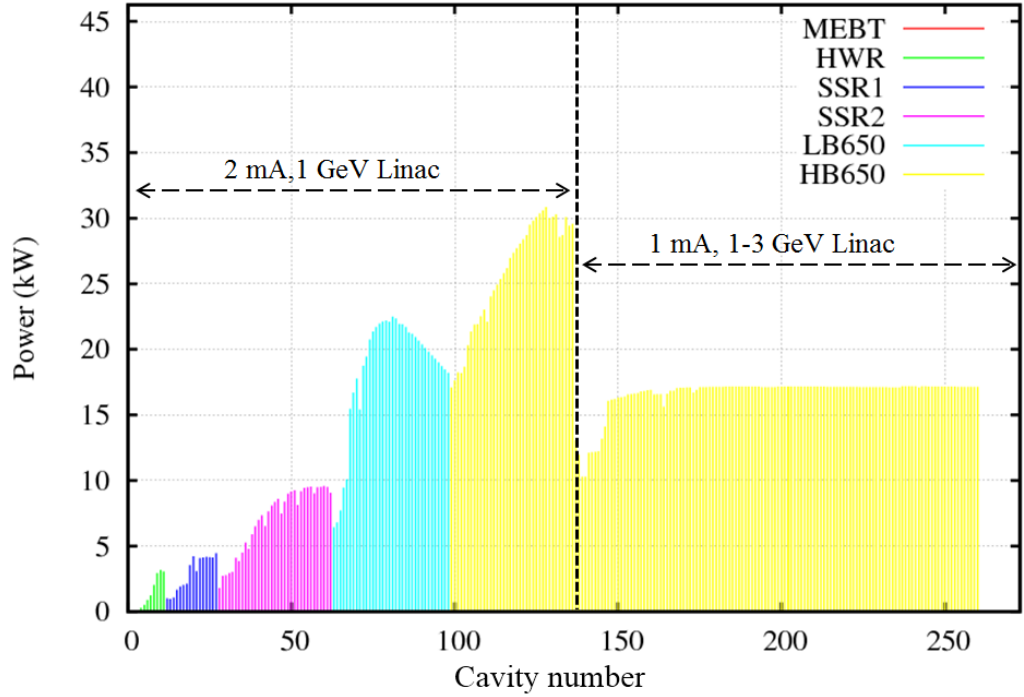

**Figure III-18**: Distribution of the RF power (per cavity) transferred to the beam; the beam current is 2 mA in 1 GeV linac and 1 mA in 1-3 GeV linac.

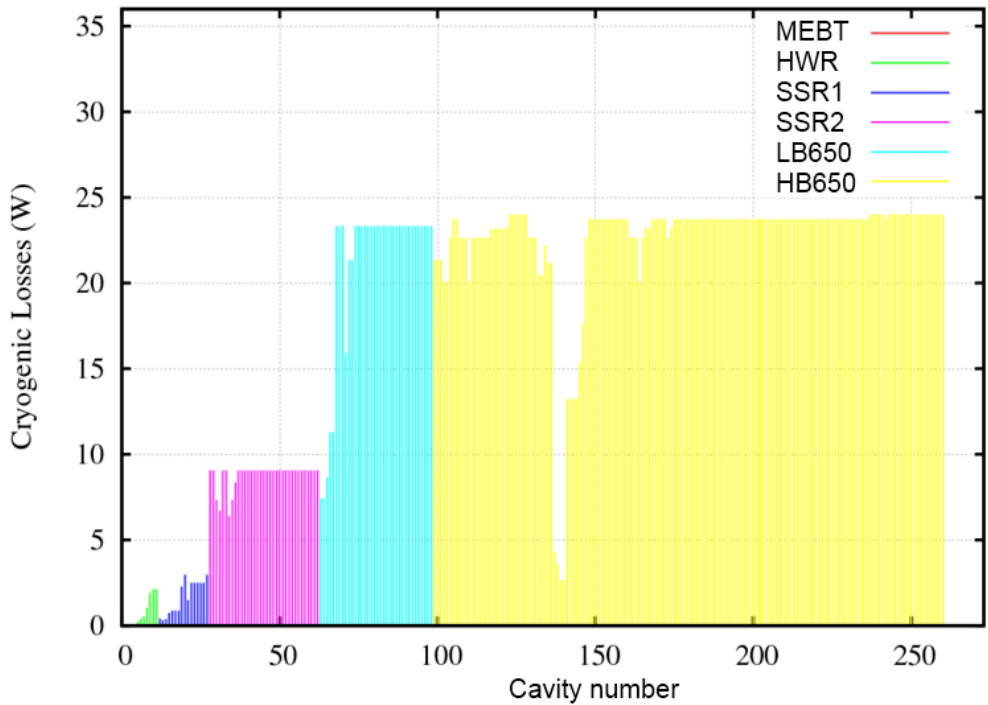

**Figure III-19**: Distribution of the cryogenic losses (dominated by cavity wall losses) per cavity in the linac.

### III.1.4 CW Linac – Beam Dynamics

The rms normalized beam emittance budget for the CW linac is established as 0.15 mm-mrad at the ion source and 0.4 mm-mrad at the exit of 3 GeV linac. The lattice design and the beam dynamics optimization are made utilizing the TRACK, TraceWin and GenLinWin codes. The results of the beam dynamics simulations are shown as an evolution of 1$\sigma$ beam envelopes through the entire CW linac in Figure III-20. The top figure shows the two transverse dimensions and the bottom figure the bunch length. Figure III-21 shows the transverse and longitudinal rms envelopes through HWR and SSR1 sections which are not clearly seen in Figure III-20. Figure III-22 displays the evolution of rms longitudinal and transverse normalized emittances through the Stage I linac. There is only modest transverse and longitudinal emittance growth. Due to diminishing space charge effects there is no significant emittance growth in the 1-3 GeV linac. However long transport through the 1 GeV arc results in a considerable emittance growth excited by beam space charge. Details are discussed in Section III.2.2. Figure III-23 displays the actual particle density distribution and Figure III-24 the evolution of beam energy through the linac. Figure III-25 shows a calculation of beam power loss (W/m) due to intra-beam stripping of $H^-$ thru through the linac. As one can see the losses due to this mechanism are below 0.1 W/m everywhere.

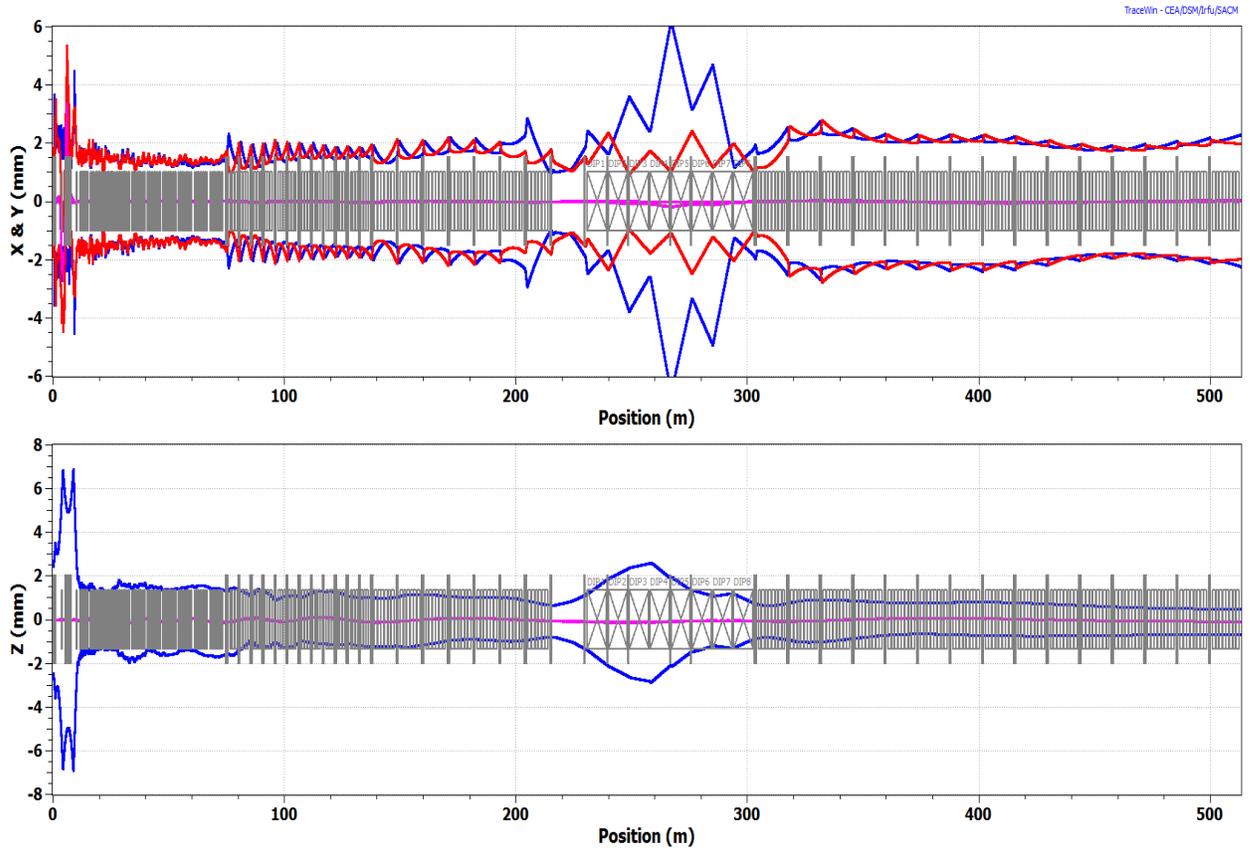

**Figure III-20**: Transverse (*x*(blue) and y (red)) rms envelopes (upper) and rms bunch length (lower) along entire CW linac (from MEBT to the end of 1-3 GeV linac) including the 180 degree bend.

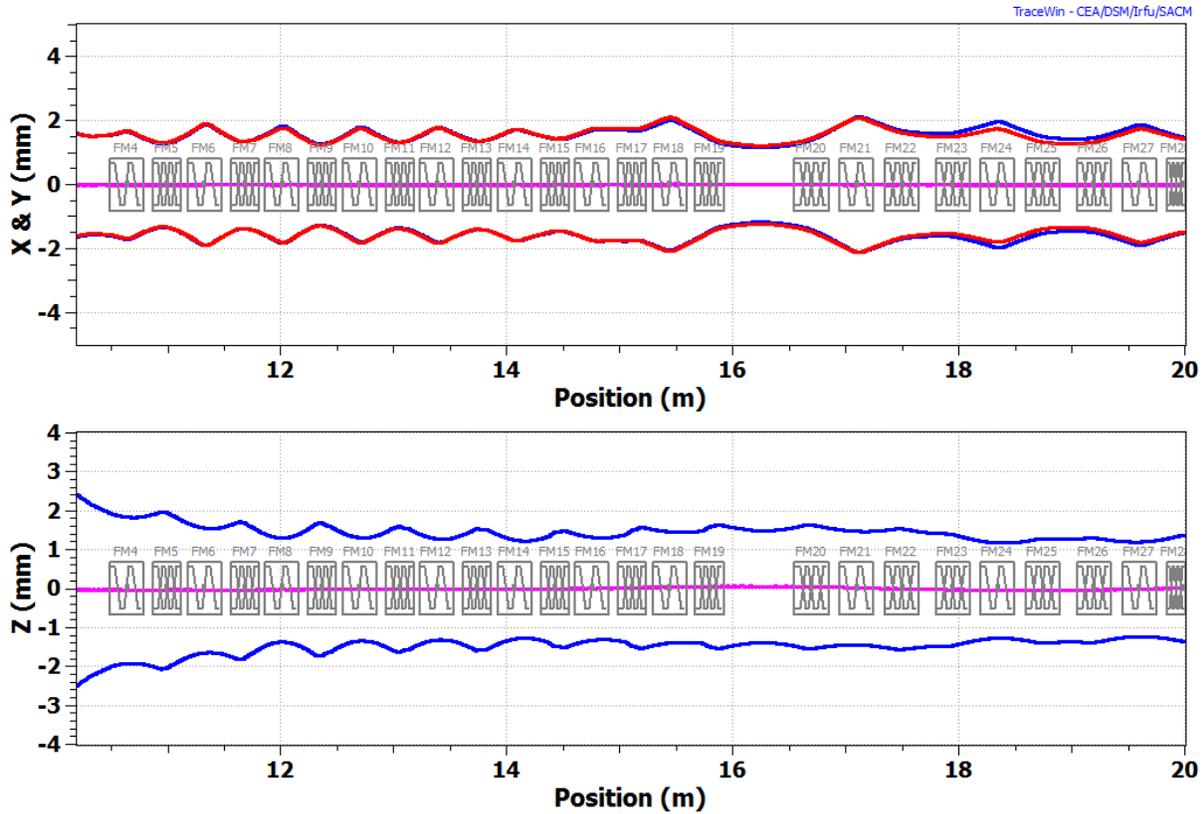

**Figure III-21**: Transverse (upper) and longitudinal (lower) rms envelopes through HWR and SSR1 sections.

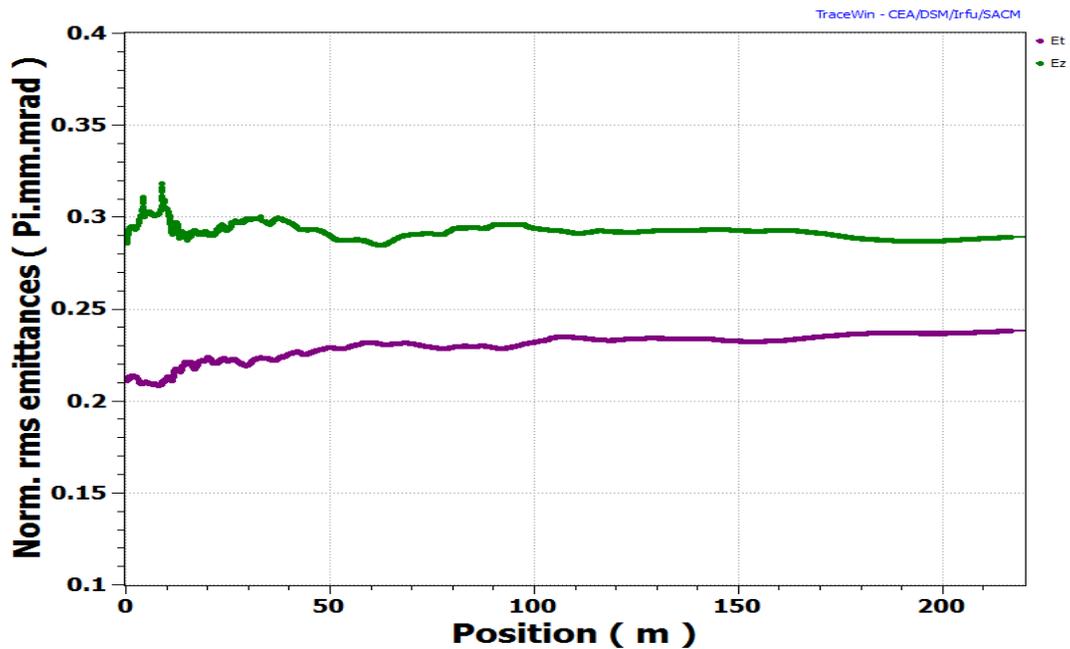

**Figure III-22**: Normalized transverse (magenta) and longitudinal (green) emittances along the linac (from MEBT entrance to the end of 1 GeV linac).

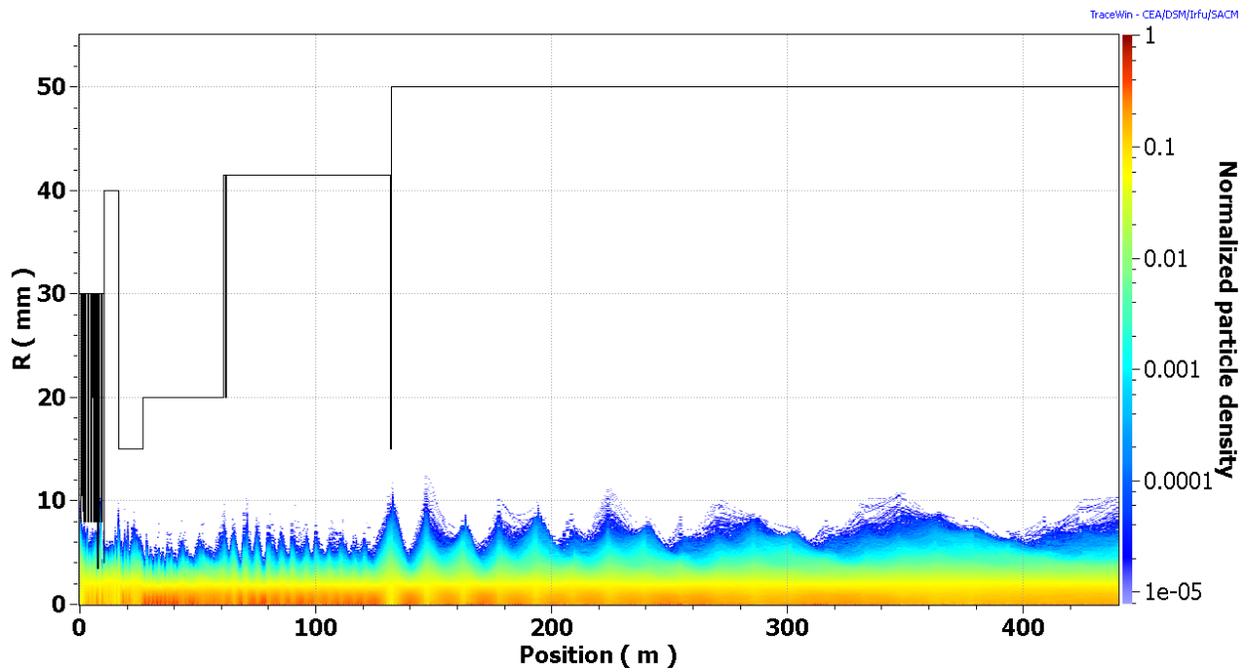

**Figure III-23**: Aperture and Particle density distribution along the linac. 100,000 particles tracked by PARTRAN. Simulations are performed without 1 GeV arc.

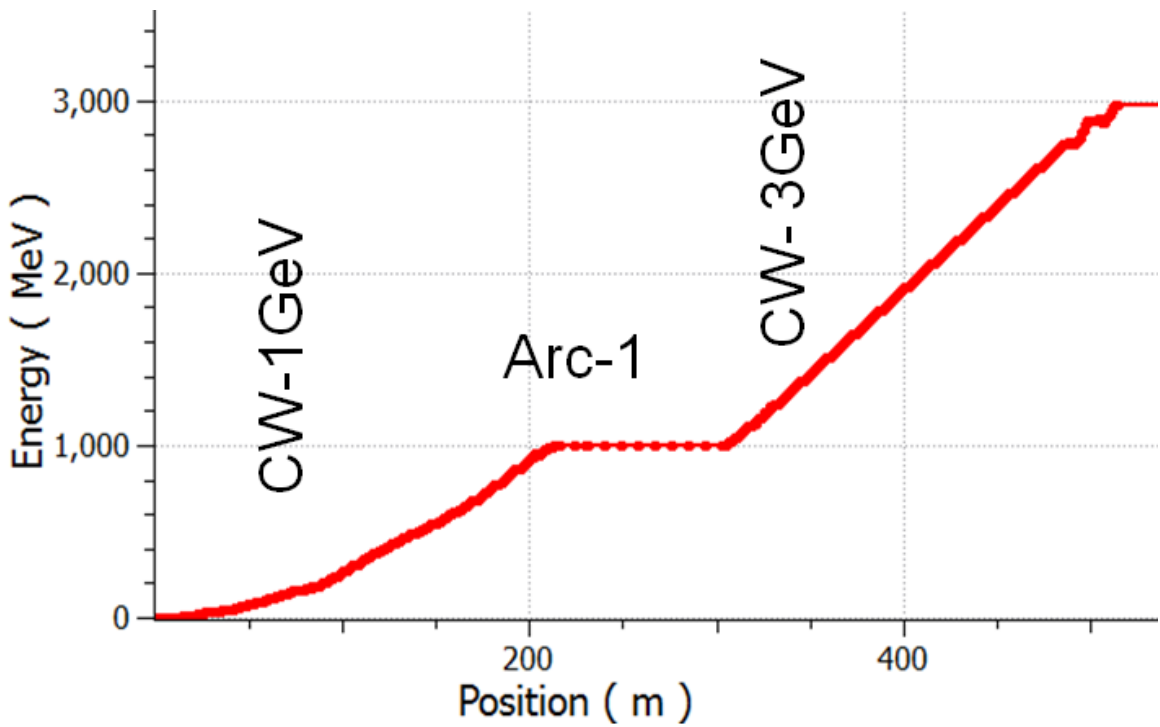

**Figure III-24**: Energy versus longitudinal coordinate.

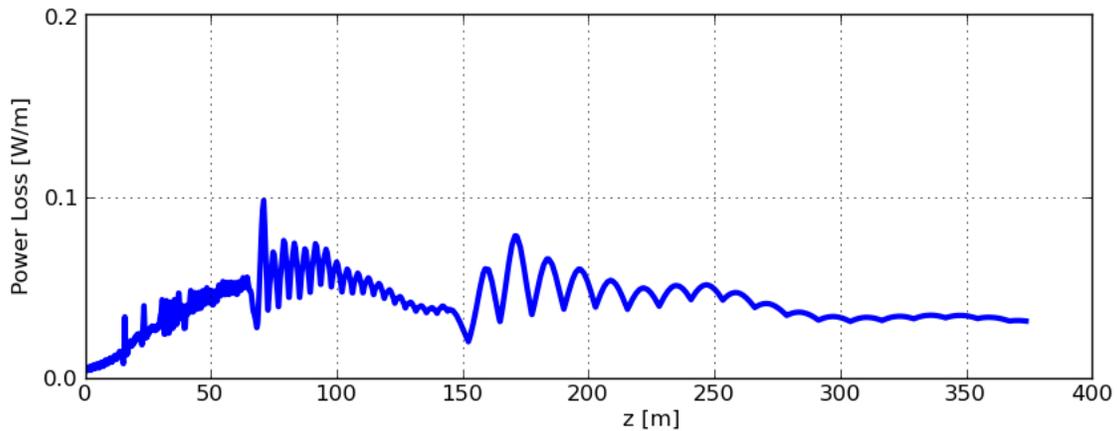

**Figure III-25**: The beam power losses per unit length caused by intra-beam stripping. Simulations are performed without 1 GeV arc which has significantly reduced intrabeam stripping due to bunch lengthening (see details in the next section).

### III.2  1 GeV Beam Handling

#### III.2.1  Transport Line Requirements and Physics Design

Stage 1 of Project X will provide up to 1 MW beam power split between the 1 GeV Experimental Area (EA) and the Muon Campus (MC), while providing 15 kW beam power to the Booster at a 15 Hz rate. The simultaneous beam delivery to the EA and MC is supported by an RF deflecting cavity that splits the EA and MC beams onto different trajectories. The final separation is produced by a three-way Lambertson magnet. Figure III-26 shows the concept for beam separation for the first two stages of Project X. As the Project matures and Stage 2 is added, it is assumed that the muon conversion experiment is relocated to the 3 GeV Experimental Area. Stage 2 provides 1 MW beam power to the 1 GeV EA and 3 MW to the 3 GeV EA. This is accomplished by increasing the 1 GeV CW linac current to 2 mA so that 1mA each can be sent to the 1 GeV and 3 GeV programs.

An RF deflecting cavity/Lambertson combination will be installed to provide the bunch structures to both the EA and MC as presented in Figure III-26. The Lambertson will provide a three-way split similar to the magnet used in the 3 GeV Experimental Area. A vertical trim dipole located near RF deflecting cavity will be used to allow for beam re-steering from the two upper to the two bottom apertures of the Lambertson. A horizontal dipole will be installed downstream of the Lambertson in the Booster/MC line to send beam to the linac dump.

The deflecting cavity deflects the EA and MC bunches in different directions. The bunches destined for the MC are given a positive deflection such that they end up going through the field region of the Lambertson while the bunches intended for the EA are given a negative deflection and end up going through the field-free region of the Lambertson. Adding a positive DC deflection equal to the deflection by the deflection cavity will offset the beam trajectory at the Lambertson such that the EA beam will once again be on the linac centerline. When Stage 2 is implemented, the trim dipole will change its polarity to offset the positive and negative kicks in the opposite direction thus splitting beam between the EA and the 1-GeV arc for the 3-GeV linac. Table III-5 summarizes the operation of the deflecting cavity/dipole for sending beam to the different destinations for both Stages.

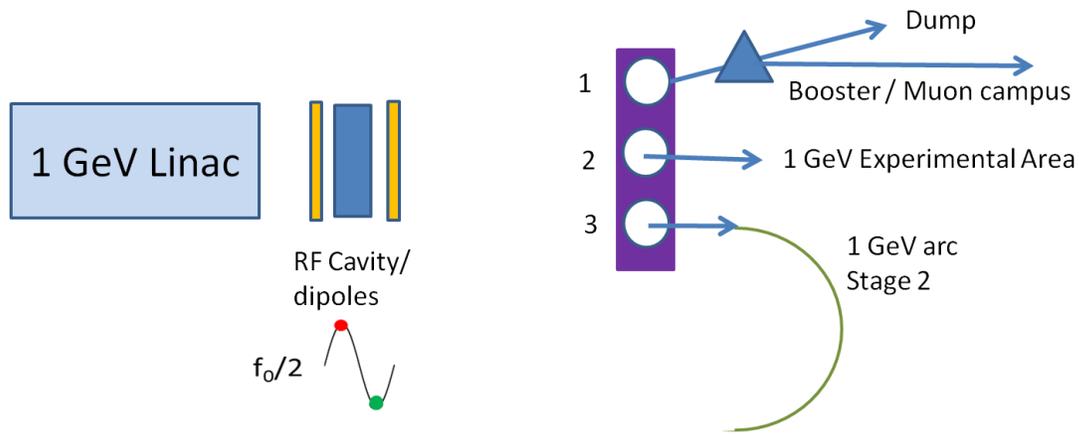

**Figure III-26**: Concepts for the 1 GeV beam separation for Stages 1 and 2.

| Stage | Destination | deflecting cavity | Dipole | DSW |
|---|---|---|---|---|
| 1 | Muon Campus | +θ | +θ | |
|  | 1 GeV EA | -θ | +θ | |
|  | Booster | off | +2θ | |
|  | Dump | off | +2θ | on |
| 2 | 1 GeV EA | +θ | -θ | |
|  | Booster | off | +2θ | |
|  | Dump | off | +2θ | on |
|  | 1 GeV arc | -θ | -θ | |

**Table III-5**: Required powering for the Deflecting cavity and dipole to send beam to the various destinations in Stages 1 and 2

Booster injection requires a 162.5-MHz beam structure which is incompatible with simultaneous operation of the 1 GeV EA and MC. This means both programs will be interrupted during Booster injections. The MEBT chopper will prepare the required bunch pattern to fill the Booster's 47 MHz buckets during the 1-ms injection period. The chopper will remove bunches that would land at the RF bucket boundaries and create a 50 to 60 ns "notch" at the revolution period to provide a gap for beam extraction. The RF deflecting cavity is turned off during Booster injections and the vertical trim dipole near splitting cavity is energized creating twice the deflection of the RF deflecting cavity. Allowing 0.5 ms for turning the cavity off and on and ramping the dipoles would create a total interruption of approximately ~2 ms out of 67 ms, or a 3% effect. The beam line (immediately downstream of the Lambertson) for Booster injections is identical to that for sending beam to the MC. A dipole switch located in F0 sector of the Tevatron tunnel will be energized (or not) to steer beam to the MC (or Booster).

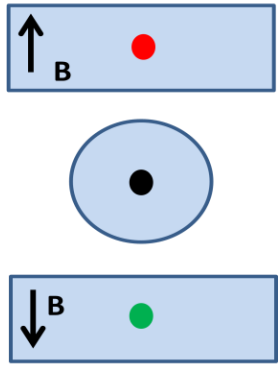

**Figure III-27**: Concept of the 1 GeV switchyard three-way Lambertson showing the beam centroids for the beam directed to the EA (black) and the MC/Booster (red) and the 1 GeV arc to the 3 GeV linac (Stage 2).

The required frequency of the deflection cavity is $f_{RFS}=(n + 1/2)\cdot 162.5$ MHz, where $n$ is an integer. The frequency of 406.25 MHz ($n = 2$) is chosen. This choice of frequency is determined by a compromise between the overall sizes of SC cavity, required voltage and a variation of the kick value due to the finite bunch length.

Both field and field-free apertures of the Lambertson should be a factor of two or three larger than the $6\sigma$ beam envelope to assure loss-free transfer. From the TraceWin output we see that both the horizontal and vertical rms sizes are approximately 1.5 mm, giving a 99% beam envelope of roughly 10 mm. A 2 kG dipole field for a meter long Lambertson will give a 35-mrad horizontal deflection, sufficient to clear the downstream magnets. This will be a symmetric Lambertson with a septum thickness of approximately 10 mm, consistent with existing designs. The septum aperture is selected to be approximately $\pm 9\sigma$ so that minimize possible beam scraping in the septum. Figure III-27 shows a conceptual design of the Lambertson aperture.

Taking into account aperture considerations, a ±20-mm displacement is required at the entrance to the Lambertson. To meet this separation, the RF deflection cavity should provide approximately ±3-mrad vertical kick. Based on the required angle, the transverse deflection voltage should be ≈ 5 MV. The deflection cavities in both the 1 and 3 GeV areas are close in frequency and maximum deflection voltage (7 MV); therefore they will have similar designs (see details in Section IV.1.7.)

### III.2.2 H⁻ Transport to 3 GeV Linac

The transport line between the 1 and 3 GeV linacs is achromatic and isochronous to suppress horizontal beam emittance growth and bunch lengthening in the course of beam transport through the arc. The arc has the total bending angle of 180 deg. Its optics is based on FODO focusing. Figure III-28 presents the beta-functions, the horizontal dispersion and beam envelopes through the line. The beamline is tuned to be isochronous at 1 GeV beam energy. If required, the isochronous energy can be varied within about ±5%, achieved by adjustments of strength of a few quadrupoles located in the large dispersion areas. In the case of small beam intensity when beam space charge can be neglected the transport through the line does not create any bunch lengthening and therefore does not produce any complications for the longitudinal beam dynamics. However for nominal intensity the space charge will result in an emittance growth as presented in Figure III-29. Although the emittance growth is not negligible it is within Project X specifications up to bunch population corresponding to 10 mA RFQ current.

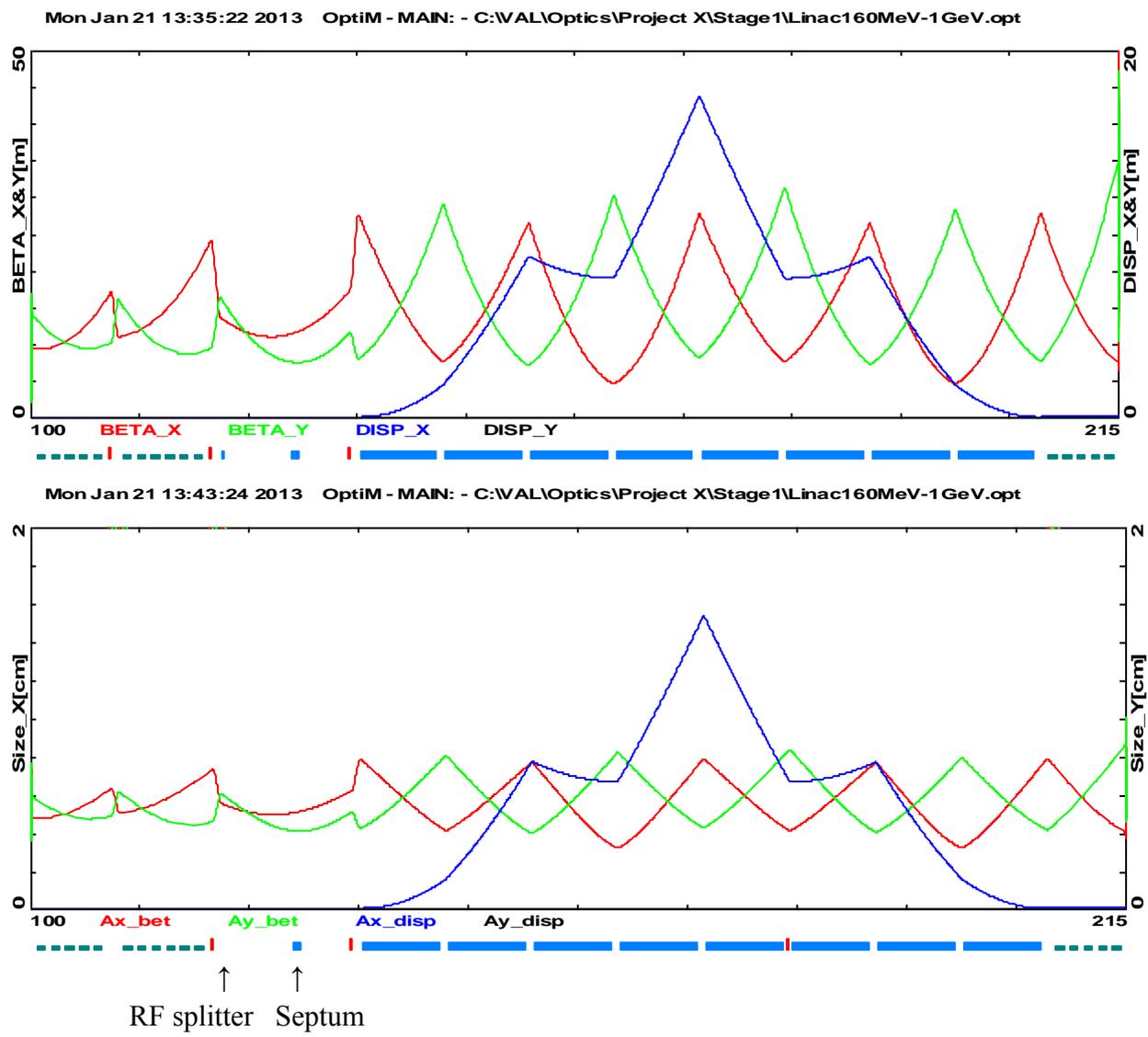

**Figure III-28**: Beta-functions and dispersion (top) and 4σ transverse beam sizes due to betatron motion and the energy spread (bottom) for the beam transport through 1 GeV arc connecting 1 GeV and 3 GeV linacs.

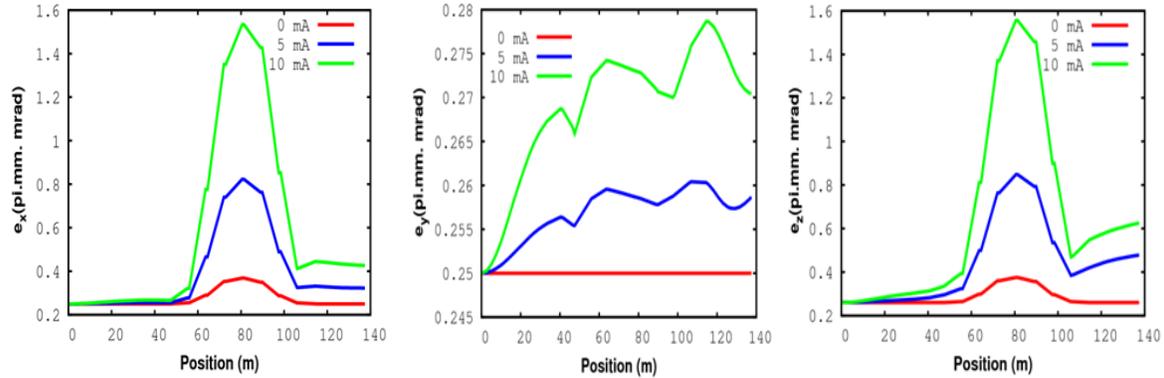

**Figure III-29**: Transverse and longitudinal emittance growths for the beam transport through transfer line between 1 and 3 GeV linacs for bunch populations corresponding to the RFQ currents of 0, 5 and 10 mA. The peaks in the horizontal and longitudinal emittances are an artifact of emittance computation related to non-zero dispersion in the arc and should be neglected. Note also that the scale for the vertical emittance is greatly expanded.

### III.2.3  H- Transport to Experimental Areas and Booster

Transport to the Booster will utilize the decommissioned Tevatron enclosure. The transport line will be at the elevation of the Tevatron (*i.e.* installed under the existing Main Ring remnant used to transport MI beam to the Muon Source and Switchyard), which will require a compact magnet design. The new beam line will enter the Tevatron enclosure upstream of F0 where a switch dipole will be installed to divert the 1 GeV beam into the Main Ring remnant, called the P2 line, for transport to F17 and the Muon campus. The transport line to Booster will follow the footprint of the decommissioned Tevatron ring till about F35 (a length of about 750 meters) where the transport line will leave the Tevatron enclosure. This will be at a point almost tangent to both the Tevatron and Booster. A new tunnel will be created for the last several hundred meters between the two enclosures. The location and length of this tunnel section will be designed to minimize interference with existing facilities and to optimize transport optics design to match into Booster injection.

A preliminary lattice for transport between the 1 GeV Linac and Booster has been designed. It matches the curvature of the Tevatron and provides an existence proof of a solution. While the design is compatible with permanent magnets, this line will use electromagnets to provide flexibility for accommodating possible energy variations in the Booster and polarity change for the proton EDM experiment (see Section III.2.5). There are four distinct sections to this transport line: 1) straight section from the Linac to the Tevatron enclosure, 2) bend to match into Tevatron F0 straight section, 3) follow the Tevatron trajectory around to F35, and 4) a straight section that transports the beam from the Tevatron tunnel to the Booster enclosure around Long 13 H- Linac Dump.

The 1 GeV linac dump will be predominately utilized for low intensity tune up. It is not expected to be used as a full intensity dump. Therefore, it will be designed for no more than 10% of full beam power, *i.e.* 100 kW. The capacity of this beam dump will need further discussion and will be optimized for price/performance and expected operational scenarios. The beam power will be controlled by duration and rep rate of the Linac. The beam will be directed toward the dump when the RF deflecting cavity is off and the dump selection dipole is powered, as shown in Figure III-25. The location of the dump on the facility site needs to be addressed. The determination of whether this will be an isolated enclosure or accessible from the transport enclosure needs to be determined by shielding considerations. The beam parameters and power range are well within the SNS operating envelope. Therefore we do not anticipate any outstanding problems for the beam dump design.

III.2.4 **Beam Loss**

Low loss beam transport is critical in the operation of MW class facility. The primary loss mechanisms for 1 GeV transport are $H^-$ intra-beam stripping, Lorentz stripping, inelastic beam-gas scattering, and scraping of beam halo on the apertures. Stripping due to black-body radiation inside the room temperature beam pipe is negligible at this energy. The overall requirement is to keep the beam loss below 100 mW/m; assuring residual radiation at below 15 mrem/hour at 1 foot – the level where the radiation does not represent a problem for machine servicing. For 1 GeV beam 100 mW/m corresponds to the fractional beam loss of less than $10^{-7}$/m.

Intrabeam stripping is the main mechanism for the beam loss. It requires, if possible, avoiding tight beam focusing. As one can see from Figure III-25 it generates a beam loss rate of about 50 mW/m at 1 GeV. This loss rate is reduced due to beam debunching in the course of beam transport through the 1 GeV transport line.

Lorentz stripping (see Section III.3.2) limits magnetic field to less than 2.9 kG, corresponding to the loss rate of about below $5\times10^{-8}$/m for 1 GeV beam. Leaving room for a possible 5% energy increase implies a maximum field of 2.7 kG in the dipoles of the 1 GeV bending arc. This generates the loss rate of about $5\times10^{-9}$/m for 1 GeV beam.

The cross section of $H^-$ stripping for 1 GeV beam on residual gas is about $10^{-19}$ cm$^2$ for molecular hydrogen and grows approximately proportional to *Z* for heavier atoms. The requirement of $2\times10^{-8}$ for partial loss rate yields a vacuum requirement of $10^{-8}$ Torr or better for H$_2$ and about an order of magnitude better for heavy molecules (hydrocarbons, water, etc.). The beam intensity transported to the Muon Campus is approximately 100 kW.

Consequently an application of vacuum practices developed in Fermilab for not-baked vacuum systems, which routinely achieve low $10^{-8}$ Torr, has to be sufficient.

The aperture, optics, and vacuum level of the transport lines will be designed with the goal of a loss free transport. This is most critical for the sections of transport line which will transport high beam power. This means that the aperture should be greater than $12\sigma$ and preferably closer to $18\sigma$ to minimize any losses to steering errors or scraping of any halo that is formed during acceleration or transport. Due to the concern of generating transverse halo by the beam space charge in a strong focusing transport line, the transverse focusing properties of the 1 GeV Linac should be continued into the beam line to the point just past the Experimental Area switch dipole to keep a more or less uniform beam size through the high intensity part of the line. At that point, we have the option to match into a FODO lattice for transport to the Booster. Due to bunch lengthening the problem of the space charge is fast diminishing with further beam transport and becomes much less significant after about first 100 m of beam transport.

### III.2.5  Provisions for sub-GeV proton based research

A proton electric dipole moment (EDM) experiment is planned as a part of the Project X physics program. This experiment will require infrequent injections (once in about 20 min.) of polarized proton beam into a newly constructed ring with electrostatic bending at beam energy of 235 MeV. A one-turn injection of 1 mA beam will take ~2 µs and deliver about $10^{10}$ protons with normalized emittance equal or below ~0.25 mm mrad. If larger intensity is required a few turn injection is possible too as well as larger current from the ion source which can be supported for a few µs time by energy stored in the SC cavities.

A few actions will be required to support sub-GeV proton operations. First, a polarized proton source is required. This source can be installed in place of the spare H$^-$ ion source. This implementation minimizes changes in the LEBT while leaving room for alternative implementations. In spite of the fact that protons are positively charged further beam acceleration can be achieved with the Project X accelerator without any significant changes from H- operations. Neither the solenoidal focusing in the HWR, SSR1 and SSR2 nor the quadrupole focusing in the LB650 requires any changes. The only change from the beam dynamics point of view will be interchanging *x* and *y*-coordinates. However, there are three actions that are required. First, some beam re-steering will be required due to imperfections of machine alignment. Normally this should not require more than 1 sec. Second, to obtain the required beam energy the phases of cavities located downstream of 235 MeV point have to be shifted to "no acceleration" and their voltage reduced to prevent longitudinal over-focusing and to reduce transverse defocusing. The low level RF controlling the cavity voltage and phases should be able to do this within 10 ms without additional modifications of the hardware. Third, the quadrupole focusing in the HB650 part of the linac has to be reduced. This action should not take much longer than 1 s for laminated quadrupoles which will be a requirement for their design if a short switching time is required. The necessity to reduce focusing in the HB650 section is related to the following: the design optics for 1 GeV linac implies gradual reduction of quadrupole focusing with energy to minimize intrabeam stripping. Therefore the quadrupole strength in the LB650 section normally accelerating beam from 177 to 480 MeV is reduced with energy and an absence of energy growth downstream of 235 MeV point does not cause over focusing. The period of quadrupole focusing is approximately doubled in the HB650 section and its focusing is set to obtain close to $90^\circ$ betatron phase advance per period at the section beginning, with beta-functions matched to the growing beta-functions of LB650 section. Therefore the reduction of energy at the HB650 entrance will result in over focusing and loss of particle motion stability. To avoid it a reduction of focusing in HB650 section is required.

If the EDM ring is located in the (former) antiproton source tunnel it can use the 1 GeV transport line directed to Muon campus. In this case the polarity of all dipoles has to be changed. If required, running the EDM experiment in normal operating conditions should not take more than 0.2% from the Project X duty cycle.

### III.3  3 GeV Beam Handling

#### III.3.1  Transport Line Requirements and Physics Design

At Stage 2, the current in the 1 GeV Linac will be increased to 2 mA. The RF splitter at the end of the 1 GeV Linac will send half of the bunches to the 1 GeV Experimental Area and the other half to the 3 GeV Linac, each with a bunch structure of 81.25 MHz. After acceleration in the 3 GeV Linac, the beam will be transported to either the 3 GeV Experimental Area, where it will be further distributed to experiments by a 3-way RF splitter/ Lambertson magnet combination, or the 3 GeV beam dump.  The selection between the Experimental Area and the dump will be controlled by a DC dipole, DS3. If it will be required in the future, additional RF splitters can be installed downstream of the primary splitter allowing an increase in number of experiments.

At Stage 3, a 3-8 GeV pulsed Linac will be added to accelerate the beam for injection into the existing Recycler. The MEBT chopper will prepare a 162.5 MHz bunch structure for acceleration in the 1 and 1-3 GeV linacs. A pulsed dipole switch, DS4, will be energized for approximately 4.4 ms to direct the 3 GeV beam into a 180 degree arc. The arc is isochronous and achromatic.  Figure III-30 shows a concept of the 3 GeV Switchyard for both Stages 2 (blue components) and Stage 3 (green components). The location of all the components for Stage 2 and upgrade to Stage 3 will be further optimized according to optical design, site conditions, and cost minimization.

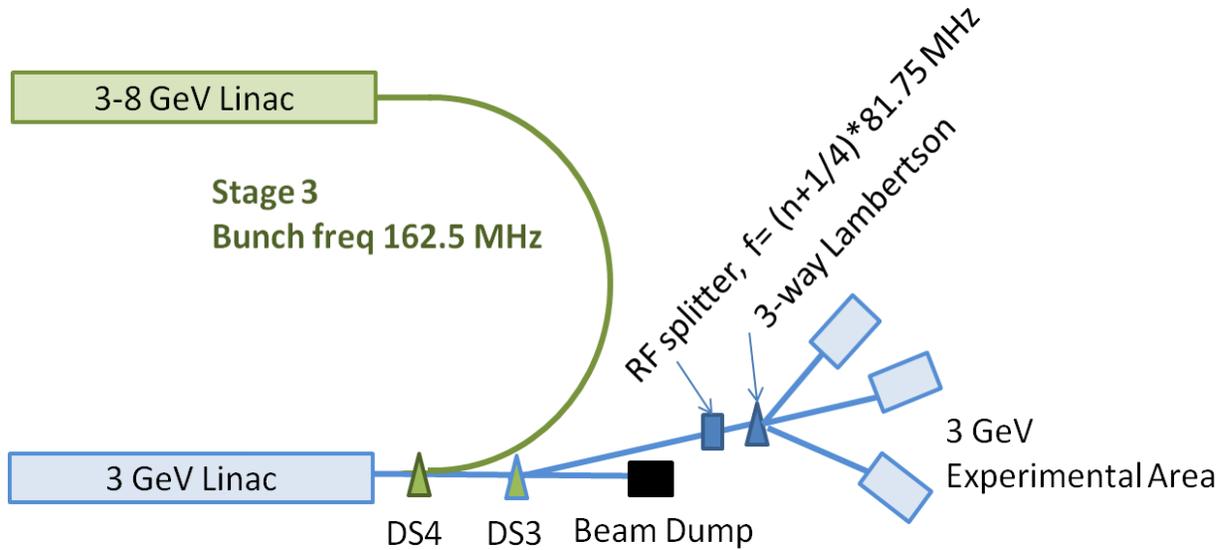

**Figure III-30**: Preliminary layout of the 3 GeV Switchyard showing the splitting to the Experimental Area (Stage 2 in blue) and the HE Linac (Stage 3 in green).

### III.3.2 Beam Loss

The configuration of transport lines supports H⁻ beam transport both to the Pulsed Linac (135 kW) and to the Experimental Area (3 MW). The requirements to the fractional beam loss are dominated by the higher intensity beam transport to the Experimental Area.

The H⁻ transport should have sufficiently small loss to minimize residual radiation in the tunnel. It is highly desirable to keep residual radiation level below 15 mRem/hr. Many facilities use the metric of 1 W/m as a limit for "hands on" maintenance, however, at 3 GeV, a 1W/m loss rate corresponds to a loss of $2.1\times10^9$ p/m/s and produces a peak contact residual dose rate of ~150 mrem/hr [22] on a bare beam pipe. This loss rate produces significantly lower residual activation on external surfaces of magnets due to shielding by the magnet core steel. However, a main concern is the residual level at magnet interfaces and instrumentation locations. These levels are based upon MARS calculations and used for order of magnitude estimations. A more accurate estimation will be required once a detailed model of the transport line is available. Setting a desirable activation level to 15 mrem/hr results in a loss goal of ~0.1 W/m, and, consequently, a fractional loss rate of $7.8\times10^{-7}$ /m and $3.4\times10^{-8}$ /m for 135 kW beam to the pulsed linac and 3 MW beam to the experiments, respectively. The single particle mechanisms contributing to the beam loss are the Lorentz stripping in dipoles, the beam stripping in the residual gas, the photo detachment by blackbody radiation in the beam pipe, and loss due to intra-beam stripping, which in a normal operational state will be a dominant contribution.

Beam motion in a magnetic field excites an electric field in the beam frame. If this electric field is sufficiently strong, it can detach the weakly bound outer electron (Lorentz stripping). Figure III-31 shows the loss rate per meter of dipole field for a 3 GeV kinetic energy H$^-$ as a function of magnetic field. One can see that a dipole field of 1.3 kG produces a fractional loss of 3.7x10$^{-8}$ /m. As the dipoles occupy a significant fraction of the beam line, in particular in the 180$^o$ arc, one would like to select a field which produces negligible loss. A dipole field of 1.2 kG is chosen as the maximum field to be used in the 3 GeV transport lines. This produces a fractional loss of ~3.4x10$^{-9}$ /m.

The loss rate due to H$^-$ scattering on the residual gas molecules is proportional to their density and the ionization cross section of the molecules present. The cross section decreases proportionally to $\beta^{-2}$ and therefore is only marginally smaller than for 1 GeV. Taking into account the three times larger beam power, one needs three times better vacuum to achieve the same radiation levels. Thus vacuum of 3×10$^{-9}$ Torr or better is required.

The contribution from the blackbody photons at 3 GeV at room temperature is 1.3x10$^{-7}$ /m as seen in Figure III-32. It is well within specifications for transport to the pulsed linac, where the beam power is limited to 130 kW at 3 GeV, however for transport of 3 MW beam power it is only slightly below the specification. Lowering the beam pipe temperature to 150 degrees K reduces the fractional loss to 5x10$^{-10}$ /m. At room temperature, one can see that the black body radiation is expected to make a major contribution.

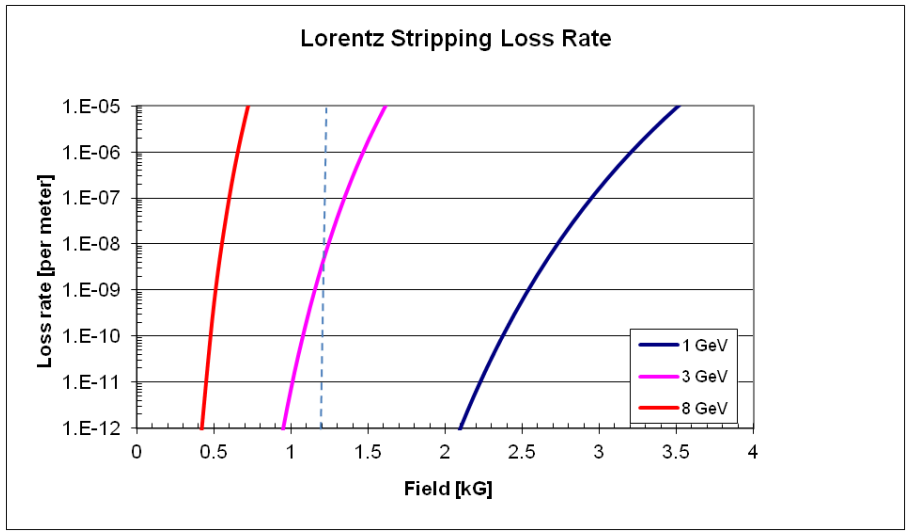

**Figure III-31**: Fractional loss for a 3 GeV H$^-$ ion traveling in a dipole field. The dashed line marks a magnetic field of 1.2 kG.

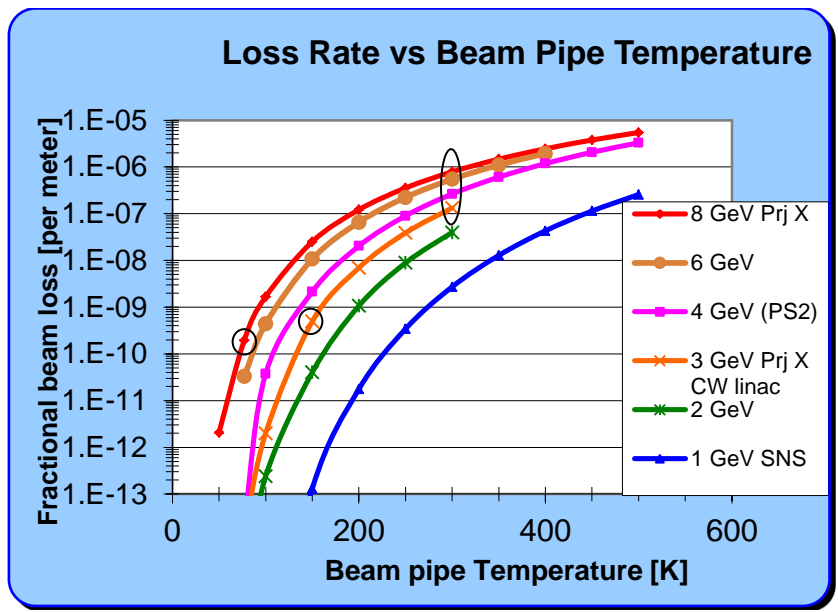

**Figure III-32**: Loss rate [m$^{-1}$] due to the interaction of blackbody photons as a function of internal beam pipe surface temperature for several values of H$^-$ kinetic energy.

Estimation of loss due to intra-beam stripping has been carried out for the entire CW linac. The loss depends on the local transverse velocity of the ions and proportional to the lattice functions. The transport line utilizes the same FODO structure as the 1.3 GHz linac so the contribution due to intra-beam stripping should be comparable to that calculated for the end of the CW linac, namely ~30 mW/m. This will be smaller in the arc due to bunch lengthening.

Table III-6 summarizes the contributions of the four single particle loss mechanisms discussed above. Note that the value for the intra-beam stripping contains only an estimated beam loss power where the value was taken from the CW Linac estimates. For the HE linac transport line, the loss rate due to Lorentz stripping with the selection of the dipole field of 1.4 kG is the dominant loss but acceptable at only 37 mW/m. Losses within the transport line to the Experimental Areas is dominated by blackbody radiation and is 0.4 W/m – on the edge of acceptability. Lowering the beam tube internal surface temperature with a beam screen to gaseous nitrogen temperature, 150°K, and improving the vacuum to $5\times10^{-9}$ Torr lowers the total beam loss to ~60 mW/m and expected residual activation on a bare beam pipe to ~10 mrem/hr.

| Loss Mechanism | CW to Pulsed Linac | | | CW to Experiments wo/ shield | | | CW to Experiments w/shield | | |
|---|---|---|---|---|---|---|---|---|---|
| | Value | Loss/m | W/m | Value | Loss/m | W/m | Value | Loss/m | W/m |
| Vacuum | 1E-08 Torr | 1.3E-08 | 0.002 | 1E-08 Torr | 1.3E-08 | 0.04 | 5E-09 Torr | 6.9E-9 | 0.021 |
| Lorentz | 1.4 kG | 2.9E-07 | 0.037 | 1.2 kG | 3.4E-09 | 0.01 | 1.2 kG | 3.4E-09 | 0.01 |
| Blackbody | 300 K | 1.3E-07 | 0.017 | 300 K | 1.3E-07 | 0.39 | 150 K | 5.0E-10 | 0.001 |
| Intra-beam | | | 0.001 | | | 0.03 | | | 0.03 |
| Total | | 4.3E-07 | 0.057 | | 1.5E-07 | 0.47 | | 1.1E-08 | 0.062 |
| Residual Activation, bare beampipe, mrem/hr | | 8.5 | | | 69.5 | | | 9.2 | |

**Table III-6**: Summary of beam losses in the 3 GeV transfer lines between the CW linac and the pulsed linac and experimental areas. The beam power within the pulsed linac transfer line is ~130 kW, while the beam power within the Experimental Areas transfer line is ~3 MW.

### III.3.3  H- Dump

A straight-ahead beam dump is provided for beam tune up. The transport line to the dump has optics similar to FODO lattice used in the linac-to-Experimental Area transport line. The beam dump is located in a separate enclosure at a distance of about 150 m from the end of the linac. Beam sweeping on the dump face reduces the power density to an acceptable level. A buried beam pipe connects the beam-line enclosure to the beam dump enclosure. For a rms emittance of $0.25\pi$ mm-mrad, the spot size on the face of the dump with a beta of 250 m is about 25 mm.

### III.3.4 H- Transport to Pulsed Linac

Similar to the 1 GeV area, the beam delivery to the next stage proceeds through 180 deg. arc. The arc is isochronous and achromatic. Its optics is built similar to the 1 GeV arc but higher beam energy requires the following changes to be taken into account. Mitigation of Lorentz stripping results in smaller field in dipoles and much larger length of the arc. An energy increase reduces the dependence of particle velocity on its momentum. It requires stronger focusing to make the linac-to-linac beam transport isochronous. The beam line Twiss functions are presented in Figure III-33. Similar to the 1 GeV transport the beam space charge results in an emittance growth. The results of simulations are presented in Figure III-34. In spite of higher energy the emittance growth is larger than for 1 GeV beam due to much larger length of the beam transport.

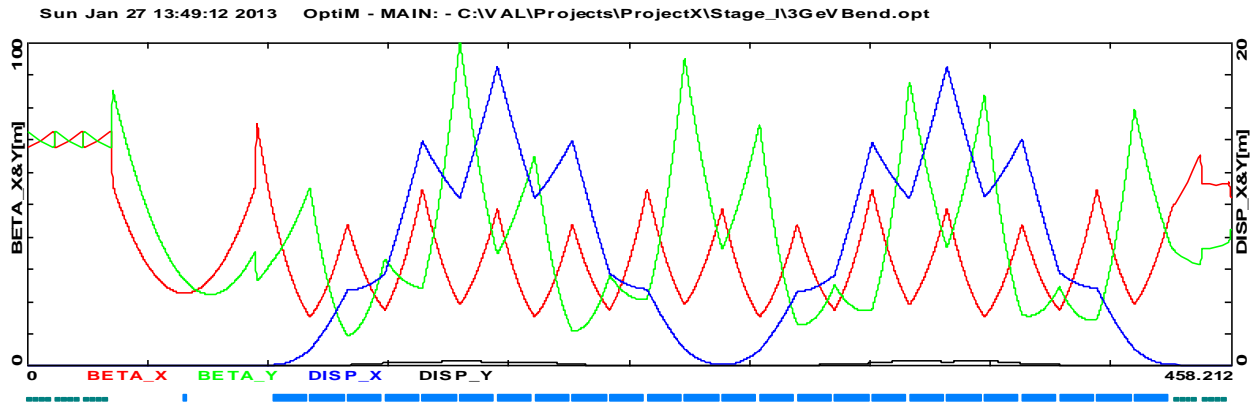

**Figure III-33**: Beta-functions and horizontal dispersion for the beam transport from 3 GeV CW linac to Pulsed linac.

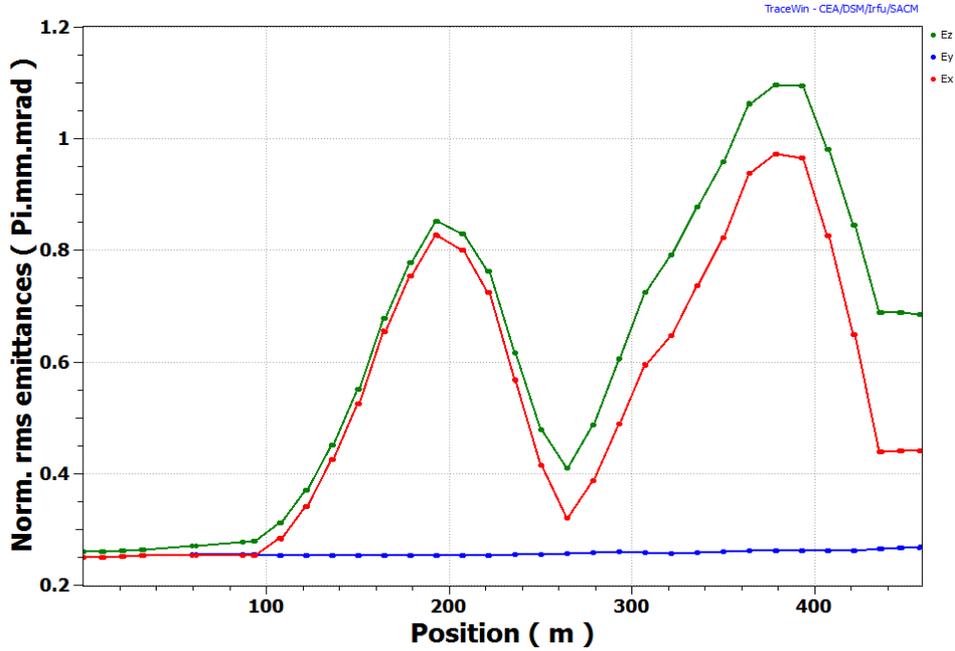

**Figure III-34**: Dependence of beam emittances on distance from the CW linac for the beam transport from 3 GeV CW linac to Pulsed linac.

### III.3.5  H- Transport to Experimental Area

If the pulsed HE linac switch, DS4, magnet is off (in Stage 3) and the DC Experimental Area switch magnet, DS3, is on, the 3 GeV H$^-$ beam will be directed toward the Experimental Area. The beam power in this transport line is up to 3 MW. It is critical to minimize single particle losses, provide sufficient aperture, and provide collimation to remove large amplitude particles that could become lost in unprotected areas.

Immediately after splitting off the dump line via a horizontal 2 dipole achromat, a section of the FODO lattice is reserved for a potential collimation system. It can increase the power handling capability of the transport line but a cooled beam screen and good vacuum are required to keep losses at acceptable level for the rest of transport line.

The beam may be directed toward any one of three experimental areas utilizing and RF splitter giving bunches (separated in time, see Figure II-3) either a positive, negative, zero vertical kick (depending on their arrival phase). A downstream three way Lambertson magnet finally separates the beams horizontally.

### III.3.6 Splitter/Lambertson Optics

Figure III-35 shows the beam positions at the face of the Lambertson magnet. The RF splitter will impart a small vertical angle based upon the phase of the beam with respect to the RF splitter phase. As this Lambertson is expected to distribute up to 3 MW beam power to the experimental facility, its apertures should not be limiting. A conservative estimate of the required aperture dictates an aperture that is at least a factor of two to three greater than the 99% beam envelope. This gives rise to a separation of $12\sigma$ to $18\sigma + \Delta s$, where $\Delta s$ is the septum thickness of the Lambertson. To avoid septum saturation at the expected field of 0.64T the septum thickness of 5 mm is required.

For a normalized rms transverse emittance of 0.37 mm-mrad at the entrance to the Lambertson, the rms vertical size varies from 1.25 mm, at the entrance, to 1.02 mm at the exit of the 2 m Lambertson. Using the $18\sigma$ figure the separation at the entrance to the Lambertson should be ±27.5 mm. The required deflection angle, $\Delta\theta$, is given by $\Delta y/(\beta_c\beta_L)\sin\varphi$, where $\beta_c$ and $\beta_L$ are the beta functions at the cavity and Lambertson, respectively, and $\varphi$ is the vertical phase advance between the two locations, is ±1.3 mrad. It requires the RF splitter voltage of 5 MV which value is well within specified 7 MV voltage for the RF splitter cavities. The kick variation during bunch passage through the splitting cavity results in an emittance growth. This effect is much stronger for the central (not deflected) bunch which sees the largest derivative of voltage change, $dV/dt$. An estimate of transverse emittance growth for this bunches shows that this is not an issue if distance between the linac and RF splitting cavity does not exceed ~100 m.

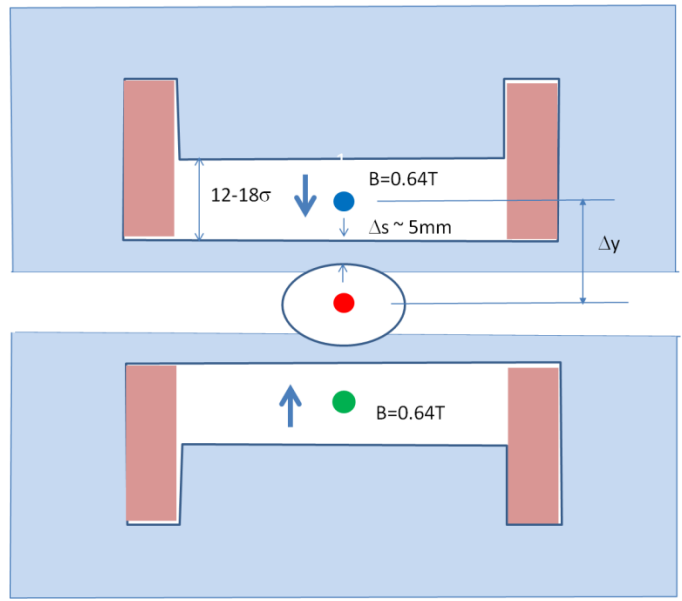

**Figure III-35**: Beam positions at the entrance of the 3 way horizontal Lambertson for μ-to-e (blue), Kaon (red), and the other experiment (green). Arrows indicate field direction.

III.4 **3-8 GeV Pulsed Linac**

III.4.1 **Linac Parameters**

The 3-8 GeV linac is a pulsed superconducting linac accelerating beam from 3 to 8 GeV. It can support a beam current of 1 mA with a pulse length of 4.3 ms and a repetition rate of 10 Hz. Figure II-1 shows the linac timeline. The bunch micro-structure (at the microsecond level) in the pulsed linac is determined by the Recycler RF bucket frequency (52.8 MHz) and the Recycler revolution frequency (90.3 kHz). The former requires bunches at the RF buckets boundaries to be removed. The latter requires removal of bunches to form a 200 ns extraction gap. The beam from the 3 GeV CW linac is directed to the pulsed linac by a pulsed dipole with a rise time of 0.5 ms. After acceleration in the pulsed linac the beam is transported to the Recycler and injected with a multi-turn strip-injection utilizing stripping foils or laser assisted stripping. Six 4.3 ms pulses are required to fill the Recycler to its maximum intensity (1.6E14) required for the neutrino program. Then, the beam is transferred to Main Injector (MI). The complete injection cycle requires 0.5 s. This time fits well within MI cycle which is of duration 1.2 s for 120 GeV operation and 0.7 s for 60 GeV operation. Additional linac pulses are available for an 8 GeV research program. The pulsed dipole and the beam line transporting the beam from the 3 GeV CW linac to the 3-8 GeV pulsed linac have been described in Section III.3.4. Table III-7 presents the basic parameters of the pulsed linac.

| Parameter | Quantity | Unit |
|---|---|---|
| Particle Species | H- | |
| Input Beam Energy | 3.0 | GeV (kinetic) |
| Output Beam Energy | 8.0 | GeV (kinetic) |
| Pulse Repetition Rate | 10 | Hz |
| RF pulse length | 8.4 | ms |
| Beam Pulse Length | 4.3 | ms |
| Average Pulse Beam Current | 1 | mA |
| 8-GeV Transverse Emittance | 0.4 | mm-mrad (rms norm) |
| 8-GeV Longitudinal Emittance | 1.6 | keV-nsec |
| 8-GeV Bunch Length | 1.0 | ps (rms) |
| Quad aperture (diameter) | 75 | mm |
| Quad strength | 1.7- 2.4 | T/m |

**Table III-7**: Pulsed Linac Parameters

### III.4.2 Accelerating Cavities and Cryomodules

The pulsed linac is based on XFEL/ILC technology: 1.3-GHz 9-cell cavities (Figure III-36) optimized for $\beta = 1$ and ILC-type cryomodules, providing a FODO focusing structure (Figure III-37). All quadrupoles are superconducting and located at the center of each cryomodule. Each quadrupole has built-in vertical and horizontal correctors and an attached Beam Position Monitor (BPM) for beam based alignment and orbit control. Basic parameters of the cavity and cryomodule are presented in Section IV.

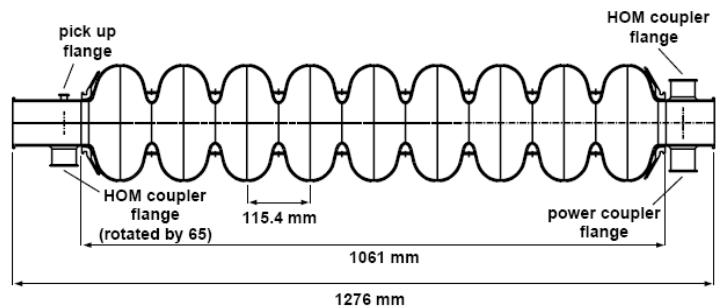

**Figure III-36**: Schematic of a 9-cell 1.3-GHz cavity

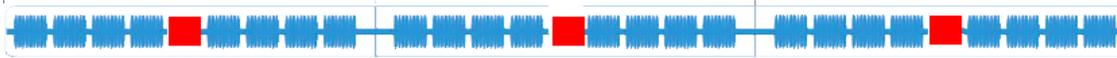

**Figure III-37**: Schematic of the linac focusing structure (red – quadrupoles, blue – SCRF cavities)

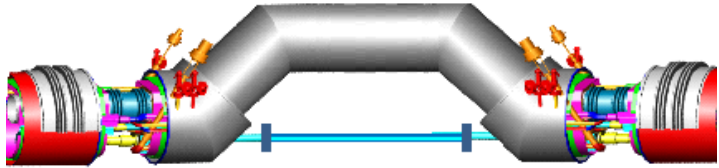

**Figure III-38**: Schematic of the warm section in between two cryogenic strings.

Each cryomodule contains 8 cavities and one quadrupole in the middle. For the average cavity gradient of 25 MV/m with acceptable gradient spread +/- 10% the linac require 28 cryomodules with 224 cavities. The total length of the linac is ~390 m. The pulsed linac has a cryogenic segmentation consisting of three cryo-strings with 10, 9 and 9 cryomodules in each string. Cryogenic strings are separated by warm sections required for beam instrumentation and collimation. In the current configuration the distance between cryo-strings is equal to the length of a cryomodule, providing ~5m of empty space for collimation and beam diagnostics. The possible solution with bypass transfer lines, designed for XFEL linac is shown in Figure III-38.

### III.4.3 Pulsed Linac – Beam dynamics

The lattice design and the beam dynamics optimization are made utilizing the TRACK and TraceWin codes. In the current lattice the gradient is set to 23.5 MV/m and synchronous phase to 8 degrees for all cavities in the linac. Over the linac energy range space-charge effects are not severe. The total number of oscillations in longitudinal phase-space is less than one over the whole linac and the longitudinal envelope is relatively smooth even without proper matching. Figure III-39 shows the simulated beam envelopes along the linac and Figure III-40 displays the actual particle density distribution through the linac. Emittance growth and structural phase advance along the Linac are shown in Figure III-41. Figure III-42 contains several plots characterizing the linac: energy gain per cavity; average cryogenic losses per cavity with $Q_0=1.10^{10}$; focusing quadrupole strengths; and beam energy.

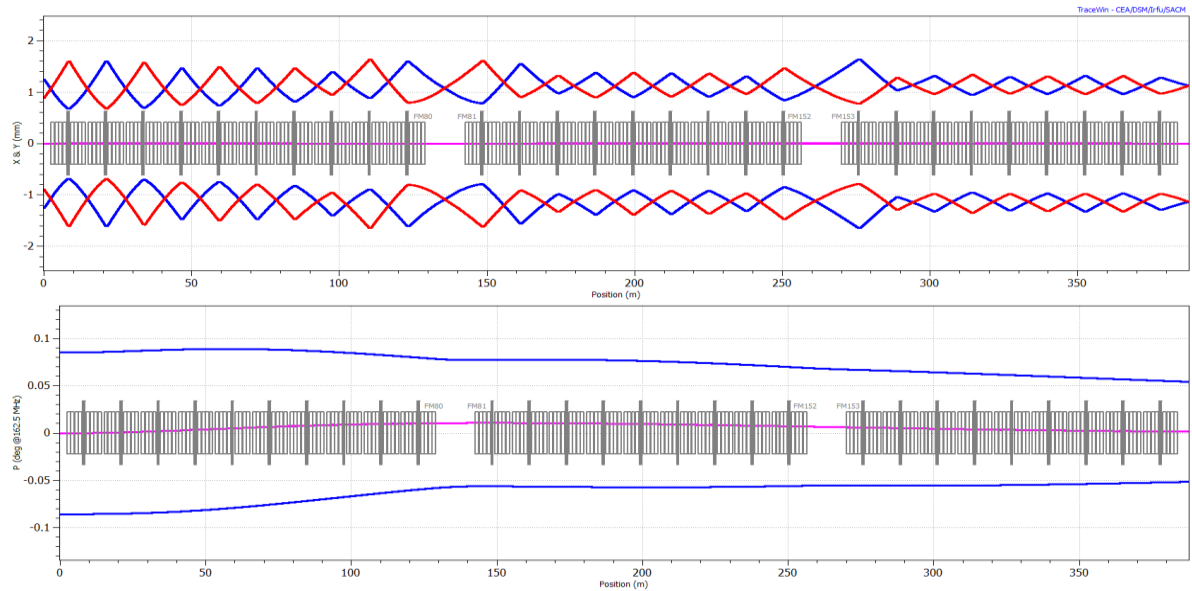

**Figure III-39**: Simulated rms beam envelopes in the pulsed linac: transverse (top) and longitudinal (bottom).

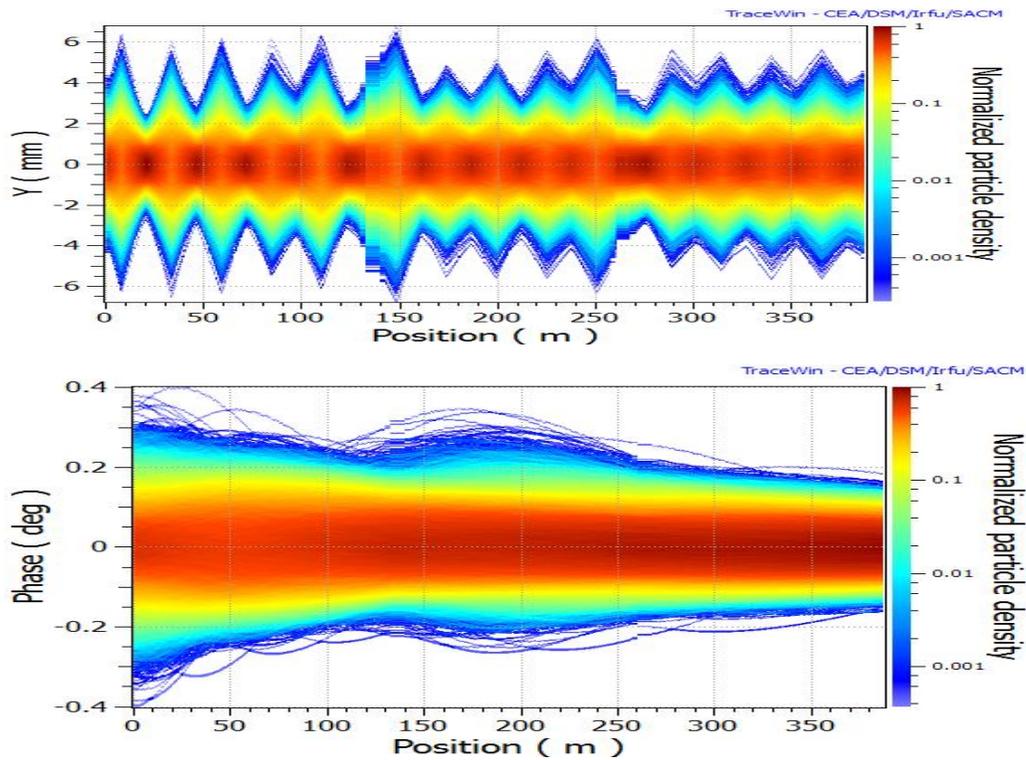

**Figure III-40**: Particle density distribution along the pulsed linac. 100k tracked by PARTRAN.

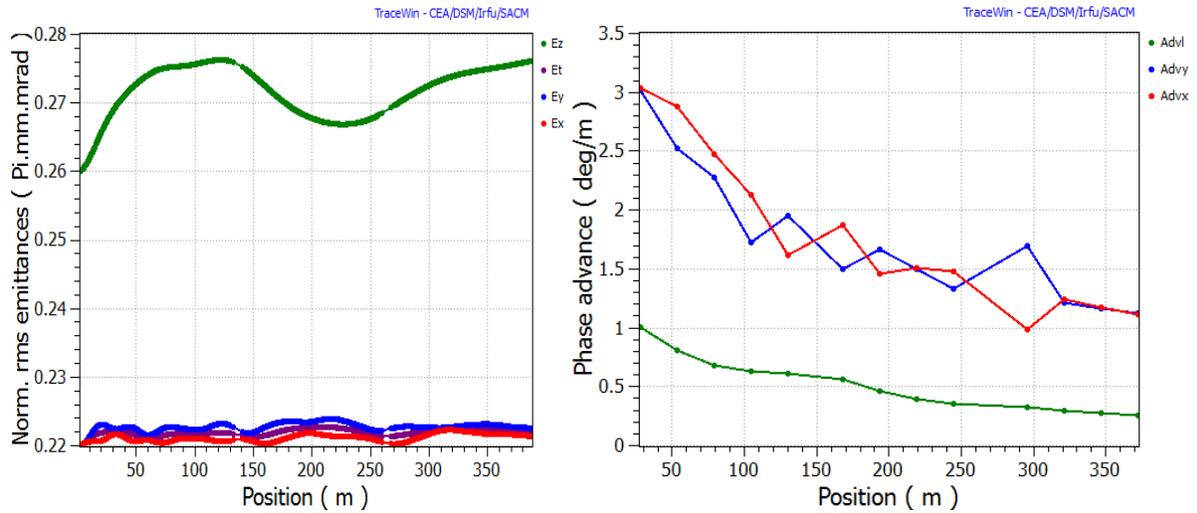

**Figure III-41:** (Left): Longitudinal (green) and transverse emittance growth along the linac. (Right): Structural phase advance along the linac: longitudinal (green), X&Y – (red/blue).

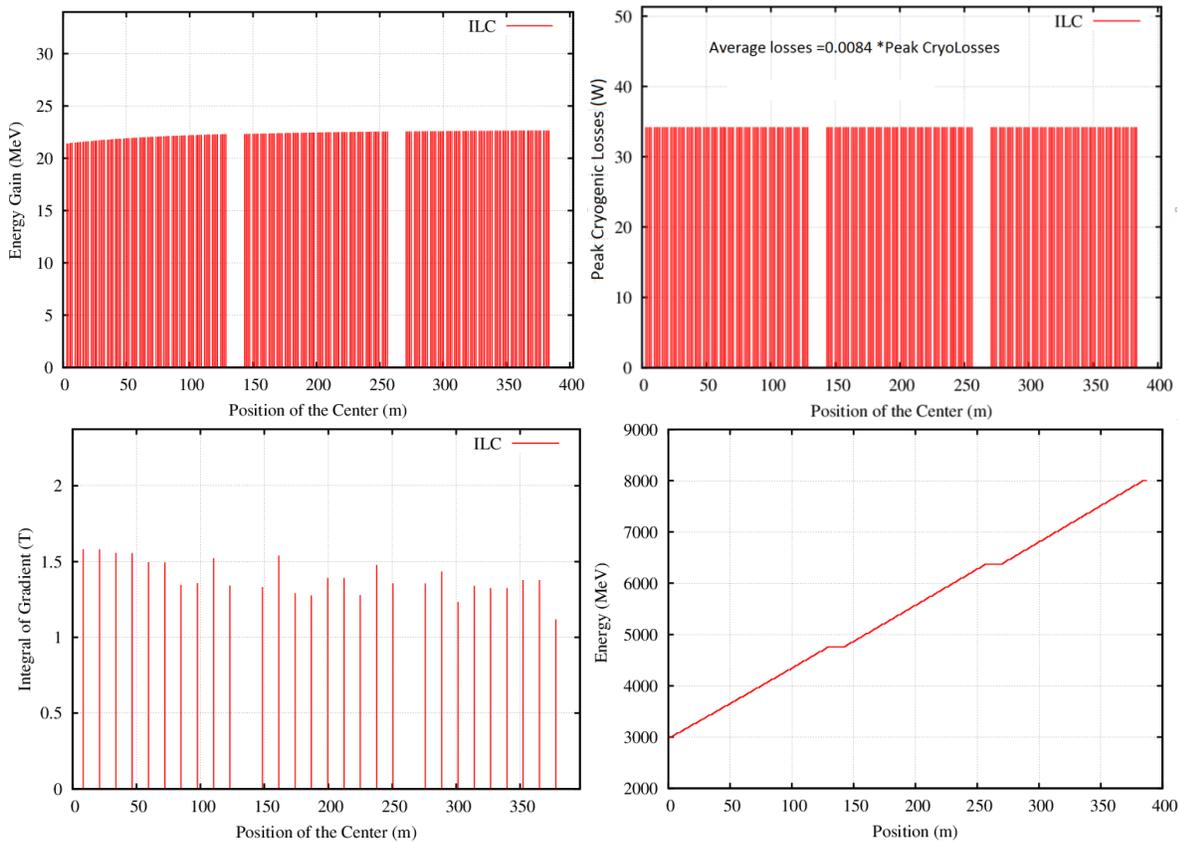

**Figure III-42**: Energy gain per cavity (top-left); Peak cryogenic losses (top-right); integrated strength of the quadrupoles (bottom-left); and Energy along the linac.

III.4.4 **RF Requirements and LLRF**

Each cavity requires 50 kW RF power taking into account the overhead needed for LLRF control and compensation of losses in the distribution system. One RF station (klystron with modulator) will feed one (or two) cryomodules. Thus, klystron power is 400 (or 800) kW and total number of klystrons in the linac is 28 (14). The vector sum (VS) signal from cavities fed by one klystron is controlled by LLRF system with required stability of ~0.5% and ~0.5 degrees in amplitude and phase. The final choice between 1 or 2 cryomodules per klystron system will be based on R&D on Lorentz force detuning compensation and errors studies.

The half-bandwidth of the cavity is 65 Hz for the designed value of $Q_L=1\times10^7$, while Lorentz forces from electromagnetic fields will detune cavity up to ~ 1 kHz at the nominal gradient of 25 MV/m. A fast piezo-tuner and dedicated Lorentz Force Detuning Compensation (LFDC) algorithm are thus necessary to achieve the required < 30 Hz frequency stability of the cavity during the RF pulse. Preliminary results from LFDC studies in long pulse operation for 2 cavities in the cryomodule tested in NML are presented in Figure III-43. These studies clearly demonstrate that the adaptive compensation algorithm is able to achieve cavity frequency detuning of ~3 Hz rms (20 Hz peak-to-peak). Note that the measured level of detuning from microphonics, 2-3 Hz rms, is comparable with total frequency detuning.

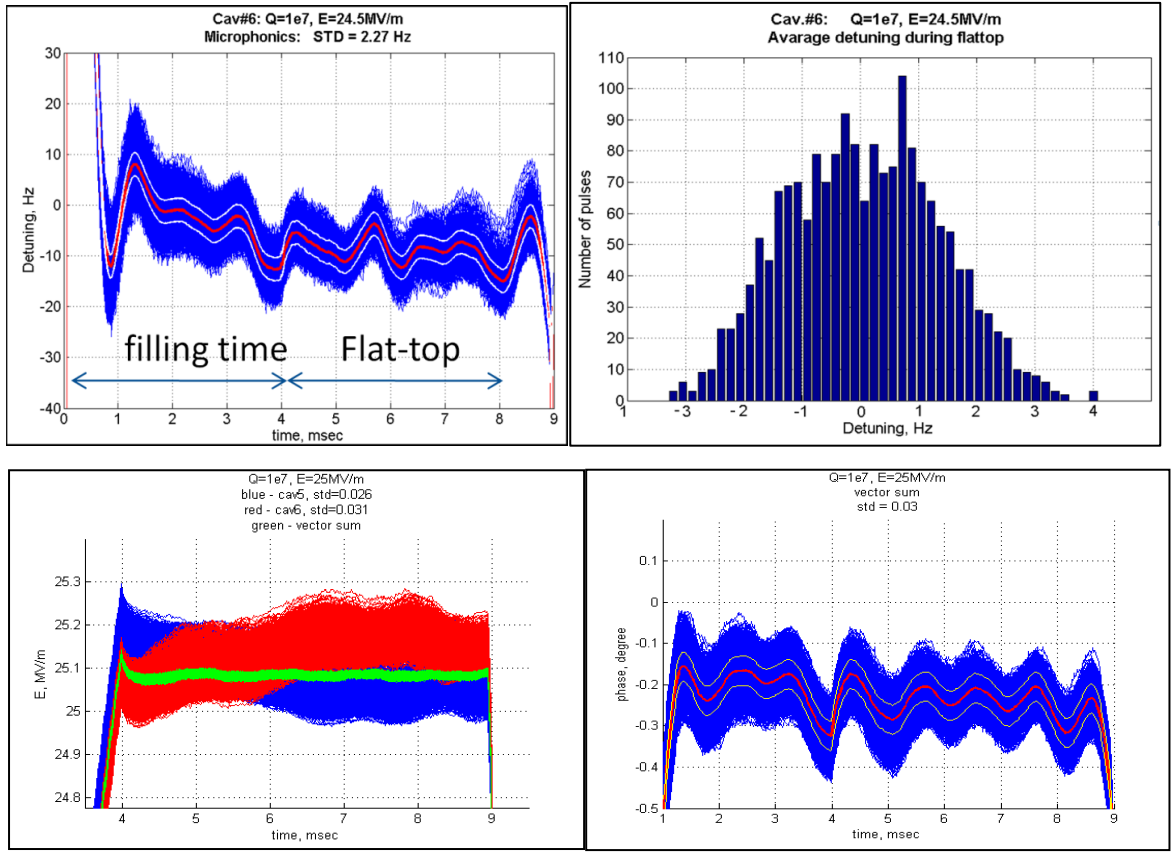

**Figure III-43**: (top-left): Residual cavity detuning after LFDC for 1800 pulses (microphonics included). (top-right): Histogram of average cavity detuning during the flat-top of the pulse. (bottom-left): Amplitude jitter in cavities (red and blue) and Vector sum signals. (bottom-right): Rms phase jitter in VS signal is ≈ 0.03 deg; rms phase jitter of individual cavities is ≈ 1 deg.

## III.5  8 GeV Transport Line

### III.5.1  Transport Line Requirements and Physics Design

Taking into account the large radius required to bend the 8 GeV H$^-$ beam (≥ 1 km), the MI-10 straight section in the Main Injector/Recycler complex is identified as the preferred injection point (see Figure I-2). This straight section will be also used for extraction from the MI for LBNE; however because injection and extraction are bound to different machines which are vertically separated this is not expected to be an obstacle.

The new Recycler injection will be via a symmetric straight section with at least 20 meters between the doublets defining the straight. The straight section has zero dispersion and will feature variable beta functions at the foil location. The condition $\alpha_x=\alpha_y\approx 0$ at the foil location will be applied to minimize the number of foil hits by the circulating beam.

The transport line will be designed to minimize losses due to all sources (discussed in the following) and maximize transmission. The transport line should match into the foil location with zero dispersion (in both planes) and should be able to adjust the spot size on the foil.

The elevation of transport line enclosure floor is expected to match the floor of the MI. The transport line must be optically matched to the new Recycler injection straight section with the additional civil constraint of minimizing the impact of new transport line enclosure on the MI and its tunnel. This is accomplished by adjusting the trajectory of the transport line to miss the "concrete nose" of the wide MI-8 tunnel intersection. The trajectory should be on the MI8 side of the "nose" which will help to minimize the impact to the MI enclosure during the tie-in of the new transport line. Figure III-44 shows the geometry.

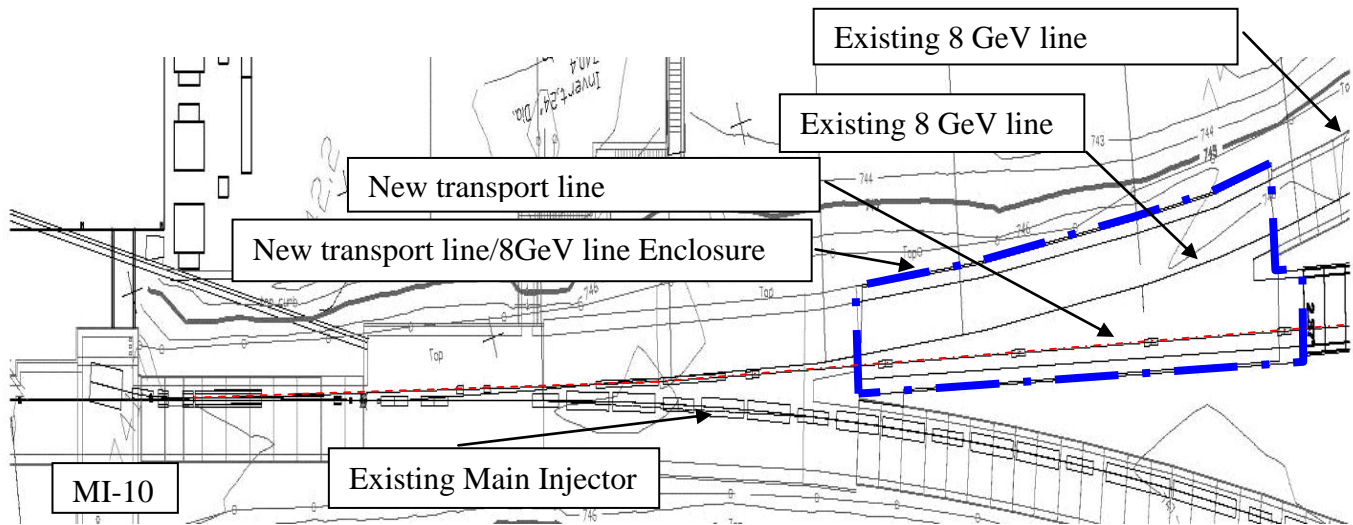

**Figure III-44**: Location of new 8 GeV H- transfer line approaching the MI-10 straight section.

### III.5.2 Beam Loss

There are two major categories of beam loss. The first is directly related to the beam optics design and aperture limitations. It can be mitigated by correct choice of phase advances per period, avoiding non-adiabatic changes of Twiss functions, maximizing apertures (as practical) and providing good orbit control. The latter requires a sufficient quantity of dipole correctors, beam position monitors and beam loss monitors. The second major category of beam loss is related to kinetic effects such as the beam-gas interactions and Lorentz stripping, stripping due to black-body radiation, and H⁻ intra-beam stripping. Each of these mechanisms has been described in detail in the previous sections (in particular see Section III.3.2.). All of these loss mechanisms may also be mitigated by design, *i.e.* by the vacuum level, limitations on magnetic fields, the interior temperature of the beam tube, and the strength of beam focusing. When operating with beam at the full 10 Hz pulsed linac capability, the beam power transported through the 8 GeV transport line will be 342 kW. Table III-8 summarizes the expected loss for the design choices of vacuum level, dipole field, beam tube temperature, and intra-beam stripping.

| Loss Mechanism | 8 GeV: 345 kW | | |
|---|---|---|---|
| | Value | Loss/m | W/m |
| Blackbody | 300 K | $8\times10^{-7}$ | 0.3 |
| Lorentz | 500 G | $5\times10^{-10}$ | 0.0002 |
| Vacuum | $1\times10^{-8}$ Torr | $1\times10^{-8}$ | 0.004 |
| Total | | $8\times10^{-7}$ | 0.3 |
| Residual activation bare beam pipe (mrem/hr) | | | 15 |

**Table III-8**: Expected Beam loss contributions at 8 GeV

### III.5.3 H⁻ Linac Dump

The design power for the H⁻ Linac dump is determined by assumptions on its utilization during the commissioning and operational phase as well as accident conditions. The initial design should be consistent with the maximum beam power expected during routine operation not only of Stage 3, but the ultimate utilization under Stage 4 as well, because once the enclosure and infrastructure are built and the dump absorber is installed and used, it becomes very difficult to make modifications to a highly radioactive component. The anticipated Stage 4 beam power is 4 MW at 8 GeV; however the design criteria for the dump is that it can absorb 10% of the total beam power implying a 400 kW capability. This is substantially less than the capability of the existing Main Injector dump. With an operational power during Stage 3 of 341 kW, this would imply that the dump absorber is capable of accommodating the entire linac beam power. However, we may wish to limit the beam directed to the dump to 34 kW based on requirements of components upstream of the absorber, for example the vacuum window at the absorber window. These evaluations will be completed as part of the more detailed dump design at the appropriate time.

### III.5.4 H⁻ Transport to Recycler

The current version of the transport line between the 8 GeV pulsed Linac and the Recycler contains a single long achromatic arc to avoid the MI65 service building. With the location of the pulsed Linac on the Tevatron infield, the transport line must cross under the Tevatron and AP1 transport line tunnels, avoid the MI65 Service building, and merge into the MI8 transport line enclosure just upstream of the MI10 straight section. The current length of this transport line is on the order of 800 meters. This length could shrink or grow as the siting of the pulsed linac becomes firmer, but this will not significantly impact the basic design of the transport line. The basic structure of the transport line is a 90° FODO cell. It contains a matching section to both the Linac and Recycler, a transverse collimation section, a momentum collimation section, and a vertical bend section to match into the Recycler elevation. All bending sections shall be achromatic. Figure III-45 and Figure III-46 show respectively the lattice functions and dispersion of the present design. Note the collimation section (made up of 4 movable stripping foils and 4 movable jaw absorbers) is at the beginning of the line right after matching. The location of the collimators will be further optimized. The section with the vertical dipoles is not shown as the elevation of the pulsed Linac has not been firmed up. We do know that this vertical offset will comprise of 2 vertical dipoles separated by 360° such that the bend is achromatic. If the elevation of the Linac and transport line are the same as the Main Injector, the required elevation change is roughly 1.5m. This requires a bending angle of approximately 12 mr which can be created with a 6 meter dipole at ~560 G. This will probably be located just before the horizontal arc.

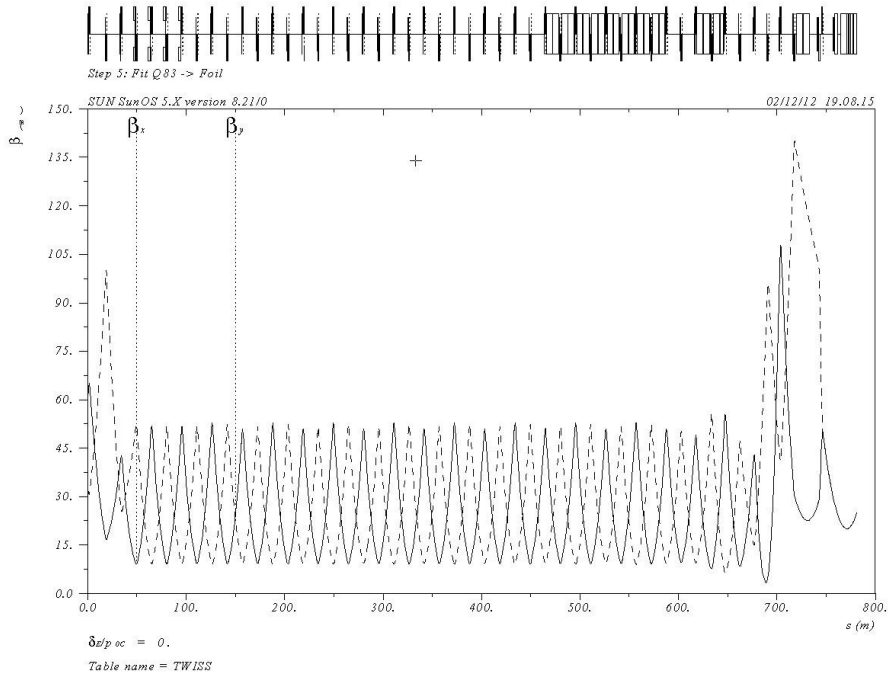

**Figure III-45**: Lattice functions of the 8 GeV H⁻ transport line directed to the Recycler.

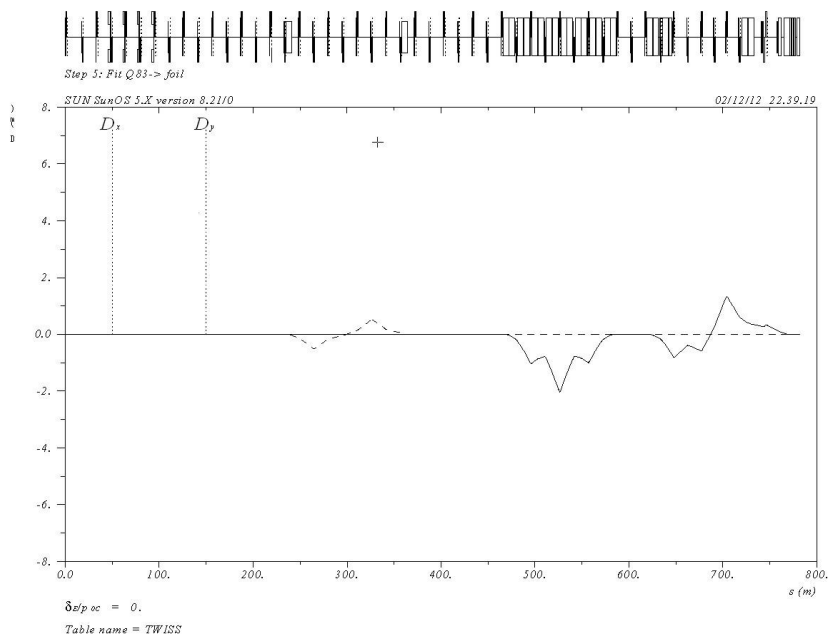

**Figure III-46**: Dispersion in the 8 GeV H⁻ transport line directed to the Recycler.

Since the transport line is matched to the permanent-magnet based Recycler it is likely that the line will be based on permanent magnets also. This is similar the existing Booster-to-MI 8 GeV transport line arrangement. The usual instrumentation and electromagnet correctors will be used.

### III.6  Recycler/Main Injector

The Fermilab Recycler Ring is a fixed energy 8 GeV storage ring constructed with strontium ferrite permanent magnets and residing in the Main Injector tunnel. For the NOvA program, the Recycler is being converted from an antiproton storage ring to a proton accumulator, providing single turn injection into the Main Injector. In Project X the Recycler serves the same purpose, although at significantly higher intensities: boxcar stacking six 4.3 ms beam pulses from the linac via a multi-turn H- stripping injection, capturing the beam in 53 MHz RF buckets, and performing a single turn extraction into the MI. The MI will receive $1.6 \times 10^{14}$ protons from the Recycler in a single turn and will accelerate them to 120 GeV in 1.2 seconds. In general, both the Recycler and Main Injector operate in a similar mode to the NOvA operation [23].

III.6.1 **Recycler Modifications**

With a new injection insert in the Recycler, we anticipate the need for more flexibility in the lattice design. The Recycler is built with permanent magnet quadrupoles, permanent magnet combined function devices, powered dipole correctors, and a tune trombone of powered quadrupoles. The installation of new powered quad elements in the injection region allows for lattice flexibility.

III.6.2 **Main Injector Modifications**

In the present Main Injector, the injection energy is 8 GeV and maximum energy 120 GeV. Transition crossing is at $\gamma_t = 21.6$ GeV. Simulation shows that emittance dilution and beam loss will occur at high intensities. Furthermore, from experience at other machines (e.g., the AGS at BNL, the CERN PS, and the KEK PS), transition crossing could become a severe bottleneck in high intensity operation.

The design being considered is a first-order system employing local dispersion inserts at dispersion-free straight sections. The normal ramp rate of the MI is 240 GeV/s. In order to have an effective $\gamma_t$-jump, the jump rate should be at least an order of magnitude higher. The system was chosen to provide a $\Delta\gamma_t$ from +1 to -1 within 0.5 ms, a jump rate of 4000 GeV/s, about 17 times faster than the normal ramp rate. Details can be found in the Proton Driver Design study document [24].

III.6.3 **Electron Cloud Mitigation**

Electron cloud induced instabilities in either the Recycler or the MI could be an important limitation to the maximum proton flux. Ongoing studies in the Main Injector are investigating the generation of electron clouds and the performance of beam pipe coatings (e.g., TiN) in reducing the secondary electron yield. Simulations are being benchmarked to measurements [25]. Based on the measurements and simulations, it appears that a combination of beam pipe coating and damper upgrades should be sufficient to mitigate electron cloud instabilities.

III.6.4 **RF System Modifications**

Upgraded RF systems are required in both the Recycler and the Main Injector. The Recycler RF system is used solely for capture of the injected beam. With the higher intensity, the Main Injector needs more power than is currently available to accelerate the beam at 240 GeV/s. To minimize changes to the existing instrumentation (specifically BPM systems), the RF will operate at the same harmonic number of 588.

The peak beam current in the Main Injector is 2.25 A and 2.7 MV/turn are necessary to reach the desired acceleration rate. With a synchronous phase angle $\phi_s = 36°$, 240 kV per cavity are required. From injection energy (8 GeV) to flattop (120 GeV), the frequency sweep is from 52.811 MHz to 53.104 MHz. A cavity design is under development, to be used in both the Recycler and the Main Injector.

To mitigate possible space charge effects, an increased bunching factor is desired. A frequency mismatch between the capture RF and the linac RF induces a parasitic longitudinal painting [26] but that is not enough. Studies have shown that a $2^{nd}$ RF system operating at the $2^{nd}$ harmonic and at half the voltage increase the bunching factor to ~0.35, within the desired range to mitigate space charge effects.

# IV  Design Concepts of Major Subsystems

## IV.1  CW Linac

### IV.1.1  Ion source, LEBT, RFQ and MEBT

The Project X front end consists of an ion source, Low Energy Beam Transport (LEBT), Radio Frequency Quadrupole (RFQ), and Medium Energy Beam Transport (MEBT). The H$^-$ beam originates from a nominally 5 mA (nominal, 10 mA peak) DC ion source and is transported through the LEBT to a CW normal-conducting RFQ, where it is bunched and accelerated to 2.1 MeV. In the MEBT a bunch-by-bunch chopper provides the required bunch patterns, removing 60-80% of bunches according to a pre-programmed timeline. To foresee possible upgrades, all elements of the front end are designed for beam currents of up to 10 mA. The beam energy of 2.1 MeV is chosen because it is below the neutron production threshold for most materials.

*Ion Source*

The ion source assembly is a DC, H$^-$ source capable of delivering up to 10 mA of beam current at 30keV to the LEBT. The ion source specifications are listed in Ref. [27]. The present scenario assumes the volume cusp ion source presently commercially available from D-Pace Inc. ([28], Figure IV-1). This source is capable of delivering up to 15 mA with a satisfactory transverse emittance of ＜0.2 μm (rms, normalized), but its life time is relatively short. The main reason is the source's filament, which needs to be replaced every ~300 hours. To improve the beam uptime, two ion sources are planned to be installed (Figure III-2). Each source can be removed for repairs, installed back, and conditioned without interrupting the operation of another source.

*Low Energy Beam Transport*

The LEBT transports the beam from the exit of the ion source to the RFQ entrance and matches the optical functions to that of the RFQ. In addition, the LEBT forms a low-duty factor beam during commissioning and tuning of the downstream beam line and interrupts the beam as part of the machine protection system.

The functional requirement specifications are listed in [8], and its schematic is shown in Figure III-2. The LEBT includes 3 solenoids (for each leg), a slow switching dipole magnet, a chopper assembly (a kicker followed by a beam absorber), and diagnostics to characterize and to tune the beam. The length of the beam line, ~3 m, insures that the gas migration from the ion source to the RFQ is tolerable.

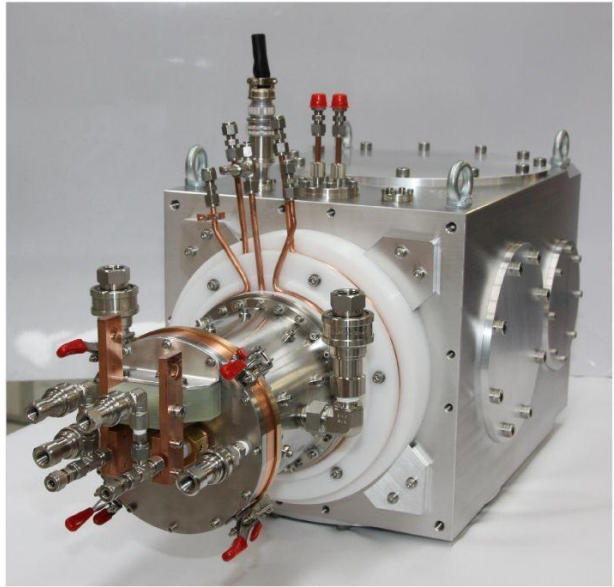

**Figure IV-1**: Photograph of the D-Pace ion source (foreground) with the vacuum chamber.

The beam diagnostics include an emittance scanner (see specifications in [29]) at the exit of each ion source, a DCCT, and a toroid. In addition, fixed electrically-insulated diaphragms are installed inside the solenoids. Moving the beam by dipole correctors, built into each solenoid and measuring the current to the downstream diaphragm allows estimating the beam size and its position. A scraper in front of the RFQ is a movable, electrically insulated, water-cooled plate that can be moved into three positions: completely removed; partially inserted so that a round opening in the plate is concentric with the RFQ entrance diaphragm; and fully inserted to completely intercept the beam. The scraper serves several purposes: the size of the opening is chosen to scrape the halo particles that otherwise would be lost in the RFQ or MEBT; the variation of beam current intercepted by the scraper while moving the beam across the opening with upstream dipole correctors gives information about the beam position and core size; and the fully inserted scraper works as a beam stop and an auxiliary beam current monitor.

Fast machine protection and pulsed beam operation are achieved via the chopper assembly, which comprises an electrostatic kicker followed by an absorber. In some scenarios, it can be used also as a pre-chopper to assist the MEBT chopping system. The primary machine protection mechanism is to disable the beam from the ion source by turning off its extraction and bias voltages, with the LEBT chopper serving as a fast beam switch during ion source turn-off.

For the Project X LEBT the optics design incorporates two regions (Figure III-3). First, the beam is nearly fully neutralized from the exit of the ion source to immediately upstream of the kicker. Further downstream, the beam can be either neutralized or un-neutralized. In the un-neutralized mode, the secondary ions created in that area are removed by a constant electric field on the kicker plates, and the ions from upstream are stopped by a positive voltage on the insulated diaphragm #2. In the neutralized mode, the kicker plates as well as the insulated diaphragm #2 have normally ground potential, while the insulated diaphragm #3 or the scraper is biased positively to prevent the ion escape longitudinally. The LEBT scheme is flexible enough to accommodate both versions by adjusting potentials and solenoid currents (see Figure III-4 for simulations of a fully neutralized transport case). The transport with an un-neutralized section is beneficial for decreasing the transition effects of the kicker pulse but results in an emittance increase. Relative benefits of each scenario can be clarified during PXIE experiments.

### *Radio-Frequency Quadrupole*

The 162.5 MHz CW RFQ will accelerate an $H^-$ ion beam with currents of up to 10 mA from 30 keV to 2.1 MeV (see Ref. 30 for specifications). Presently the Project X RFQ is assumed to be identical to that being developed for PXIE by LBNL [31]. This 4.45-m long, four-vane structure consists of four longitudinal modules and uses a 60 kV vane-to-vane voltage. Most of the RF input power is dissipated in the cavity walls to establish the needed RF field with ~12% beam loading. A series of 32 water-cooled pi-mode rods provides quadrupole mode stabilization, and a set of 80 evenly spaced fixed slug tuners is used for the final frequency adjustment and local field perturbation corrections. The design incorporates selected portions of the technology validated by the Spallation Neutron Source (SNS) Front End RFQ [32] designed and constructed at LBNL. An overall view of the full four-module RFQ is shown in Figure IV-2.

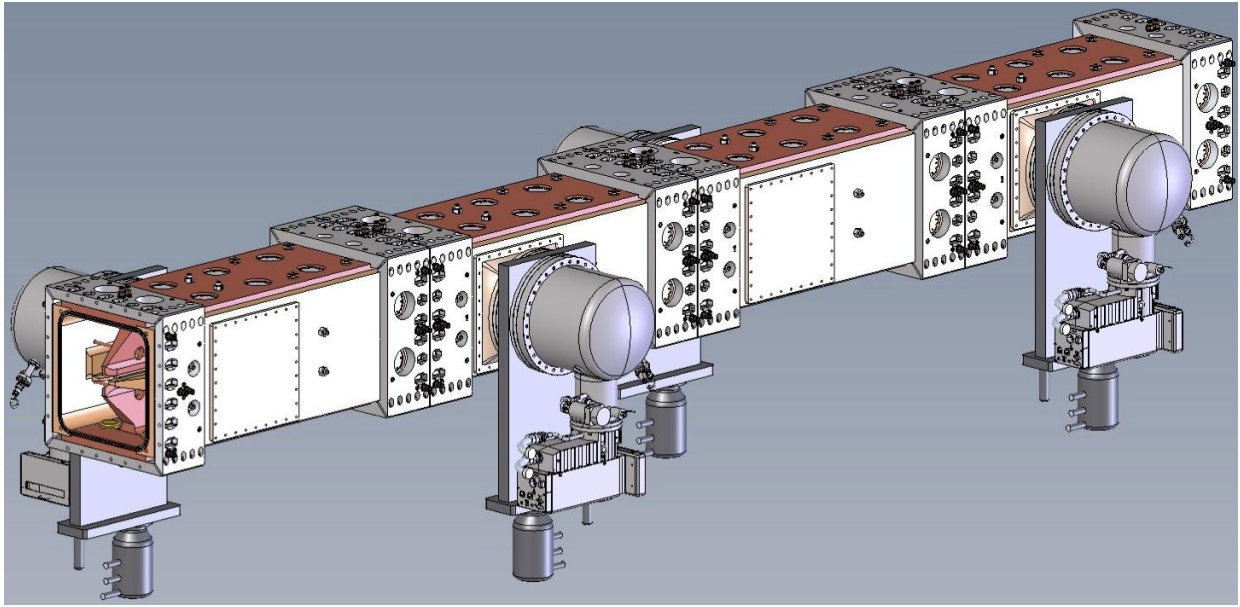

**Figure IV-2**: CAD model of the full four-module RFQ.

The beam dynamics design of the RFQ provides over 96% transmission for beam current from 1 to 15 mA. At 5 mA nominal current, 99.8% beam capture is achieved with transverse and longitudinal emittance (rms, norm) of 0.15π-mm-mrad and 0.64 keV-nsec, respectively. The RF design studies [33] include mode stabilization, field flatness, radial matching, and entrance and exit terminations. Table IV-1summarizes the RF design results.

| Parameter | Value |
|---|---|
| Frequency, MHz | 162.493 |
| Frequency of dipole mode, MHz | 181.99 |
| Q factor | 14660 |
| Max power density, W/cm$^2$ | 7.9 |
| Total power loss, kW | 74.6 |
| Beam power @ 5mA, kW | 10.5 |
| Total RF power from source, kW | 150 |

**Table IV-1**: Main parameters of the Project X RFQ electromagnetic design

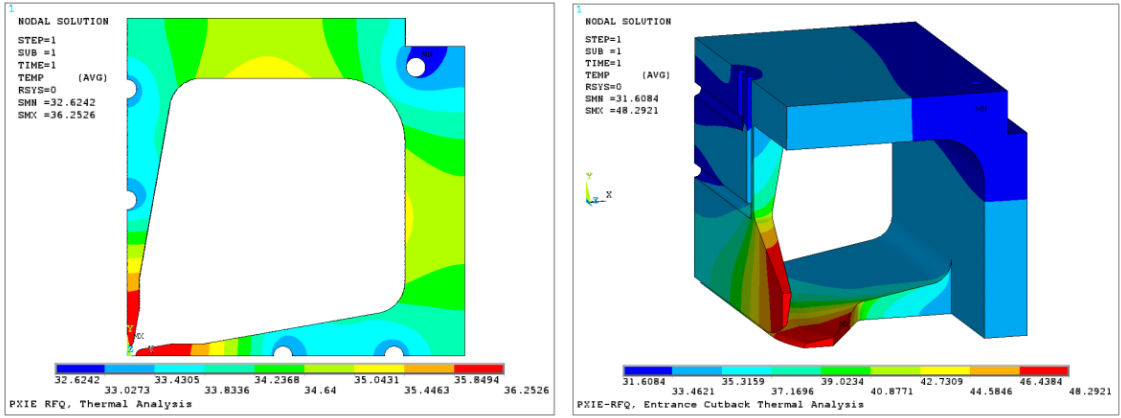

**Figure IV-3**: Temperature distribution in one RFQ quadrant body (left) and cut-back (right). The color scheme (degrees C) is at the bottom of each plot.

A series of RF and thermal finite-element models of the RFQ have been developed using ANSYS®. An example of the temperature contour plots for the cavity body and vane cutback region is shown in Figure IV-3. From the RF analysis, the average linear power density was determined to be 137 W/cm with a peak heat flux on the cavity wall of only 0.7 W/cm². With the 30°C water in the vane and wall cooling passages, the resulting temperature profile in the cavity body ranges between 32 and 37°C at full RF gradient.

Additional modeling that has been carried out includes stress and displacement analyses, thermal analyses of the tuners, pi-mode rods and vane cutbacks, and prediction of the frequency shift of the RFQ cavity due to thermal loading and changes in the cooling water temperature.

The RFQ cooling scheme will use differential water temperature control in the vane and wall passages. This technique provides active tuning of the RFQ by holding the wall water temperature constant and adjusting the vane water temperature up and down. The frequency of the RFQ can be shifted by -16.7 kHz for every 1°C rise in the vane cooling water temperature. For uniform water temperature control, the shift would only be -2.8 kHz/°C.

### *Medium Energy Beam Transport*

The MEBT layout is given in Figure III-8 (reproduced in Figure IV-4). The MEBT generally provides beam transport and focusing (both transverse and longitudinal), beam chopping, chopped beam absorption, vacuum pumping, and diagnostics. Transverse focusing is provided primarily by equidistantly placed quadrupole triplets; the only exception is two doublets at the RFQ exit. Each triplet or doublet is followed by a pair of dipole correctors.

The specifications for the quadrupoles and correctors are listed in [34]. The spaces between neighboring triplets or doublets are referred to as MEBT sections, with each section providing a particular service. The section separation in the regular part of the MEBT is 1140 mm, which leaves a 650-mm long (flange-to-flange) space for various equipment (350 mm in the section between doublets labeled #0).

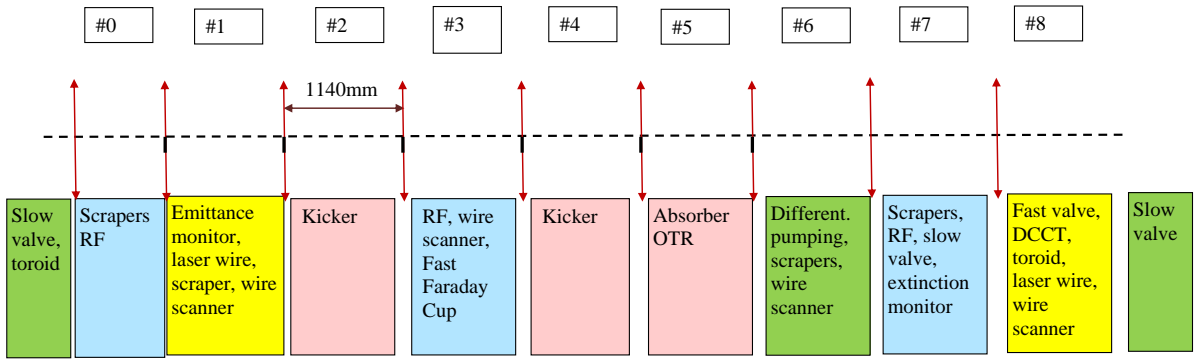

**Figure IV-4**: The MEBT structure. Sections are colored according to their main functions. The red vertical arrows schematically show the transverse focusing (doublets or triplets) elements.

The undesired beam bunches will be removed in the MEBT by a chopping system, represented in the figure by red boxes. The system consists of two identical 50 cm long kickers separated by a 180º transverse phase advance and an absorber (90º from the last kicker). In the broadband, travelling-wave kicker, the transverse electric field propagates with the phase velocity equal to the speed of H⁻ ions (~ 20 mm/ns, $\beta$= 0.0668) so that the ion vertical velocity changes sufficiently to push the ion bunches, designated to be removed, to the absorber. The separation between the kicker plates is 16 mm. The aperture is limited by protection plates on both sides of the kicker to 13 mm so that in the case of a mismatched transport, the intercepted bean current on these protection plates would trigger the beam turn-off. Detail specifications for the kicker can be found in Ref. [35]. The simulated transverse beam envelopes in the MEBT for both passing and chopped bunches are presented in Figure III-9.

Presently two versions of the kicker, which differ by the structure's impedance, are being investigated [16]. In the 50 Ohm version, the kicker plates are connected in vacuum by cable delay lines (Figure IV-5, purple loops). Compared to the helical structure discussed below, this arrangement considerably reduces coupling between neighboring turns and consequently dispersion. Each kicker is driven by two commercially available linear amplifiers. Signal distortion caused by the imperfections of the amplifier characteristics, cabling, and dispersion in the structure are corrected by the corresponding pre-distortion of the amplifier's input signal.

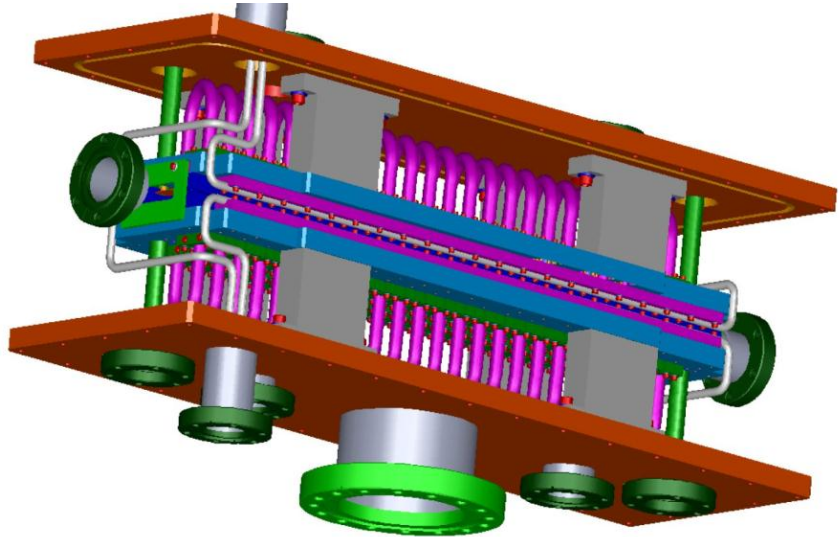

**Figure IV-5**: 3-D model of the 50-Ohm kicker structure (side walls of the vacuum box are removed for presentation purpose).

The 200-Ohm structure is comprised of two helical windings around grounded cylinders with plates attached to windings (Figure IV-6). In this scheme, the kicker driver is a fast switch being developed in-house. In addition, this scheme requires custom made feed-throughs, transmission lines, and current loads. While the 50 Ohm version has many elements commercially available and the design of its structure is more mature, the 200 Ohm is potentially less costly and has a lower power loss in the structure. The selection of the kicker technology for Project X will be based on beam tests at the PXIE.

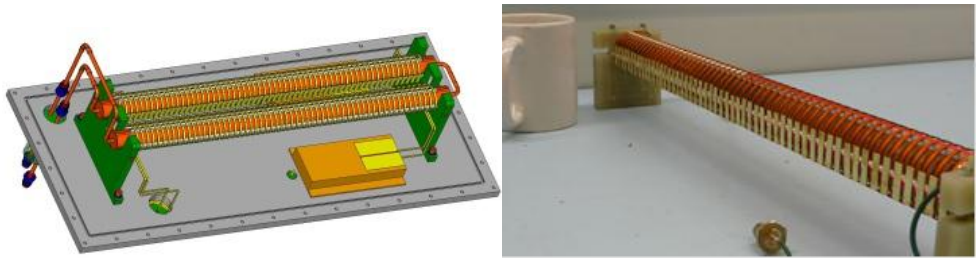

**Figure IV-6:** Conceptual design (left) and photograph (right) of a single-helix model of the 200 Ohm dual-helix kicker.

The undesired bunches are directed to an absorber, displaced vertically from the beam trajectory. To accommodate the entire beam that the upstream system is capable to deliver, the absorber is being designed for the maximum beam power of 21 kW that corresponds to a 10-mA beam completely diverted to the absorber (see specifications [36]). The power density in the beam with a ~2 mm rms radius exceeds by an order of magnitude what is technically possible to absorb without melting the surface. To decrease the surface power density, the absorber is positioned with a small angle, 29 mrad, with respect to the beam (Figure IV-7).

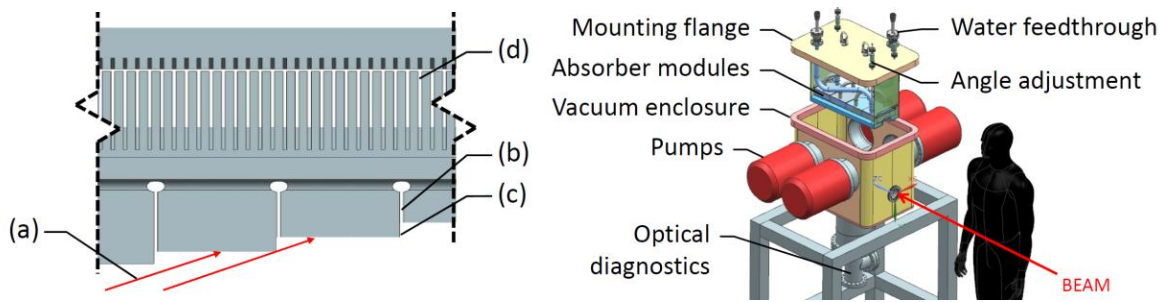

**Figure IV-7**: A conceptual design of the MEBT absorber. Left: a side-view of the absorber showing (a) beam incident on surface, (b) axial stress relief slits, (c) shadowing step increment (magnitude exaggerated), (d) 300μm wide by 1mm pitch water cooling channels. The horizontal scale is exaggerated. Right: an exploded view.

Challenges presented by the absorber design include maintaining vacuum quality, managing surface effects (sputtering and blistering), containing secondary particles, accommodating radiation effects, spreading energy deposition, and the survival at high temperatures and temperature-induced mechanical stresses. Presently the design choice is a monolithic absorber made from the molybdenum alloy TZM.

To keep the beam properly bunched and to match the longitudinal phase space to the first superconducting cryomodule, the MEBT includes 3 identical (room temperature) bunching cavities [37]. Each cavity is a quarter-wave 162.5 MHz resonator with the nominal accelerating voltage of 70 kV (at β=0.0668). A conceptual view of the cavity being designed for PXIE and some results of its simulations are shown in Figure IV-8.

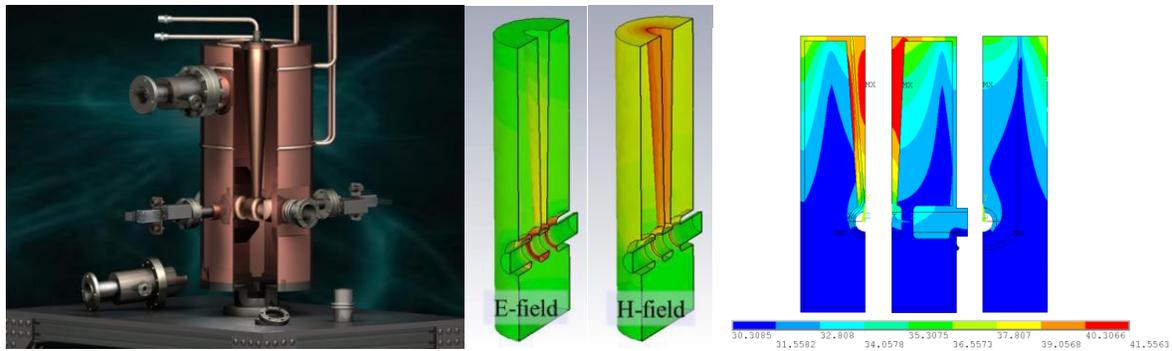

**Figure IV-8**: A conceptual design of the MEBT bunching cavity (left), simulated surface field distributions (center), and the temperature distribution (right) (from Ref. 38).

Each of the sections #0, 1, 6, and 7 in Figure IV-4 contains a set of 4 scrapers. Each of these 16 scrapers is envisioned as an electrically insulated, 100W-rated plate precisely movable across the half-aperture [39]. The scrapers will be used for several purposes: for beam halo measurements and removal; protection of downstream equipment from a beam loss caused by beam envelope and trajectory mismatches; as an auxiliary beam density distribution diagnostics in the pulse mode; and formation of a pencil H- beam for measurements downstream (in the pulse mode). The scraper sets in the upstream and downstream pairs are separated by ~90$^o$ of transverse phase advance to insure an effective removal of particles with large transverse actions.

The vacuum requirements for the MEBT are determined by the electron detachment in H- beam and by the necessity to have a low gas flow into the HWR cryomodule. Obviously, the electron detachment results in a loss of H$^-$ beam intensity - an additional restrictive effect creating a flux of neutral hydrogen atoms that may reach the SRF cavities. For the conceptual design of the vacuum system it is convenient to choose a reasonable limit for the integral of the pressure over the distance along the MEBT axis. The selected value of $1\times10^{-6}$ Torr·m corresponds to a relative loss of ~$10^{-4}$ and to an additional ~0.1 W heat load to the SRF by neutral atoms.

Gas flow from the room-temperature MEBT to the 2K HWR cryomodule can cause a gas deposition on the cryogenic surfaces, which negatively affects the cavity performance. To limit this effect to safe levels, the vacuum pressure upstream of the HWR cryomodule is specified to be below $1\times10^{-9}$ Torr (hydrogen). To ensure this level, with a quite high gas load coming from the absorber, ~1 mTorr·l/s, the absorber section is followed by a differential pumping section and is separated from the SRF by two additional sections.

Most of diagnostics are listed in Figure IV-4 and will allow measuring beam properties at the RFQ exit (before sending it into the HWR cryomodule. In addition, each triplet and doublet has a BPM inserted between quadrupoles. The BPMs will measure the transverse beam position and the bunch phase; functional requirements [40] foresee a capability of gating on a single bunch in the bunch train.

### IV.1.2 CW Accelerating Structures Requirements

The parameters and requirements associated with all accelerating structures and cryomodules within the CW linac have been summarized in Table III-2, Table III-3, and Table III-4. This section describes design concepts for the cavity types required in the CW linac, and the associated cryomodules.

### IV.1.3 Low-beta section (2.1-10 MeV, 162.5 MHz)

The original design of Project X included 325 MHz Single Spoke Cavities of type 0 (SSR0), to accelerate the H$^-$ beam from 2.1 to 10 MeV. To maintain high beam quality, an adiabatic increase of the accelerating gradient in SSR0 cavities was necessary, and satisfying the adiabaticity condition required 3 cryomodules comprising 24 SSR0 cavities. After careful considerations, a design based on 162.5-MHz Half-Wave Resonator (HWR) cavities was selected instead. This design has several substantial advantages as compared to the 325 MHz SSR0 option:

- Only 8 HWRs are required to accelerate the beam to ~10 MeV while maintaining high beam quality.

- Reduced RF defocusing due to both the lower frequency and the lower synchronous phase angle results in a much faster energy gain without emittance growth.

- It opens the possibility to use 162.5 MHz re-bunchers in the MEBT to allow for longer drift spaces for the fast beam choppers.

- Significant cost reduction due to the reduced component count.

The beam dynamics optimization determines that a cavity beta of $\beta_{OPT}$=0.112 is optimal. The cavity design is based on recent advances in SRF technology for TEM-class structures being developed at ANL. Highly optimized EM parameters which maximize the real-estate gradient while maintaining low dynamic cryogenic loads and peak surface fields were achieved using a conical shape for both the inner and outer conductors. A "donut" shaped drift tube in the center conductor has been developed to minimize the undesirable quadrupole component of the electric field as is shown in Figure IV-9. Utilization of the HWR requires two major sub-systems: a 10 kW RF coupler and a slow tuner. A capacitive adjustable 10 kW RF coupler prototype has been designed, constructed, and is being prepared for testing. It will provide RF power through the port which is perpendicular to the beam axis in the center of the cavity (Figure IV-10). A pneumatically actuated mechanical slow tuner which compresses the cavity along the beam axis is located outside of the helium vessel and will be attached to the SS beam port flanges shown in Figure IV-10. A fast tuner is not required for operations with a 4.9 kW RF power source (Stage II). This power supports 2 mA beam loading and has sufficient margin for amplifier control and suppression of microphonics (mainly related to helium pressure fluctuations). The required cavity bandwidth (loaded) is 49 Hz. Stage I has a 1 mA beam current. Consequently, the power can be reduced to about 3 kW. The main parameters of the HWR are shown in Table IV-2.

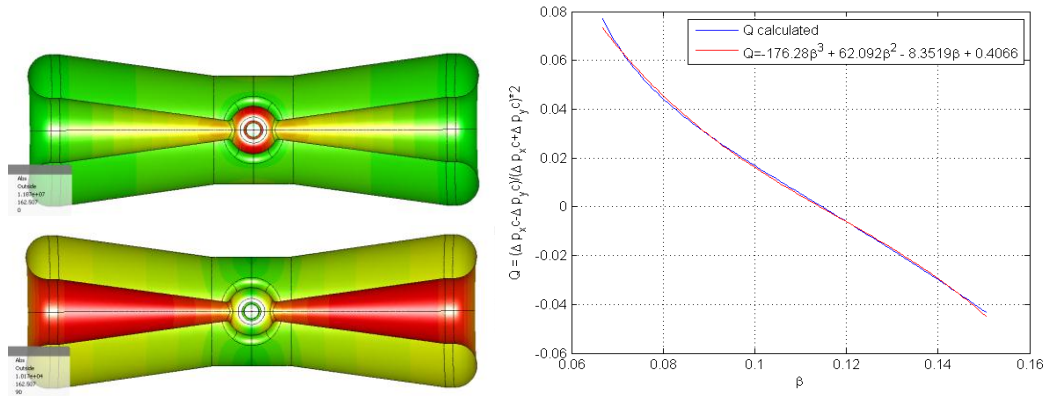

**Figure IV-9**: Left - Half-wave resonator model in Microwave Studio (MWS). The picture shows electric (top) and magnetic field (bottom) distributions on the surface. Red is high intensity and green is zero. Right - Quadrupole effect in the HWR cavity versus the particle velocity $\beta$ in operating domain; red and blue line show simulation and approximation.

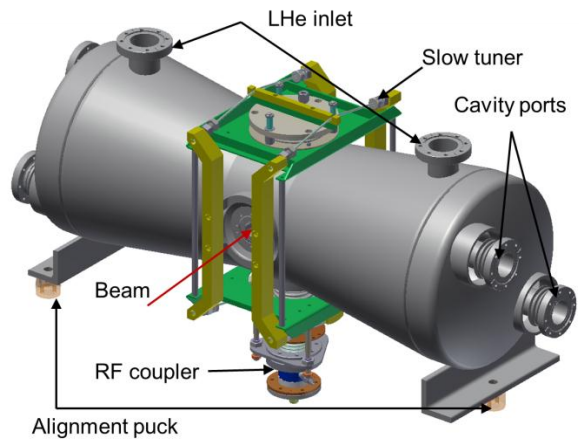

**Figure IV-10**: HWR cavity 3D model in INVENTOR

Extensive finite element analysis of the cavity included simulations to evaluate the integrity of the cavity per the Fermilab ES&H manual. The simulations include protection against plastic collapse, local failure, buckling, ratcheting and fatigue failure to ensure that the operating loads are below the maximum allowable limits. The maximum structural load is determined by the pressure set by the operation of the cryogenic system and the safety pressure relief valve. The evaluation was performed for 30 psig at room temperature and 60 psig at 2 Kelvin in the helium space of the cavity in compliance with the Fermilab requirements. In general, the over pressure condition could occur during the initial cryogenic cooling with the cavity structure at or near room temperature; since the room temperature strength limits (i.e., yield and ultimate) are lower than for cryogenic temperatures and the operating margin is smaller here, the room temperature limits were studied in more detail. The stress analysis was performed in the presence of the slow tuner and other appurtenance loads. The final design exceeds all evaluation criteria for the niobium and the stainless steel (SS) parts respectively.

| HWR Cavity | Design Value | |
|---|---|---|
| Frequency | 162.5 | MHz |
| Geometric Beta ($\beta_G$) | 0.094 | |
| Optimum Beta ($\beta_{OPT}$) | 0.112 | |
| Aperture (diameter) | 33 | mm |
| $L_{EFF} = \beta_{OPT}\lambda$ | 20.7 | cm |
| $R/Q_0$ | 275 | $\Omega$ |
| $G = Q_0 R_S$ | 48 | $\Omega$ |
| $E_{max}/E_{acc}$ | 4.65 | |
| $B_{max}/E_{acc}$ | 5.0 | mT/(MV/m) |
| Gradient | 8.2 | MV/m |
| Peak Surface Electric Field | 38 | MV/m |
| Peak Surface Magnetic Field | 41 | mT |
| $Q_0$ | 0.5 | $10^{10}$ |
| Operating Temperature | 2 | K |

**Table IV-2**: HWR cavity requirements

Two methods have been studied for minimization of the cavity frequency sensitivity to fluctuations of the helium pressure: (1) adding gusseting to reduce the cavity deflections in the high magnetic and electric field regions and (2) varying the depth of the flat dish located opposite to the RF coupler port. The results of these studies showed that no gusseting is required; a minimal value of 1.4 kHz/atm was achieved by optimizing the dimensions of the flat dish penetration. Simulations of the slow tuner were performed by applying a force to the SS flanges of the helium jacket. For example, a 10 kN force results in a frequency shift of -120 kHz.

The primary operational parameters for the HWR are contained within Table IV-2, and are based on our recent experience with the ATLAS energy upgrade cryomodule and its long term operation. As is shown in Figure IV-11, recent tests of new 72 MHz Quarter Wave resonators (QWR) show just 2 n$\Omega$ residual resistance at 41 mT which readily supports design parameters of the HWRs. The peak fields shown in the table were taken directly from the CST MWS screen in simulations with ~200K tetrahedral mesh cells.

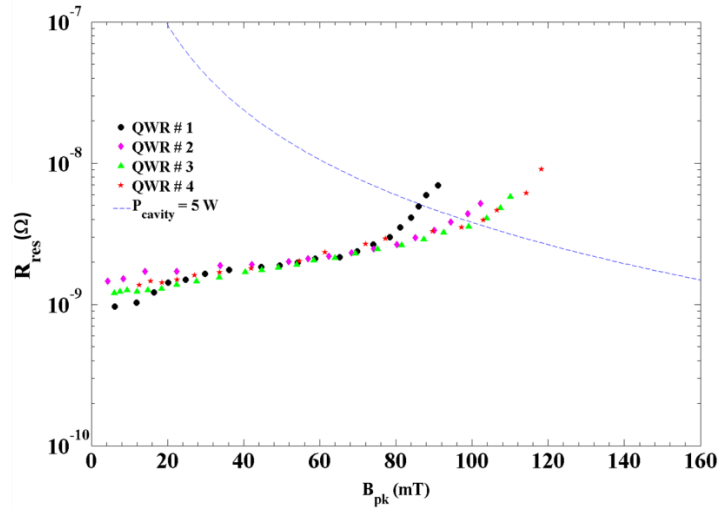

**Figure IV-11**: Cavity residual resistance measured in the ANL Intensity Upgrade QWR

The cryomodule designs all build upon past ANL experience with box cryomodules. Here the cryomodule is much wider due to the half-wave cavities being mounted on their sides. To keep the half-cylinder bottom would make the vacuum vessels unacceptably tall. We have arrived at making the vacuum vessel a box which appears to be a good compromise between fabrication cost, structural integrity and minimizing cryostat height. The radii of the rounded corners were chosen to fit the contents of the box minimizing overall height including the depth of the required gussets. Figure IV-12 shows the results of ANSYS calculations of the structural deformations due to vacuum being pulled on the inside. Notice that the structure pulls in about 0.25" on average due to evacuation, the maxima are between 0.5" and 0.67". Motion of the vacuum vessel wall moves the internal magnetic shielding and stresses the baton points which may degrade performance. Reducing the maximum displacement to less than 0.25" will avoid this but it adds the cost of additional gusseting. Future tests are planned here to evaluate the magnetic shielding.

The cryomodule houses 8 sets of identical components. Each set forms a focusing period and includes a resonator, a SC solenoid with 4 dipole coils and a Beam Position Monitor (BPM). Beam dynamics requires the solenoids to be aligned to better than ±0.5 mm peak transversely with ±0.1$^0$ for all of the rotation angles with similar constraints on the cavities. The beam-line string length is 6 meters and will be supported and aligned on a cryomodule spanning titanium rail system, called the strong-back as shown in Figure IV-13. The strong-back is composed of 2 inch × 8 inch grade 2 titanium plates formed into a box and supported by titanium hangers. Each component is mounted on top of the strong back with its own independent kinematic-alignment hardware.

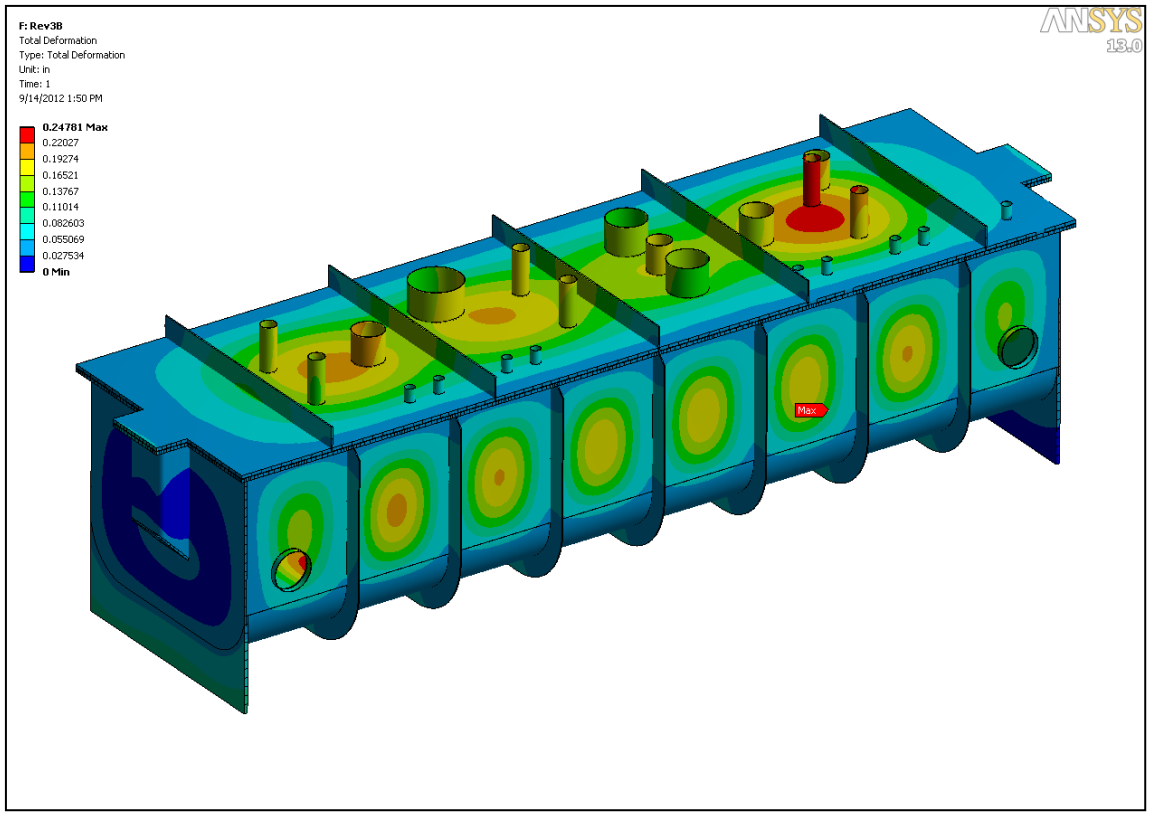

**Figure IV-12**: ANSYS results of the vacuum vessel deformation due to a 14.7 psi static pressure gradient across the walls. The red color corresponds to displacements greater than 0.6 inches with the maximum being 0.67 inches.

Table IV-3 summarizes the estimated static and dynamic heat loads to each temperature level in the cryomodule assembly (Figure IV-14) from all sources. The following sources were included in the calculation of 2K heat load: cavities, RF couplers, helium manifold, radiation from 70K to 2K, instrumentation, high current leads, strongback hangers, cavity and solenoid cooldown lines, vacuum manifold, slow tuners, and gate valves. Changing the operating voltages by + and -20% will result to 28W and 21 W total 2K heat load respectively. Currently two HWR prototypes are being fabricated. In addition, a high-power RF coupler, a BPM and SC solenoid have been built and are being cold tested now (spring 2013).

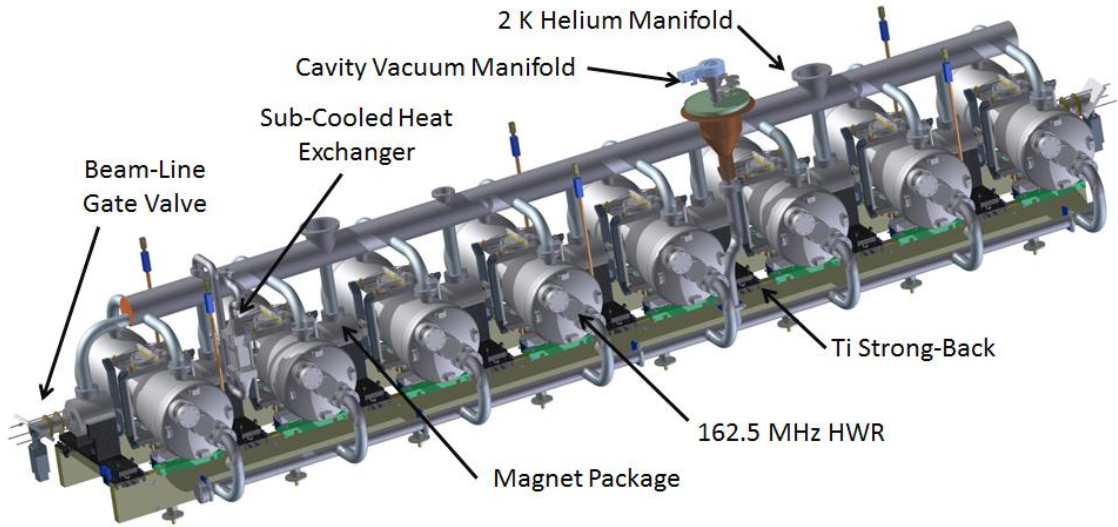

**Figure IV-13**: HWR Cavity String Assembly

| Temperature | Load, W |
|---|---|
| 2 K, static | 14 |
| 2 K, dynamic | 10* |
| 5 K | 60 |
| 70 K | 250 |

*This value takes into account actual voltage distribution on the HWR cavities

**Table IV-3**: HWR Cryomodule Heat Load Estimate

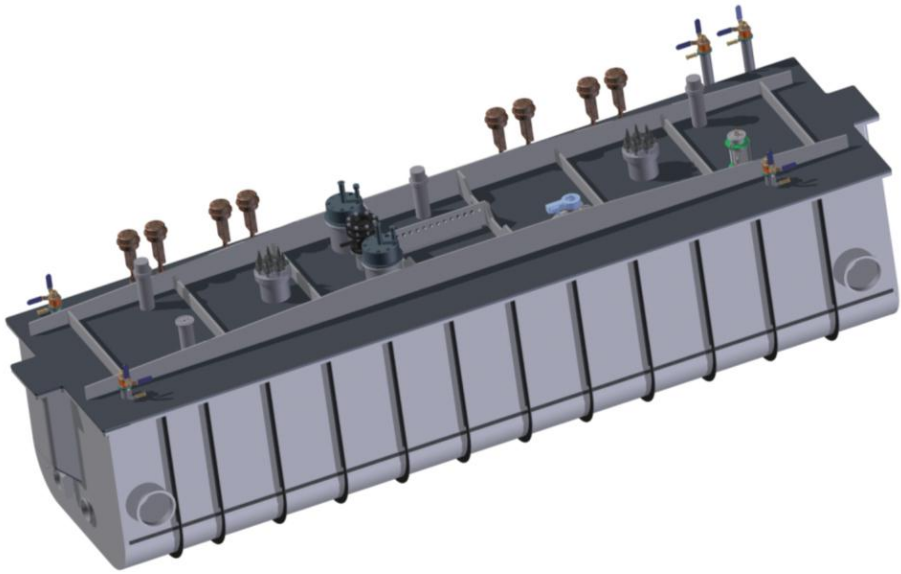

**Figure IV-14**: HWR cryomodule assembly.

### IV.1.4 Low-beta section (10-160 MeV, 325 MHz)

Two cavity types are required to accelerator beam from 11 to 177 MeV ($\beta$=0.15 to 0.61). The general requirements on the sections containing these cavities are listed in Table III-3. All cavities are of the single spoke resonator (SSR) type. During the R&D phase, studies will be completed comparing the relative benefits of these and other possible low beta configurations.

*SSR1 Cryomodule*

Acceleration from 11 to 38 MeV utilizes superconducting SSR cavities with $\beta_{opt}$=0.22 (SSR1). The cavity has geometrical and electro-magnetic parameters shown in Table IV-4. A SSR1 cavity matching these requirements has been designed, fabricated, and tested with RF power as part of the HINS program. The mechanical design, including focusing elements, is displayed in Figure IV-15.

| SSR1 Cavity | Design Value | |
|---|---|---|
| Frequency | 325 | MHz |
| Geometric Beta ($\beta_G$) | 0.186 | |
| Optimum Beta ($\beta_{OPT}$) | 0.222 | |
| Aperture (diameter) | 30 | mm |
| $L_{EFF} = \beta_{OPT}\lambda$ | 20.5 | cm |
| $R/Q_0$ | 242 | $\Omega$ |
| $G = Q_0 R_S$ | 84 | $\Omega$ |
| $E_{max}/E_{acc}$ | 3.84 | |
| $B_{max}/E_{acc}$ | 5.81 | mT/(MV/m) |
| Gradient | 10 | MV/m |
| Peak Surface Electric Field | 38 | MV/m |
| Peak Surface Magnetic Field | 58 | mT |
| $Q_0$ | 0.5 | $10^{10}$ |
| Operating Temperature | 2 | K |

**Table IV-4**: Requirements of the low-beta single spoke resonator (SSR1) cavities

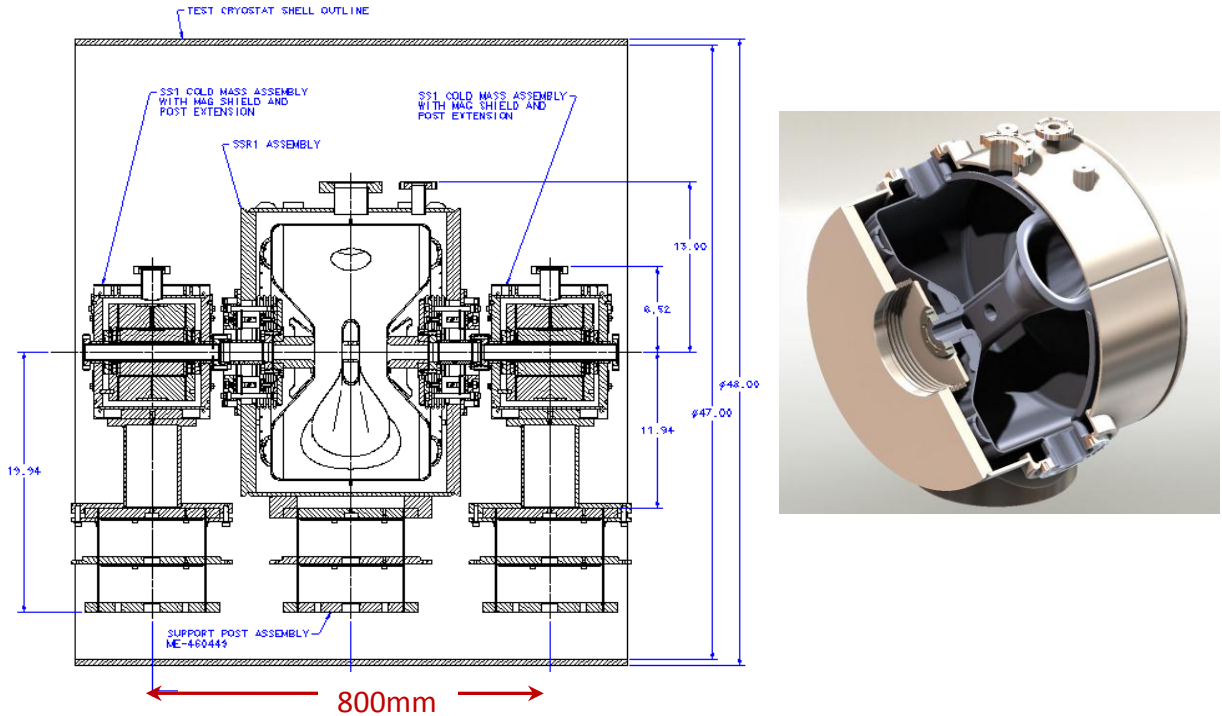

**Figure IV-15**: SSR1 cavity mechanical design and cutaway view.

Figure IV-16 shows the first( SSR1-02) cavity fabricated as part of the HINS program. The left photograph shows the bare cavity, the right a "dressed" cavity encased in its He jacket with ancillary slow and fast (piezo) tuners. To date an additional ten bare cavities were fabricated and delivered to Fermilab, (SSR1-05 – SSR1-14). Six have been tested with parameters suitable for operation in Project X. The measured performance at 2K of the bare cavity in a vertical test is displayed in Figure IV-17. Note that the cavities are made of Cabot niobium, which is not certified for high-gradient ILC operation, and demonstrated higher losses than material from certified vendors. However, all the cavities show a $Q_0 > 0.7 \times 10^{10}$ at the 2K at the operating gradient of 12 MeV/m, which is well above required value $> Q_0 > 0.5 \times 10^{10}$. Note that the cavity SSR1-02 made of certified material demonstrated a $Q_0 = 1.1 \times 10^{10}$ at 2K at the operating gradient. The measured surface resistance of this cavity as a function of temperature is shown in Figure IV-18. The cavity operational and test requirements are summarized in Table IV-5.

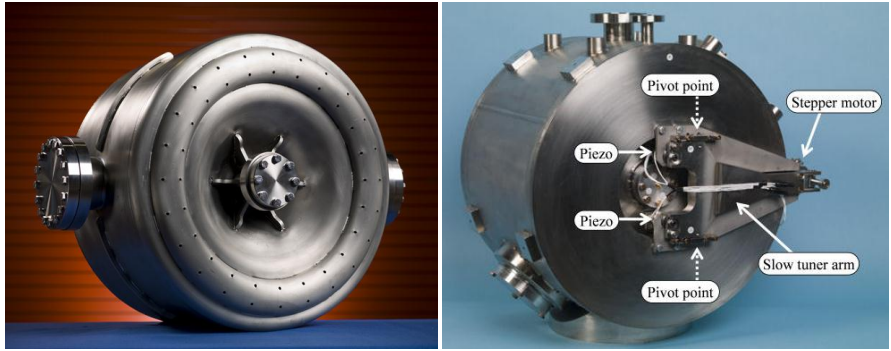

**Figure IV-16**: Photographs of the bare and dressed prototype SSR1 cavity

| Parameter | Requirement |
| --- | --- |
| Max leak rate (room temp) | < $10^{-10}$ atm-cc/sec |
| Operating gain per cavity | 2.0 MeV |
| Maximum gain per cavity | 2.4 MeV |
| Max. power dissipation per cavity at 2 K | 5 W |
| Sensitivity to He pressure fluctuations df/dP | < 25 Hz/Torr |
| Field flatness | Within ±10% |
| Multipacting | None within ±10% of operating grad. |
| Operating temperature | 1.8-2.1 K |
| Operating pressure | 16-41 mbar differential |
| MAWP | 2 bar (RT), 4 bar (2K) |
| RF power input per cavity | 6 kW (CW, operating) |

**Table IV-5**: SSR1 cavity operational and test requirements.

A spoke cavity has no axial symmetry. Therefore its quadrupole component cannot be compensated over the entire range of cavity operation. Figure IV-19 presents the dependence of the quadrupole effect on the beam velocity. Due to engineering limitations, mainly related to the RF couplers, the cavities are rolled by 45°; consequently, their quadrupole field is also rolled and is equivalent to a skew-quadrupole field. The cavity skew-quadrupole fields will be compensated by correction coils located inside nearby focusing solenoids capable to create dipole and skew-quadrupole fields.

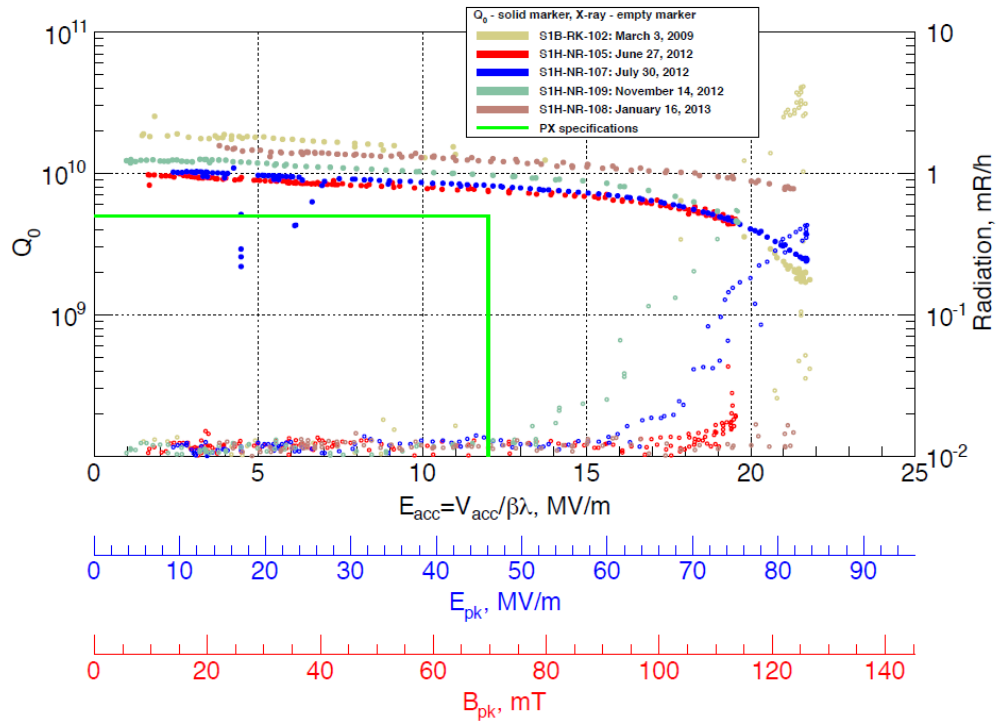

**Figure IV-17**: $Q_0$ vs. acceleration gradient from the cold test of the SSR1-02, SSR1-05, SSR1-07, SSR1-08, and SSR1-09 single-spoke cavities ($\beta = 0.22$). Maximal gain is 4.2 MeV @ 2K.

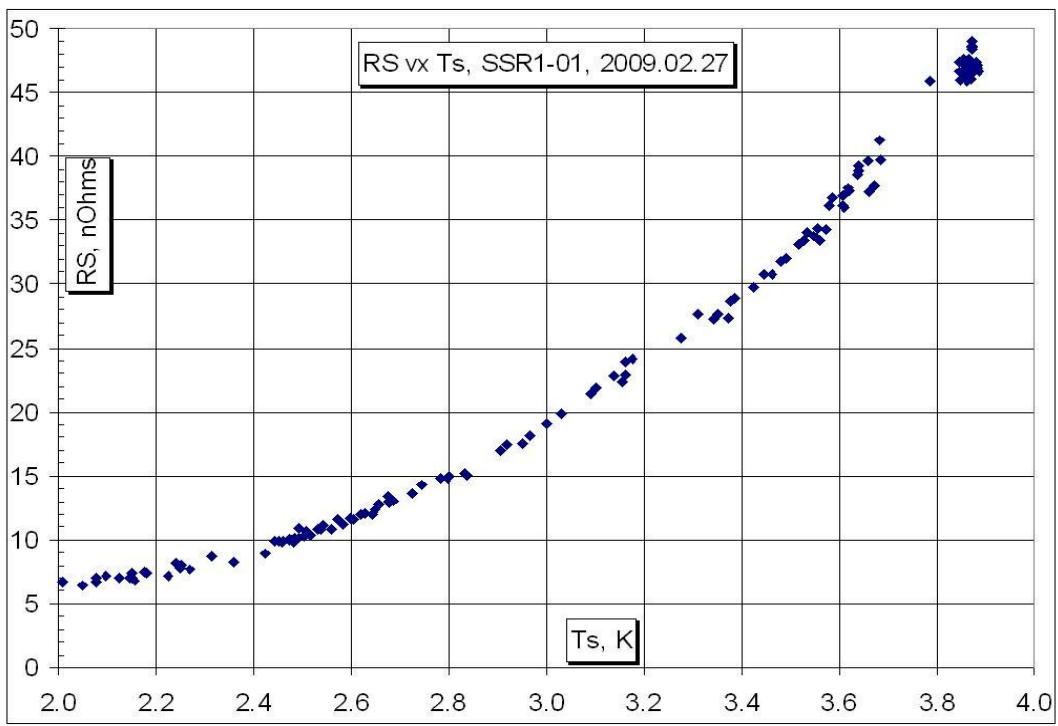

**Figure IV-18**: Temperature dependence of the surface resistance for SSR1 cavity.

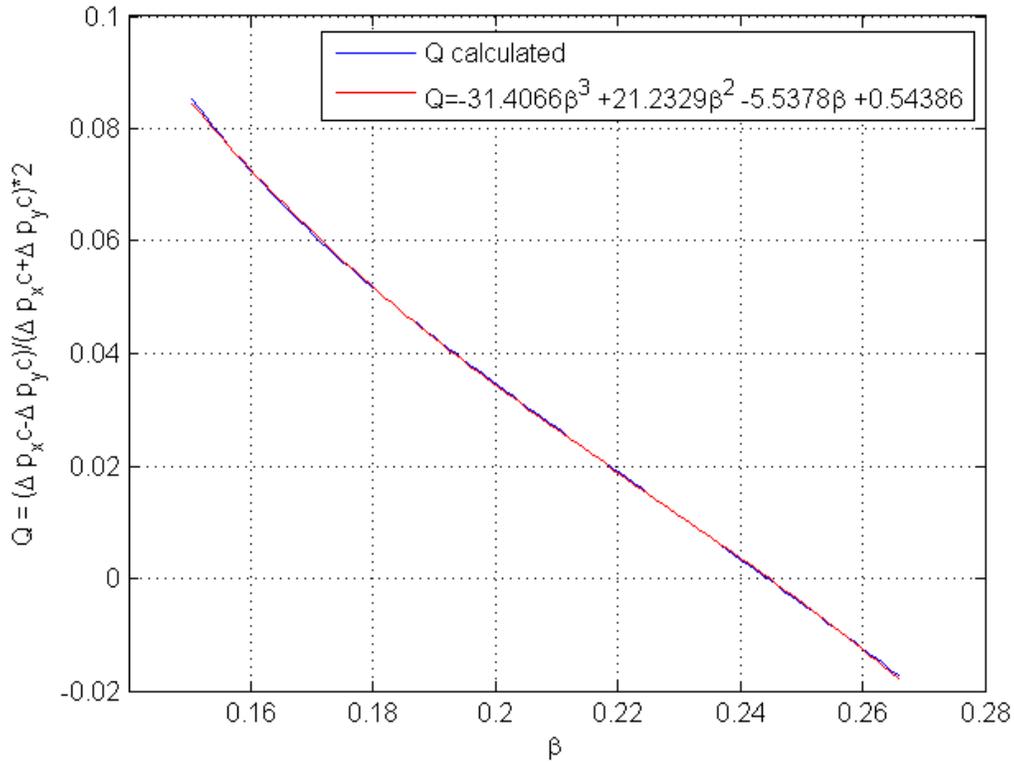

**Figure IV-19**: Quadrupole effect in SSR1 cavity versus the particle velocity β in the operating domain; blue and red line present simulation and approximation.

In order to attain the requirements for frequency range and resolution (Table IV-6), the tuning systems for cavities of narrow bandwidths such as SSR1 typically integrate a coarse and a fine mechanism engaged in series. The first normally utilizes a stepper motor with large stroke capability and limited resolution, the latter usually contains piezo-electric actuators with limited stroke but virtually infinite resolution.

|  | Requirement |
| --- | --- |
| Coarse frequency range | 135 kHz |
| Coarse frequency resolution | 20 Hz |
| Fine frequency range | 1 kHz |
| Fine frequency resolution | ≤2 Hz |

**Table IV-6**: SSR1 tuning system requirements

The coarse tuner is predominantly used to achieve consistently the resonant frequency during the cool-down operations. The range necessary to compensate for the cool-down uncertainties is estimated to be 50 kHz. In the event that a cavity must be detuned as a result of a malfunction, the coarse tuning system must be able to shift the frequency away from resonance by at least 100 bandwidths which equal to ≈ 10 kHz, so that the beam is not disturbed. The requirement on the range was set arbitrarily considering a safety margin of 2.7. The requirement on the resolution of the coarse tuning system is set to a value that would allow operation in the event of a failure of the fine-tuning system. Based on other applications, it is believed that such resolution can be achieved with a coarse tuning system.

It is conservatively assumed that the coarse system cannot be operated during beam acceleration; it is thought that the vibration of a stepper motor may induce vibrations in the cavity severe enough to disrupt the operation. Thus, fine tuners shall be designed to compensate, at a minimum, the frequency shifts of the cavity induced by fluctuations of the helium bath pressure. The use of fine tuners will reduce considerably the hysteresis of the system by limiting the elements in motion during the tracking of the frequency. A particular design effort shall be dedicated to facilitate the access to all actuating devices of the tuning system from access ports on the vacuum vessel. All actuating devices must be replaceable from the ports, either individually or as a whole cartridge.

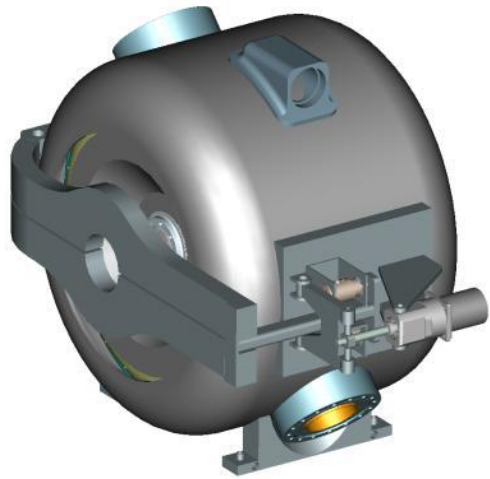

**Figure IV-20**: SSR1 cavity, helium vessel, and tuner

The Helium vessel will be fabricated from a non-magnetic stainless steel (e.g. 316L) designed to house a 2 K helium bath sufficient to remove up to 5 watts average dissipated power, with appropriately sized supply and return piping. It must meet the requirements of the Fermilab ES&H Manual for cryogenic pressure vessels and be rated at an MAWP (Maximum Allowable Working Pressure) of no less than 2 bar at room temperature and 4 bar at 2 K. Every effort should be made to minimize the weight and physical size of the helium vessel in all dimensions. The cavity vessel with tuner system is shown in Figure IV-20.

*SSR1 Input Coupler*

The design of the input coupler is based on a 105 Ω coaxial line. It will be used for both SSR1 and SSR2 cryomodules. Its maximum power of 30 kW CW is determined by requirements for an eventual upgrade of Project X to 5 mA average beam current [7]. The coupler contains a single warm ceramic window that provides separation of the warm and cold coupler sections. During cryomodule fabrication, the cold section can be installed on the cavity in the cleanroom prior to assembly of the string. The warm section can then be installed from outside the vacuum vessel during final assembly. The inner conductor is solid copper with phosphor bronze bellows to accommodate motion due to misalignment and thermal contraction. The cold end of the outer conductor is 316L-stainless steel. The warm end is copper with phosphor bronze bellows. Heat load estimates don't suggest a significant penalty for not copper plating the outer conductor. A forced-air cooling tube is inserted into the inner conductor after assembly that supplies air to cool the coupler tip. The coupler parameters are shown in Table IV-7. Figure IV-21 shows the current coupler design.

|  | Requirement |
|---|---|
| CW Power | 30 kW |
| Multipactor threshold | 25 kW (Trav. Wave) |
| Passband | 50 MHz |
| Input | 3.125'' coaxial |
| Input impedance | 50 Ω |
| Output | 3''× 0.5'' coaxial |
| Output impedance | 105 Ω |

**Table IV-7**: Design parameters of SSR1 input coupler.

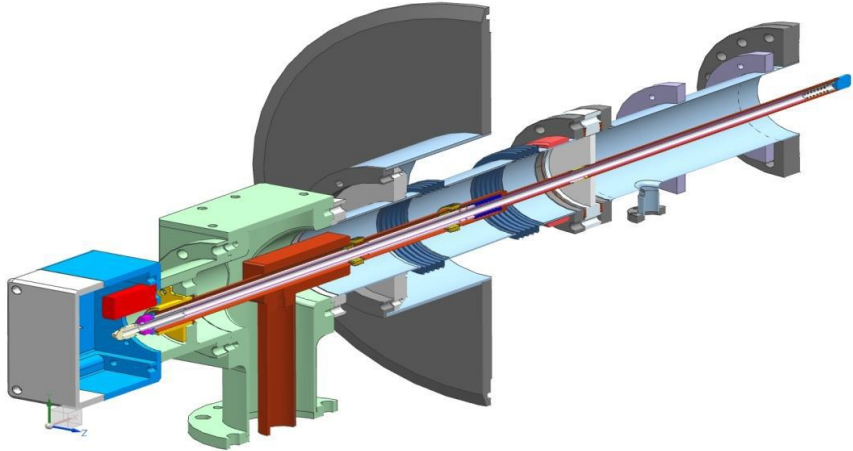

**Figure IV-21**: SSR1 input coupler.

### SSR1 Current Leads

Each focusing element package contains five coils: the main solenoid, operating up to 100 A, and four coils which can be combined to serve as both *x* and *y* steering and skew-quadrupole correctors. Each coil can operate up to 50 A. A conduction cooled current lead design modeled after similar leads installed in the LHC at CERN is being developed for use in the SSR1 cryomodule. Thermal intercepts at 45-80 K and at 5 K help reduce the heat load to 2 K, nonetheless, these current leads represent a significant source of heat at the low temperature end. There will be one lead assembly for each magnetic element.

### SSR1 Solenoid and Beam Position Monitor

The four magnet packages in the cryomodule each contain a focusing solenoid (lens) and four corrector coils all operating in a helium bath at 2 K. The general design requirements for the lenses in the SSR1 cryomodule are summarized in the list below:

Requirements essential for the beam dynamics in the linac:

- Integrated focusing strength of the lens must be not less than 4 $T^2$m;
- Each lens must contain 4 coils which can be combined into two dipole correctors; bending strength of each corrector must be not less than 0.0025 T-m;
- Clear aperture in the lens must be not less than 30 mm;

- Uncertainty of the location of the effective magnetic axis in the focusing solenoid of the lens relative to reference points on the outer surface of the device must be better than 0.1 mm rms.

Requirements essential for proper functioning of the cryomodule:

- Maximum current in the solenoid must be less than 100 A;

- Maximum current in the dipole correctors must be less than 50 A;

- LHe vessel must be used for cooling the windings down to 2 K;

- The lenses must be quench-protected; the energy deposited in the lenses after quenching must be as low as reasonably achievable;

- The LHe vessel must meet the requirements of the Fermilab's ES&H manual chapters for pressure vessel;

- The design of the LHe vessel must ensure reliable and reproducible mechanical connection to the alignment fixture of the cryomodule;

- Maximum magnetic field generated by lenses in the cryomodule in the area near the surface of the SSR1 superconducting cavities must not exceed the level that would result in more than two-fold reduction of the intrinsic quality factor after quench event at any point on the surface of the cavity.

The Project X lattice, especially the low-beta section, provides limited space along the beamline for beam diagnostics either inside individual cryomodules or between adjacent modules. In order to conserve axial space along the beamline a button-type beam position monitor (BPM) has been chosen for installation in the SSR cryomodules. For a non-relativistic beam they also generate larger signal than the strip-line BPMs. A total of four BPMs will be installed in the cryomodule, one near each magnetic element. These devices are compact and lend themselves well to incorporation right into the solenoid magnet package as shown below in Figure IV-22. The bellows in either end of the beam tube allow independent adjustment of each magnet.

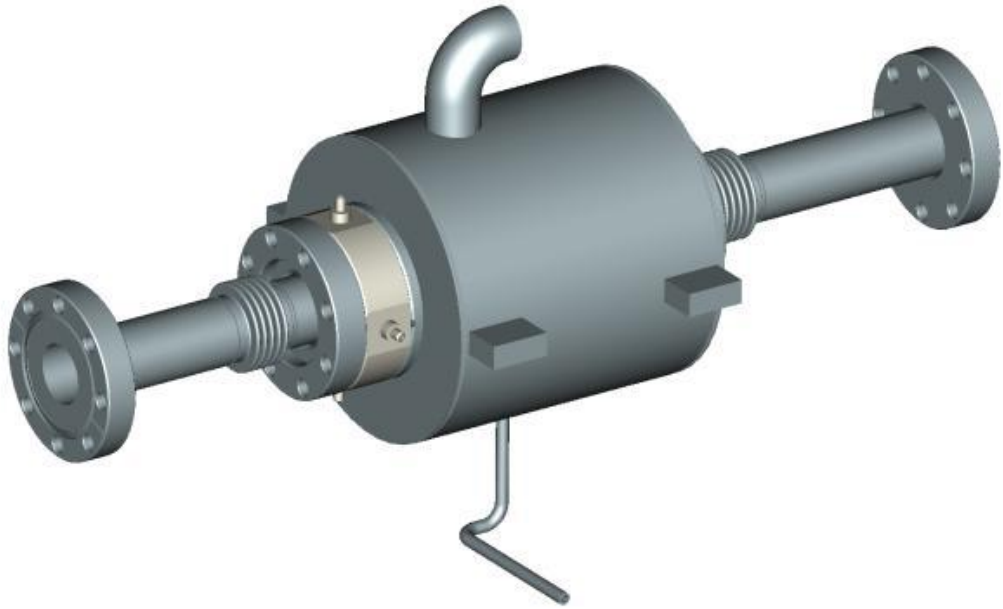

**Figure IV-22**: Solenoid and BPM assembly.

### Final Assembly

The final assembly of the SSR1 cryomodule for SSR1 is shown in Figure IV-23 and Figure IV-24. Figure IV-23 shows the cavity string consisting of the cavities, solenoids, beam position monitors, and internal piping mounted on support posts that are in turn mounted to the strongback. Figure IV-24 shows the entire cryomodule assembly.

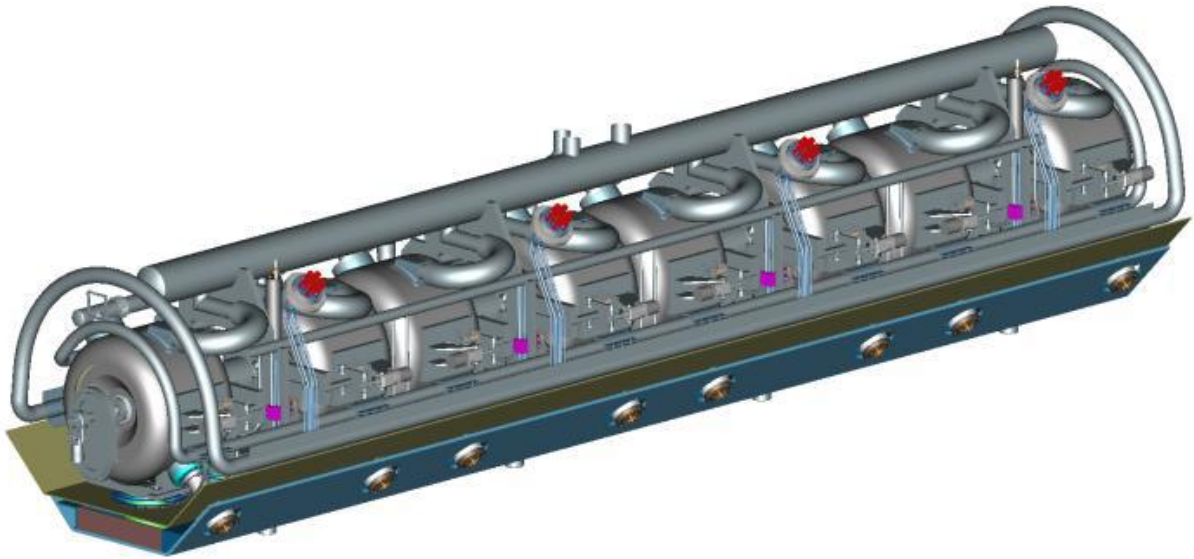

**Figure IV-23**: SSR1 cavity string assembly.

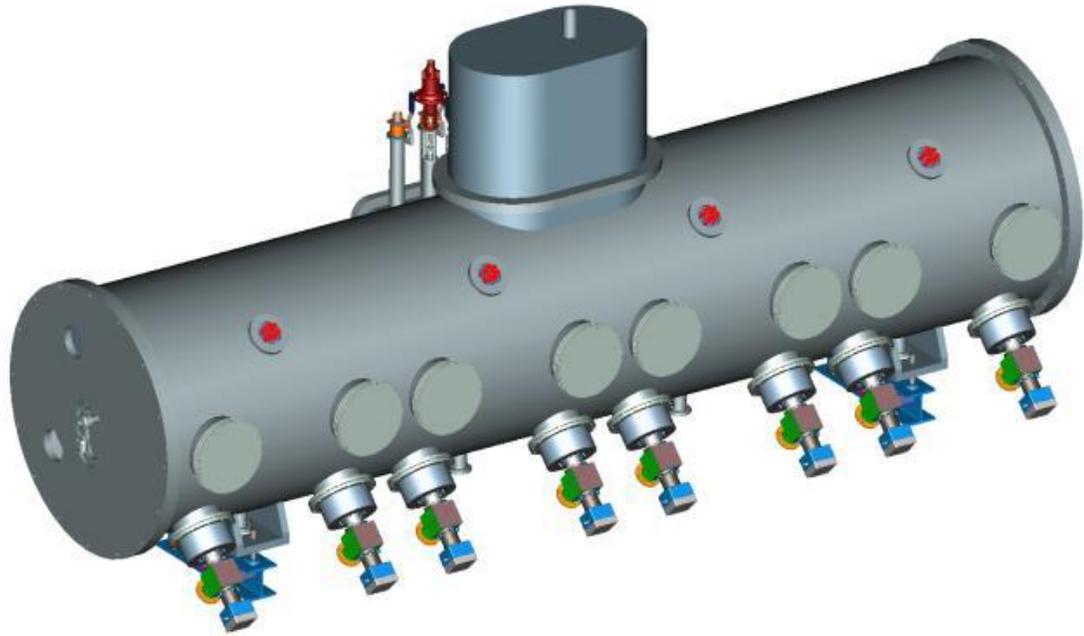

**Figure IV-24**: SSR1 cryomodule assembly

### SSR1 Heat Load Estimate

Table IV-8 summarizes the estimated static and dynamic heat loads at each temperature level in the cryomodule assembly from the primary sources. As mentioned earlier, the nominal 80 K thermal shield and intercepts may operate anywhere between 45 and 80 K.

|  | Per Unit (W) | | | Units | Total (W) | | |
| --- | --- | --- | --- | --- | --- | --- | --- |
|  | 80 K | 5 K | 2 K |  | 70 K | 5 K | 2 K |
| Input coupler, static | 5.4 | 2.8 | 0.5 | 8 | 43 | 23 | 4 |
| Input coupler, dynamic | 0 | 0 | 0.25 | 8 | 0 | 0 | 2 |
| Cavity, dynamic | 0 | 0 | 1.8 | 8 | 0 | 0 | 14 |
| Support post | 2.8 | 0.4 | 0.05 | 12 | 33 | 4 | 0.6 |
| Conductor Lead Assembly | 36.8 | 13.2 | 1.2 | 4 | 147 | 53 | 5 |
| MLI (70 K + 2 K) | 30.5 | 0 | 1.4 | 1 | 31 | 0 | 1 |
| Cold-warm transition | 0.7 | 0.1 | 0.01 | 2 | 1 | 0.2 | 0.02 |
| TOTAL |  |  |  |  | 255 | 80 | 27 |

**Table IV-8**: SSR1 Cryomodule Heat Load Estimates

*SSR2 Cavities and Cryomodule*

Acceleration from 38 to 177 MeV utilizes superconducting SSR cavities with $\beta_{OPT} = 0.51$ (SSR2). The cavity geometrical and electro-dynamic and mechanical design parameters are listed in Table IV-9 and Table IV-10. The cavity layout is shown in Figure IV-25. The electromagnetic optimization as well as mechanical design is complete, including the piezo tuner and helium vessel.

As in the SSR1 cryomodule, the quadrupole field is compensated by corrector coils which have independent leads. The SSR2 cavity has the same type of helium vessel as the SSR1 cavity, and the same type of tuners - coarse and fine. The input coupler is the same as for SSR1 cavity.

The SSR2 cryomodule design contains eight identical slots, which can accommodate both SSR2 dressed cavities and a focusing solenoid with a corrector package and BPM. The SSR2 cryomodule comprises 5 dressed cavities and 3 solenoids. The overall cryomodule length will be approximately 6.5 m. The vacuum vessel diameter will be 1.22 m (48 inches). Each cryomodule will be configured as a stand-alone unit, i.e. the vacuum vessel ends will be closed and cryogenic connections will be made at each module. Connections for cryogens and cryogenic control valves will be located in a mid-span vacuum vessel extension. The only module-to-module connection will be the beam line. The only beam instrumentation internal to the cryomodule assembly will be BPMs.

| SSR2 Cavity | Design Value | |
|---|---|---|
| Frequency | 325 | MHz |
| Geometric Beta ($\beta_G$) | 0.431 | |
| Optimum Beta ($\beta_{OPT}$) | 0.515 | |
| Aperture (diameter) | 50 | mm |
| $L_{EFF} = \beta_{OPT}\lambda$ | 47.5 | cm |
| R/Q | 275 | Ω |
| $G = Q_0 R_S$ | 118 | Ω |
| $E_{max}/E_{acc}$ | 3.53 | |
| $B_{max}/E_{acc}$ | 6.25 | mT/(MV/m) |
| Gradient | 11.2 | MV/m |
| Peak Surface Electric Field | 40 | MV/m |
| Peak Surface Magnetic Field | 70 | mT |
| Max Energy Gain | 5.3 | MeV |
| Operating Temperature | 2 | K |
| $Q_0$ at 2 K | 1.2 | $10^{10}$ |
| $P_{diss}$ at Max Energy Gain | 8.6 | W |

**Table IV-9**: Electro-dynamical requirements of the low-beta spoke cavities (SSR2)

| | Requirements | |
|---|---|---|
| **Mechanical** | Radius, Length | 280 mm, 540 mm |
| | He Vessel Material | Stainless Steel |
| | Maximum Allowable Pressure, MAWP | 2 bar RT, 4 bar CT |
| | df/dp | ≤ 25 Hz/mbar |
| **Coupler** | Max. design forward power | 30 kW |
| **Tuning** | Coarse tuning range | 135 kHz |
| | Fine tuning range | 1000 Hz |

**Table IV-10**: Mechanical parameters of the low-beta spoke cavities (SSR2)

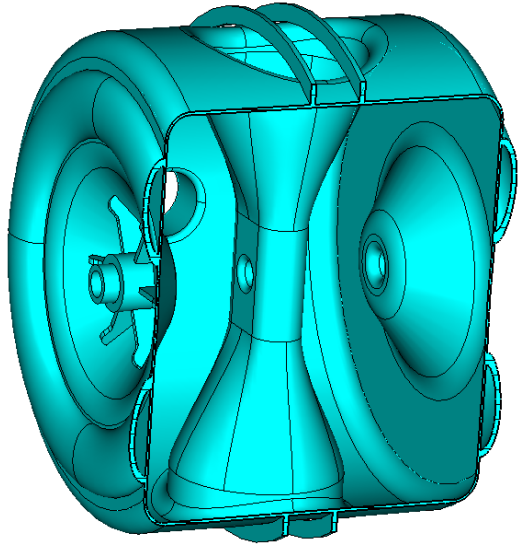

**Figure IV-25**: SSR2 cavity layout.

IV.1.5 **Medium-beta section (160 – 3000 MeV, 650 MHz)**

Acceleration from 177 MeV to 3 GeV will be provided by two families of the 5-cell elliptical cavities operating at 650 MHz and designed to $\beta_G$ =0.61 and $\beta_G$ =0.9. The cavity shape is optimized to decrease the field enhancement factors (magnetic and electric) in order to improve the interaction between the beam and the cavities. In order to do this, the cavity aperture should be as small as possible subject to the following considerations:
- field flatness,
- beam losses,
- mechanical stability,
- reliable surface processing.

The working gradient is chosen to provide the peak surface magnetic field that allows operation below high-field Q-slope, see Figure III-15. For a frequency of 650 MHz the peak magnetic field should be not greater than ~70 mT. In addition we require that the peak surface electric field be lower than 40 MV/m in order to avoid the risk of strong field emission.

Linear perturbation theory indicates that for a given relative error in the frequencies of cavity cells the field flatness is determined mainly by the distance, $\delta f$, between the operating frequency and the frequency of the neighboring mode, $\pi(n-1)/n$. In terms of the coupling parameter, $k$, between the cavity cells and the number of cells one obtains:

$$\delta E/E \sim f_\pi/|f_\pi\text{-}f_{\pi(n-1)/n}| \equiv f_\pi/\delta f \approx (n-1)^2/k.$$

Thus, a cavity with fewer cells allows a smaller coupling coefficient, $k$, for a given field flatness. For example, the 9-cell ILC cavity has $\delta f/f_\pi$ of $6\times10^{-4}$ ($k = 1.87\%$) and for the 5-cell 650 MHz cavity one can take the same $\delta f/f_\pi$ at least, yielding $k > 0.5\%$.

The aperture selected for the cavity represents a trade-off between requirements related to the cell-to-cell coupling and beam loss. The 805 MHz superconducting section of the SNS proton linac, which is close to Project X linac in average current, operates with cavities that have an aperture of 83 mm for low-beta part and 100 mm in high-beta part. Their experience is that these cavities operate with tolerable beam loss at these apertures. Thus, we have adopted these dimensions for the 650 MHz cavities of Project X. In addition, it appears that these apertures will also facilitate the required surface processing.

The 650 MHz cavities require sufficient wall thickness to minimize sagging caused by the overall weight. Figure IV-26 shows results of a simulation of the cavity sag caused by its weight as a function of wall thickness for the 650 MHz cavities and the ILC (1300 MHz) cavity. Note that stiffening rings are used for both the ILC and 650 MHz cavities to increase the rigidity of cavities. Requiring a maximum cavity sag of 120 μm (the same as ILC) results in a 4 mm wall thickness. A small cavity wall slope (designated by $\alpha$ in Figure IV-27) gives more freedom to decrease the field enhancement factors. However, the slope is limited by surface processing and mechanical stability requirements. For beta=0.9 we select the slope to be 5°. For beta=0.61 the slope is reduced to 2°, in order to maintain an acceptably low field enhancement.

Optimization of the two 650 MHz cavity shapes was done based on the constraints discussed above. Cavity performance parameters are summarized in Table IV-11. The physical description of the cavity shapes is displayed in Figure IV-27 and Table IV-12. Requirements for maximum cavity detuning amplitude and cavity sensitivity versus helium pressure fluctuations are discussed further in [41]. Note that the 650 MHz cavities have small beam loading, and thus microphonics mitigation is essential. In order to do this, the cavities are over-coupled; both active and passive means for microphonic compensation are planned to be used also [42]. Neither HB605 or LB650 cavities contain HOM dampers – they are not necessary for required beam current (see for example [43]).

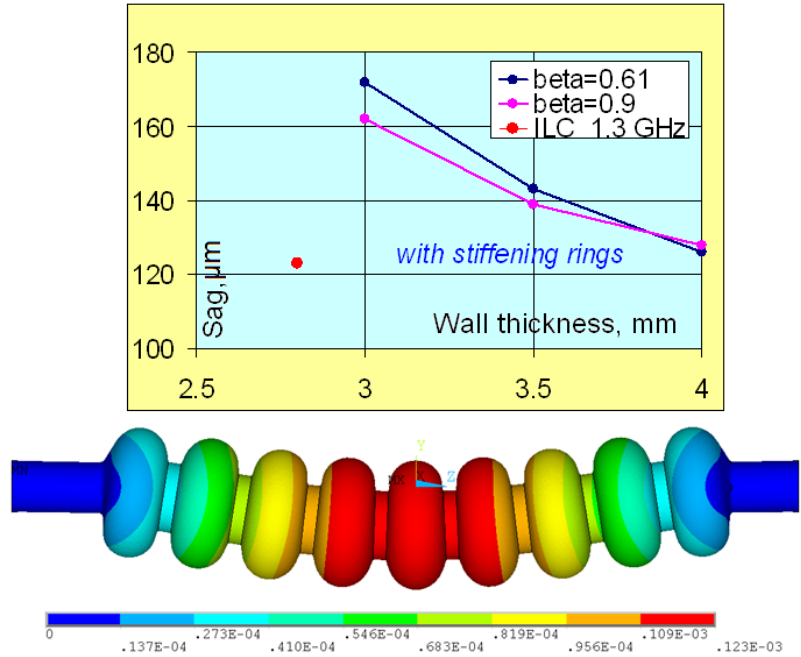

**Figure IV-26**: The cavity sag versus the wall thickness.

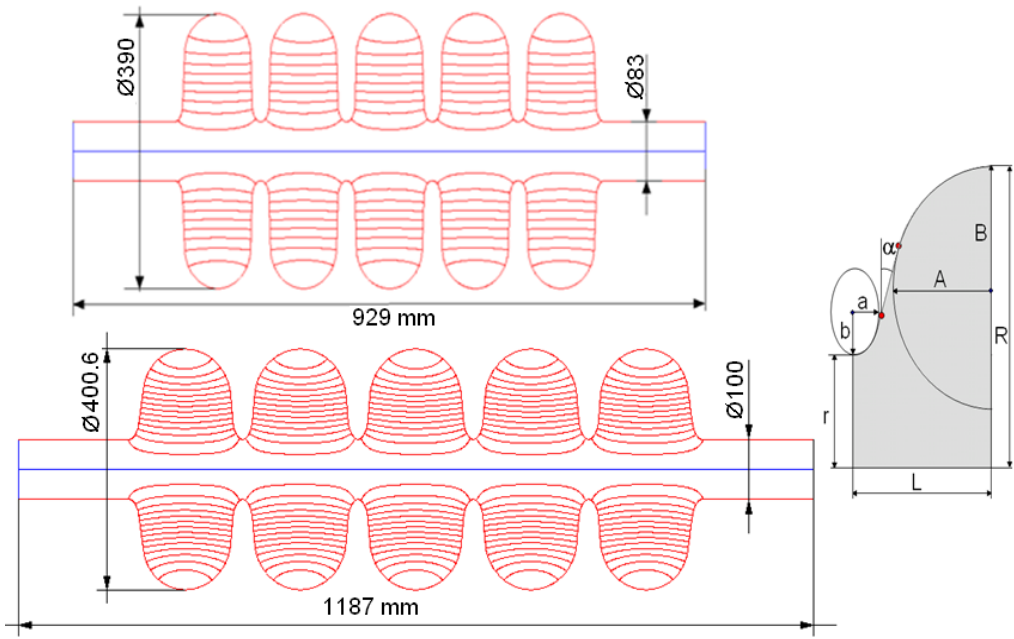

**Figure IV-27**: Layout of 650 MHz cavities. Beta=0.61(top) and beta=0.9 (bottom).

| 650 MHz Cavities | Design Value | | |
|---|---|---|---|
| | Low Beta | High Beta | |
| Frequency | 650 | 650 | MHz |
| Geometric beta ($\beta_G$) | 0.61 | 0.90 | |
| Optimum Beta ($\beta_{OPT}$) | 0.64 | 0.95 | |
| Aperture (diameter) | 83 | 100 | mm |
| Cells per Cavity | 5 | 5 | |
| Cavity Length | 70 | 104 | cm |
| $R/Q_0$ | 378 | 638 | $\Omega$ |
| $G = Q_0 R_S$ | 191 | 255 | $\Omega$ |
| $E_{max}/E_{acc}$ | 2.27 | 2.07 | |
| $B_{max}/E_{acc}$ | 4.25 | 3.78 | mT/(MV/m) |
| Gradient* | 16.5 | 17 | MV/m |
| Peak Surface Electric Field | 37.5 | 35.2 | MV/m |
| Peak Surface Magnetic Field | 70 | 64 | mT |
| Max. gain per cavity, MeV(on crest) | 11.6 | 17.7 | MeV |
| Operating temperature | 2.0 | 2.0 | K |
| $Q_0$ at 2 K | 1.5 | 1.5 | $10^{10}$ |
| Cryogenic losses per cavity at 2 K | 24 | 24 | W |
| Peak detuning amplitude ♦ | <20 | <20 | Hz |
| Bandwidth † | 45 | 45 | Hz |
| Sensitivity to He pressure fluctuations df/dP | <25 | <25 | Hz/mbar |
| Beam current (Stage 1/Stage 2) | 2 | 2/1 | mA |
| Power consumed by the beam ♦ | 23 | 35/17.5 | kW |
| Maximal input RF power† | <34 | <50/31 | kW |

\* The gradient is taken for the optimal beta and normalized on the effective cavity length

♦ The peak microphonics amplitude plus static detuning errors.

† Beam power plus the power overhead to account for waveguide losses, cavity detuning, etc.

**Table IV-11**: RF parameters of the 650 MHz cavities.

| Dimension | Beta=0.61 | | Beta=0.9 | |
|---|---|---|---|---|
| | Regular cell | End cell | Regular cell | End cell |
| r, mm | 41.5 | 41.5 | 50 | 50 |
| R, mm | 195 | 195 | 200.3 | 200.3 |
| L, mm | 70.3 | 71.4 | 103.8 | 107.0 |
| A, mm | 54 | 54 | 82.5 | 82.5 |
| B, mm | 58 | 58 | 84 | 84.5 |
| a, mm | 14 | 14 | 18 | 20 |
| b, mm | 25 | 25 | 38 | 39.5 |
| α,° | 2 | 2.7 | 5.2 | 7 |

**Table IV-12**: 650 MHz cavities dimensions (see Figure IV-27 for definition of dimensions).

A possible bunch structure required for muon, kaon, and nuclear experiments running in parallel at 3 GeV is shown in Figure IV-28. Average beam current in the 1-3 GeV linac in this mode is 1 mA. Figure IV-29 shows the spectrum for this beam structure, assuming very short bunches of equal charge and in the absence of timing jitter. The spectrum and *(R/Q)* values of the $\beta_G$ =0.9 650 MHz cavity are shown in Figure IV-30.

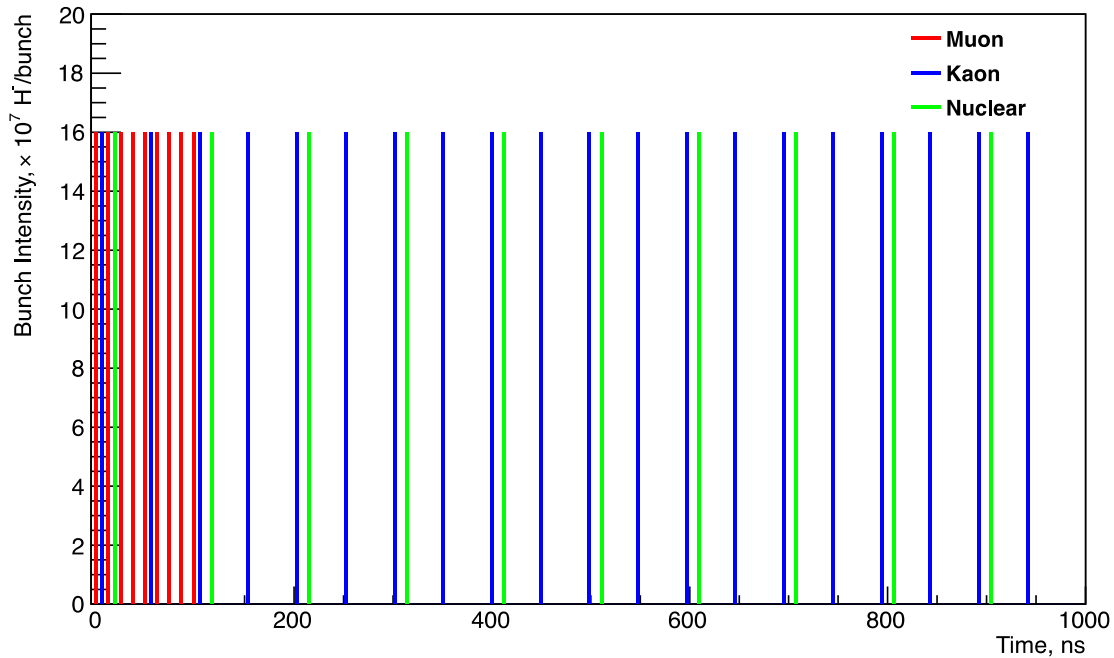

**Figure IV-28**: Beam structure for 3 GeV program.

The amplitude of an excited monopole HOM depends on the amplitude of the nearest beam spectrum line, $I$, the detuning $\delta f$, and the distance between the HOM frequency $f$ and the beam spectrum line frequency. It can be estimated for a high $Q$ resonance (assuming $\delta f/f \ll 1/Q$) as the following:

$$U_{HOM} \approx \frac{I(R/Q)}{4\sqrt{2}\,\delta f/f}.$$

If the high order mode is exactly at resonance,

$$U_{HOM} = \frac{1}{2}I(R/Q)Q_L,$$

where $Q_L$ is the loaded quality factor of the mode. The cryogenic losses depend on square of HOM amplitude:

$$P_{loss} \approx \frac{U_{HOM}^2}{(R/Q)Q_0}.$$

Requiring that $P_{loss}$ is much smaller than the sum of static heat load and cryogenic losses due to accelerating mode (20 W), and assuming that the intrinsic quality factor is $Q_0 = 5 \times 10^9$, the maximum allowable value of the monopole HOM loaded quality factor is obtained: $Q_L \ll 6 \times 10^7$.

Similarly, requiring that excitation of a monopole mode does not increase longitudinal emittance, $\varepsilon_z \gg U_{HOM}\sigma_z/c$, an estimation of the safe frequency detuning yields:

$$\delta f \gg f\frac{I(R/Q)\sigma_z}{4\sqrt{2}\varepsilon_z c}.$$

Here $\sigma_z$ is the bunch length, and $c$ is the speed of light. The worst case is at the beginning of $\beta_G = 0.9$ section, where the bunch length is at maximum ($\sigma_z/c = 7.7 \times 10^{-3}$ ns). Assuming the second pass-band monopole HOM (1241 MHz and $R/Q$ = 130 Ohm) is the nearest beam spectrum line ($I$ = 1 mA), and an emittance of $\varepsilon_z$ = 1.5 keV ns, one obtains the following estimation for frequency detuning: $\delta f \gg 140$ Hz.

A more accurate estimation of coherent HOM excitation in the Project X linac is performed using statistical analysis based on the expected spread of data for the HOM parameters (frequency, impedance and quality factor). Errors of cavity shape introduced in manufacturing are taken into account by allowing random variations of cavity profile within 0.2 mm of ideal shape. In order to estimate probability of cryogenic losses, and relative change of longitudinal emittance, $10^5$ random linacs were generated using predicted deviations of frequency, loaded quality factor and impedance values of monopole HOMs. It was found that the probability to have losses above 0.1 W per cryomodule is extremely small: $10^{-4}$ for an average beam current of 1 mA.

The beam structure, shown in Figure IV-28 consists of three main sub-components (1 MHz, 10 MHz and 20 MHz). The phase of the voltage of an HOM excited by the resonance with one of the beam components is random with respect to two other components of the beam. In case of a high-$Q$ resonance such a HOM may introduce a significant energy variation and longitudinal emittance growth along the beam train. Results of statistical analysis show, that the probability of the emittance to double is $10^{-3}$ for the beam current 1 mA. Based on this analysis the conclusion is made that HOM couplers are not needed in 650 MHz cavities. More details can be found in Ref. [43].

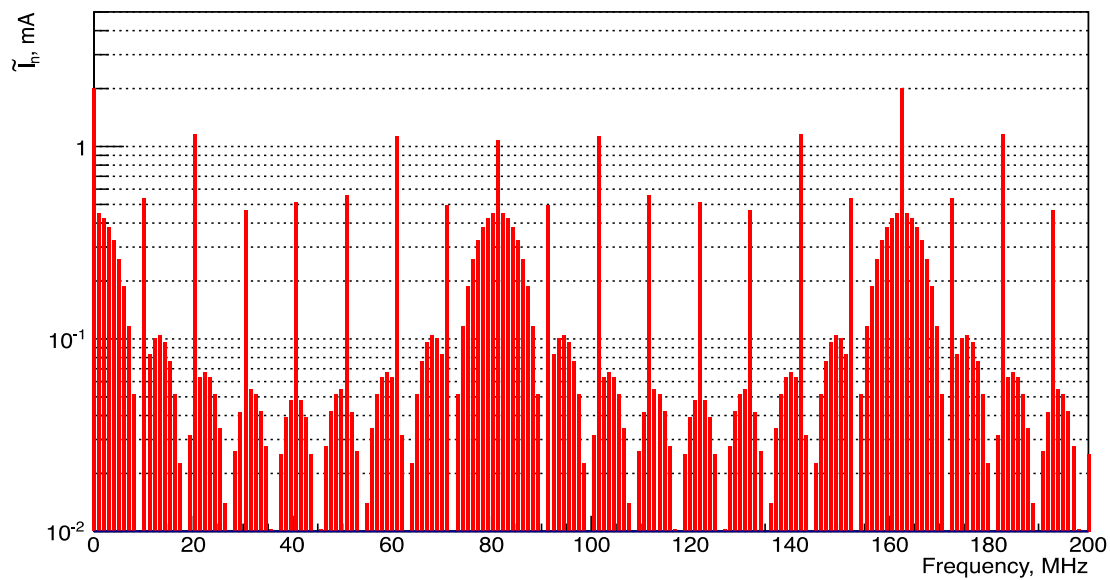

**Figure IV-29**: Beam spectrum of 3 GeV program.

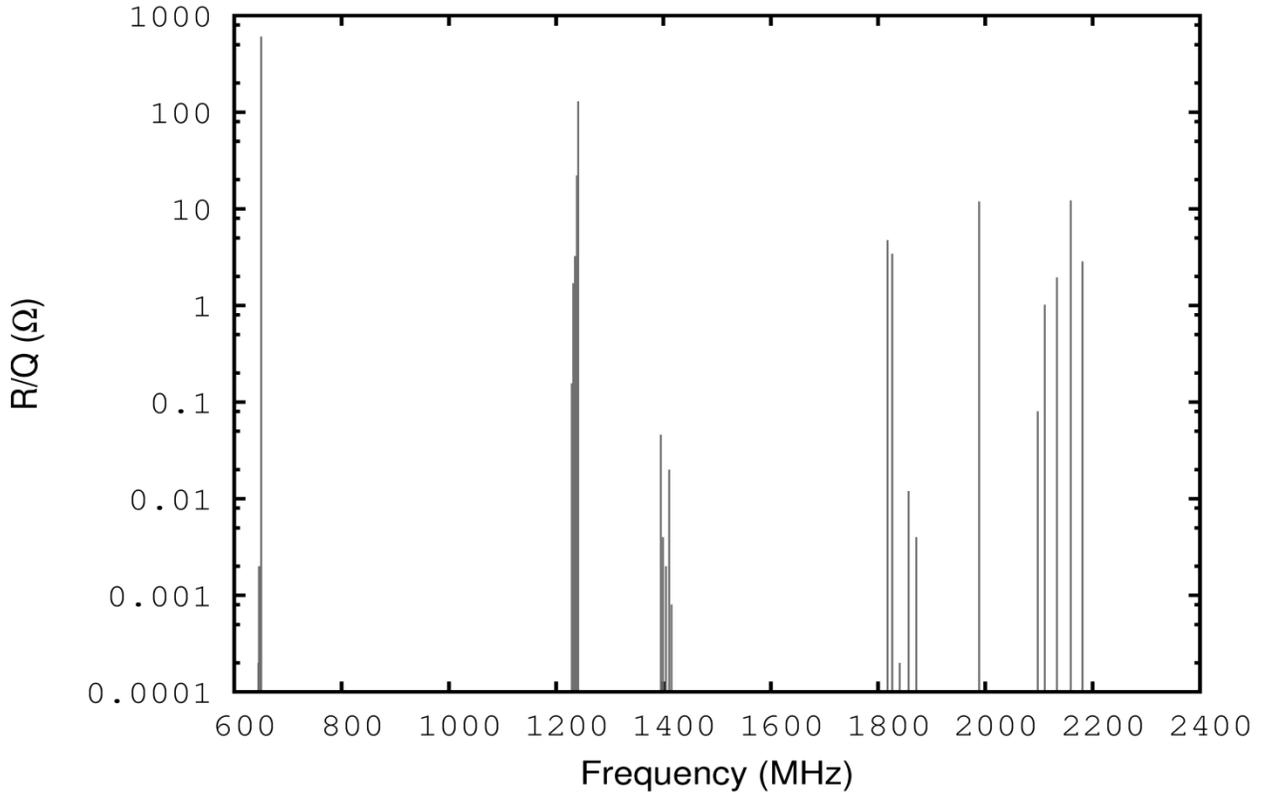

**Figure IV-30**: (R/Q) of the monopole HOMs in $\beta_G$ =0.9 section of 650 MHz cavities.

Cryogenic losses in the cavities are determined by the R/Q value, G-factor and surface resistance. The surface resistance is in turn is a sum of residual resistance and BCS resistance. Modern surface processing technology may provide a residual resistance of ~5 nΩ [44]. BCS resistance as a function of the frequency $f$ and temperature $T$ may be estimated using the formula

$$R_{BCS} = 2.\,10^{-4} \frac{1}{T}\left(\frac{f}{1.5}\right)^2 e^{-17.67/T}$$

which represents a good approximation to the value of BCS resistance. For 650 MHz one obtains ~3 nΩ for BCS and, thus, ~8 nΩ total. Assuming a medium field Q-slope at the peak field of 70 mT of about 30%, this gives a target for $Q_0$ value of the 650 MHz cavity of ~2×10¹⁰, and losses at the operating gradient (17.7 MeV/m) of ~30 W/cavity, or < 250 W/cryomodule. The preliminary mechanical design of the beta=0.9 cavity is shown in Figure IV-31.

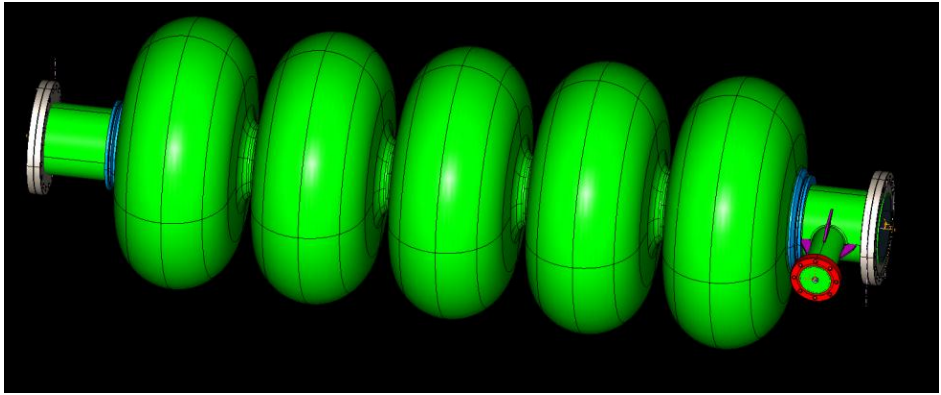

**Figure IV-31**: Preliminary mechanical design of the beta=0.9 cavity.

IV.1.6  **RF power**

There are more than 250 RF systems included in the CW linac. With the exception of the RFQ they are deployed as one amplifier per cavity. All RF systems – RFQ, buncher, half wave resonator (HWR), single spoke resonator (SSR1 & SSR2), low beta and high beta elliptical (LB650 & HB650) – will utilize continuous wave (CW) amplifiers.  There are three frequencies of operation, 162.5 MHz (RFQ & HWR), 325 MHz (SSR1 & SSR2), and 650 MHz (LB650 & HB650.  Power levels range from 4 kW to 150 kW. RF source requirements for the superconducting portion of the linac are listed in Table IV-13.

| Section | Freq MHz | Microphonic amplitude Hz | Minimal bandwidth Hz | Maximal loaded Q | Max power per cavity kW |
|---|---|---|---|---|---|
| HWR | 162.5 | 20 | 49 | $3.3 \times 10^6$ | 4.9 |
| SSR1 | 325 | 20 | 56 | $5.8 \times 10^6$ | 5.5 |
| SSR2 | 325 | 20 | 45 | $7.2 \times 10^6$ | 17 |
| LB | 650 | 20 | 45 | $1.4 \times 10^7$ | 34 |
| HB1 | 650 | 20 | 45 | $1.4 \times 10^7$ | 50 |
| HB2 | 650 | 20 | 45 | $1.4 \times 10^7$ | 31 |

**Table IV-13**: Maximum power consumption per cavity, tolerable microphonic amplitude, loaded Q and bandwidth.

The RFQ has two input ports and is driven by two 75 kW CW solid-state amplifiers. The beam energy at the RFQ output is 2.1 MeV. Three room temperature buncher cavities and one cryomodule containing eight superconducting HWRs at 162.5 MHz will have one solid-state amplifier each operating at power levels of 4 to 6 kW. Two cryomodules at 325 MHz will be populated with eight SSR1 cavities powered by 6 kW solid-state amplifiers. 7 kW solid-state amplifiers power seven SSR2 cryomodules with 5 cavities each. Five LB650s with 30 cavities and fifteen HB650s with 120 cavities will each be powered by a 35 kW IOT amplifier. Seven HB650s with 42 cavities in the 1 GeV linac will each be powered by a 50 kW IOT. It is possible that a solid-state or injection locked magnetron could be used at 650 MHz as R&D for those technologies mature.

The low level RF (LLRF) system will provide a drive signal on the order of 0 to +10 dBm for each RF power source. The amplifier(s) will provide sample signals of the pre-driver and final outputs. All amplifiers will be self-contained units complete with integral power supplies, protection circuits, and control interface. The RF distribution system for the CW linac will utilize rigid coax commensurate with system power levels, 6-1/8", 3-1/8", or 1-5/8" EIA flanged sections. The final connection to the cryomodules will utilize a section of flexible transmission line to minimize connector location tolerances. Each RF system will have a circulator and load to isolate the cavity from the power amplifier. This level of protection is essential in SRF systems due to full power reflection from the cavity in the absence of beam. Cavity and drive sample signals will be provided to the LLRF for vector regulation and frequency control of the cavities. All of the RF amplifiers will be water cooled to minimize the heat load to the building HVAC system.

While each amplifier has built in protection which includes, water flow, water temperature, pressure differential, and reflected power monitoring; a global interlock and hardware protection system will need to be designed for all RF systems. This will include water flow to loads and circulators, spark detection on cavity couplers, and RF leakage detection.

IV.1.7 **RF Splitter**

RF splitters are used to support a quasi-simultaneous beam delivery to multiple users. Presently we expect that two different splitter types will be used in the Project X. The first will split the "beam RF buckets" into two equal parts for bunch frequencies of 162.5 MHz. This requires operations at frequencies equal to $(n+1/2)\times 162.5$ MHz, where $n$ is an integer. The splitter of the second type, to be used in Stage 2 at 3 GeV, splits the beam with 81.25 MHz bunch frequency into 3 parts, so that the non-deflected beam (crossing the splitter at the RF wave form zero crossing) would have 2 times larger number of buckets than two other beams deflected in different directions (the corresponding bunches cross the splitter at maximum and minimum of the waveform). The frequency of this RF splitter has to be $(n+1/4)\times 81.25$ MHz.

The operating frequency of the deflecting RF structure is limited (i) by the bunch length – at high frequency and (ii) by the cavity transverse size – at low frequency. The cavity should have a reasonable aperture (a compromise between the deflecting properties and possible beam loss heating the cavity). The required deflecting kick is $\Delta pc/e \sim 7$ MV for both splitter types. It yields a total deflection angle of approximately ± 4 mrad at 1 GeV and ± 2 mrad at 3 GeV. A choice of RF frequency in vicinity of 400 MHz satisfies the above mentioned requirements and allows one to have very close designs for both splitters. The corresponding frequencies are 406.25 MHz = (2+1/2) )×162.5 MHz and 426.5625 MHz = (5+1/4) )×81.25 MHz. The cavities of these frequencies have sizes which fit existing cryogenic test facilities.

In order to provide this transverse kick, a two-cell version [45] of a ridge deflecting cavity [46,47] is suggested, see Figure IV-32. Parameters of the cavities are specified in Table IV-14.

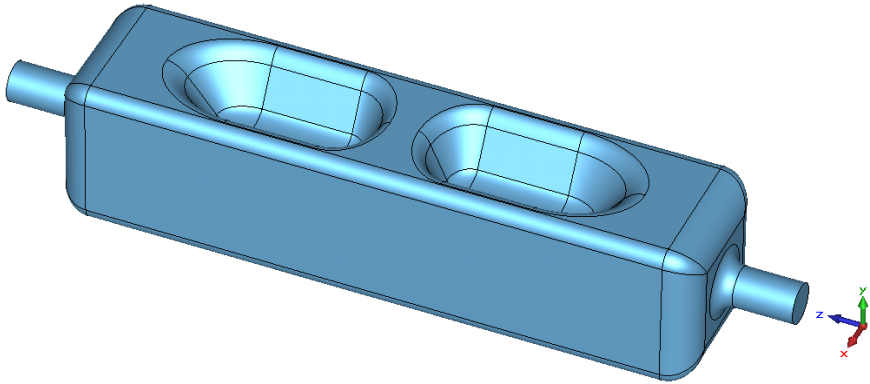

**Figure IV-32**: Two-cell ridge deflecting cavity.

| Stage | I | II |
|---|---|---|
| Operating frequency, MHz | 406.25 | 426.5625 |
| Number of cells | 2 | 2 |
| Optimal beta | 0.87 | 0.92 |
| Transverse kick, MeV | 7 | 7 |
| Maximal surface electric field, MV/m | 36 | 37 |
| Maximal surface magnetic field, mT | 50.5 | 52 |
| R/Q*, Ohm | 485 | 510 |
| G-factor, Ohm | 115 | 115 |
| Dimensions, mm$^3$ | 270×270×1200 | 260×260×1150 |
| Aperture, mm | 70 | 65 |

*$R/Q = V_{kick}^2 / 2\omega W$, where W is the stored energy.

**Table IV-14**: Parameters of deflecting cavities

Input power requirements are determined mainly by the amplitude of microphonics $\delta f$. $P_{inp} < V_{kick}^2 / [(R/Q) \cdot f / \delta f]$. For a cavity operating at 7 MeV and for $\delta f$=15 Hz, $P_{inp}$ <4 kW for both cases. Consequently, the cavity bandwidth should have the same order, or ~15 Hz. The microphonics amplitude is determined by expected helium pressure variations and $df/dp$ value. Minimization of $df/dp$ is one of the major goals during the cavity mechanical design. The helium pressure fluctuation budget is limited to 0.2 mbar rms. Thus, $df/dp$ is to be about ~20 Hz/mbar taking into account the safety factor of 4.

## IV.2 Collimators and Beam Dump

### IV.2.1 8 GeV Collimators

A transverse collimation system is employed in the 8 GeV transfer line from the pulsed linac to the injection foil in the Recycler. Each collimation station system consists of a "thick" stripping foil, located just upstream of a quadrupole, and a downstream absorber. Large amplitude H- ions that strike the foil are converted into protons which are defocused (in the collimation plane) by the downstream quad while the H- remain focused. The protons receive a kick proportional to their offset in the quadrupole field such their amplitude is increased prior to being intercepted by an absorber placed just upstream of the next quadrupole to intercept these particles. This technique has been successfully implemented at SNS. There they use a fixed aperture absorber. The Reference Design employs a movable jaw absorber to increase the efficiency of absorbing the stripped protons. This technique has been simulated in TRACK and the movable jaw absorber has been simulated in MARS. The technique is described in [48]. A conceptual description of the system is shown in Figure IV-33. The phase advance per cell in the 8 GeV transport line will be 90º, and therefore four stations (two in each plane) are required to cover the full phase space.

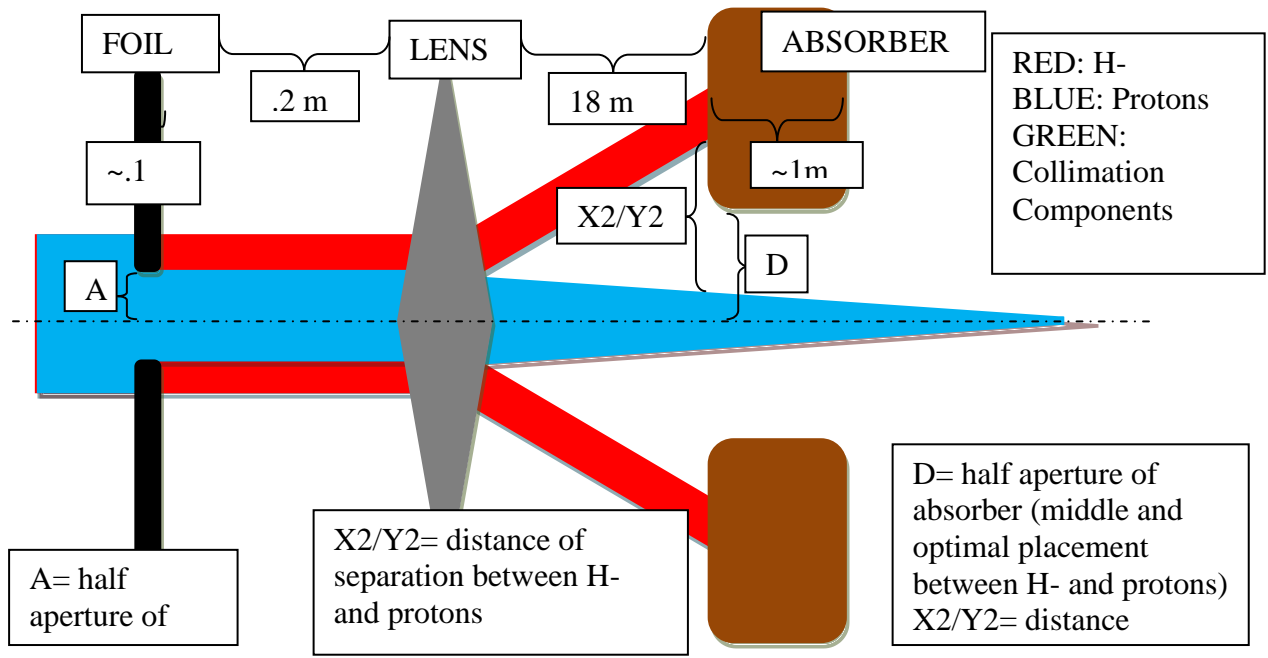

**Figure IV-33**: Concept of a horizontal H- transverse collimation station. A thin foil (black) intercepts large amplitude H- ions (blue). The focusing quadrupole (grey) defocuses the converted protons (red). The variable aperture absorber (brown) is adjusted to intercept the protons at an impact parameter that minimizes out-scattering.

Figure IV-34 displays the TRACK output for beam transport from the pulsed linac to the Recycler. The entrance to the transport line is on the left, and the figure ends with the injection stripping foil in the Recycler on the right. In this simulation 1 million particles were tracked through the beam line and the rms and 100% amplitudes are displayed: horizontal (blue) and vertical (red). The top figure displays the amplitudes when the collimation system is <u>not</u> engaged. The bottom figure displays the amplitude of the particles when the collimation system was set to intercept 1% of the particles. This particular simulation is based on a prior version of the beamline containing six collimating stations (three horizontal and three vertical) and 60º cells. The effect of the six stations is readily seen in the maximum particle amplitudes on the left hand side of the plot. As can be seen the collimation system is effective in reducing the maximum particle amplitudes, with essentially no impact on the rms beam size. Figure IV-35 shows the resultant x-y beam distribution at the Recycler stripping foil location. In this figure, the width of the uncollimated beam distribution (shown in red) is almost a factor of three larger than the collimated distribution (shown in green). It is clear that this collimation scheme reduces the required injection foil size substantially, as well as minimizing the number of H⁻ in the tails that miss the injection foil. This will, in return, help reduce the activation in the injection area of the Recycler.

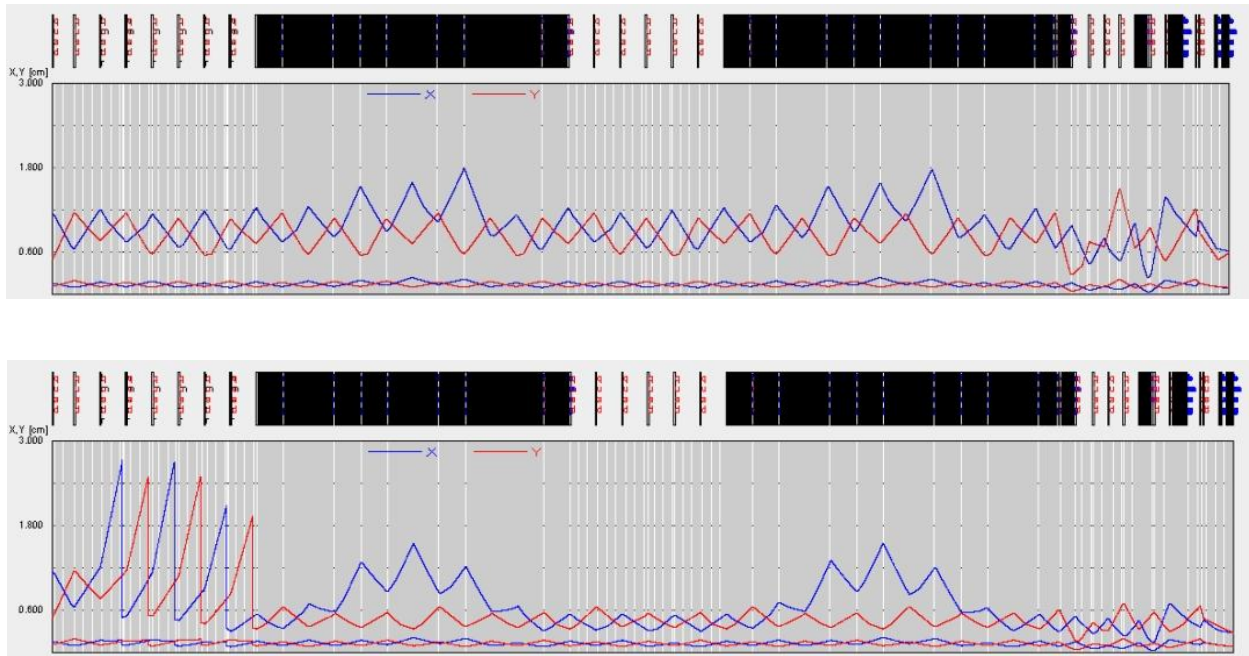

**Figure IV-34**: Comparison of maximum (100%) and rms particle amplitudes for no collimation (top) and collimation (bottom).

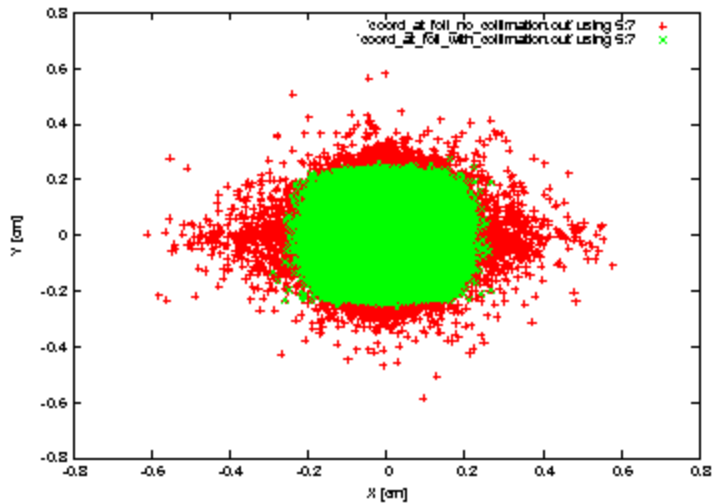

**Figure IV-35**: Beam distribution at the Recycler injection foil location, showing the impact of transverse collimation. Red is uncollimated and green is the collimated distribution.

The movable jaw absorbers were modeled in MARS to determine the requirements on the collimation system shielding which will meet Fermilab radiological requirements. Figure IV-36 shows the layout of the front face and top view of a horizontal absorber. The absorber consists of a 14×18×1 cm stainless steel vacuum chamber with two movable tungsten collimator jaws (magenta). The vacuum chamber is surrounded by 45 cm of steel and 10 cm of marble. The absorber jaws are 10 cm inside the front and rear face of the collimator. Also included in the model, but not shown, are the holes through the shielding for the motion control of the collimator jaws.

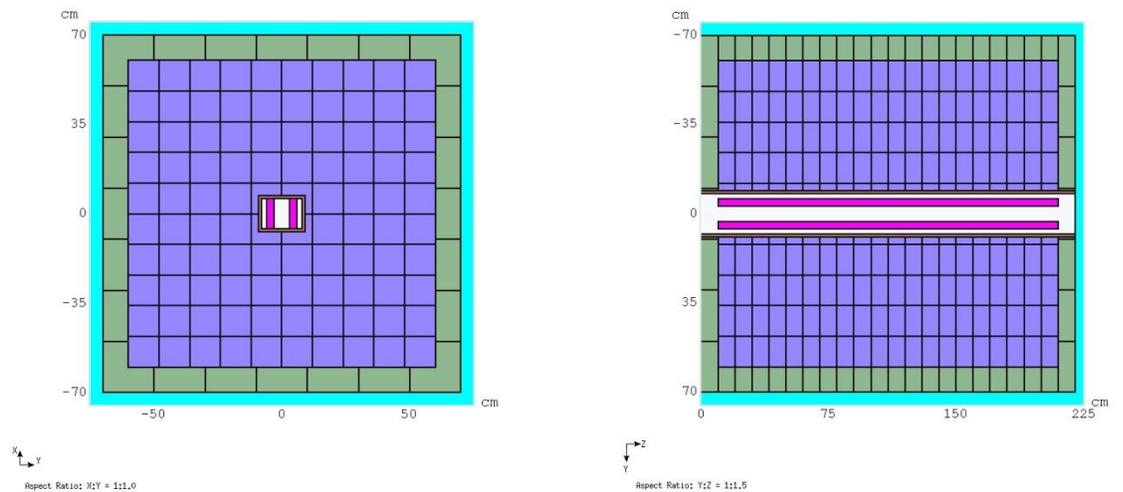

**Figure IV-36**: Geometry of the MARS model for a movable jaw transverse collimator.

Although this simulation is for 8 GeV transport, the same systems could also be installed in both 1 and 3 GeV transfer lines if it were deemed necessary. The expectation is that these will be installed in a straight section located in the 1 GeV transport line between exiting the Tevatron enclosure and entering the Booster enclosure. The utilization in the 3 GeV line requires further study.

### IV.2.2 Recycler Injection Beam Absorber

The Recycler accumulates six pulses (spaced at 100 ms) containing $2.6\times10^{13}$ protons each, every 1.2 seconds for the Main Injector 120 GeV neutrino program. This leaves six additional 10 Hz linac pulses that could be utilized for an 8 GeV program based off the pulsed linac or Recycler. The total injection beam power for the 120 GeV program is about 170 kW. If we assume all 10 Hz linac pulses were injected into the Recycler, the maximum delivered beam power from the pulsed linac would increase to approximately 350 kW.

The beam power delivered to the injection absorber under nominal operating conditions is due to 1) unstripped or partially stripped ions exiting the foil and 2) H⁻ ions missing the foil. Although the foil thickness would be sized to give greater than 99% stripping efficiency, we assume for these calculations that 2% of the injected H⁻ passing through the foil exit the foil as unstripped H⁻ or neutrals. Depending on the transverse tail distribution of the incoming H- ions, a conservative approach is to assume that up to 3% of the beam in the tail could miss the foil and be directed to the absorber. Therefore, we assume 5% of the injected beam as a load on the injection absorber under normal operating conditions. The beam absorber is designed conservatively to accept 10% (which is consistent with designs at other facilities, in particular SNS) of the maximum beam power deliverable, i.e. 35 kW to allow for commissioning activities and abnormal injection conditions.

The design of the Recycler injection optics yields itself to a short dump line to the injection absorber. The current concept is to install the absorber in the MI-10 alcove and provide sufficient shielding to satisfy ALARA requirements. Although the goal is to utilize the existing alcove, it is recognized that some civil construction may be required to place the absorber farther from the ring centerline. The geometry of the absorber is shown in Figure IV-37. This particular layout is for a Main Injector absorber – Figure IV-38 compares the elevation of the absorber as utilized in the Recycler.

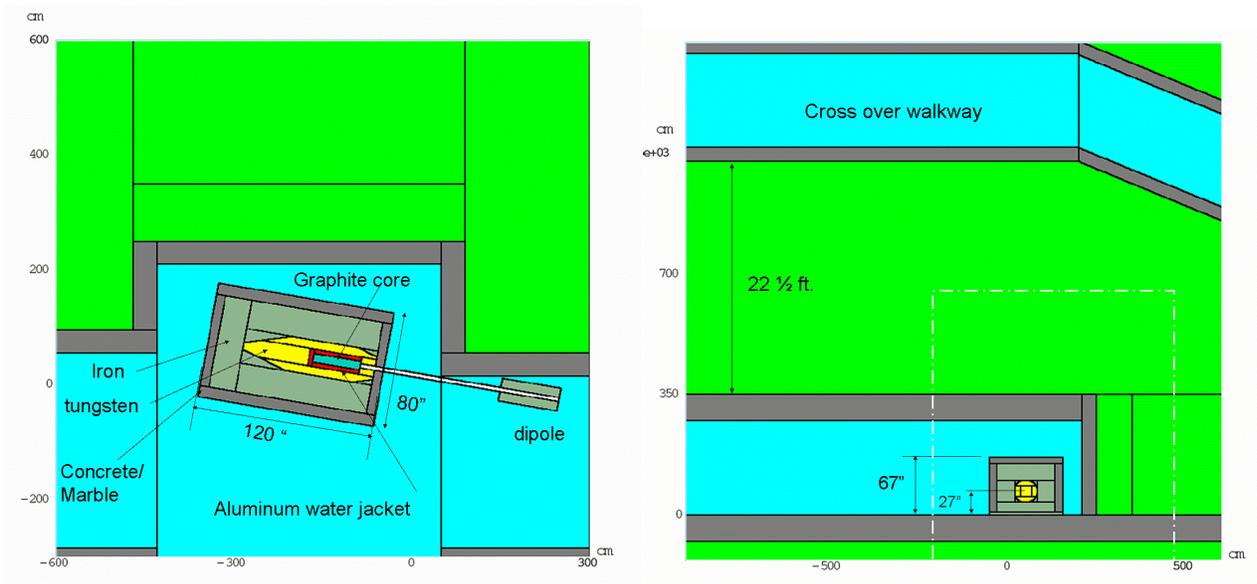

**Figure IV-37**: Mars model for injection absorber in the Main Injector.

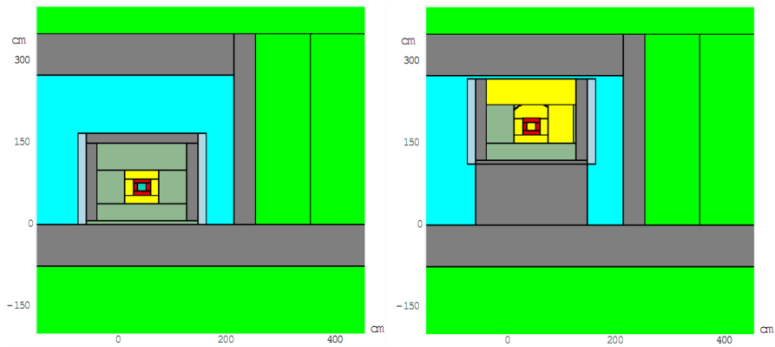

**Figure IV-38**: Comparison of potential absorber elevations for MI injection (left) and Recycler injection (right).

Much of the initial shielding design work was done with previous incarnations of Project X with the absorber located at the Main Injector elevation (Figure IV-37) with different injection scenarios and expected beam powers. These investigations showed that the prompt dose in the cross over walkway was ~0.002 mrem/hr for 6.5 kW on the beam absorber. The same simulation showed residual dosages (with 30-day irradiation and 1-day cool down) on the outside of the marble to be less than 40 mrem/hr and the star density (for sodium and tritium production estimation) in the unprotected soil at the location of the sump drains to be a factor 10 less than the allowed levels [49]. Therefore, the conceptual design described in the reference easily met the radiological requirements for a beam power of 6.5 kW on the absorber.

With the current maximum beam power of 35 kW into the injection absorber, the previous results can roughly be scaled by the absorbed power. Using this scaling, the prompt dose rate in the crossover walkway is still 0.01 mrem/hr, still well below the limit of 0.05 mrem/hr for unrestricted access. Scaling the residual dose on the outside of the marble yields an unacceptable level of over 200 mrem/hr which will require additional attention. Scaling the star density in the soil still meets the requirements. Additional simulations for the Recycler installation and current expected injection scenario and beam powers are required.

With the LBNE plans to use the Main Injector straight section for extraction, the elevation of the injection absorber is raised to the Recycler elevation. It will be necessary to investigate changes in shielding for prompt dose in the walkway and star density in the unprotected soil. This analysis was completed in 2008 for the then assumed beam power of 10% of 1MW into the injection absorber. Figure IV-39 shows the star density surrounding the absorber in this configuration. This simulation showed that for 100 kW into the absorber the star density under the enclosure was at the limit for tritium and sodium production. Scaling these results to the present configuration, the star density is reduced to 35% of the limit.

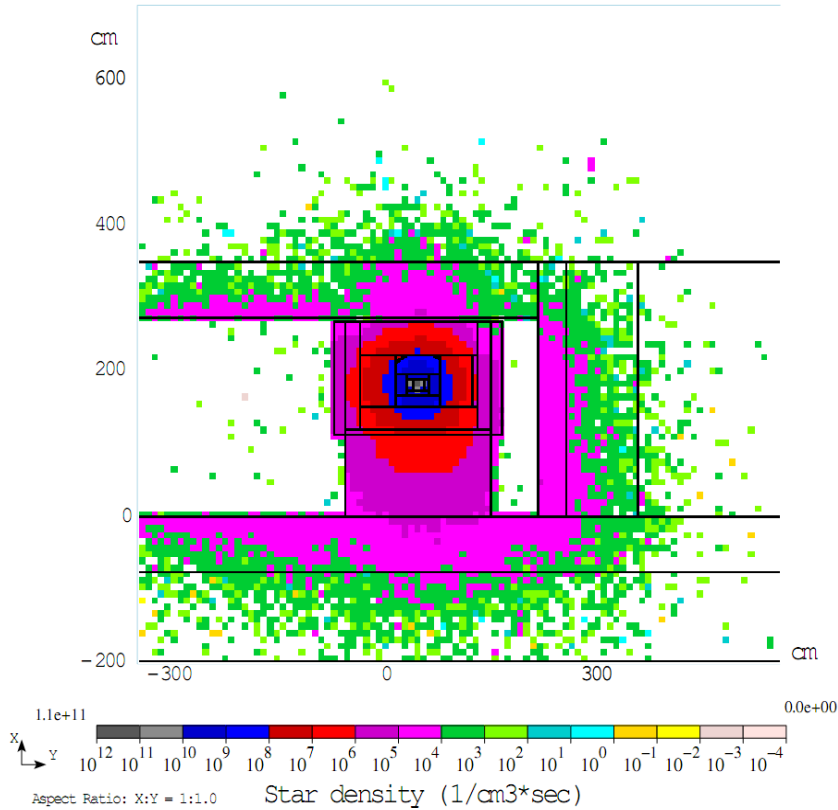

**Figure IV-39**: Star density around a Recycler injection absorber model with 100 kW into the absorber shows that the star density is at the radiological limit for surface water.

The initial injection absorber core design was patterned after the MI abort core, namely a graphite core surrounded by an aluminum water jacket. Simulations showed the maximum energy deposition localized outside the aluminum water jacket leading to heating of the shielding. In an effort to keep all the energy deposition within the aluminum water jacket (without lengthening the core) a sandwich core was investigated. This injection dump core design is shown in Figure IV-40. The core is a sandwich of alternating graphite and tungsten disks inside a sealed aluminum water jacket. The concept of alternating density of the core allows the energy deposition to be spread out over a larger distance thus reducing peak internal temperatures and forces. Further optimization of disk thicknesses is expected to reduce peak temperatures even further.

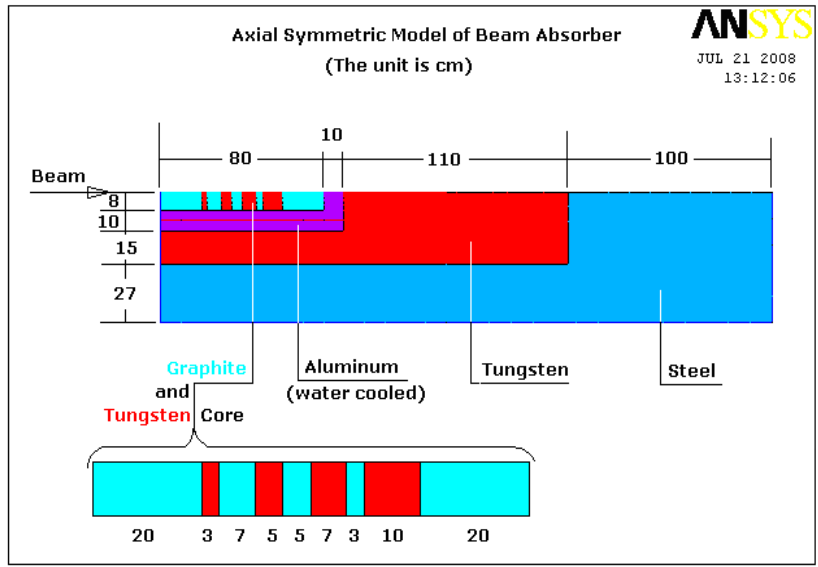

**Figure IV-40**: Geometry of the ANSYS model of Recycler injection absorber.

The thermal and stress simulations were completed at 36 kW. The design is robust against several failure modes (i.e. loss of cooling for 10 minutes and 2 full intensity pulses of 360 kW). Figure IV-41 and Figure IV-42 show results for 36 kW of steady state input power. Spreading out the energy deposition over a larger distance allows the absorber shielding to remain at room temperature. In addition, the stresses induced in the core from the increase in temperature, are well below the yield points of any of the materials. These studies indicate that under assumed operating conditions in the Reference Design the injection absorber provides a performance margin of between six and twelve. While these simulations indicate that the absorber concept is viable, more detailed simulation with more accurate assumptions on beam power will be pursued in the future.

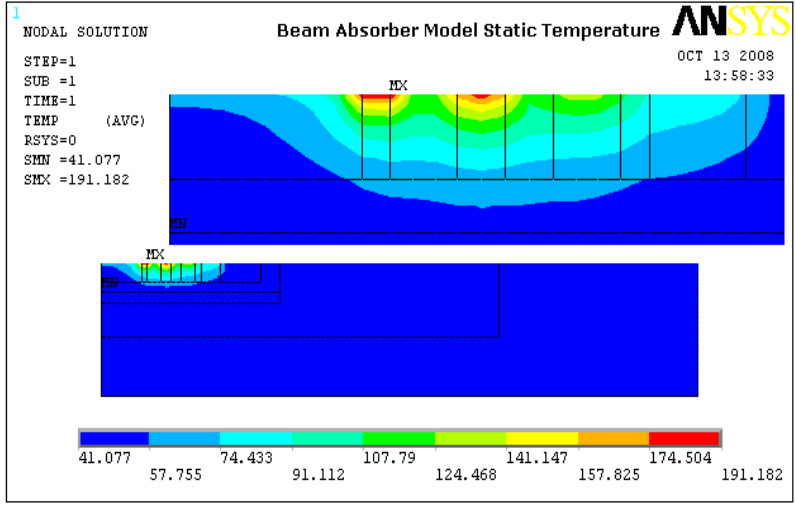

**Figure IV-41**: Steady state temperature in the absorber core for 36 kW incident beam power.

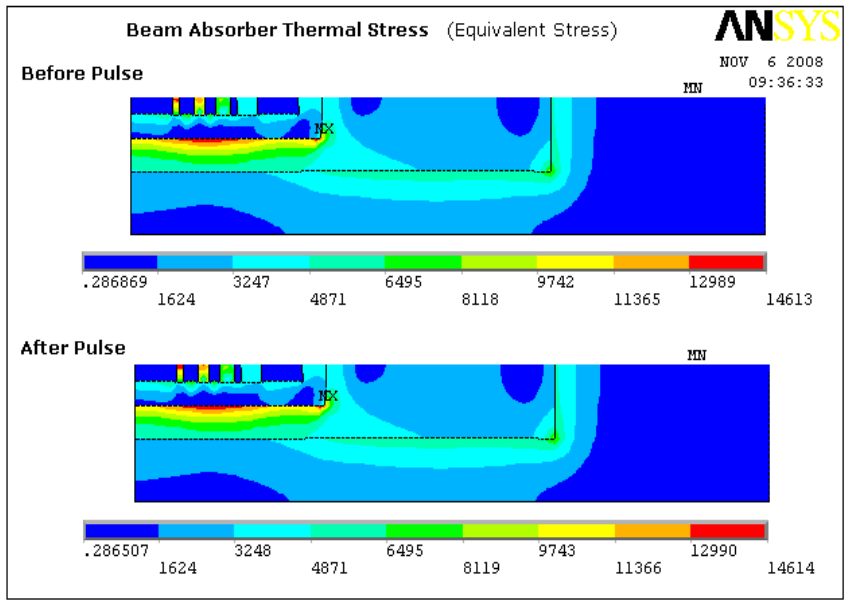

**Figure IV-42**: Thermal stress in the absorber core due to beam heating at 36 kW.

## IV.3 3-8 GeV Pulsed Linac

### IV.3.1 Accelerating Structures

Acceleration from 3 to 8 GeV in pulsed linac is provided by a β=1.0, 9-cell cavity operating at 1300 MHz. A photograph of the ILC 9-cell cavity is given in Figure IV-43 . Performance requirements for the 1300 MHz cavities are summarized in Table IV-15.The gradient is chosen to be 25 MV/m. For this gradient we expect a $Q_0$ that is at least $1.0 \times 10^{10}$ – see Figure IV-44 where the test results for various TESLA/ILC type 9-cell cavities produced and tested at DESY and Americas region are summarized.

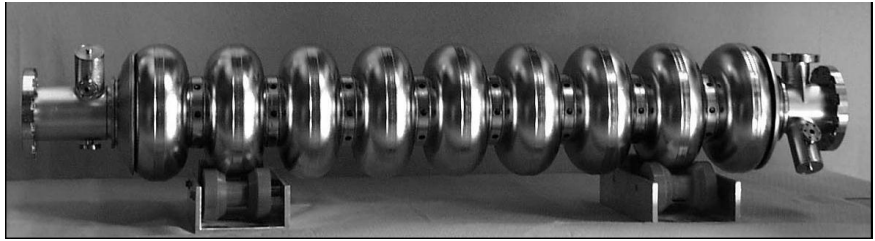

**Figure IV-43**: ILC 1300 MHz accelerating cavity

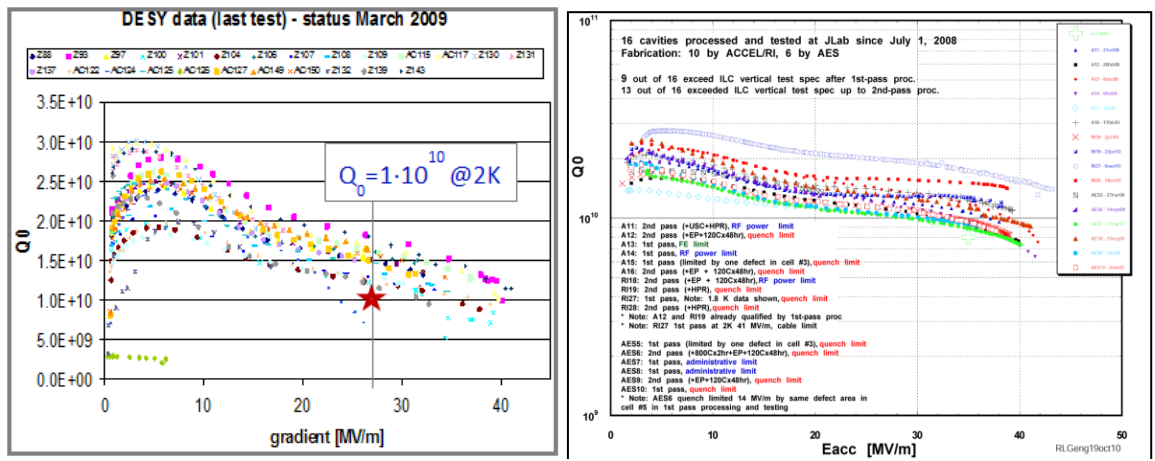

**Figure IV-44**: Data of Q measurements in latest cavities achieved at DESY (left) and Americas region (right)

| 1.3 GHz Cavity | Requirement | |
|---|---|---|
| Frequency | 1300 | MHz |
| Geometric Beta ($\beta_G$) | 1.0 | |
| Optimum Beta ($\beta_{OPT}$) | 1.0 | |
| Iris diameter | 70 | mm |
| Cells/cavity | 9 | |
| Cavity Length | 103.8 | cm |
| R/$Q_0$ | 1036 | Ω |
| G = $Q_0 R_S$ | 270 | Ω |
| $E_{max}/E_{acc}$ | 2.0 | |
| $B_{max}/E_{acc}$ | 4.2 | mT/(MV/m) |
| Gradient | 25 | MV/m |
| Peak Surface Electric Field | 50 | MV/m |
| Peak Surface Magnetic Field | 106 | mT |
| Operating Temperature | 2 | K |
| $Q_0$ at 2 K | 1.0 | $10^{10}$ |
| Losses@2K | 5 | W |
| Pulse length | 7.4 | ms |
| Flat-top | 4.3 | ms |
| Rep. rate | 10 | Hz |

**Table IV-15**: RF parameters of the ILC-type 1.3 GHz cavities.

The accelerating gradient and $Q_0$ shown in Table IV-15 are quite modest for that frequency. Recent progress in 9-cell 1.3 GHz cavity production and surface treatment supported by ILC and XFEL R&D programs push SRF technology up. As today more that 80% of cavities produced in Americas region demonstrated the accelerating gradient above 25 MV/m at the first test. Cavities not passing through 1st test were re-processed and tested again. After second test yield was increased up-to 90% as shown in Figure IV-45.

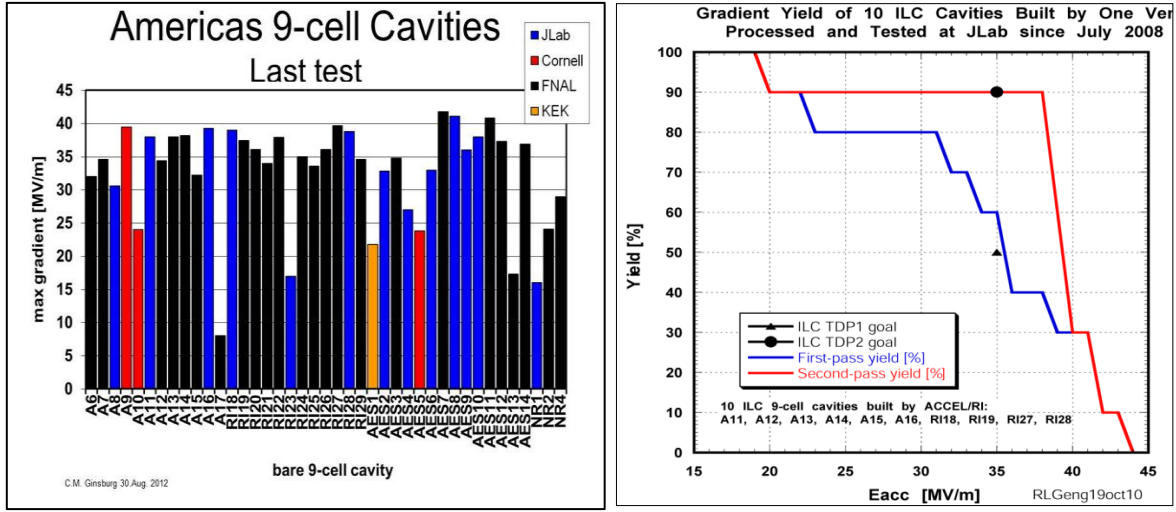

**Figure IV-45**: Maximum accelerating gradient achieved in American region (left) and gradient yield for 10 cavities build by one vendor (right).

A total of 224 cavities are required in the 1300 MHz section. These cavities are contained within 28 cryomodules. Each cryomodule contains 8 accelerating cavities and one focusing quadrupole. Figure IV-46 shows the ILC Type-4 cryomodule and schematically accompanied by the disposition of focusing elements within neighboring cryomodules. The requirements are summarized in Table IV-16.

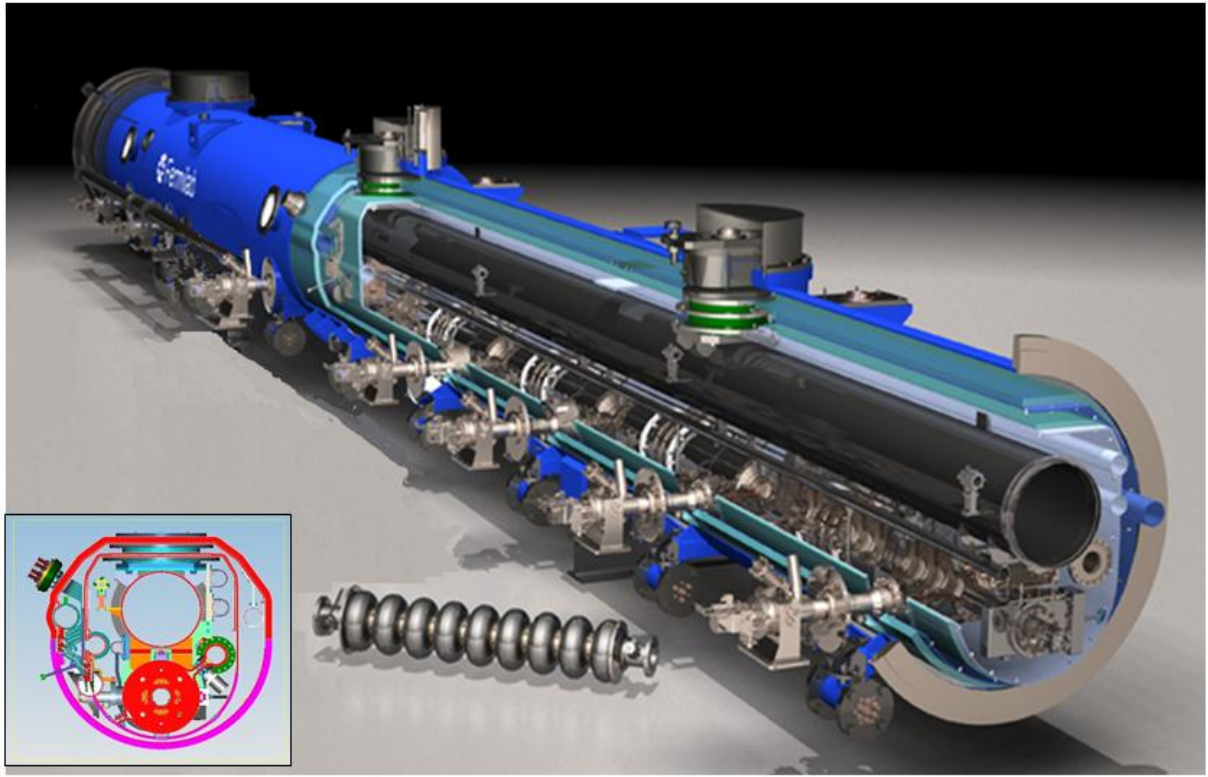

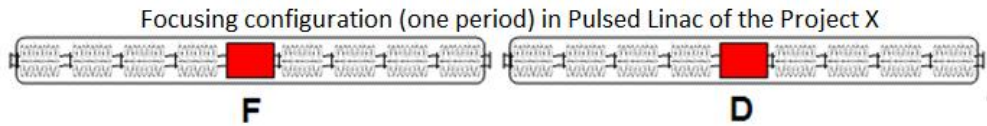

**Figure IV-46**: Type-4 ILC cryomodule (upper) and focusing schematics (lower).

|  | ILC-1.3 GHz | Units |
|---|---|---|
| Number of cavities | 224 |  |
| Number of SC Quadrupoles | 28 |  |
| Number of Cryomodules | 28 |  |
| Total Length | ~390 | m |
| Period Length | 25.23 | m |
| Number of FODO periods | 14 |  |
| Energy | 3 - 8 | GeV |
| Range of β | 0.973 - 0.99 |  |

**Table IV-16**: Summary table of 1.3 GHz cryomodules.

IV.3.2 **RF Power**

There are a total of 28 RF systems in the pulsed 3-8 GeV linac. These are distributed over 224 cavities in 28 cryomodules. The beta of the beam at the entrance to the pulsed linac is close to one; hence a single klystron power source can drive 8 cavities (one cryomodule). The RF power required for the cavity depends on loaded Q and cavity detuning as shown in Figure IV-47. The loaded Q of the cavity was chosen $Q_L=1\times10^7$, as a compromise between required power and cavity passband (Note: for an ideally tuned cavity the minimum power corresponds $Q_L=2.5\times10^7$ for 1 mA and 25 MV/m). Assuming control of microphonics and uncompensated Lorentz forces to limit detuning to less than 30 Hz maximum, the required RF power to a cavity is 35 kW. Taking into account the required power overhead for LLRF control and compensation of RF losses, the set power is 50 kW/cavity.

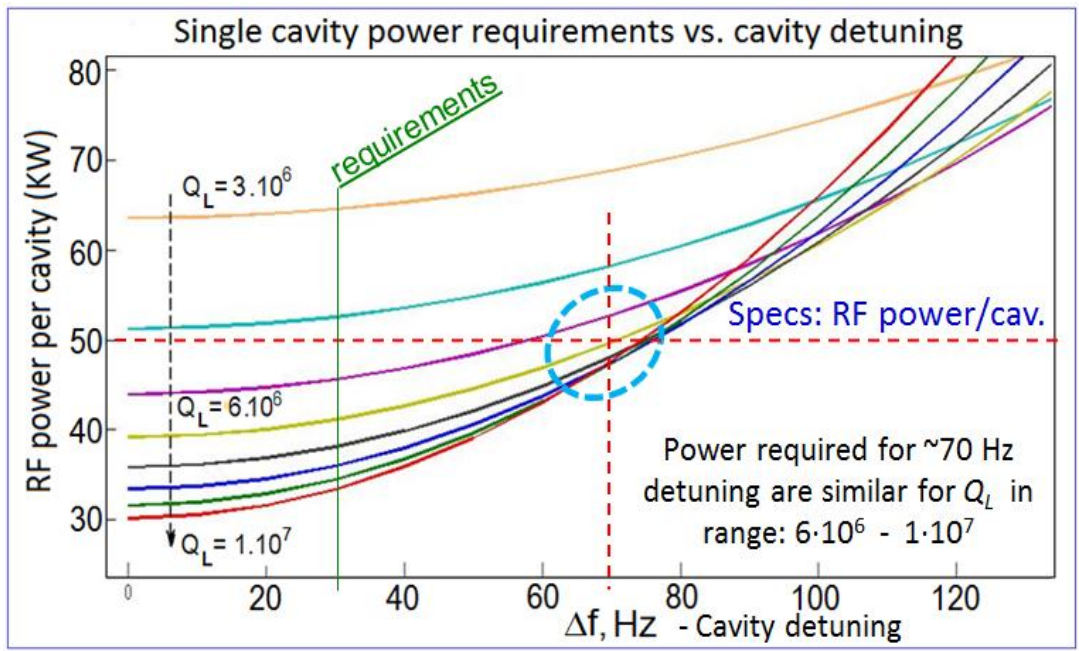

**Figure IV-47**: Required cavity RF power as function of cavity detuning for different loaded Q-values. Parameters: Gradient=25 MV/m, Beam current 1 mA.

In baseline configuration one RF station will feed one or two cryomodules. The RF power distribution scheme would be very similar to that of the XFEL, with a vector-sum feed-back control providing both phase and amplitude control. The required RF power source parameters for one cryomodule are shown in Table IV-17. Two possible types of power sources are considered: single-beam klystron (SBK) and multi-beam klystron (MBK). Klystron parameters are shown in Table IV-18.

| Parameter | Recycler/MI | Direct Injection to MI | Units |
|---|---|---|---|
| Frequency | 1.3 | 1.3 | GHz |
| Loaded Q | 1.e7 | 1.e7 | |
| RF pulse width | 7.4 | 30 | ms |
| Cavity Gradient | 25 | 25 | MV/m |
| Beam current | 1 | 1 | mA |
| Repetition rate | 10 | 10 | Hz |
| Cavity RF power | 32 | 32 | kW |
| Cavity power + losses + regulation +EOL | 50 | 50 | kW |
| Power per Cryomodule | 400 | 400 | kW |

**Table IV-17**: RF power requirements for a Pulsed Linac cryomodule

| | Klystron per 1CM | | Klystron per 2 CM's | | Units |
|---|---|---|---|---|---|
| Klystron type | SBK | MBK | SBK | MBK | |
| Klystron power | 400 | 400 | 800 | 800 | kW |
| Voltage | 54 | 33 | 71 | 44 | kV |
| Current | 12 | 20 | 19 | 30 | A |
| Efficiency | 60-62 | 62-64 | 60-62 | 62-64 | % |
| Average Power | 32 | 32 | 64 | 64 | kW |
| Number of klystrons/linac | 28 | 28 | 14 | 14 | |

**Table IV-18**: Klystrons parameters for the Pulsed Linac. Only the 800 kW SBK would require a modulator with oil.

One possible candidate for the pulsed linac power source is a MBK under development by Omega-P as a Small Business Innovative Research (SBIR) project (shown in Figure IV-48 with list of parameters in Table IV-19). As alternative solutions we considering source based on Injection locked 2-stage magnetron (SBIR proposal by Muons Inc). The schematic of magnetron station for 1 cryomodule is shown in Figure IV-49 with two 200 kW magnetrons per 1 cryomodule.

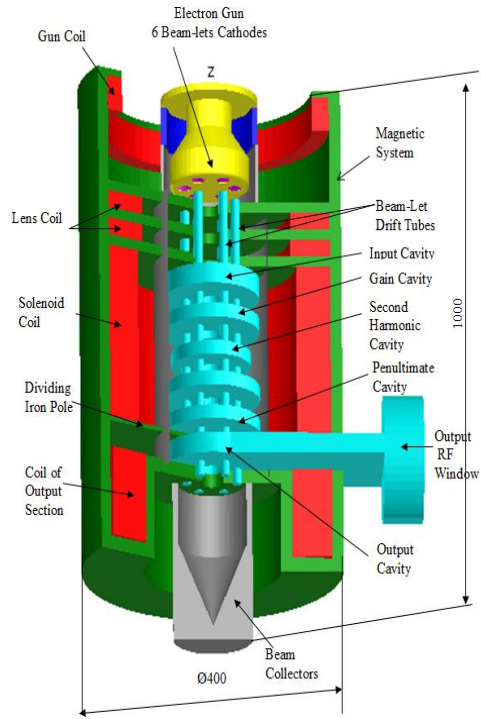

**Figure IV-48**: Multi-beam klystron proposed by Omega-P (SBIR project)

| 500 kW, 1.3 GHz MBK parameters | | |
|---|---|---|
| Operating Frequency | GHz | 1.3 |
| Beam Voltage | kV | 30 |
| Number of Beam-lets |  | 6 |
| Beam-let / total Current | A | 4.4 / 26.4 |
| Beam-let / total perveance | $10^{-6}$ | 0.847 / 5.08 |
| Cathode Loading | A/cm$^2$ | 2 |
| Beam / RF Power | kW | 792 / 515 |
| Saturated Efficiency | % | 65 |
| Saturated Gain | dB | 46 |
| Pulse length | ms | 8.4 / 30 |
| Rep. rate | Hz | 10 / 2 |
| Brillouin / operating B$_{solenoid}$ | Gs | 400 / 800 |
| Power of Magnetic System | kW | 2.5 |
| Max. collector power | kW | 60 |

**Table IV-19**: Multi-beam klystron Parameters

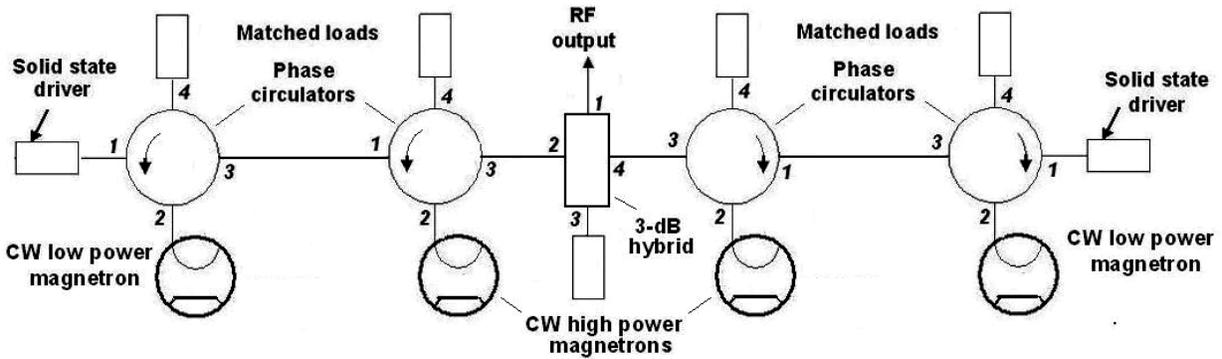

**Figure IV-49**: Schematic of a magnetron based RF source (SBIR proposal by Muons Inc.)

The low level RF (LLRF) system will provide a drive signal on the order of 0 to +10 dBm for each RF power source. The amplifier(s) will provide sample signals of the pre-driver and final outputs. All amplifiers will be self-contained units complete with integral power supplies, protection circuits, and control interface. The RF distribution system for the pulsed linac will use WR650 waveguide. The final connection to the cryomodules will utilize a section of flexible transmission line to minimize connector location tolerances. Each RF system will have a circulator and load to isolate the cavity from the power amplifier. This level of protection is essential in SRF systems due to full power reflection from the cavity in the absence of beam. Cavity and drive sample signals will be provided to the LLRF for vector regulation and frequency control of the cavities. All of the RF amplifiers will be water cooled to minimize the heat load to the building HVAC system.

While each amplifier has built in protection which includes, water flow, water temperature, pressure differential, and reflected power monitoring; a global interlock and hardware protection system will need to be designed for all RF systems. This will include water flow to loads and circulators, spark detection on cavity couplers, and RF leakage detection.

## IV.4 MI/RR

The increased intensity provided by the Project X linac requires upgrades to several areas in the Recycler and Main Injector in order to achieve the performance goals listed in Table II-1. These upgrades are described below.

IV.4.1 **H- Injection**

Injection into the Recycler for the Neutrino program requires the accumulation of about 26 mA-ms. With the average pulsed Linac current of 1 mA the required injection time would be 26 ms. The current concept is to use six injections each of 4.3 ms in duration. With the revolution period of the Recycler being 11.2 μs, this corresponds to approximately 388 turns injection time. With the Linac rep rate of 10 Hz, the total injection time into the Recycler is 0.6 seconds or 1 injection every 100 ms. There are two potential techniques that are under discussion for accomplishing the multi-turn H- injection into the Recycler, a standard carbon foil charge exchange technique and a new Laser Assisted H- stripping injection technique being pioneered at SNS. R&D for both techniques is being pursued and the salient features of each will be addressed. Currently, the carbon foil charge exchange is the default technique. Independent of stripping technique, transverse phase space painting must be utilized due to the difference in transverse emittance between the linac beam and final emittance of the circulating beam in the Recycler. Longitudinally, we will be utilizing micro bunch injection where individual linac micro bunches will be injected into already formed Recycler RF buckets. This will require accurate control of the phasing between the linac and Recycler RF.

The Recycler straight section 10 will be modified to be a symmetric straight for the purposes of H- injection. This has been discussed previously. Figure IV-50 and Figure IV-51 show the lattice functions of the modified Recycler and the layout of the injection chicane used for merging the H- onto the recycler closed orbit and transverse phase space painting. Both have zero dispersion and D'=0 to eliminate targeting position of the H- on the foil due to changes in momentum.

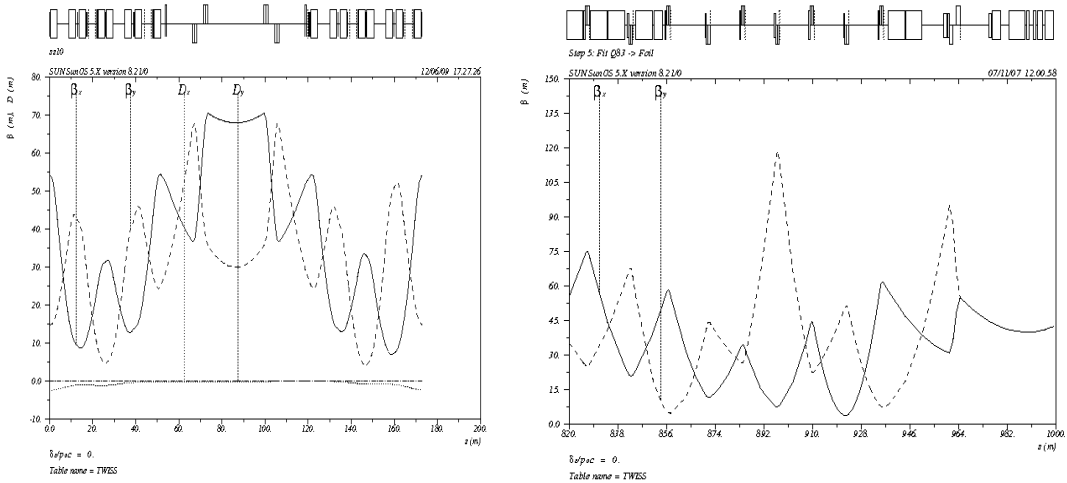

**Figure IV-50**: Lattice functions for of the modified Recycler straight section 10 for H- multi-turn injection (left) and the lattice functions of the end of the transport line (right). Note that both the ring and transport line have α=0 at the foil location.

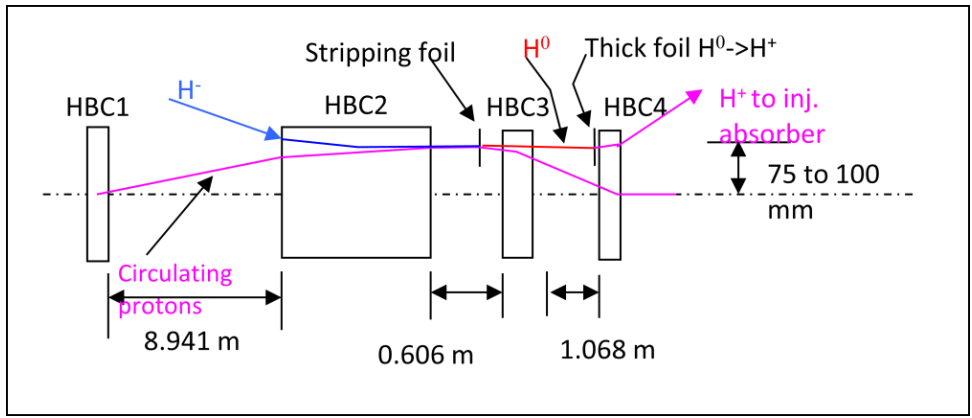

**Figure IV-51**: Layout of the recycler injection chicane showing the location of the "stripping foil" and the incoming H-, circulating proton, and waste beam trajectories.

In Figure IV-51 the "stripping foil" is the location where the H⁻ is converted into protons once the H⁻ have been merged onto the proton closed orbit. This conversion process could utilize either carbon foil or laser method of converting the H⁻ into protons. Any H- that miss the foil or $H^0$ created in the foil will be completely stripped to protons for loss free transport to the injection absorber. The entire injection chicane is located within the doublets which form the symmetric straight section. Since the Recycler is used only as an accumulation ring and the accumulated charge will be transferred to the Main Injector immediately after the sixth injection, it is not critical that the injection chicane magnets be ramped. It is envisioned that they are still electromagnets.

Transverse phase space painting involves the systematic moving of the ring closed orbit about the fixed "conversion point" of the incoming H⁻ ions. Figure IV-52 shows an example of this motion in the horizontal plane for a particular painting scenario. Here the green ellipse represents an empty circulating proton phase space determined by the final painted ring emittance. At the start of the painting injection process the proton closed orbit determined by the fast painting magnets (red) is bump to the outside of the ring and centered on the injection point for the H-. As the injection process proceeds the amplitude of the closed orbit is slowly collapsed with the painting magnets to the point of the red ellipse. This painting distance is equal to 3σ of the final painted ring emittance. In the figure a second step for removal of the circulating beam is shown. Here, the painting magnets are fully collapsed. It should be noted that the location of the purple ellipse is determined by the chicane dipoles.

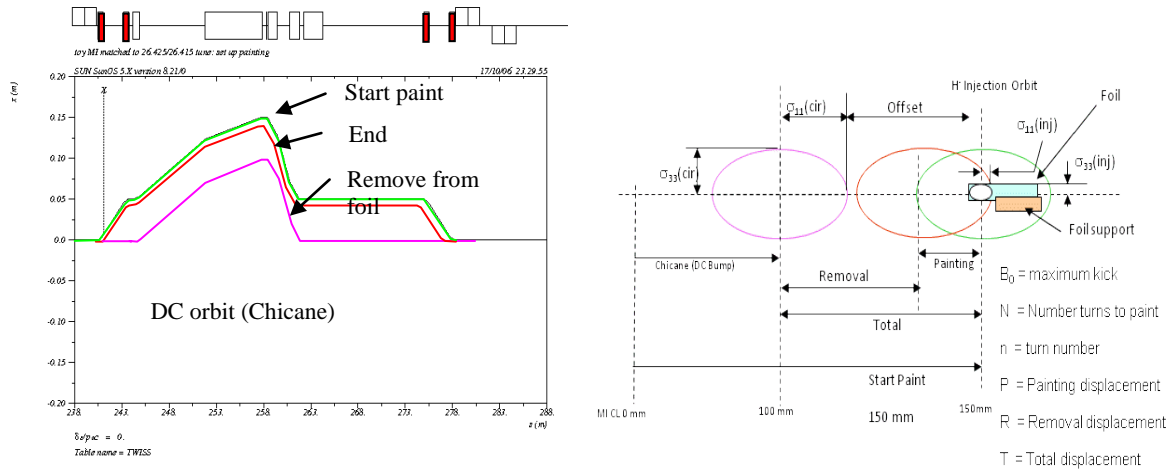

**Figure IV-52**: Example of movement of the circulating proton phase space for horizontal painting.

There are numerous transverse phase space painting scenarios that can be investigated, each with their own advantages and disadvantages. The injection chicane design would support any proposed painting scheme. The scheme considered at this stage is one that has been adopted by JPARC for their carbon charge exchange injection where they utilize an anti-correlated horizontal painting in the ring with vertical steering from the beam line (fixed y and varying y'). In this scheme, the foil is placed on the median plane where secondary electrons are collected by an electron catcher on the side of the aperture. This painting scheme has been detailed elsewhere and results in a uniform distribution in x and y.

When considering foil injection, five main issues need to be addressed: 1) foil temperature and lifetime, 2) handling of secondary electrons, 3) minimizing secondary foil hits by circulating protons, 4) losses due to $H^0$ excited state produced by the incomplete stripping process and interactions with the foil, and 5) beam power directed to the injection absorber. Many of these can be addressed by design.

In injection the scenario considered here where the incoming H- beam current is low which requires longer injection periods (number of turns), the largest contribution to foil heating comes from circulating beam. This is in contrast to the injection of higher beam current and fewer turns, where the incoming beam is the major contribution to foil heating. Preliminary simulations [50] for six injection painting into the Recycler utilizing the J-PARC scenario for various injection times show peak temperatures approaching 2000°K during the last injection. Figure IV-53 shows peak temperature verses time. In addition, a single injection with a duration of approximately 26 ms is shown. Here, the peak temperature is close to 2500°K and the foil would not survive 1 pulse. A potential technique for reducing the peak temperature during the injection time is to move the foil during the injection time so the energy deposition is spread out over a larger area.

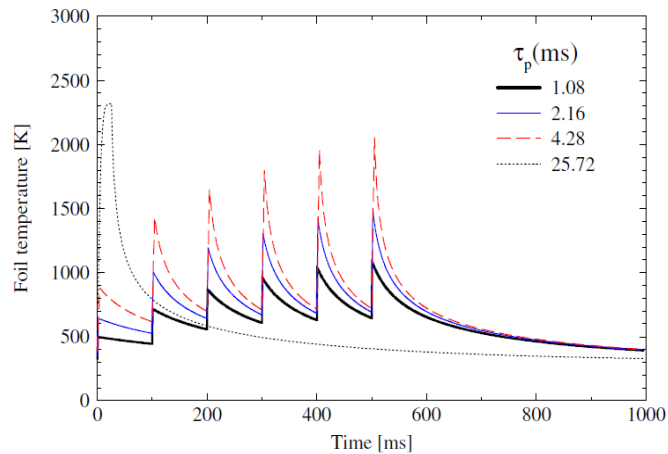

**Figure IV-53**: Peak temperature of foil during the injection of 26 mA-ms charge into the Recycler during various injection times.

A simulation was carried out in STRUCT to simulate a "rotating" foil by moving a linear foil in one dimension and the hit densities along the foil motion were accumulated. This hit density as a function of position and time was used to calculate a peak foil temperature. The results of a stationary foil and two rotation periods were calculated. The results are shown in Figure IV-54.

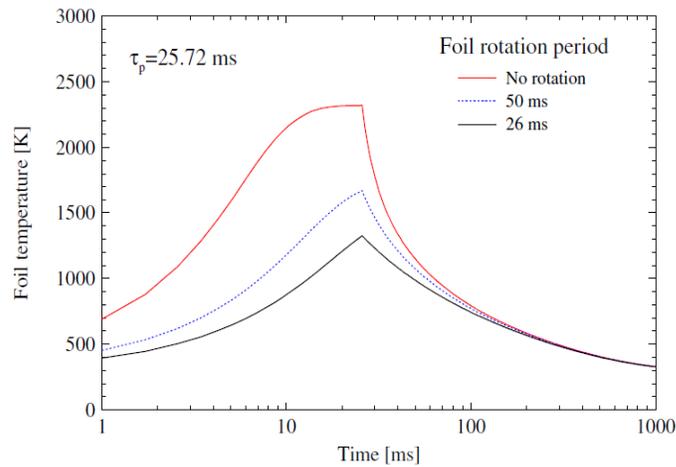

**Figure IV-54**: Temperature evolution during a single 26 ms injection period for a stationary foil (red) and two rotation periods.

Here it is clear that matching the rotation speed to the injection period reduced the peak hit density and thus the peak temperature. The amount of reduction is dependent not only on the speed but the area over which the energy is deposited. It should be clear that a moving foil does not address any issues with losses due to the interaction of the circulating beam with the foil. The concept of a rotating foil is depicted in Figure IV-55. Here, a thin Ultra Nano-Crystalline Diamond film is grown on a Silicon wafer. The outer 1 cm of wafer is removed exposing the UNCD diamond used for stripping. In figure, the green, red, and magenta ellipses represent the start and end of painting and circulating beam, respectively. This concept is being pursued to determine if this concept in feasible. Initial ANSYS simulations indicate that the rotational forces on the foil should be much smaller than the yield strength of the UNCD film. A thermal model was also modeled using ANSYS and indicates the same sort of temperature reductions as the STRUCT simulation. The first stage of the R&D program is to determine if this configuration can be commercially made and the steps involved in producing the foil. To this end we have contracted with several vendors to look at a UNCD and a graphene foil and make us some samples than can be tested. A rotational set up in the laboratory will be used to spin the foil at up to 2000 rpm and measure any deformations. A small spot on the foil will be heated with a laser or micro-torch while the foil is rotating to test the robustness of the foil and determine any deformation. Ultimately, a beam test and temperature measurement are planned. Currently SNS has a foil testing facility that could be potentially utilized for temperature and lifetime tests.

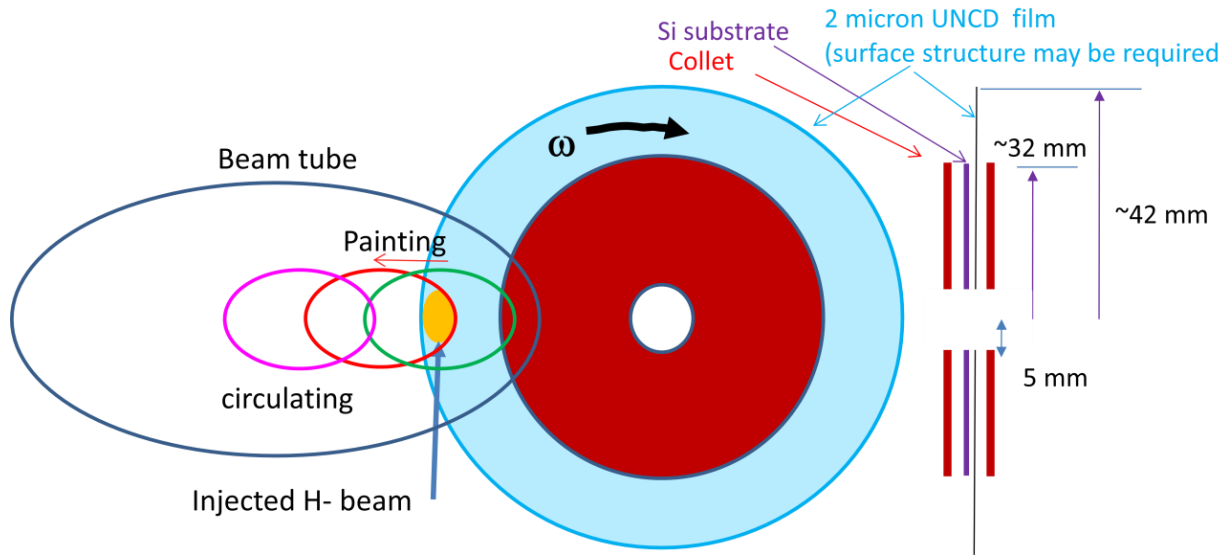

**Figure IV-55**: Concept for a rotating foil.

In addition to the development of a rotating foil system, we are collaborating with SNS on their next phase of laser stripping experiment. The proof of principal stripped a single 400 MHz bunch. The next stage will be to demonstrate stripping for a 10 us beam pulse. Final operational requirements require stripping a 1 ms beam pulse. SNS continues to develop the laser system required, the H- transport optical system, and the interaction chamber and hardware, for the next phase test. Due to the beam energy of 1 GeV at SNS, the energy of the photons needs to be 255 nm (UV). On the other hand FNAL injection into the Recycler at 8 GeV allows the use of 1064 nm (IR) photons. Many of the technical issues associated with UV are not present in the IR.

### IV.4.2 Main Injector $\gamma_t$-jump

Designs for a $\gamma_t$-jump in Main Injector have been studied for the last 15 years. Details can be found in the Proton Driver Design Report [24]; a brief summary is included here. The system consists of 8 sets of pulsed quadrupole triplets. Each triplet has two quads in the arc and one of twice the integrated strength in the straight section, with a phase advance of $\pi$ between each quadrupole. The perturbation to the original lattice is localized. In particular, the dispersion increase during the jump is small ($\Delta D_{max} \approx \pm 1$ m), which is the main advantage of a first-order jump system. Each triplet is optically independent from the others and provides roughly 1/8 of the total required jump amplitude (i.e., $\Delta \gamma t \approx \pm 0.25$ per triplet).

The power supply uses a GTO as the fast switch and a resonant circuit with a 1 kHz resonant frequency. The beam pipe is elliptical and made of Inconel 718. It has low electrical conductivity σ and high mechanical strength so eddy current effects are relatively small. The eddy current effects scale as σd, where d is the pipe wall thickness. The σd value of Inconel 718 is about four times lower than that of stainless steel.

The 8 pulsed triplet locations are summarized in Table IV-20. Since the original study was done, there have been changes to the Main Injector and these locations need to be revisited. A set of magnet design parameters has been developed and modeled (see Table IV-21).

| Pulsed Triplet | Quad Locations |
| --- | --- |
| 1 | 104, 108, 112 |
| 2 | 226, 230, 302 |
| 3 | 322, 326, 330 |
| 4 | 334, 338, 400 |
| 5 | 404, 408, 412 |
| 6 | 526, 530, 602 |
| 7 | 622, 626, 630 |
| 8 | 634, 638, 100 |

**Table IV-20**: Candidate $\gamma_t$ quad triplet locations

|  | Requirement |
| --- | --- |
| Integrated Gradient | 0.85 T |
| Vacuum pipe cross section (elliptical) | 2.4 x 1.125 in |
| Field Quality, 1 inch radius | 2% |
| Maximum length | 17 in |
| Maximum Current | 200 A |
| Maximum Voltage | As low as possible |

**Table IV-21**: Pulsed quadrupole magnet parameters

### IV.4.3 Electron Cloud Mitigation

Electron cloud generation could be a possible instability source for the intensities in the Recycler and Main Injector.

The best approach is to mitigate the generation of the cloud itself. There have been a series of measurements in the Main Injector, looking at secondary electron yield and cloud generation. A dedicated measurement setup now exists at MI-52, with newly developed RFA detectors. Both TiN and C coated beam pipes have been installed and measurements made. Both coatings show significant reductions in secondary electron generation when compared to an uncoated stainless steel pipe. VORPAL simulations are being benchmarked against these measurements. Plan to install a SEY (Secondary Emission Yield) stand in MI in order to measure the effect of scrubbing in situ for different kinds of stainless steel. A picture of the SEY measuring stand is shown in Figure IV-56.

Research continues into the coating process. The Main Injector beam pipe is captured in the dipole magnets, so coating needs to take place in situ. We set up a coating (sputtering) facility in E4R and successfully coated with TiN a 6m long piece of round MI pipe and measured the coating thickness. It will be used to coat test coupons for SEY measurements in MI. The experience from our coating facility will be used to estimate the effort required to in-situ coat the MI beam pipe with TiN.

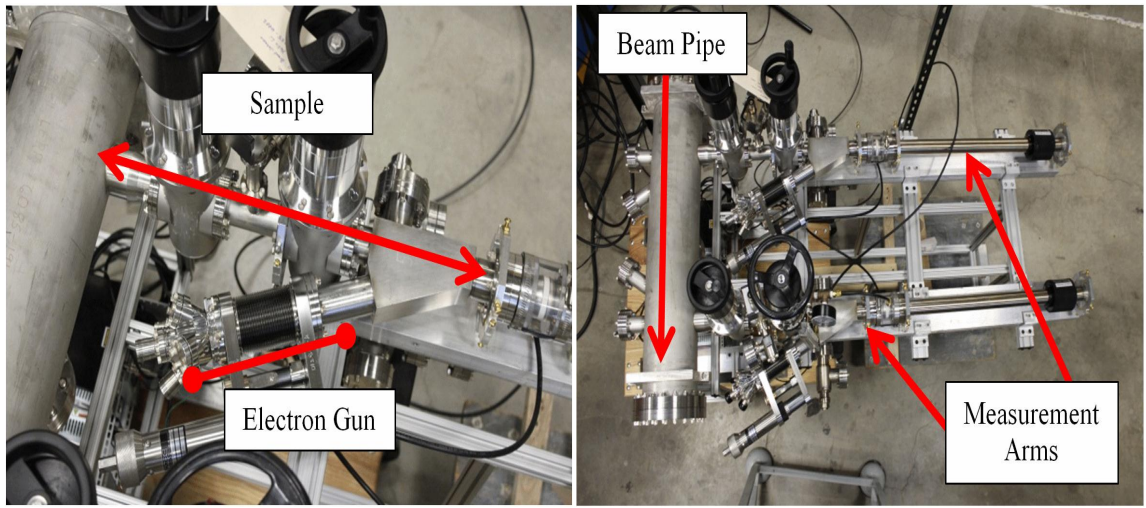

**Figure IV-56**: SEY Measuring Stand

IV.4.4 **RF Systems**

The Recycler and Main Injector need new $1^{st}$ and $2^{nd}$ harmonic RF cavities. The same cavities will be used in both machines. A cavity design has been developed, with perpendicular biased tuners and R/Q ~ 30 Ω. A mechanical drawing of the cavity is in Figure IV-57. The cavity parameters are shown in Table IV-22. Higher Order Mode (HOM) Coaxial dampers for the 53 MHz cavities have been designed. The effect of the HOM dampers on the first 2 monopole cavity modes is shown in Figure IV-58. A mock-up of the first harmonic cavity has been constructed and a set of low level RF measurements were taken. A preliminary design of the second harmonic cavity that is a scaled down version of the first harmonic has been completed.

The power source needs to provide greater than 550 kVA of total power and 4.65 A of current. To simplify operation and maintenance, the source should have enough bandwidth to power both the 1$^{st}$ and 2$^{nd}$ harmonic cavities. The EIMAC 8973 power tetrode amplifier has a maximum operating frequency of 110 MHz, output power capabilities greater than 1 MW, and plate dissipation of 1 MW. An 8973 tube has been purchased and a power test stand is being developed.

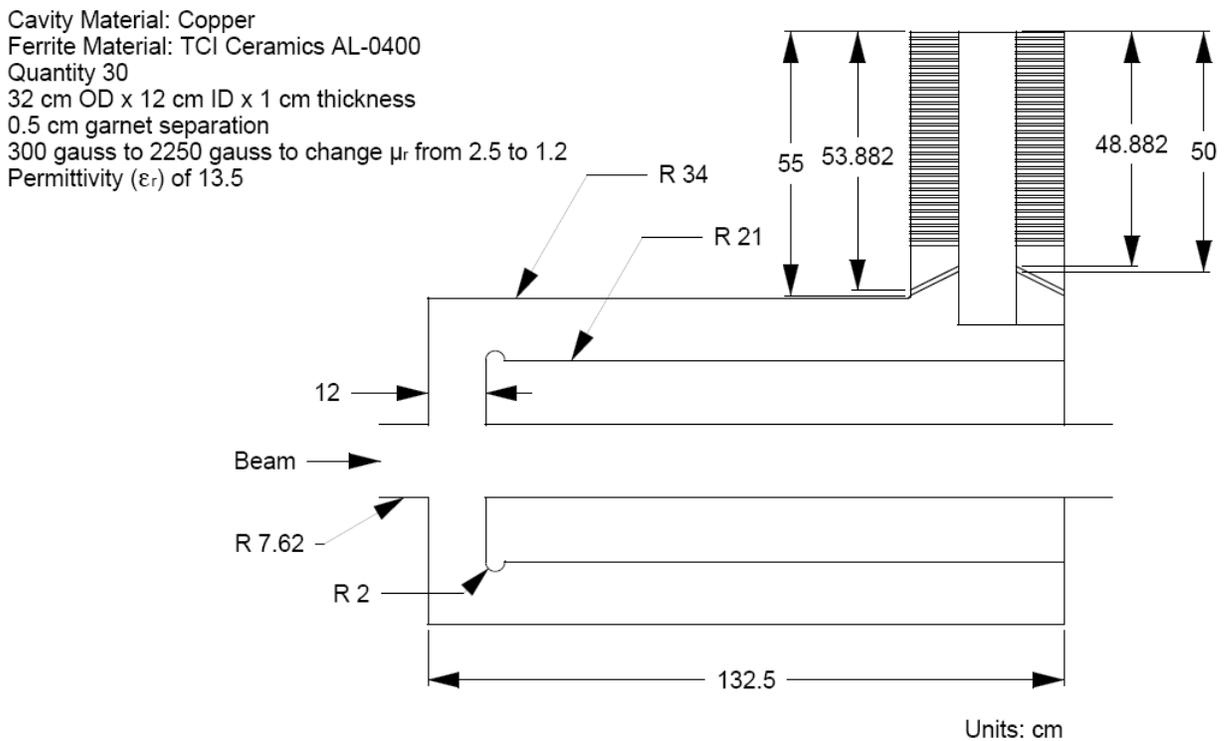

**Figure IV-57**: Mechanical dimensions of the 53 MHz cavity.

| Tuner Intrusion 75 mm @ $V_{pk}$=240 kV | $\mu_r$=1.2 | $\mu_r$=2.5 | |
|---|---|---|---|
| R/Q | 36.4 | 31.5 | Ω |
| $Q_0$ | 12244 | 12023 | |
| F | 53.3047 | 52.6152 | MHz |
| $P_{wall}$ | 64 | 76 | kW |
| $P_{ferrite}$ | 6 | 42 | kW |
| $P_{ceramic}$ | 0.2 | 0.6 | kW |

**Table IV-22**: Parameters of the new 53 MHz Cavity as a function of ferrite permeability.

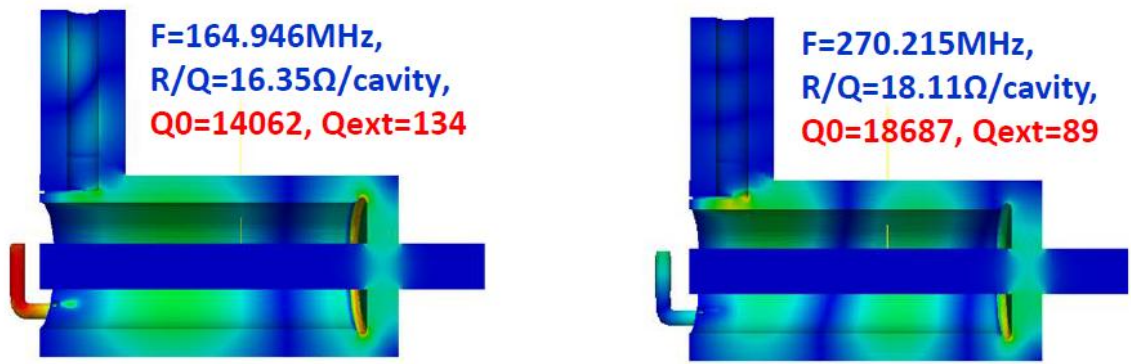

**Figure IV-58**: Effect of the 53 MHz cavity HOM dampers on the first two monopole modes.

### IV.5 Cryogenics

The cryogenic system scope for Project X includes a new cryogenic plant, a cryogenic distribution system, and the necessary ancillary systems (purification system, cryogenic storage, etc.) to support the plant. A conceptual layout for the cryogenic system is shown in Figure IV-59.

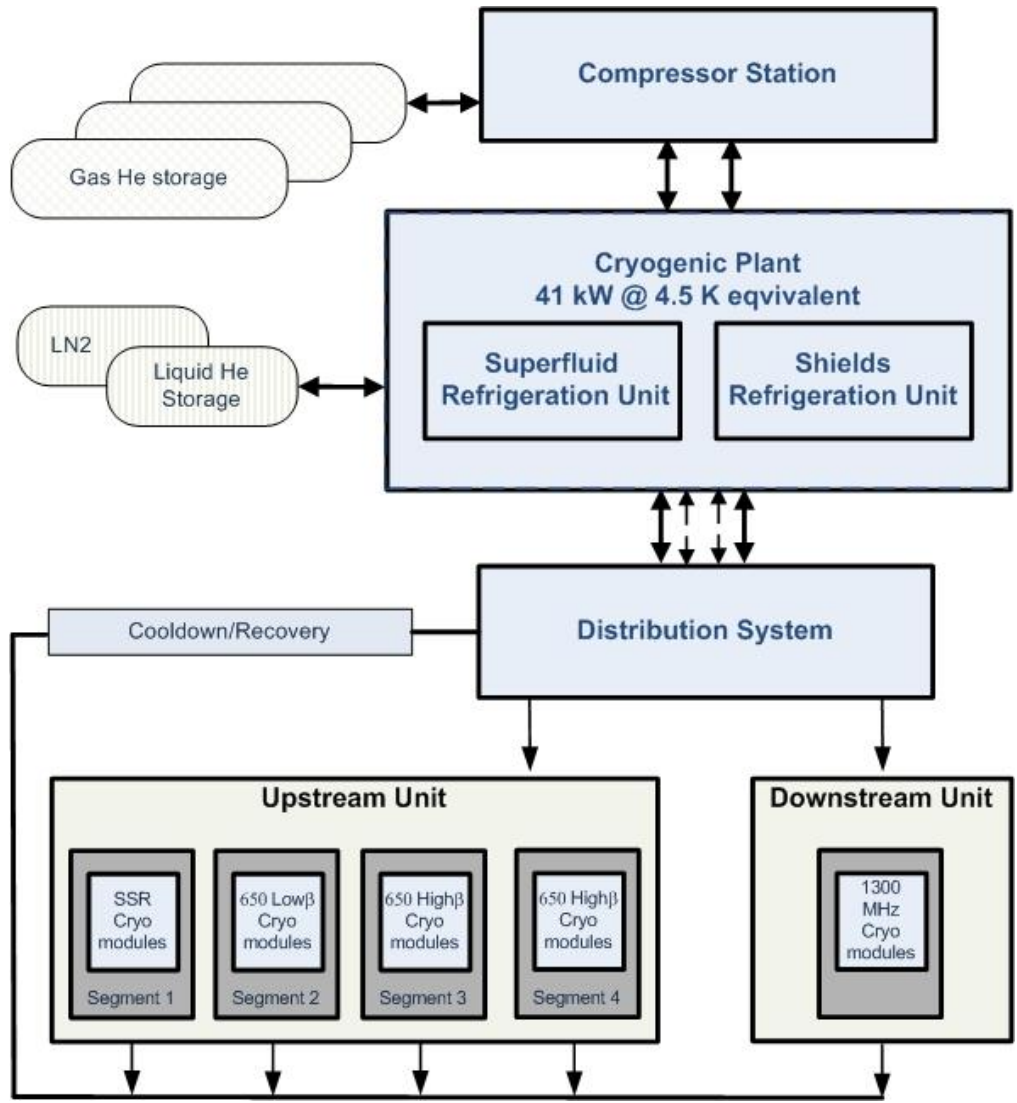

**Figure IV-59**: Conceptual Layout for the Project X Cryogenic System

The cryogenic distribution system accommodates a range of steady state and transient operating modes including RF on/RF off, cool down, and warm-up and fault scenarios. The system includes feed boxes, cryogenic transfer lines, bayonet cans, feed and end caps, string connecting and segmentation boxes, gas headers, etc. It will be capable of supporting operation of the linac within cool down and warm-up rate limits and other constraints imposed by accelerating SRF components. Protecting the superconducting RF cavities from over pressurization beyond the component's maximum allowable working pressure during fault conditions will be accommodated by the cryomodules and cryogenic distribution system. Components of the cryogenic distribution system contain cryogenic control and isolation valves, cryogenic instrumentation, safety valves, etc.

It is assumed that on the time scale of the Project X, a large portion of the Tevatron ancillary cryogenic components will be available for use by the project. These components include cryogenic nitrogen and helium dewars, gas storage tanks, purifier compressors, cryogenic transfer line and parts of the inventory management system.

SRF cavities are cooled by two-phase superfluid liquid helium. The cavities are grouped in 62 cryomodules of six different types, HWR, SSR-1, SSR-2, 650-low β, 650-high β, and 1300 MHz cryomodules. For cost estimating purposes, all cryomodules are divided into two cryogenic units. The cryogenic feed point is located between the upstream and downstream units. The upstream cryogenic unit consists of HWR, SSR-1, SSR-2, 650-low β and 650-high β, while the downstream unit contains 1300 MHz cryomodules.

The upstream unit contains 34 cryomodules and is subdivided into four (4) distinct cryogenic segments. The first segment contains ten cryomodules (one (1) HWR, two (2) SSR-1 and seven (7) SSR-2 cryomodules). The second segment comprises of five (5) 650 MHz Low β cryomodules. The third and fourth segments each contain eleven (11) 650 MHz High β cryomodules. A cryogenic transfer line runs along the upstream cryogenic unit and is used to distribute superfluid helium and helium gas for the thermal shields and intercepts to each segment.

The downstream unit contains twenty eight (28) 1300 MHz cryomodules divided into four segments. The TESLA concept is utilized for connecting 1300 MHz cryomodules within a segment.

Operation of Project X in a continuous wave mode results in very high dynamic heat loads to the cryogenic system. For SRF components, dynamic heat load (due to RF power dissipation) on average is an order of magnitude greater than the static heat load (due to conduction and thermal radiation).

A preliminary heat load estimate has been performed. For the design study, an additional 50% margin is applied to the estimated heat loads to ensure the system could meet all operational requirements. With this factor, the total equivalent design capacity at 4.5 K for the entire linac is approximately 41 kW. The physical size of a plant of this capacity is too large to house in a single cold box. As a result, two cold boxes are envisioned; one to support the superfluid loads and one to support the thermal shield loads.

A wide range of possible cryogenic plant design solutions that satisfy all requirements and constraints for Project X will be studied further. Combining effective use of the existing Fermilab infrastructure with commercially available components requires further study. The final solution will be based on a cycle with either cold compression alone or utilizing a hybrid approach (both cold and ambient temperature compression). Ease of operation, reliability, ability to operate efficiently over a wide range of loads, capital and operational costs, and other factors will be considered in selecting the technology for the Project X cryogenic plant.

A study will be performed to ensure that the Project X cavity operating temperature and segmentation layout are optimized to minimize the combined capital and operating cost while maintaining system reliability.

## IV.6 Instrumentation

Various beam instrumentation and diagnostics systems are necessary to characterize the beam parameters and the performance in all Project X sub-accelerators. For startup and initial beam commissioning we need to provide, at a minimum, beam instruments in order to observe:
- Beam intensity
- Beam position / orbit
- Transverse beam profiles
- Beam phase / timing

The high beam intensity / power and the presence of superconducting technologies also require a reliable, fail safe machine protection system (MPS) to prevent quenches in cryogenic elements or damage due to an uncontrolled loss of the high power beam. This system will be based on beam loss monitors (BLM) and other beam intensity monitors (e.g. toroids).

Beside these core beam instrumentation systems, more specialized beam diagnostics need to be provided, e.g. to characterize the longitudinal bunch profile and tails, transverse beam halo, and more advanced beam emittance measures. Many types of beam monitors (e.g. BPMs, toroids, etc.) can be standardized. However many areas in Project X, such as the front-end ($H^-$ source, RFQ and MEBT), the synchrotron, and injection/ejection sections, will demand dedicated beam diagnostics (Allison scanner, fast Faraday cup, e-beam scanner, vibrating-wire, etc.).

In order to develop this beam instrumentation, a complete set of "beam measurement requirements" has to be established. Each sub-accelerator (linac, transport lines, MI, Recycler) needs to address the operating modes with the nominal, as well as non-standard beam parameters, and all requirements for the different beam measurements (resolution, precision, dynamic range, etc.). We foresee the following general detectors and systems for beam instrumentation and diagnostics:

### Beam Position Monitors

The beam orbit monitoring (i.e. transverse position and longitudinal phase) is the most fundamental measurement and the most powerful diagnostics tool in an accelerator. Project X requires a large number (~100) of new warm and cold beam position monitors (BPM), thus making it a complex and expensive measurement system. BPM pickups need high quality RF cables to transmit their low-power signals to the read-out hardware outside the accelerator tunnel. This requirement may impact the arrangement or layout of some conventional facilities. It is necessary to locate BPM pickups in cryogenic SCRF sections of the machine, which needs extra care to meet UHV, cryogenic and clean room requirements simultaneously.

### Beam Monitoring in the SCRF Linac

The beam monitoring within the cryogenic environment is probably limited to beam orbit monitoring with BPMs. Other measurements such as beam profiling and other beam monitors, that may include moving parts, cannot operate in a cryogenic environment. Similar measurement systems would require a "warm" diagnostics section within the SCRF linac.

### Beam Profile Monitors

Profile monitors are required in the transfer line for measuring emittance and matching between the linac and transfer line and the Recycler. Options for transverse profile monitors in the H- sections of Project X are the standard multi-wire monitor and the newer laser profile monitor. Laser-based profile monitors are intended as the primary technology choice with standard multi-wires as a fallback technology. . In addition, profile measurements in the rings will be made using ionization profile monitors and electron wire profile monitors.

### Beam Loss Monitors

Typical fast ionization chambers with a large dynamic range will be utilized for most loss measurements. However, there may be instances where measurements of thermal neutrons or machine activation during cooldown periods are desired. The loss monitors will be incorporated in a machine protection system.

### Beam Current Monitors

To determine transport and injection efficiency, a combination of DCCTs and beam toroids will be used to obtain an accurate measure of beam current throughout the linac, transfer line, injection sections.

### Special Monitors

Several types of special beam monitors and diagnostic tools are required to verify the beam quality and minimize beam losses. These include the monitoring of the transverse beam halo (e. g. vibrating wires, laser wires) and the detection of longitudinal tails (e. g. using optical sampling techniques with mode-locked laser wires). A list of special beam monitors and diagnostic tools is not complete but design of generic instrumentation ports in diagnostic sections will allow future instruments to be installed.

### Data Acquisition and Timing

Most beam monitoring systems will use digital signal conditioning and processing methods to extract the wanted beam parameter(s). The generated output data needs to be "time stamped" with respect to the beam event, so beam and other recorded data can be cross-correlated throughout the entire Project X complex. This cross correlation will simplify diagnostics and trouble-shooting on the day-to-day operation.

Internal calibration systems, as well as data acquisition and transport cannot make use of a long pause in the beam pulse (there is none in CW operation). The integration times of some of the measured parameters, e.g. beam intensity, and their time stamping has to be discussed in detail.

### Instrument Physical Space Issues

Sufficient physical space will be made available to accommodate the required beam detection elements. At some critical, real-estate limited locations, e.g. LEBT, MEBT, injection / extraction, and SRF areas, a compromise has to be worked out, which enables a decent way to sense the beam without compromising its quality in the diagnostic sections.

## IV.7 Controls

The control system is responsible for control and monitoring of accelerator equipment, machine configuration, timing and synchronization, diagnostics, data archiving, and alarms. Project X will use an evolution of the Fermilab control system ACNET [51], this is the system that is used in the main accelerator complex and also at the NML/ASTA and PXIE test facilities [52]. ACNET (Figure IV-60) is fundamentally a three tiered system with front-end, central service, and user console layers. Front-end computers directly communicate with hardware over a wide variety of field buses. User console computers provide the human interface to the system. Central service computers provide general services such as a database, alarms, application management, and front-end support. Communication between the various computers is carried out using a connectionless protocol over UDP. Subsystems developed by collaborators based on the EPICS control system can be integrated into the main system.

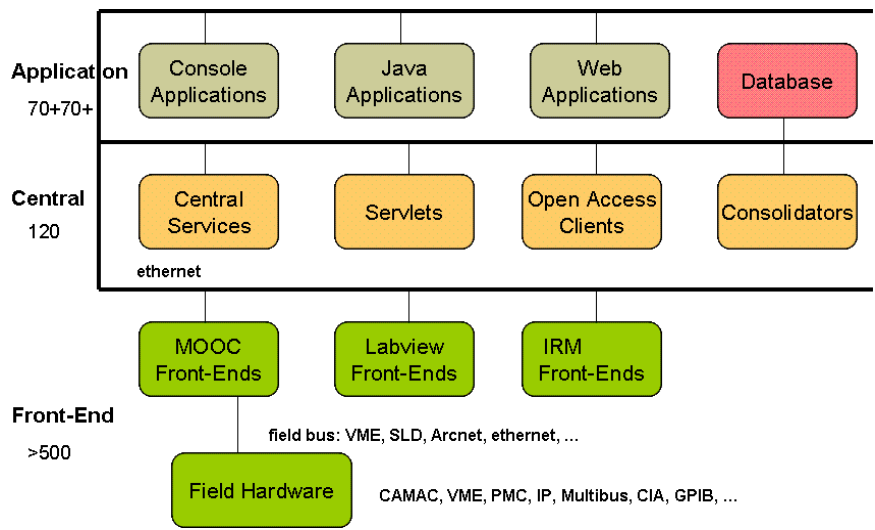

**Figure IV-60**: Project X Controls System architecture.

The scale of the control system is expected to be similar to that of the complex when the Tevatron was operating. The system should support up to 1M device properties. Time stamping must be provided so that all data from the CW linac and the pulsed linac can be properly correlated with that from the existing complex. The control system should contribute less than 1% to operational unavailability. The high beam power implies the need for a sophisticated machine protection system to avoid damage to the accelerator due to errant beam pulses. Furthermore, to minimize routine losses and activation of accelerator components a fast feedback system will be required to stabilize beam trajectories.

A new timing system will be developed that is a major enhancement over the TCLK and MDAT links in the main complex. A prototype has been developed based on a 1 Gbps data link that adds a data payload and cycle stamp to each clock event transmission. The latter will allow reliable correlation of data across different front-ends. The prototype design will be updated based on requirements for the CW linac.

It is highly desirable to have s single control system operating the entire complex rather than separate systems for the new linac(s) and older parts of the system. There should only be a single copy of core services such as alarms and data archiving. Software applications should be able to access any device in the system. This model simplifies development and operation and reduces long term maintenance costs of the complex. The Fermilab control system will be updated for the NOvA era and to support the NML and PXIE test facilities. This upgrade will include modernizing the application software environment as well as replacing obsolete hardware. Upgraded timing and machine protection systems will be developed for the Project X linac that will accommodate the legacy hardware in the existing parts of the complex. These systems as well as linac control software will be prototyped at NML and PXIE.

It is recognized that some equipment will be developed outside of Fermilab by institutions with expertise in the EPICS control system. Also it may be appropriate in some cases to use commercial hardware that comes with EPICS software. It is planned to support integration of EPICS front-ends and some core applications in the Fermilab control system. This has been demonstrated in several different ways at the HINS and NML test facilities. A gateway process between ACNET and EPICS based front-ends has been developed. The console environment also supports EPICS displays based on the EDM display manager.

The control system will specify standard interfaces between its internal components as well as with technical equipment. This will make integration, testing, and software development easier and more reliable and reduce the long term maintenance load. Also standard interfaces allow parts of the system to be more easily upgraded if required for either improved performance or to replace obsolete technologies. Only portions of the system need be changed while the core architecture of the control system remains the same.

## IV.8 LLRF

The Low Level RF system encompasses the programming and regulation of the cavity field amplitude and phase as required by longitudinal beam dynamics in the machine. It also controls or interfaces to the ancillary equipment that is involved in the generation of RF. Hardware and software modules include Cavity Field Controllers, Resonance Controllers,

Master Oscillator, Phase Reference Line, LO distribution, transfer synchronization and the interface to interlocks, timing systems and the control system. This section will cover first the 3 GeV CW linac section operating at the frequencies; 162.5 MHz, 325 MHz and 650 MHz, followed by the pulsed 3-8 GeV 1300 MHz linac.

### IV.8.1 LLRF for the CW 3 GeV SRF LINAC

The high level RF configuration for the CW linac provides a single power amplifier for each of the cavities. The beam current, cavity field gradients and worst-case microphonics determine the loaded cavity Qs and bandwidths. Analysis of the 1mA beam current and cavity gradients show that loaded Qs may be kept in the low $10^7$ range where control of cavity microphonics will not create excessive RF power demands, nor large field disturbances. In addition, except for turn on, CW operation eliminates the dynamic effects of Lorentz force detuning and modulator droop. In many ways the CW linac will operate in a regime similar to the 12GeV upgrade of CEBAF. The largest disturbance to the cavity fields will be the beam structure produced by the beam chopper, particularly when the beam is turned off while ramping for injection into the Recycler. The LLRF system will provide the chop pattern for the MEBT and will therefore have the information required for accurate beam loading compensation.

For cost savings and hardware simplicity, the LLRF controllers will be grouped into stations that will cover two cryomodules (nominal 16 cavities) and the associated RF equipment. Fermilab has been extensively involved in SRF LLRF Control Systems for the ILC and ASTA. Two types of digital controllers have been developed and are in use, each of which could be modified to meet Reference Design requirements. RF and microwave hardware has also been developed for HINS and ILCTA that directly apply to Project X. Ongoing R&D will determine the details of the beam-based calibration and of the phase reference distribution system.

### IV.8.2 LLRF for the Pulsed Linac

The 1300 MHz pulsed linac is implemented with cryomodules containing eight, nine-cell cavities. An RF station will contain one or two cryomodules with RF distributed from a single klystron. With a single RF source, individual cavity fields will vary within the group due to cavity resonance errors generated by Lorentz force detuning, microphonics and static cavity tuning errors. At 3 GeV and above, the beam is relativistic and longitudinally stiff so that field errors in individual cavities in a cryomodule will not cause energy errors, as long as the cavity field vector sum is maintained. The LLRF creates the vector sum of the calibrated baseband analytic signals from each cavity. This vector sum is then regulated by the digital controller with a control bandwidth in the range of 100 kHz. The digital controller actuators are both the klystron RF drive signal and the piezo tuners on each cavity. Slow drifts in the resonant frequency errors are corrected by the motor tuner. There are a few noteworthy parameter differences of the pulsed linac section from ASTA or the ILC design such as an increase in Q from 3E6 to 1E7 and an increase in flattop length from 1 to 4 ms. Both of these differences are more challenging for Lorentz force compensation and therefore are part of ongoing studies at the ASTA facility.

## IV.9 Safety and radiation shielding

Design requirements and radiation limits for accelerators and beam transport lines are provided by the Fermilab Radiological Controls Manual (FRCM). The manual requires that shielding designs shall be well-engineered to maintain occupational and environmental radiation exposures as low as reasonably achievable (ALARA) and compliant with applicable regulations and DOE Orders. The first choice for accelerator shielding designs are to be passive shielding elements designed to achieve areas external to shielding to be classified as minimal occupancy. Minimal occupancy is defined to mean any area which is not normally occupied by people more than 1 hour in 8 consecutive hours. Dose rates for potential exposure to radiological workers in areas without continuous occupancy shall be ALARA and such that individuals do not receive more than 20% of the applicable limits. The design goal for dose rates in areas of continuous occupancy is to be less than an average of 0.05 mrem/hr and as far below this and as low as is reasonably achievable (ALARA). Reliance on active systems such as radiation safety interlocks and/or beamline instrumentation to achieve radiation safety goals should be chosen only if passive elements cannot, in view of planned accelerator operations, reasonably achieve the level of protection required by the FRCM. Discharges of radioactive liquid to the environment should be kept ALARA. Materials and components should be selected to minimize the radiological concerns, both occupational and environmental. Where removable contamination might be associated with accelerator operations, provisions should be made in facility designs for the containment of such material. Internal radiation exposure is to be minimized in accordance with ALARA principles by the inclusion of engineered controls such as ventilation, containment, filtration systems, where practicable and with appropriate administrative procedures. Efficiency of maintenance, decontamination, operations, and decommissioning should be maximized.

The FRCM specifies the manner in which radiological posting requirements are to be determined. The maximum dose is that which can be delivered under the worst credible accident in that area, taking into consideration circumstances and controls, which serve to limit the intensity of the maximum beam loss and/or its duration. Some examples of accident scenarios are (1) beam intensity significantly greater than the nominal beam intensity; (2) unanticipated beam losses; and (3) single pulse full machine loss on an element. The maximum dose is to be determined through the safety analysis, which shall document calculations and measurements of possible radiation exposures, radiation shielding, beam optics and other relevant information. The safety analysis must be reviewed and approved by the SRSO prior to construction and/or operation of the beam.

The FRCM specifies requirements for controls. Accelerator/beam line areas shall be posted and controlled for the normal operating conditions when the safety analysis documents that delivering the maximum dose to an individual is unlikely. Accelerator/beam line areas shall be posted and controlled for accident conditions when the safety analysis documents a scenario in which it is likely that the maximum dose may be delivered to an individual.

The application of the FRCM design requirements to Project X is described in the remainder of this section.

### IV.9.1 Radiation Limits

*Safety Analysis*

Project X will be capable of the production and delivery of multi-megawatt beam power. Accelerator components such as cryomodules and beam pipes can be destroyed very quickly by beam power at this level. The Project X accelerators and associated beam lines require unprecedented control of beam orbit, beam optics, and beam losses in order to provide decades of safe operation for experimental programs. Consequently, the control of beam loss through a machine protection system is a primary design consideration for Project X.

Principal design features required for the control of beam loss include precision alignment of all accelerator components, precise control of beam focusing, elimination of RF jitter, and precision control of beam orbit. Operation of Project X accelerators and beam lines without precision controls could easily result in beam losses on the order of 100 W/m. Machine protection systems are required to ensure that all beam control features are functional and operating as intended. The loss of any precision control feature will cause the machine protection system to inhibit beam acceleration at the ion source, LEBT, or MEBT. The machine protection systems will be capable of limiting or stopping machine operation within a few microseconds of sensing an abnormal condition.

Project X accelerators and beam lines accelerate and transport $H^-$ beams. Principal beam loss mechanisms are related to stripping electrons from $H^-$ ions; the causes of stripping include $H^-$ ion collisions with residual gas, blackbody photon interactions, Lorentz force (magnetic stripping), and interbeam scattering. Beam loss due to these various mechanisms is dominated by interbeam scattering and has been determined to be about 0.1 W/m. Losses from the remaining mechanisms are << 0.1 W/m.

### *Facility Design Beam Loss Level*

The conclusion of the Safety Analysis is that the average beam loss under normal conditions will be of the order of 0.1 W/m. Machine protection systems will monitor the performance of beam focusing, beam orbit, RF stability, and machine alignment. Machine protection systems will reduce accelerator beam power or inhibit accelerator operation in the event precision control of the accelerator control is lost. The reaction time of machine protection systems under consideration is on the order of a few microseconds. Therefore, only operation under normal conditions should be possible. For purposes of the facility design, it is assumed that the peak average beam loss will be 1 W/m, a factor of 10 higher than what is expected during nominal beam operating conditions.

### *Facility radiation level design goals*

The design goals for Project X accelerators and beam lines are:

1. Permit unlimited occupancy for all service buildings, shielding berms, parking lots, control rooms, and associated areas. By design, radiation levels are to be kept below 0.05 mrem/hr in all accessible locations outside of the beam enclosures for normal operating conditions, based upon an assumed continuous beam loss of 1 W/m. The actual nominal beam loss condition described in the Safety Analysis is expected to be about 0.1 W/m.

2. Permit inspection and maintenance activities within tunnel enclosures while maintaining personnel radiation exposure due to residual activation of accelerator components and beam enclosures at levels as low as reasonably achievable.

3. Limit radiation exposure due to air activation both within the beam enclosure during inspection and maintenance activities and at the site boundary.

4. Limit ground water and surface water activation to levels well below regulatory standards.

5. Prevent the activation of beam component surfaces to avoid the generation of removable radioactivity.

6. Minimize the activation of accelerator components which can impact their useful service life

## IV.9.2 Radiological Design Requirements and Consequences of Project X Radiation Limits

Nominal beam loss throughout the Project X accelerator and beam lines is expected to be about 0.1 W/m. Machine Protection Systems will limit or inhibit beam operations within microseconds of sensing a machine fault. The design requirements for radiation shielding discussed below are based upon an assumed continuous beam loss of 1 W/m. The consequences of the activation of accelerator components, enclosure structures, air, water, and removable contamination are discussed in terms of the expected nominal beam loss of 0.1 W/m as defined in the Safety Analysis.

### *Radiation Shielding*

An early conceptual design of a Project X accelerator enclosure is shown in Figure IV-61. An enclosure height of 16 feet is indicated along with passive shielding of 24.5 feet. An option to transport 1 GeV beam from the 1 GeV Linac to the existing Booster accelerator is shown in Figure IV-62; a remnant of the Main Ring tunnel is used as a beam transport line. The tunnel height is 8 feet and the shield is approximately 20 feet. At this time, details of the Project X layout and facility design have not been finalized. It is necessary that the accelerator design precede the shield design, but some shielding design concepts for Project X are considered here.

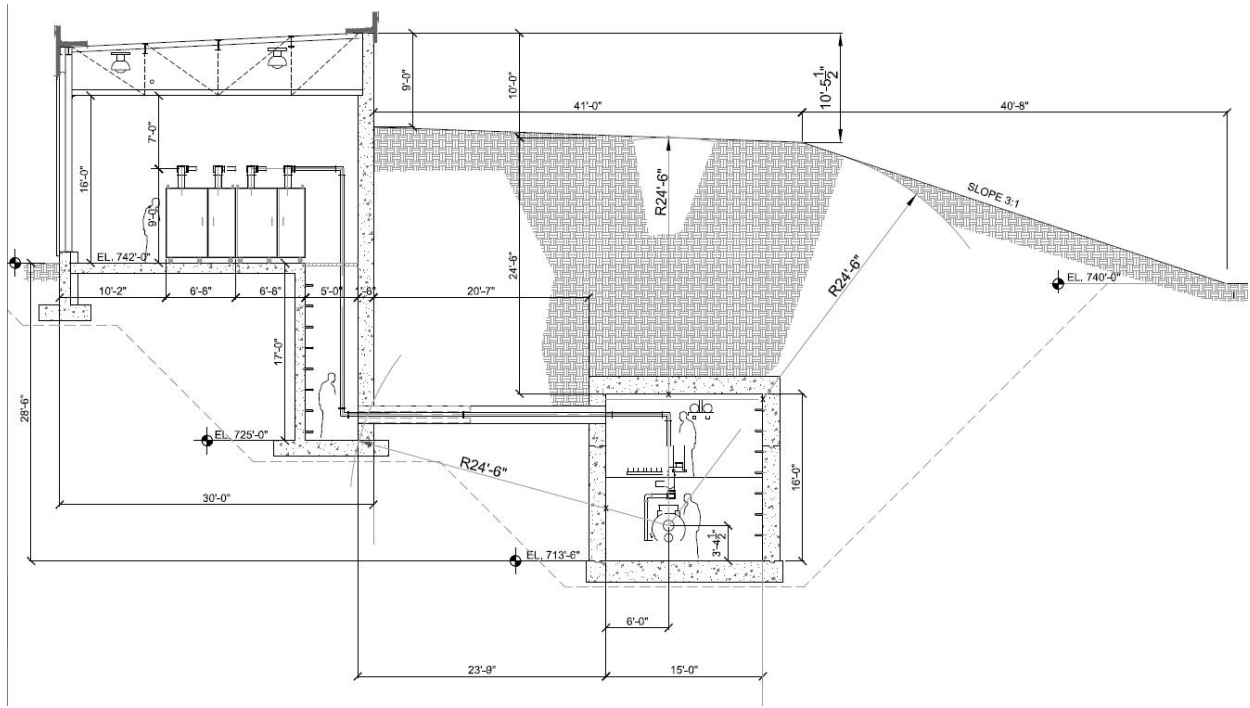

**Figure IV-61**: Early concept cross section of Project X linear accelerator enclosure

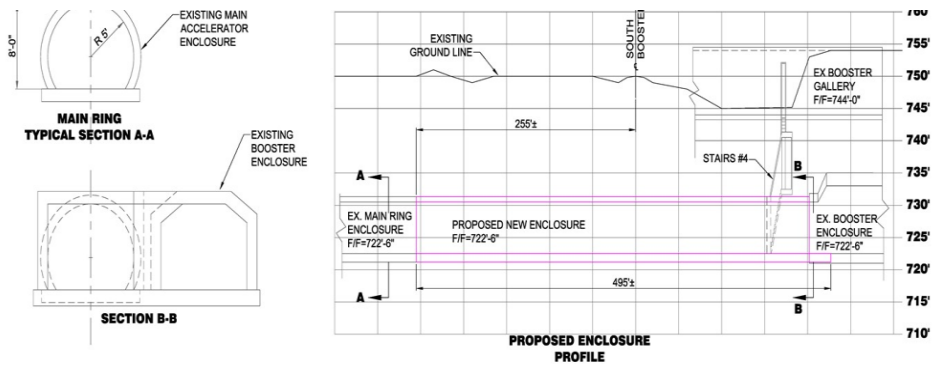

**Figure IV-62**: 1 GeV transport line to the Booster accelerator

An established parameterization [53] is used to determine the radiation dose equivalent rate as a function of energy (GeV), distance (feet), and angle with respect to incident beam direction (degrees) from a low energy proton beam (<1 GeV) incident upon a target:

$$S(E, r, \theta_s) = 2 \times 10^{-5}(1 + E^{0.6}) \left( \frac{1 - e^{-3.6 E^{1.6}}}{\left(0.3048\, r\, \left(\theta_s + \frac{40}{\sqrt{E}}\right)\right)^2} \right)$$

For neutron energies below 100 MeV, the attenuation length in concrete is significantly shorter than that for neutrons considered in higher beam energy based assessments. For example, for high energy shielding problems, 3 feet of concrete provides a reduction factor of 10 in radiation dose rate. The mean free path of low energy neutrons relative to the high energy asymptote has been parameterized [53]:

$$\frac{\lambda_{LE}}{\lambda_{HE}} = 1 - 0.8\, e^{-5*E}$$

The reduction in radiation dose rate as a function of energy (GeV and concrete shield thickness (feet) is:

$$A(E, T_{conc}) = 10^{\frac{-T_{conc}}{3}/(1 - 0.8 e^{-5E})}$$

In the dose rate calculations, the peak neutron energy E (GeV), is taken to be equal to the beam energy. This simplification is conservative in that the actual neutron energies are necessarily lower and hence lead to better attenuation provided by the concrete shielding than indicated by calculations. In addition, the dose equivalent per neutron conversion factor is taken as a constant value of 40 fSv/n over the range of the neutron spectrum. The resulting shielding calculations are implicitly conservative.

Radiation shielding required to limit radiation dose rates to 0.05 mrem/hr for a 1 GeV linac and beam transport line assuming various beam levels of beam loss is shown in Figure IV-63. The shielding requirement varies with beam energy with the assumed maximum beam power loss for normal and accident conditions. The ultimate shielding thickness chosen will depend up a number of factors including the confidence level given to the Safety Analysis including consideration of the projected loss mechanisms and the machine protection system.

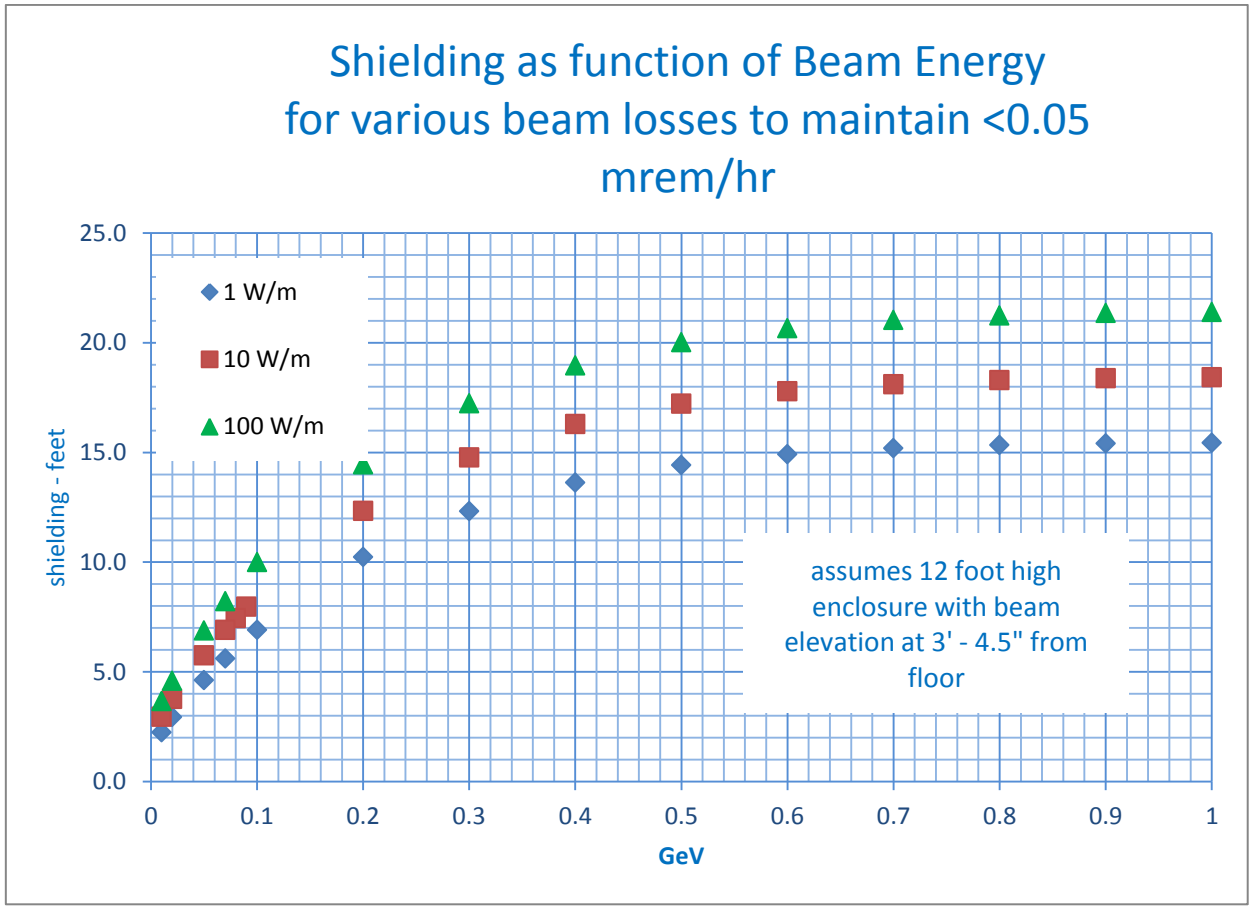

**Figure IV-63**: Radiation shielding requirements as a function of beam energy and beam power loss for a Project X beam enclosure

An active protection system, the Total Loss Monitor (TLM), currently under development, could be used to guarantee the limitation of any given beam power loss. The use of a TLM system could help to fix the level of beam power loss, and as a consequence, fix the amount of radiation shielding required.

The TLM is an argon gas filled ion chamber of variable length with an applied bias voltage. A beam loss in the vicinity of the ion chamber produces a charge whose magnitude is proportional to the amount of beam loss. The TLM response to an 8 GeV proton beam loss made under controlled conditions measured over a wide range of bias voltage and over two decades of beam intensity has been determined as shown in Figure IV-64. The response has been shown to be independent of the TLM length. At the nominal bias of 500 volts, the TLM response to 8 GeV proton beam loss is about 3 nC/E10 protons. Preliminary scaling laws, to be verified in further TLM development work, can be used to predict TLM response at other energies.

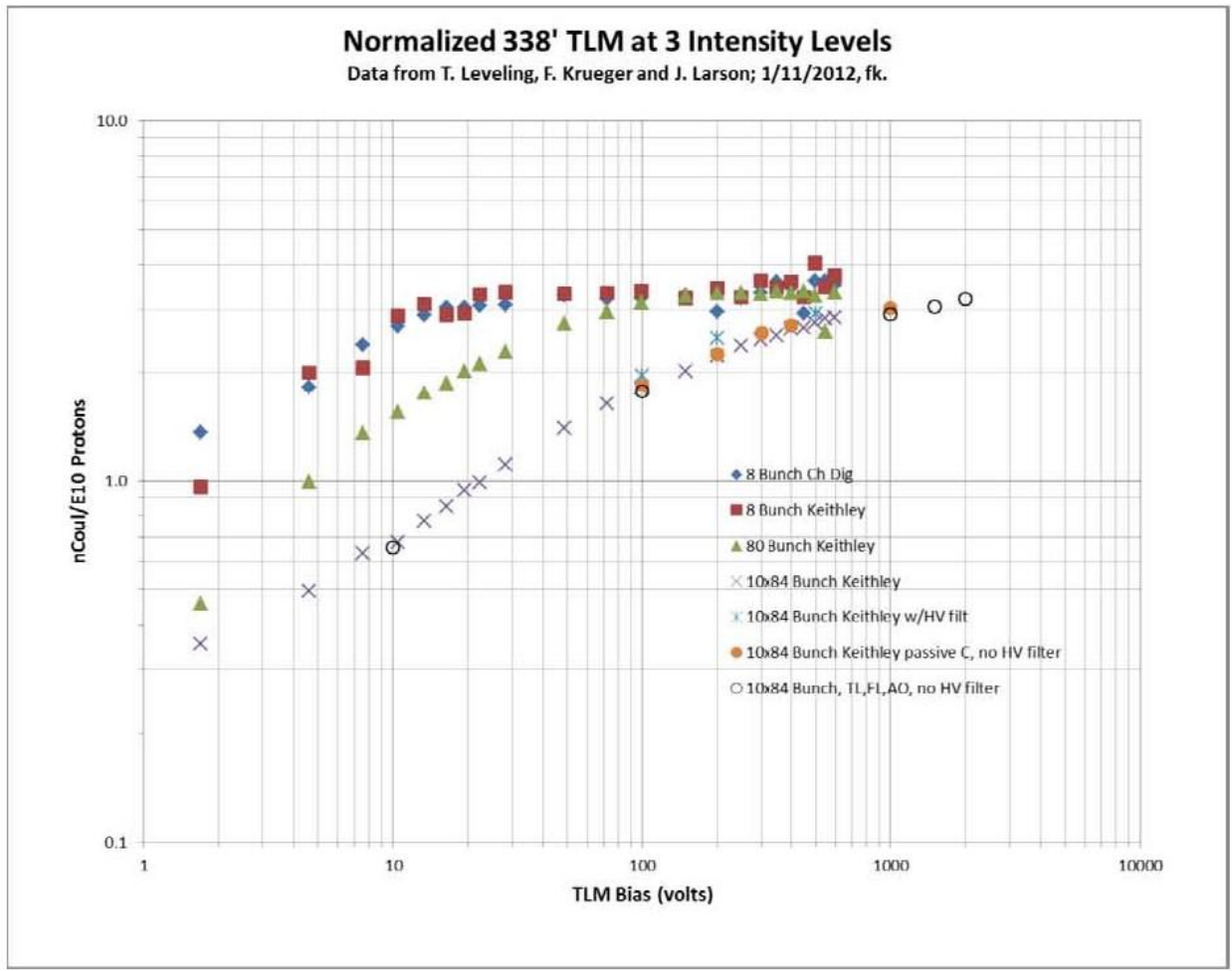

**Figure IV-64**: Response of 338 foot (103 meter) TLM as a function of applied bias voltage over 2 decades of beam intensity

The response can be scaled to beam energy down to 1 GeV by the relationship:

$$3\frac{nC}{E10} protons * \left(\frac{E}{8\,GeV}\right)^{0.8}$$

The response to beam energy below 1 GeV remains to be determined.

A feature of the TLM system is that an interlock trip level can be established to limit beam loss to 1 W/m or virtually any beam power loss. A TLM as presently conceived, does not distinguish between distributed losses and single point beam losses. The process to set TLM trip levels consists of two steps: 1. Establish the total charge to be collected for a distributed loss, e.g., 1 W/m. 2. Evaluate the shielding considering that the total charge can be deposited at a single location. If the shielding is sufficient for the maximum charge collection rate at any location, then the TLM can effectively limit both the distributed beam loss and worst case single point beam loss.

Based upon preliminary TLM work, it should be possible to limit Project X beam loss with a TLM system beginning at the HW cryomodule and continuing through the entire accelerator and beam transport chain.

*Residual Activation of accelerator components and structures*

Residual radiation levels in beam transport lines and accelerators due to operational beam losses must be controlled in order to conduct maintenance activities while keeping personnel radiation exposure as low as reasonably achievable (ALARA). For 2-3 MW beam power, small fractions of a percent loss would result in very high residual radiation levels which would render beam enclosure access difficult and maintenance at loss points extraordinarily difficult. A sensitive machine protection system which inhibits beam operation when significant losses are present will be required to allow access and maintenance activity modes historically enjoyed at Fermilab.

For design purposes, a loss rate of 3 to 10 watts/meter results in a dose rate of about 100 mR/hr at one foot from beam line components such as magnets and accelerating cavities following a 30 day irradiation period and 1 day of cool down. A loss of 0.25 watts/meter results in a dose rate of about 100 mR/hr at one foot from low mass components such as beam pipes for the same irradiation/cooling period. Radiation levels are typically at least a factor of 5 less than these levels. For example, for a typical magnet beam loss location at 2 watts per meter, the fractional beam power loss is 1 ppm. A sensitive machine protection system will be required to quickly identify and suspend operation in the event such losses occur.

In the Safety Analysis, projected normal losses due to interbeam scattering and other loss mechanisms are about 0.1 W/m. The machine protection system as presently conceived should limit beam loss to < 1 W/m. Consequently, residual activation of the accelerator, beam line components, and tunnel structures should be comparable to or less than levels tolerated in existing and previous machines. While the machine protection system would serve to protect the accelerator and beam line components, the TLM system would serve in a parallel role as a personnel safety system to limit residual activation of accelerator components.

*Air activation*

Air activation must also be characterized due to projected Project X operations. Based upon the anticipated losses described in the safety analysis, the combination of anticipated normal beam loss and the machine protection system should serve to limit the total beam loss levels at or below those produced at existing facilities. Based upon projected losses from the Safety Analysis, no significant air activation is anticipated. While the machine protection system would serve to protect the accelerator and beam line components, the TLM system would serve in a parallel role as a personnel safety system to limit air activation within accelerator enclosures.

*Water activation*

The site chosen for the new Project X accelerator and beam line enclosures is presently at the center of the Tevatron beam enclosure. In order to evaluate surface and ground water activation, a geological survey (core borings) will be required to understand ground water migration rates at this site since no data presently exists. An estimate of surface and ground water activation is necessary in order to ensure compliance with regulatory requirements for surface and ground water. However, based upon losses projected by the Safety Analysis, no significant surface water or ground water activation is anticipated. The machine protection system would serve to limit the total beam loss that would also determine the level of surface water and ground water activation. The TLM system would serve a parallel, redundant role to also limit surface and ground water activation.

*Radioactive surface contamination*

Radioactive surface contamination results coincidentally with the activation of accelerator and beam line components. Maintenance activities are rendered more complicated when radioactive surface contamination is present due to prescriptions for the use of personnel protective equipment including coveralls, gloves, shoe covers, and other protective measures. It would be possible in megawatt beam power machines to produce very significant levels radioactive surface contamination at beam loss locations. However, as indicated in the Safety Analysis, nominal beam power losses are expected to be approximately 0.1 W/m, about a factor of 100 below the beam power loss required to produce the onset of measurable radioactive surface contamination. Consequently, radioactive surface contamination on accelerator, beam line components, and tunnel structures should be comparable to or less than levels tolerated in existing and previous machines.

*Lifetime of machine components*

Based upon the level of beam loss projected by the Safety Analysis and also upon experience with existing accelerator and beam line facilities, machine component lifetimes, in general, should be on the order of many decades.

## IV.10 Machine Protection System

The Project X linac will accelerate 1mA of beam current at a CW duty cycle. While the peak current is relatively low compared to most HEP linacs, the average current will be equal or greater than any other HEP hadron linac. In order to protect the machine and associated diagnostics from beam induced damage and excessive radiation damage will require a robust Machine Protection System (MPS) that provides the requisite amount of protection for the entire system. The main goals of the MPS will be the following:
- Protect the accelerator from beam induced damage and operator damage.
- Manage and monitor the beam intensity.
- Safely switch off the beam in the case of failures.
- Determine the operational readiness of the machine.
- Manage and display alarms.
- Provide a comprehensive overview of the machine status.
- Provide high availability.
- Provide fail safe operation where possible.
- Provide post mortem analysis.

Several signals from devices or systems will be monitored and utilized as actuators to inhibit the beam at various stages of the accelerator. The main actuator for beam is the ion source power supply itself. In addition, signals from the LEBT/MEBT choppers, the Radio Frequency Quadrupole Amplifier (RFQ), cavity power amplifiers as well as beam stops and gate valves status will all contribute as additional control devices. A comprehensive overview of the entire machine will be obtained by carefully monitoring all relevant inputs from machine diagnostics and critical systems affecting safe or fail safe operation.

The protection system is modeled based on experience gained from commissioning and operating the SNS accelerator. Their peak current specifications are about 20× higher than the Project X peak beam current, but their average beam current is equivalent to Project X specifications due to the lower duty factor. Their copper to SC cavity transition occurs at 187 MeV. Above 200 MeV the Project X MPS hardware design and placement can be modeled after the SNS system. The Project X MPS system will not need response times as stringent as SNS because of our low peak currents. The challenge for the Project X MPS comes from the low energy cryomodule protection (2.1 MeV – 150 MeV) where beam losses have difficulty penetrating the cryomodule and beam pipe.

### IV.10.1    MPS Configuration

The MPS will be considered to be the collection of all subsystems involved in the monitoring and safe delivery of beam to the dump and not limited to any particular subsystem or diagnostic device. It has connections to several external devices and sub-systems. Figure IV-65 shows a conceptual overview diagram of the MPS. The top layer comprises signal providers such as beam loss monitors, beam position monitors, magnet power supplies etc. Systems at this level send alarms or status information to the MPS logic subsystems (permit system) which issues a permit based on the comprehensive overview of all inputs and request. Only simple digital signals (e.g. on-off, OK-alarm) are transmitted. All devices or subsystems that are determined to be pertinent to protecting the machine or necessary for machine configuration are included. The permit system layer of the MPS will be FPGA based and is thus fully programmable and handles complex logic task. The logic here will be designed to ensure safe operating conditions by monitoring operational input, chopper performance, the status of critical devices and by imposing limits on the beam power.

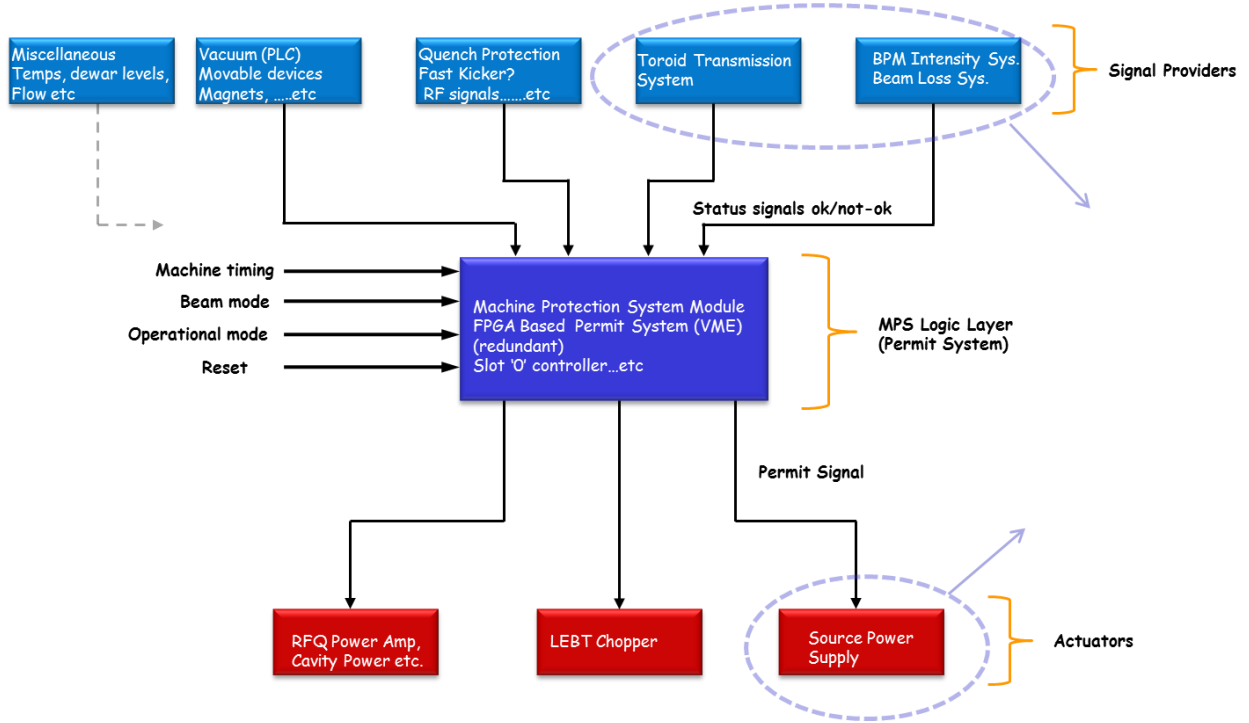

**Figure IV-65**: MPS Conceptual Layout

The final layer of the system shows the main actuators. This will comprise of all points where the MPS logic may act on the operation of the machine to prevent beam from being produced or transported. The entire protection system interfaces with the accelerator control system and machine timing system for configuration management, timing and post mortem analysis as shown in Figure IV-66. The operational modes, operational logic, reaction time and complexity of inputs will differ based on the machine configuration and damage potential at various stages of the accelerator complex.

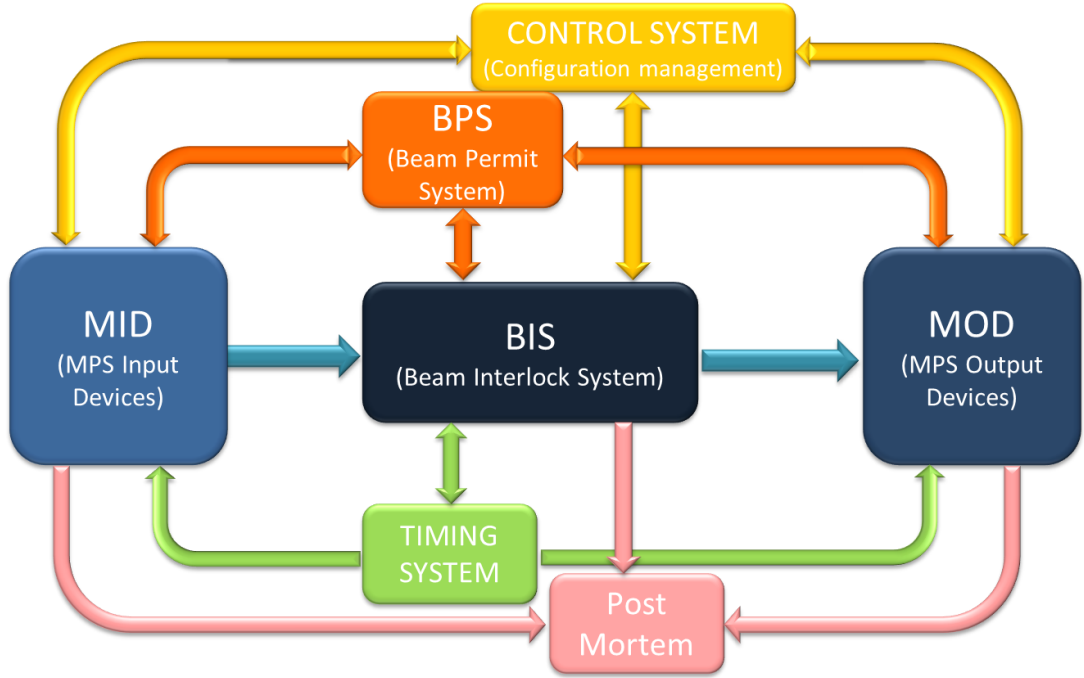

**Figure IV-66**: Conceptual layout integrated with control system

IV.10.2      **Protection System R&D**

Protecting the superconducting cavities from low energy protons losses where the particle energies are too low to produce significant detectable radiation will be a major part of the developmental work needed to effectively inject beam without quenches. To achieve this we will need to research sensitive means for measuring these losses and develop an effective feedback for machine protection. In addition we plan to achieve the following goals as a result of designing, constructing and operating the PXIE MPS:

- Understand and verify acceptable loss rates in the room temperature sections.
- Develop a strategy to monitor chopped beam from the MEBT.
- Estimate particle shielding effect of superconducting cavities and cryomodules.
- Develop effective algorithms for the FPGA based logic system.

- Demonstrate effective integration with controls/instrumentation and all subsystems.
- Understand dark current effects as it relates to protection issues.

In order to protect the accelerator from damage as the beam transitions from the room temperature sections of the machine to the superconducting sections, some specialized instrumentation may be developed at PXIE. Developing an effective algorithm to monitor the beam position as a feedback to machine protection will be of interest for both PXIE as well as Project X.

# V Siting and Conventional Facilities

## V.1 Siting Options

Ample space exists on the Fermilab site to construct all Project X related facilities within the site boundaries. Selection of the specific facility layout is based on an optimization of functionality, performance, interfaces with existing facilities, expandability, environmental impacts, and costs among other factors. In accordance with the latest funding guidance, a staged approach has been developed whereby the first 1 GeV is constructed in Stage 1, followed by 1-3 GeV in Stage 2 and finally 3-8 GeV in Stage 3. The layout of the facility must accommodate this staged approach and minimize cost in Stage 1.

The fundamental requirement of Stage 1 is to deliver 1 GeV beam to the existing Booster, Muon Campus (under construction), and a new Spallation Facility simultaneously. However, the Stage 1 layout must also consider future stages and their constraints. An important constraint is the location for 8 GeV injection into the Recycler (RR) for Stage 3. Transfer line radii become larger with energy and therefore there are limited options for routing the 8 GeV transfer line to the Main Injector (MI) tunnel. For this reason the layout designs originate at MI/RR and work backwards toward the front end. The favored location for injection is at MI-10 since the other symmetric 4 or 8 half-cell straight sections (MI-30, 40, 60) all include other major equipment (collimation and MI/RR transfer line, abort, RF). Since the beam is counter-clockwise in MI, the beam from Project X approaches MI-10 from the east and drives the location of the 3-8 GeV linac to the area inside the Tevatron ring.

Initial designs for a "one-shot" Project X in which all three stages are constructed concurrently had each accelerator in a linear arrangement as shown in Figure V-1. This is still a viable option should funding projections change but has some disadvantages in a staged scheme. An investment in conventional facilities would be required for all envisioned accelerators through 8 GeV to allow a path from the Stage 1 section to the Booster and Muon Campus. Also, it would not allow for concurrent construction of Stages 2 and 3 while Stage 1 is operating since they would share tunnels.

The layout designed for this report is shown in Figure V-2. It shows the Stage 1 1 GeV accelerator as the southernmost section, followed by the 3 GeV section to its north and finally the 3-8 GeV pulsed section. Locating the Stage 1 section south of F0 provides paths to the Booster by way of the Tevatron tunnel and to the Muon Campus through AP0 (see section III.2 for 1 GeV beam handling at F0). Connection from Tevatron tunnel to Booster is shown in Figure V-3. This layout allows for construction of Stages 2 and 3 concurrently with operation of Stage 1 and also provides flexibility for expansion beyond Project X. Figure V-4 presents notable existing and planned facilities relevant to the construction of Project X Stage 1. The 3-8 GeV pulsed section could further be expanded to a muon accelerator as shown in Figure V-5.

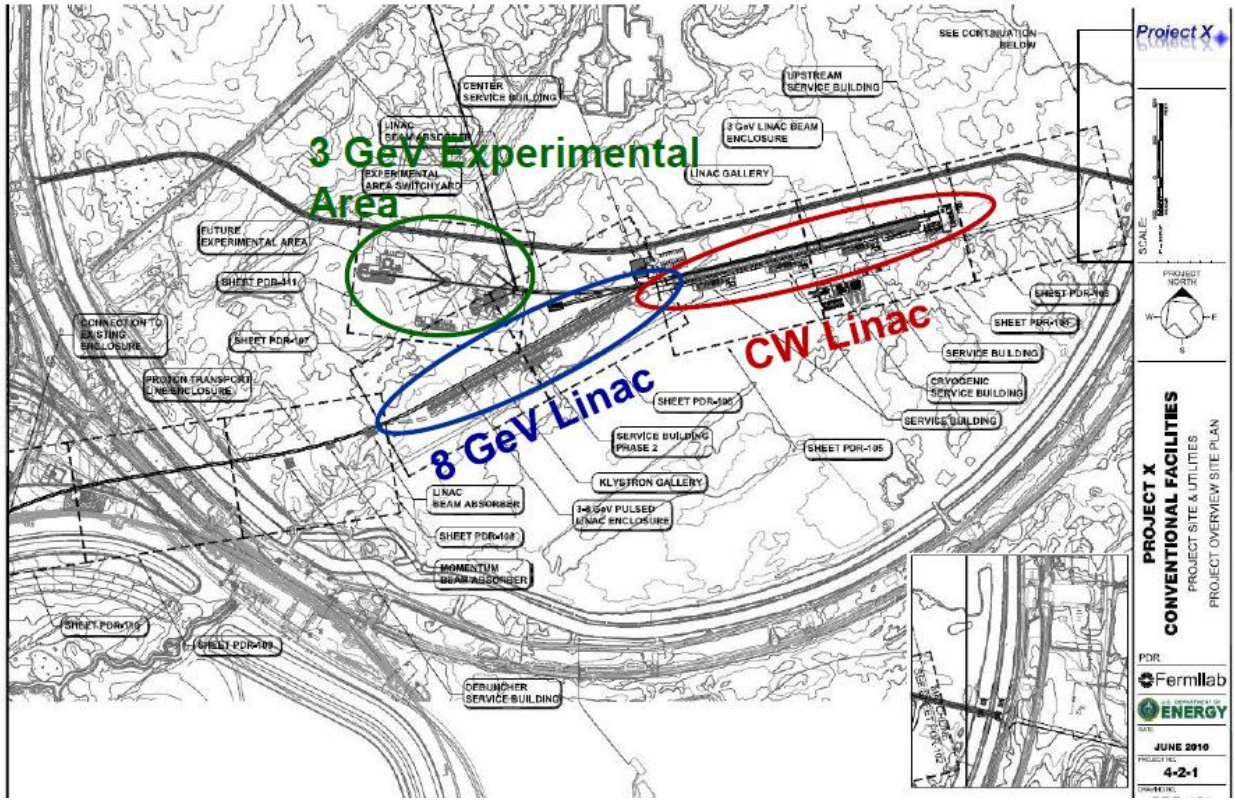

**Figure V-1**: "One-shot" Project X layout from 2010 Reference Design Report.

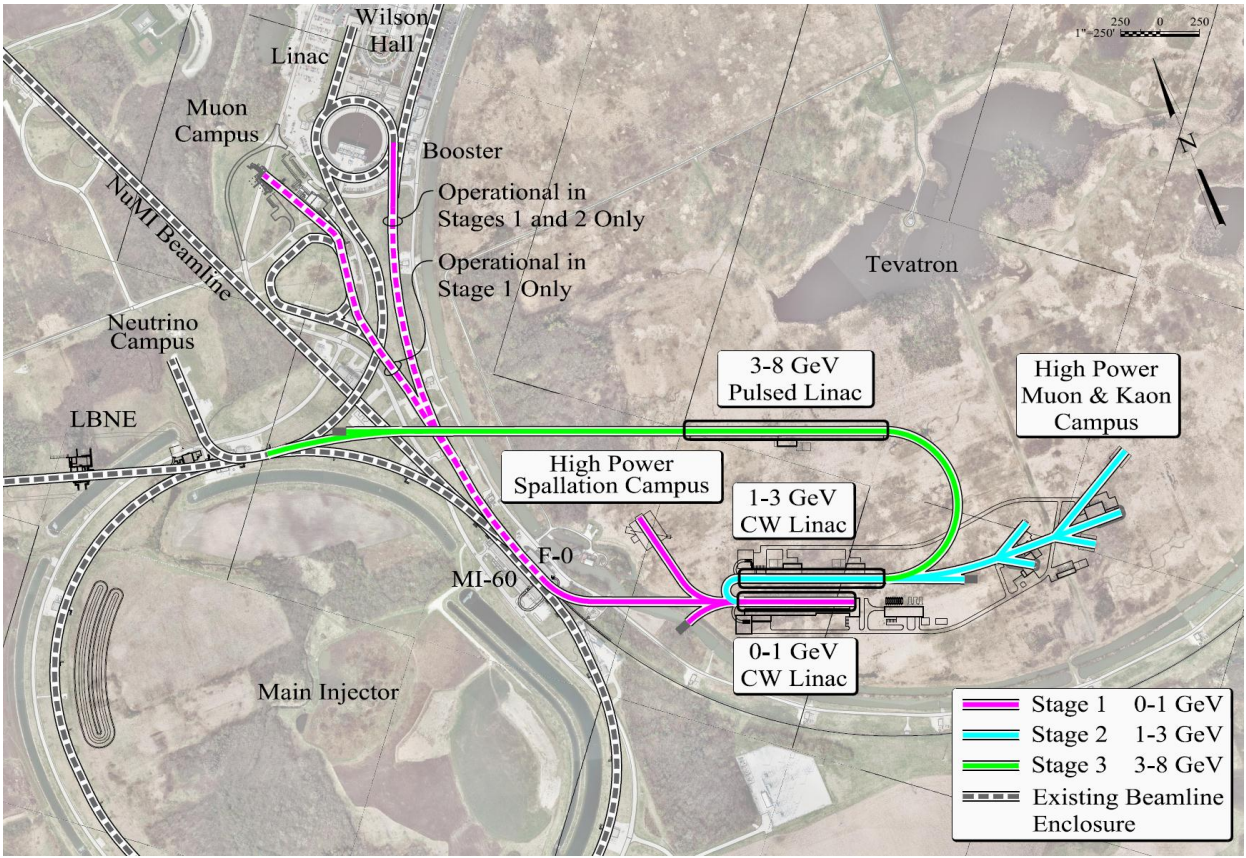

**Figure V-2**: Staged layout selected for the Reference Design

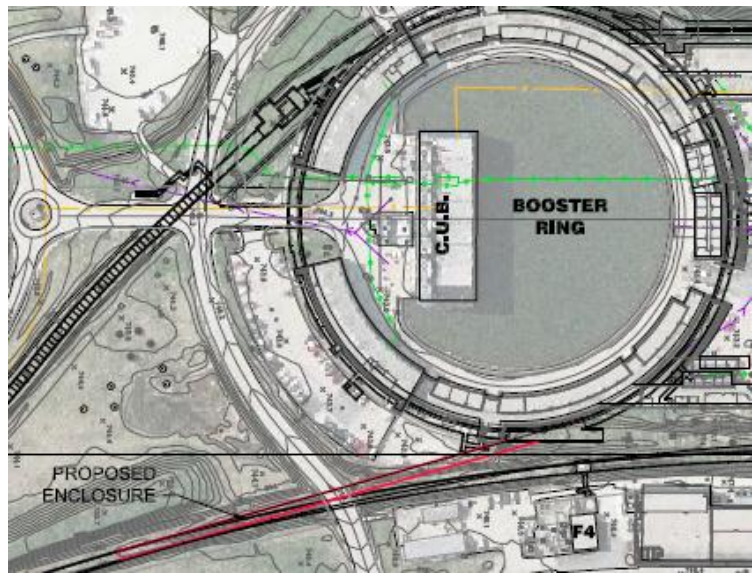

**Figure V-3**: New ~ 500 ft. tunnel section from Tevatron tunnel to Booster for Stage 1.

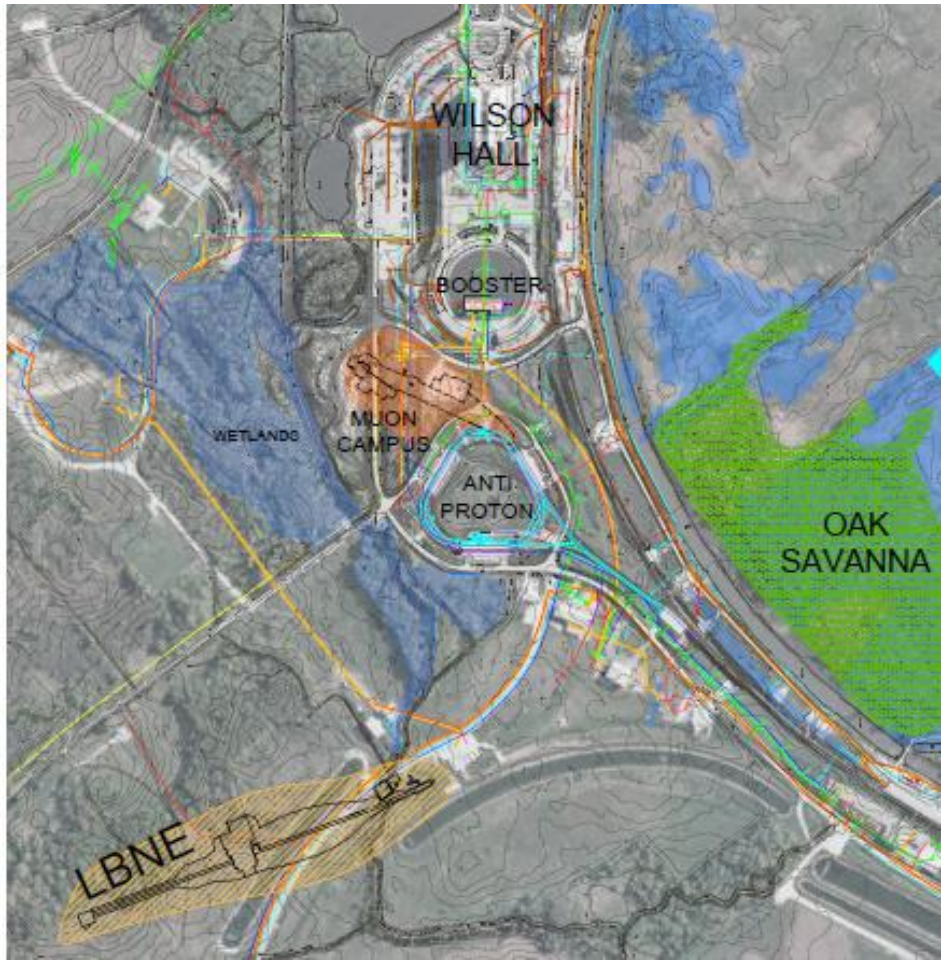

**Figure V-4**: Other notable existing and planned facilities relevant to Project X Stage 1 include the Booster, Muon Campus, and LBNE beamline.

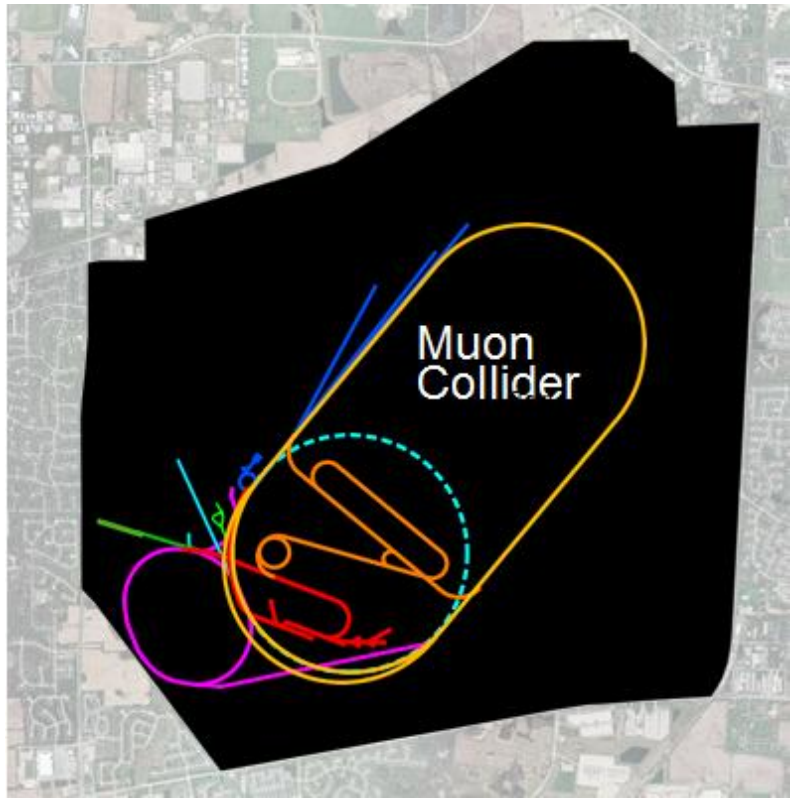

**Figure V-5**: Potential future facilities and their relation to Project X. Project X facilities in red, Muon Accelerator in orange, Muon Collider in yellow.

## V.2  Conventional Facilities

This section outlines the conventional facilities required to house and support the 0 to 1-GeV CW Linac (Stage 1), 1 to 3-GeV CW Linac (Stage 2) and the 3 to 8-GeV pulsed Linac (Stage 3).  Civil construction for the facility includes all below-grade beam-line enclosures.  All above-grade buildings, roads, parking, utilities and services to accommodate the equipment for the operation of the facility on the Fermilab site are also included.

Construction of the below-grade linac(s) and beam transport lines as well as the above-grade service buildings are similar to previously utilized and proven construction methods previously executed at Fermilab.  Construction of all below-grade enclosures consists of conventional open-cut type construction techniques.  The architectural style of the new buildings reflects, and is harmonious with, the existing buildings.  Currently, the layout has been optimized for the accelerator requirements.  Future layouts will consider existing topography, sustainability, watersheds, vegetation, natural habitat, and wetlands.  All the aspects will be thoroughly addressed in the Environmental Assessment for this project.

Site Construction
 1. Site work

a. <u>Site Drainage</u> will be controlled by ditches and culverts, preserving the existing watershed characteristics both during construction and subsequent operation.
b. <u>Road Construction</u> includes a new temporary construction road providing access to Butterfield Road. This road will provide direct access for construction traffic during construction only – roadway will be restored to original condition upon completion of the project. New Service Road will provide permanent access to all service buildings and utility corridor.
c. <u>Landscaping</u> includes the restoration of disturbed areas. Construction yards and stockpile areas will be removed after completion of the construction phase of the project. All disturbed areas will be returned to a natural state or landscaped in a similar manner as found at other Fermilab experimental sites. Erosion control will be maintained during all phases of construction.
d. <u>Wetlands Mitigation</u> includes the avoidance or minimization of adverse impacts to wetlands in the project area. Environmental consultants would delineate wetlands, and a Clean Water Act permit application prepared for submittal to U.S. Army Corps of Engineers for impacts that cannot be completely avoided. Compensatory mitigation would be provided according to terms and conditions of the permit. This may be in the form of purchased wetland bank credits, restoration or enhancement of existing wetlands on site, or creation of new wetland areas. The permit would dictate the amount and type of mitigation, which must be in place prior to the initiation of construction. A Floodplain/Wetland Assessment pursuant to 10 CFR 1022 would be incorporated into the Environmental Assessment.

2. <u>Utilities</u>
The following utilities are required to support the operation of the facility. The list incorporates current assumptions and will require further refinement as the design process progresses.
a. <u>Electrical Power</u> includes new duct banks and utilization of existing duct banks from two sources including Kautz Road Substation (KRS) and Master Substation (MSS). Separate high-voltage feeders with backup will be provided for conventional, machine and cryogenic power.
b. <u>Communications</u> include new duct banks tied into the existing communication network along Kautz Road.
c. <u>Chilled Water (CHW & CHWR)</u> for machine and building cooling will be supplied via new supply and return lines from the existing Central Utility Building (CUB).
d. <u>Low Conductivity Water (LCW)</u> for machine cooling will be supplied via new supply and make-up water from the existing Main Injector ring LCW system.
e. <u>Industrial Cooling Water (ICW)</u> for fire protection will be supplied via new supply and return lines from the existing site wide ICW system.

  f. <u>Domestic Water Supply (DWS)</u> for potable water and facilities will be supplied via new supply line from the existing site wide DWS system.
  g. <u>Sanitary Sewer (SAN)</u> for facilities will be supplied via new sewer main and lift station from to the existing site wide sanitary sewer system.
  h. <u>Natural Gas (NGS)</u> for building heating will be supplied via new supply lines from the existing site wide NGS system.
3. <u>Facilities Construction</u>

Conventional facilities will be constructed with future upgrade capabilities considered in the initial design phase. Equipment galleries, enclosures and surface buildings will be designed to accommodate future expansion of the technical components of the facility. See

Figure V-2 for a site map and facilities locations. The major elements for the conventional facilities are as follows:
 a) Below-Grade Construction
  0 to 1-GeV Continuous Wave (CW) Linac Enclosure
  1 to 3-GeV Continuous Wave (CW) Linac Enclosure
  3 to 8-GeV Pulsed Linac Enclosure
  3-GeV Experimental Area Switchyard
  Various Transfer Line Enclosures
 b) Above-Grade Construction
  0 to 1-GeV Linac Gallery
  1 to 3-GeV Linac Gallery
  3 to 8-GeV Linac Gallery
  Cryogenic Service Building

## V.3 Site Power Requirements

An estimate of site power requirements for the linacs and beam transfer lines is given in Table V-1. Included on the various lines are:

<u>CW Linac RF</u>: RF power sources to accelerate 2 mA of beam to 1 GeV and 1 mA of beam from 1 GeV to 3 GeV. Also included are LLRF, protection circuits, RF controls, filament and bias supplies.

<u>Pulsed Linac RF</u>: RF sources, including modulators, to accelerate 1 mA of beam from 3 GeV to 8 GeV with a 4.4% duty factor. Also included are LLRF, protection circuits, RF controls, filament and bias supplies.

<u>Cryogenic Systems</u>: Based on the estimated cryomodule dynamic and static heat loads through both the CW and pulsed linacs, and system operation at 2 K. Because the heat load is overwhelmingly dynamic, the cryogenic system power associated with the CW linac is about 10.2 MW, while the pulsed linac is only about 1.0 MW.

<u>LCW</u>: The primary load is cooling of the RF sources.

ICW: Utilized for heat removal in the cryogenic systems.

HVAC: The primary loads are the 20% of RF power not removed by the LCW system and the removal of heat from equipment galleries.

Conventional Systems: Power required for other linac/beamline components (magnets, vacuum pumps), and for occupied spaces.

| System | Wall-plug Power (MW) |
| --- | --- |
| CW Linac RF | 16.4 |
| Pulsed Linac RF | 2.1 |
| Cryogenic Systems | 11.2 |
| Low Conductivity Water | 1.8 |
| Industrial Chilled Water | 1.0 |
| HVAC | 1.0 |
| Conventional Systems | 3.0 |
| TOTAL SITE POWER | 36.5 |

**Table V-1**: Estimated site-wide power requirements for the Project X Linacs

The total site-wide power of 36.5 MW supports operations with 1 MW of beam power at 1 GeV, 3 MW of beam power at 3 GeV, and 350 kW of beam power at 8 GeV. Thus the total system efficiency (beam power/wall-plug power) is about 12%.

# Appendix I: Staging Scenarios

As described in the body of this report, the Reference Design provides a facility that will be unique in the world with unmatched capabilities for the delivery of very high beam power with flexible beam formats to multiple users. Financial and budgetary constraints have led to available funding have led to consideration of a staged approach to Project X. Development of a staging plan for Project X is based on application of the following principles:

- Each stage should have a cost significantly below $1B.
- Each stage should present compelling physics opportunities.
- Each stage should utilize existing elements of the Fermilab complex to the extent possible.
- At the completion of the final stage the full vision of a world leading intensity frontier program at Fermilab should be realized.

## I.1 Staging Scenario

A three stage approach to the Reference Design consistent with the above principles has been developed and is described below. The siting plan consistent with this approach has been presented in Figure I-1.

*Stage 1*

Stage 1 of Project X is comprised of a 1 GeV, CW, superconducting linac operating with an average current of 1 mA. A small fraction (~2%) of the beam will be injected into the existing Booster. Injection at 1 GeV is projected to result in a 50% increase in the per pulse proton intensity delivered from the Booster to the Main Injector complex, relative to current operations. Stage 1 thus establishes the potential for delivering up to 1200 kW onto a long baseline neutrino target (either NuMI or LBNE). Depending upon the operating energy of the Main Injector and the allocation of the Main Injector timeline between neutrino production and a possible rare kaon experiment, significant power could also be devoted a program based on 8 GeV protons. The balance of available linac beam can be delivered to the Muon Campus currently under development, providing a factor of ten increase in beam power available to the μ-to-e experiment, and/or to a newly developed experimental programs devoted to nuclear electric dipole moments (edm), ultra-cold neutrons, and possibly nuclear energy applications. An additional substantial benefit of Stage 1 is that the existing 400 MeV linac will be retired from service, removing a substantial operational risk within the Fermilab proton complex.

*Stage 2*

Stage 2 is based on extension of the CW linac to 3 GeV, with an average current of 1 mA accelerated to 3 GeV. Stage 2 thus provides 3 MW of beam power at 3 GeV, with the capability of delivering flexible beam formats to multiple experiments. It is anticipated that a Main Injector based kaon experiment would be relocated to the 3 GeV linac in Stage 2. Also accommodated are any number of muon and nuclei based experiments. Injection into the Booster at 1 GeV is retained, as will simultaneous delivery of 1 MW to the 1 GeV program established in Stage 1.

To support the Stage 2 performance the initial 1 GeV of the linac will be upgraded to 2 mA capability, with 1 mA available at 1 GeV and 1 mA transmitted into the 1-3 GeV linac section. In addition the Booster will be upgraded to 20 Hz capability.

*Stage 3*

Stage 3 completes the Reference Design via construction of the pulsed linac for acceleration of beam from 3 GeV to 8 GeV. This beam is delivered to the Recycler/Main Injector complex in support of the long baseline neutrino program. At Stage 3 >2MW of beam power is available at any energy between 60-120 GeV. Upgrades to the Recycler/Main Injector are required to support the increased beam power. Enhanced capability for delivery of 8 GeV beam, directly from the pulsed linac, is also created at this Stage. Beam capabilities at 1 and 3 GeV remain as in Stage 2.
In addition, with the completion of Stage 3 the existing 8 GeV Booster can be retired from service, taking the second substantial operating risk in the current program along with it.

*Explanation of Tables*

The tables below summarize the performance at all available beam energies at each stage. The organization of the tables is as follows:
- Each table describes beam performance in support of a particular program by Stage: Long Baseline Neutrino Program (Main Injector Fast Spill); 8 GeV Program (Booster in Stages 1 and 2, pulsed linac in Stage 3); 3 GeV Program; 1 GeV Program.
- Under each Stage there are two entries, corresponding to operations of the Main Injector at 120 GeV or at 60 GeV.
- There is a trade-off (proton economics) between beam power available for the Long Baseline Neutrino and 8 GeV program. The tables present a self-consistent set, based on the maximum beam power achievable in the Long Baseline Program and the corresponding minimum in the 8 GeV program. Subsequent to the tables is a set of figures displaying the trade-offs between the two programs.
- The beam formats for the 3 GeV and 1 GeV programs are flexible, subject to certain constraints that are described at the end of this document.

**Long Baseline Neutrino Program**
**(Main Injector Fast Spill)**

|  | Stage 1 | | Stage 2 | | Stage 3 | | |
|---|---|---|---|---|---|---|---|
|  | 120 | 60 | 120 | 60 | 120 | 60 | |
| Maximum Beam Power* | 1200 | 900 | 1200 | 1200 | 2450 | 2450 | kW |
| Protons per pulse | $7.5\times10^{13}$ | $7.5\times10^{13}$ | $7.5\times10^{13}$ | $7.5\times10^{13}$ | $1.5\times10^{14}$ | $1.5\times10^{14}$ | |
| Pulse length | 9.5 | 9.5 | 9.5 | 9.5 | 9.5 | 9.5 | µs |
| Number of bunches | 504 | 504 | 504 | 504 | 504 | 504 | |
| Bunch spacing | 18.9 | 18.9 | 18.9 | 18.9 | 18.9 | 18.9 | ns |
| Bunch length (FWHM) | 2 | 2 | 2 | 2 | 2 | 2 | ns |
| Pulse repetition period | 1.2 | 0.8 | 1.2 | 0.6 | 1.2 | 0.6 | s |

**8 GeV Program**

|  | Stage 1 (Booster) | | Stage 2 (Booster) | | Stage 3 (Pulsed Linac) | | |
|---|---|---|---|---|---|---|---|
|  | 120 | 60 | 120 | 60 | 120 | 60 | GeV |
| Minimum Beam Power* | 42 | 0 | 84 | 0 | 172 | 0 | kW |
| Protons per pulse | $6.6\times10^{12}$ | $6.6\times10^{12}$ | $6.6\times10^{12}$ | $6.6\times10^{12}$ | $2.7\times10^{13}$ | $2.7\times10^{13}$ | |
| Pulse length | 1.6 | 1.6 | 1.6 | 1.6 | 4300 | 4300 | µs |
| Number of bunches | 81 | 81 | 81 | 81 | 140,000 | 140,000 | |
| Bunch spacing | 18.9 | 18.9 | 18.9 | 18.9 | 30 | 30 | ns |
| Bunch length (FWHM) | 2 | 2 | 2 | 2 | .04 | .04 | ns |
| Pulse repetition rate | 15 | 15 | 20 | 20 | 10 | 10 | Hz |

| 3 GeV Program | Stage 1 | | Stage 2 | | Stage 3 | | |
|---|---|---|---|---|---|---|---|
| | 120 | 60 | 120 | 60 | 120 | 60 | GeV |
| Beam Power | NA | NA | 3000 | 3000 | 2870 | 2870 | kW |
| Protons per second | NA | NA | $6.2\times10^{15}$ | $6.2\times10^{15}$ | $6.2\times10^{15}$ | $6.2\times10^{15}$ | |
| Pulse length | NA | NA | CW | CW | CW | CW | µs |
| Bunch spacing** | NA | NA | Programmable | Programmable | Programmable | Programmable | ns |
| Bunch length (FWHM) | NA | NA | .04 | .04 | .04 | .04 | ns |

| 1 GeV Program | Stage 1 | | Stage 2 | | Stage 3 | | |
|---|---|---|---|---|---|---|---|
| | 120 | 60 | 120 | 60 | 120 | 60 | GeV |
| Beam Power | 984 | 984 | 990 | 990 | 1000 | 1000 | kW |
| Protons per second | NA | NA | $6.2\times10^{15}$ | $6.2\times10^{15}$ | $6.2\times10^{15}$ | $6.2\times10^{15}$ | |
| Pulse length | NA | NA | CW | CW | CW | CW | µs |
| Bunch spacing** | NA | NA | Programmable | Programmable | Programmable | Programmable | ns |
| Bunch length (FWHM) | NA | NA | .04 | .04 | .04 | .04 | ns |

**Notes**

* Beam Power available from the Main Injector and at 8 GeV are dependent upon the disposition of protons provided at 8 GeV and the operational energy of the Main Injector. It is assumed that the disposition of protons will be a program planning decision based on the physics opportunities at the time. Below are presented the dependence upon available beam power at 8 GeV as a function of beam power available from the Main Injector, at each stage and for each of two operating energies.

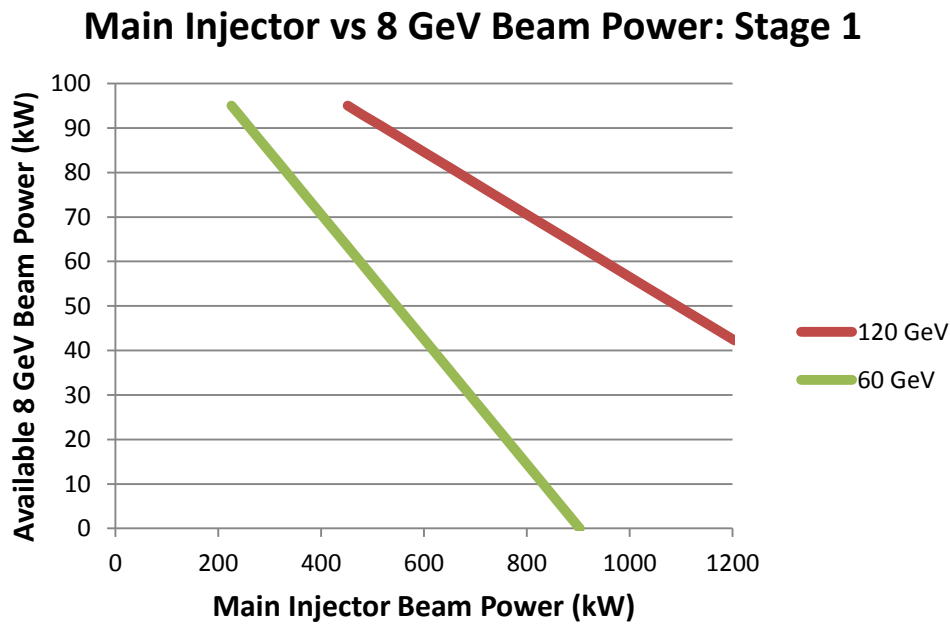

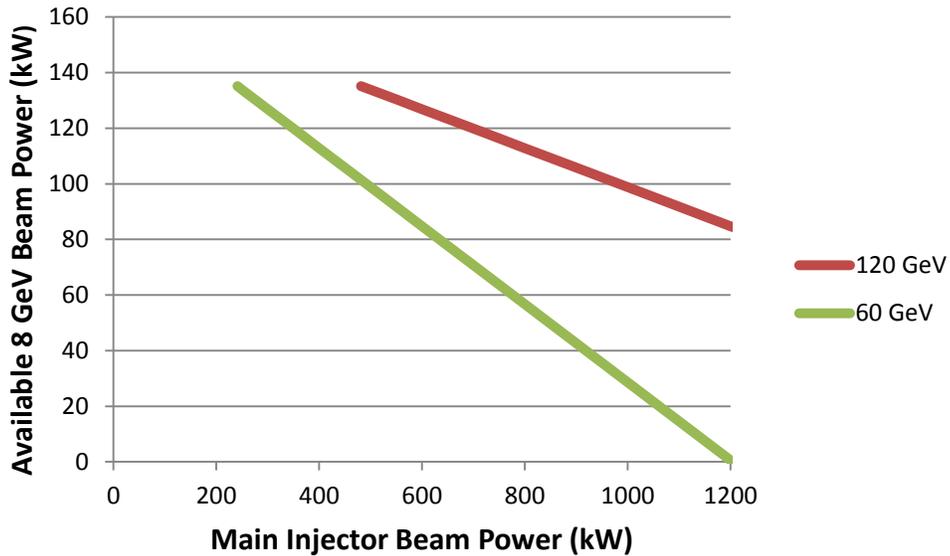

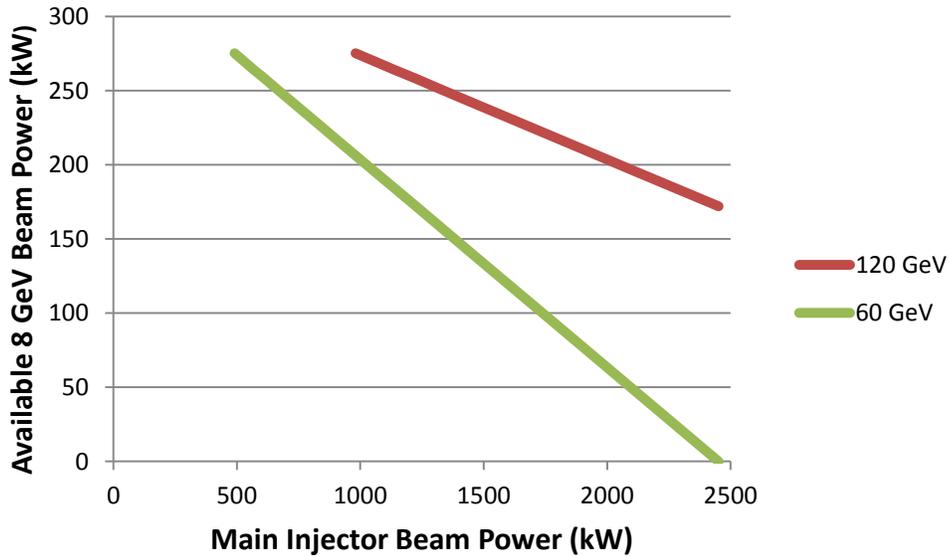

\*\* Independent bunch structures can be provided from the 1 and 3 GeV linac to three experimental areas simultaneously. The bunch pattern in any particular area must conform to the following requirements:

- Each bunch contains up to $1.9 \times 10^8$ particles (H- ions);

- Bunches in each experimental area must be separated by either 12.4, 24.8, 49.6, or 99.2 ns (80, 40, 20, 10 MHz);
- The bunch patterns must repeat every 1.0 μsec;
- The total current, summed over the three experimental areas, must be 1 mA averaged over the 1.0 μs period.

An example is given in the body of this report (Figure II-2 and Figure II-3).

## I.2 Booster Upgrades for Stage 1

The Booster synchrotron will require modification for its role in Project X Stage 1 where it acts to accelerate the 1 GeV beam to 8 GeV so it can be injected into the Recycler and ultimately the Main Injector. Major items of consideration are:

- The ultimate intensity limitations within the Booster.
- The new transfer line and injection insertion for bring the 1 GeV H⁻ beam into the Booster.
- Changes to the RF system to accelerate the increased beam current.
- Beam loss control and shielding to allow operation.
- Consideration of replacement of various components as they become aged or outdated.

The Booster accelerated its first beam to 8 GeV on May 21, 1971. When Project X begins operation it is likely that the Booster will be at least 50 years old. Many items in the Booster have been replaced over time, but several significant legacy systems remain including: the main magnets, much of the RF cavity system, the support girders, and the enclosure itself. Each of these items is an asset when considering the future, but also a liability.

The Project X Booster upgrades rely on previous upgrade programs to be completed successfully, the Accelerator and NuMI Upgrades (ANU) of the NOvA Project, and the Proton Improvement Plan (PIP). ANU is nearly complete (as of spring 2013) and concentrates on the Main Injector and Recycler. PIP is primarily concerned with the Proton Source, including the Booster and Linac. The PIP upgrades are a vital prerequisite for the Project X upgrades. PIP will act to significantly increase the repetition rate and intensity limits in the Booster; it will also address legacy issues, particularly the main RF cavities.

### Booster Intensity Limitations

The ultimately achievable Booster beam intensity depends on several factors. The first and most apparent is the space charge limitation. The Booster beam experiences a force from its own electromagnetic field, distorting the optics in an intensity-dependent way. The primary impact is to cause an amplitude-dependent defocusing that results in a tune spread of the beam particles. This can be represented by a tune spread $\Delta \nu$

$$\Delta \nu = -\frac{\pi \lambda r_0 R}{2 \varepsilon_N \beta \gamma^2}$$

where $\lambda$ is the linear charge density, $r_0$ is the classical proton radius, $R$ is the radius of the accelerator, $\varepsilon_N$ is the normalized emittance of the beam, and $\beta$ and $\gamma$ are the relativistic parameters. The tune shift rapidly decreases with beam energy, so that a correspondingly larger line density (beam intensity) will result in the same tune shift. From this relationship we find that an intensity increase of a factor of 2.57 (157%) would result in the equivalent space charge tune shift at 1 GeV as at 400 MeV.

However, the above ratio does not fully determine the intensity limit. A larger tune shift at 1 GeV would also produce a larger tune shift through the remainder of the cycle, including crucial times such as transition crossing. Furthermore, the effects of space charge presently lead to beam loss at low energy which cannot be withstood at the Stage 1injection energy. An equivalent relative loss with Project X would occur at a higher energy and greater proportional intensity. We believe these effects balance out with an increase of Booster intensity of 50% above current operations.

**Existing Booster Injection Straight section**

The Booster lattice contains 24 periods and can be described as a FoDoDoFo lattice utilizing gradient magnets with long straight sections (5.66 meters) between the defocusing gradient magnets and short straight sections (0.5 meters) between the F and D gradient magnets. Horizontal beta function varies from about 6 meters in the long straight to 33 m in the short straights while the vertical beta function varies from 20 m in the long straights to ~5.3 m in the short straights. The horizontal dispersion varies between approximately 1.8 (in the long straights) and 3.6 meters around the ring. The ring optics are shown if Figure 1.

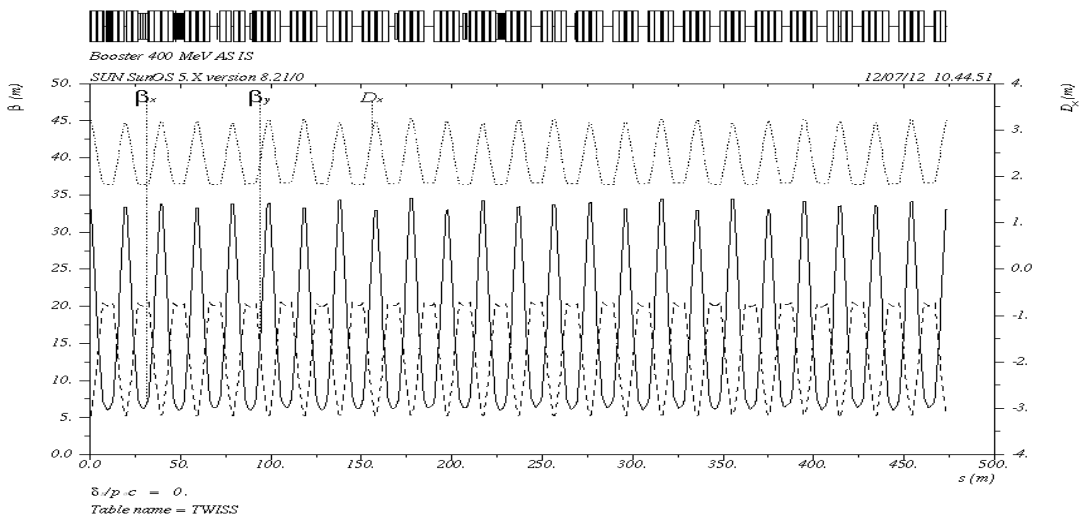

**Figure 1**: Lattice functions of the current Booster ring (MAD)

The injection straight section is located in the long straight section 01. A new injection design using a three dipole orbit bump magnet system (ORBUMP) was installed in 2006. The lattice functions and horizontal dispersion are shown in Figure 2 and listed in Table 1. The gradient magnets, ORBUMP magnets, and correctors are shown at the top of the figure. This new design, as compared to the original injection system with a septa and 4 chicane magnets, reduced the injection angle, the number of required dipoles, and the required strength of the dipoles by about a factor of two. The center dipole of this insert used to merge the incoming H- on the closed orbit produced by the three dipoles and runs at twice the field (current) as the outer dipoles. The injection foil is located immediately after the middle ORBUMP magnet. The current angle produced by the center dipole is approximately 44 mr which corresponds to an integrated field of 1.4 kG-m. Lorentz stripping in these magnets is not an issue at 400 MeV.

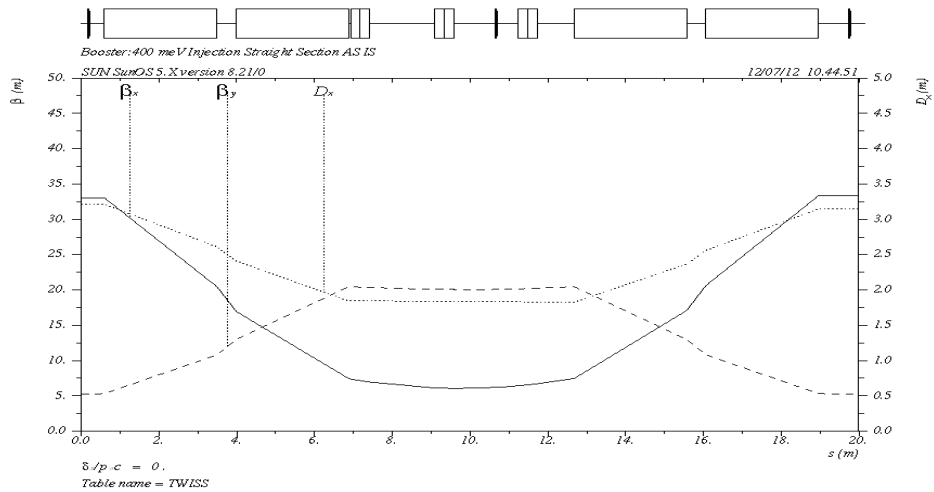

**Figure 2**: Lattice functions of the current Long 01 injection straight.

| | |
|---|---|
| $\beta_x$ | 4.88m |
| $\alpha_x$ | 0.049 |
| $\beta_y$ | 18.52m |
| $\alpha_y$ | 0.011 |
| $D_x$ | 1.73 |
| $D'_x$ | -0.003 |

**Table 1**: The current lattice functions at the foil.

New pulsed ORBUMP magnets were designed to be capable of continuous operation at 15 Hz and at a peak/integrated dipole field of up to (at 15 kA excitation) 3 kG/1.676 kG-m. At 400 MeV this would produce a bend angle of 52.6 mr. A cross section of the magnet is shown in Figure 3.

The ORBUMP displaces the closed orbit during injection by approximately 45 mm at the location of the foil. The edge of the foil is placed at approximately 32 mm. Figure 4 shows a 6σ beam envelope for a normalized emittance of 16 π-mm-mr. In the current design, the unstripped H-, protons, and neutrals follow the same trajectory until they enter the third ORBUMP magnet. This dipole places the protons on the Booster central trajectory while the $H^0$ continue undeflected until they hit the radial outside of the second downstream gradient magnet. The H- are initially deflected inward by the third dipole, but then deflected radially outward, crossing the closed orbit by the gradient magnets and are also deposited on the radially outside of the second gradient magnet.

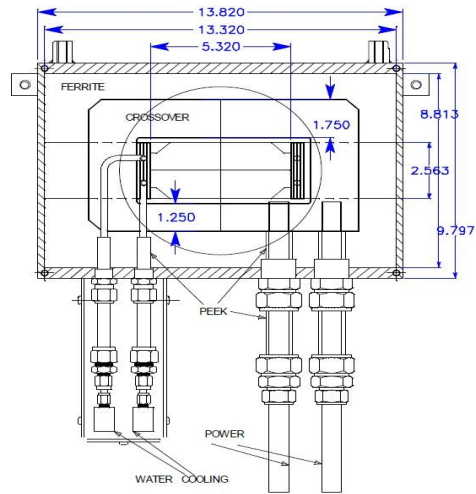

**Figure 3**: Cross section of the Booster pulsed ORBUMP magnet.

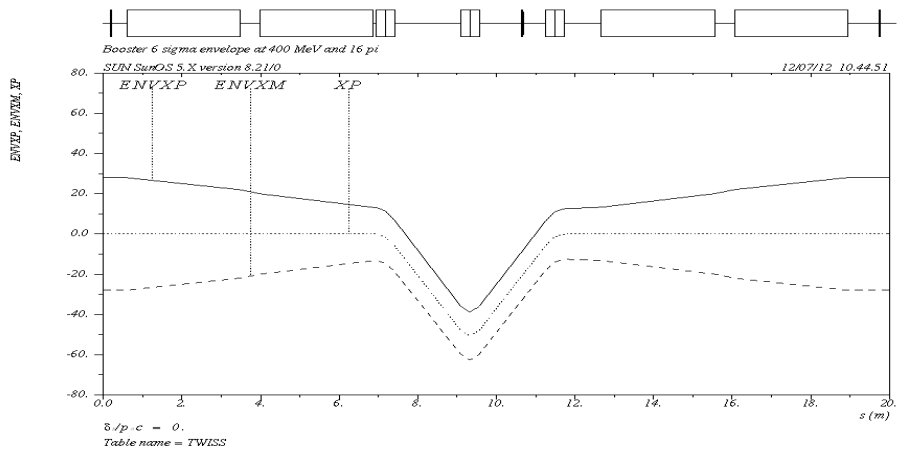

**Figure 4**: The horizontal 6σ beam envelope for a 16 π-mm-mr 95% emittance showing the closed orbit due to the ORBUMP dipoles. The injection foil is located approximately 4.5 inches downstream of the middle bump magnet.

The injection foil changer, pictured in Figure 5, has the capability of holding up to 8 foils.

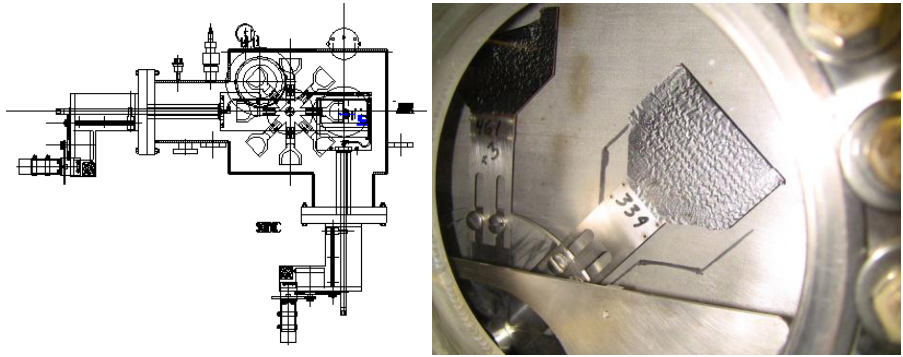

**Figure 5**: Current Booster injection foil changer and foil.

The expected stripping efficiency for the production of protons at 400 MeV, according to scaled Gulley parameters, for a carbon stripping foil of thickness 380 μg/cm$^2$ (~1.15 μm) is 99.9% which means that only 0.1% of the beam exits the foil as excited states of H$^0$. The majority of these states will be in the lowest states, will not be stripped, and end being lost in the second downstream gradient magnet on the aisle side. In addition, any H$^-$ leaving will get bent to the radial inside of the ring by the third ORBUMP magnet and any that enter the downstream gradient magnet will get bent to the outside of the ring and be lost in the downstream gradient magnets. Two loss points, one for H$^-$ and the other for H$^0$ have been identified on the aisle side of the second gradient magnet downstream of the injection straight. With an average injection intensity of 5E12 ions/cycle at a 7.5 Hz repetition rate the injected beam power is 2.43 kW. If 0.1% of the beam is lost due to incomplete stripping, this would correspond to approximately 2 W of beam loss at these two locations. Radiation surveys of these locations upon turning the Booster off show on the order of a few R "on-contact". At the 1 GeV injection energy and a 15 Hz repetition rate and 7E12 injected/cycle, the injected beam power rises to 16.8 kW and for 99% stripping efficiency and one is looking at a loss of 170 Watts or a factor of 8.5 increase in "on-contact" residual activity. A better solution for handling and safely disposing of the un-stripped injected H- will be incorporated into the new design.

The current injection into Booster utilizes a linac pulse length equivalent to 1-10 turns (~2.2 us/turn) and does not utilize any transverse phase space painting. The transport line is "matched" to the ring lattice functions and therefore the linac transverse emittance defines the "base" emittance of the beam in Booster. To this is added any emittance growth due to lattice, dispersion, or trajectory mismatch and scattering in the foil. The average scattering angle is only 33 μr. The current 95% normalized emittance of the Booster beam at extraction is on the order 12 to 16 π-mm-mr.

**Injection at 1 GeV**

Stage 1 of Project X will increase the injection energy of the current Booster to 1. The average current of the new CW linac is 1 mA over a period of 1ms (Stage 1, rising to 2 mA over 0.5 ms in Stage 2).

To meet the proton intensity out of the MI/Recycler of 7.5E13/cycle requires Booster to provide 6.5E12/15Hz cycle, assuming a 95% slip stacking / acceleration efficiency in MI/RR. Assuming a 90% injection and acceleration efficiency in Booster, the linac must inject 7.3E12/15Hz cycle. If we assume a chopping factor of 50%, the peak linac current will be required 2 mA.

*Booster Geometry for 1 GeV*

We are currently evaluating utilization of the present long straight section in the Booster for injection at 1 GeV. It should be noted that the 1 GeV injection straight section for SNS is 12.5 m between quadrupoles and they utilize a 4 dipole chicane for injection. A secondary foil is utilized to convert any neutrals or H- that miss the foil to protons and transport them to an external dump. As noted above, the current Booster long straight section is 6 m between gradient magnets and does not contain an external dump for the neutrals or H- that miss the foil.

Two possible geometries for the injection chicane are a three dipole chicane, as in the current Booster implementation, and a four dipole chicane as utilized at SNS and the default design for H- injection into the Recycler. In addition there are two potential injection orientations, horizontal and vertical.

To determine if the existing ORBUMP magnets could be utilized for 1 GeV operation we scale the central ORBUMP magnetic field by the ratio $\beta\rho_{1G}/\beta\rho_{0.4G}$, and the required integrated dipole field of the central magnet becomes ~2.5 kG-m for the existing magnet steel length of 0.4823m (effective length of 0.5585m). This would correspond to a field of ~4.5 kG, about 50% greater than the current magnet design field. This indicates that the current three bump design would not work with the existing ORBUMP magnets. Increasing the (effective length) to roughly 1 m (physical length 0.86 meters) or using 2 of the existing ORBUMP magnets for the central bend brings the magnet strength back down to ~2.5 kG, well within the design parameters of the existing magnets. During the construction of the chicane magnets the asymmetry in stripline leads contribute to a significant quadrupole term such that the project should consider the construction of new dipoles.

A four-dipole design reduces the required strength of the center dipole as the chicane essentially consists of two "dog-leg" bumps with the stripping foil located between the two central dipoles. If the injection offset remains at 45 mm, then the required angles for all chicane magnets is roughly 22 mr. At 1 GeV this corresponds to an integrated strength of ~1.25 kG-m or a field of roughly 2.2 kG, well within the design of the existing magnets. In this scenario, the $H^0$ would pass undeflected through the third chicane dipole and the unstripped H- would be deflected in the opposite direction of the protons. This geometry requires the incoming H- trajectory to be at an angle of 22 mr with respect to the Booster straight section. This injection trajectory of the H- entering the second ORBUMP magnet (for merging the H- onto the circulating orbit) would need to miss the upstream gradient magnet which has a half-width of roughly 9 inches. Since this second magnet is only a few meters downstream of the gradient magnet a new strong septum magnet would be required for the injected beam to miss this gradient magnet.

Another consideration regarding the geometry of the injection is which injection plane would provide the fewest parasitic hits from the circulating beam on the foil. Current Booster injection is in the horizontal plane where the beta function is a factor of 3.8 smaller than that of the vertical plane. This will lead to a smaller orbit movement during painting injection, thus allowing more time for the already injected protons to impact the foil. In addition, the vertical size of the upstream gradient magnet is only 2/3 that of the horizontal which makes it easier to clear the upstream gradient magnet, in the case of a four dipole design. The selection of the injection plane would also impact the transport line design.

### *Magnetic Field and Lorentz stripping*

At 1 GeV, dipole fields in the 2.5 to 3 kG range would lead to a loss rate due to Lorentz stripping in the range of $10^{-7}$ to $10^{-9}$/m, a negligible loss for the expected injection power of 16.8 kW. Figure 6 shows the loss rate/meter due to Lorentz stripping for 400 MeV and 1 GeV kinetic energies.

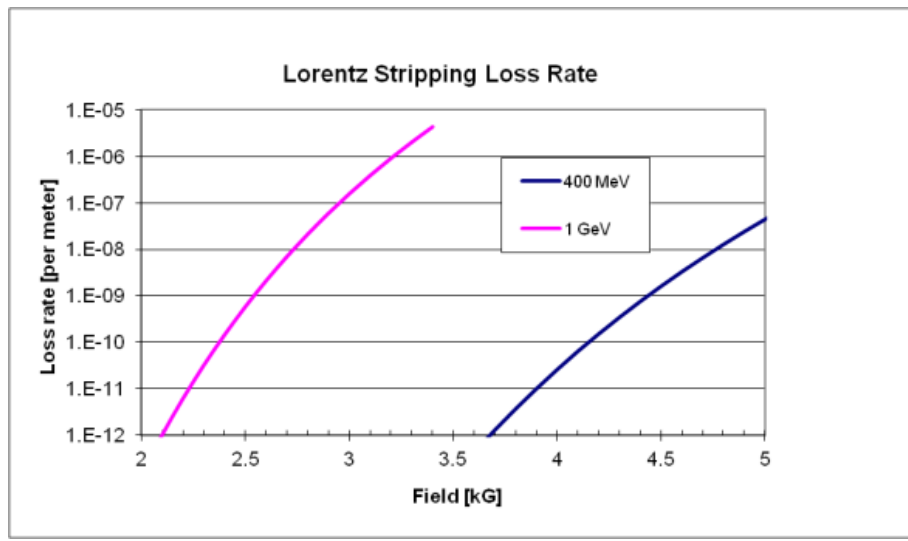

**Figure 6**: Loss rate due to Lorentz stripping for 400 MeV and 1 GeV H- ions.

*Booster straight section modifications*

In order to minimize the activation of the downstream magnets by the unstripped H⁻ and H⁰, it is advantageous to install a well shielded absorber after the injection stripping foil, between the last ORBUMP dipole and the first gradient magnet. Currently, this is not feasible as the space is only ~ 1/3 meter in length. However, if the defocusing gradient magnets on either side of the injection straight can be shortened and the space occupied by ORBUMP can be slightly reduced, space can be opened up in which an absorber could be installed.

In order to create additional room for either injection scenario, the two gradient magnets were shortened by 25% (from 2.8896m to 2.167m) and the dipole field and gradients were increased by 25%, while keeping the bend center at the same point. This will increase the straight section by a modest 0.72m. This modification introduced an RMS variation in $\beta_x$ ~0.6%, $\beta_y$ ~4%, and $D_x$ by ~3%. The lattice with this modification is shown in Figure 7.

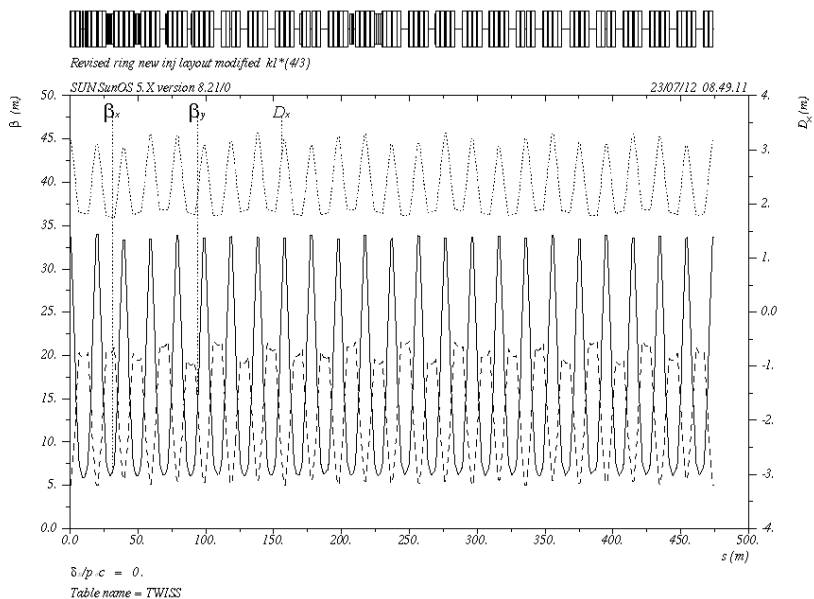

**Figure 7**: Lattice distortion due to shortening the two D gradient magnets on either side of the straight to ¾ length and increasing gradient by 4/3 keeping bend center fixed.

These errors in $\Delta\beta_x$, $\Delta\beta_y$, $\Delta D_x$ may be corrected to a ~1% level by utilizing five quad correctors around the straight section, requiring only a few amps in each, however the correction at high field pushes the correctors close to their maximum. Another technique is to modify the gradient slightly in the two shorter gradient magnets. Reducing this to 98% of the scaled value corrects distortions in both beta functions, but there is a small residual dispersion distortion. These corrections are a first pass solution. Figure 8 shows the result of the lattice with only the ratio of the dipole and gradient adjusted.

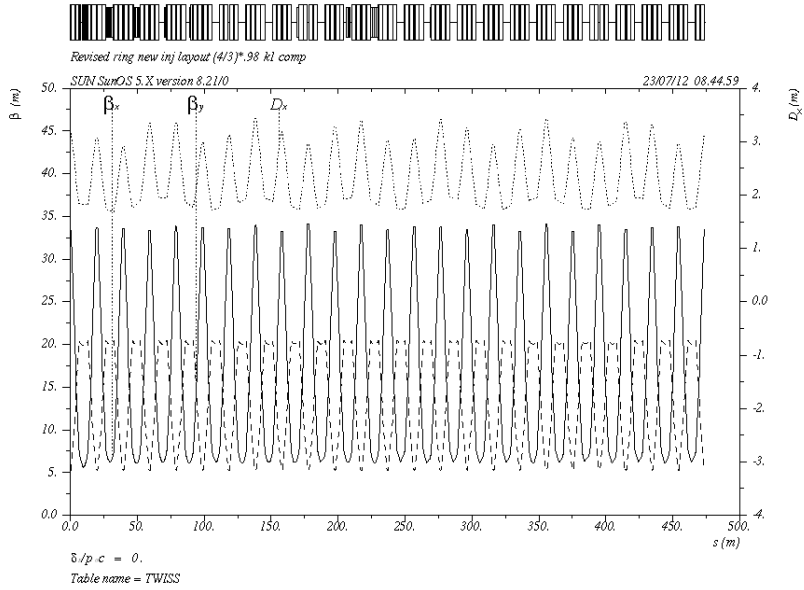

**Figure 8**: Lattice error from shortened magnet corrected by reducing gradient

Shortening the ORBUMP insert and shortening the adjacent gradient magnets allows the insertion of an absorber capable of shielding the downstream magnets from the unstripped neutral hydrogen atoms and the H⁻ missing the foil. Figure 9 shows an example of the new layout with the 6σ beam envelope for a 20π normalized emittance of a 1 GeV beam. The face of the absorber is approximately 0.95 m (~3 ns) downstream of the last ORBUMP magnet. The absorber with some downstream shielding is shown as a blue block in the figure. The $H^0$ and $H^-$ trajectories are shown in blue.

According to PSTAR tables the projected range of 1 GeV protons in tungsten is 6.020 g/cm² which for a density of 19 g/cm³, yields a stopping range of 31.7 cm or 12.5 inches. The current Booster injection straight section was modified in MAD to reduce the adjacent gradient magnet length by 25% and shorten the ORBUMP insert which would allow the insertion of an absorber between the last ORBUMP magnet and the first gradient magnet. This absorber would have the dimensions of approximately 5 cm H x 6 cm W x 35 cm L with the inside edge at about 1.8 cm from the Booster centerline.  This would still allow additional 30 cm shielding between the absorber and gradient magnet. Further refinement of the absorber geometry and a MARS simulation for the absorber assembly needs to be done.

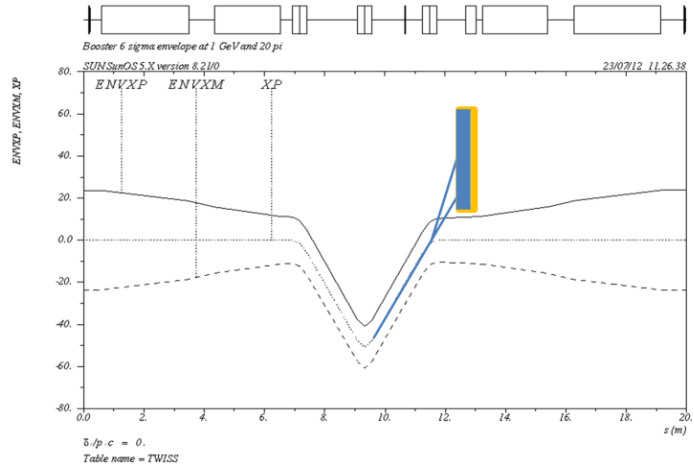

**Figure 9**: The potential location of an absorber for un-stripped neutrals and H- missing the foil located between the last ORBUMP magnet and the first gradient magnet.

The efficiency for converting H$^-$ ions into protons depends on the ion energy (velocity) and foil areal density (thickness). Figure 10 shows the yield of protons and the first two states of neutral hydrogen for 800 MeV and 1 GeV. The 1 GeV estimate is scaled from the 800 MeV predictions. The red curve contains the yield for the first three states of H$^0$. Table 2 summarizes the values for a 500 μg/cm$^2$ carbon foil. The yield of states n $\geq$ 4 is not tabulated here, but the n=4 state is roughly a factor of 2 less than n=3 and the yield of the higher order states drop rapidly. So, essentially only the first few excited states are of interest. All of these states are stable in the absence of an external magnetic field. In the proposed configuration the stripping foil is after ORBUMP 2 magnet, in the decaying field. The third ORBUMP magnet has a peak field of roughly 1.5 kG. Figure 11 shows the lifetime of the Stark States of neutral hydrogen as a function of magnetic field. We see that the field of 1.5 kG is in the middle of the n=5 stark states. All lower states will not be stripped. With the position of the absorber it will intercept all H$^0$ and H$^-$ (that miss the foil).

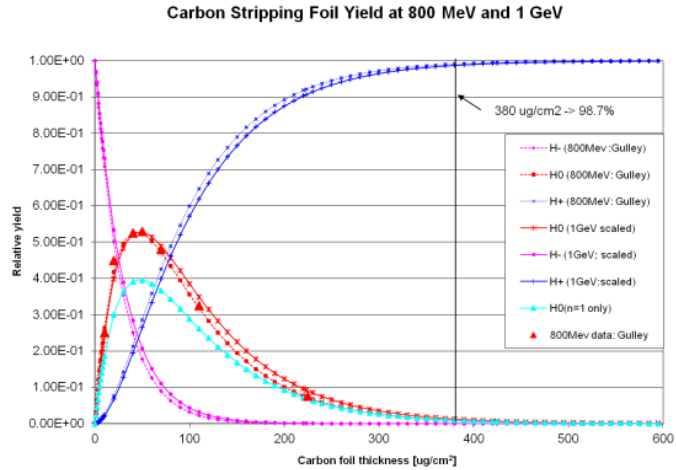

**Figure 10**: Stripping yield as a function of carbon foil thickness for 800 MeV data and scaled to 1 GeV.

| State | Fraction |
|---|---|
| $H^+$ | 99.7% |
| $H^-$ | 9E-6% |
| $H_0$ (n=1) | 0.2% |
| $H_0$ (n=2) | 0.07% |
| $H_0$ (n=3) | 0.02% |

**Table 2**: Estimates for production of the first few excited states of hydrogen from 1 GeV H- interacting with a carbon foil with thickness of 500 μg/cm².

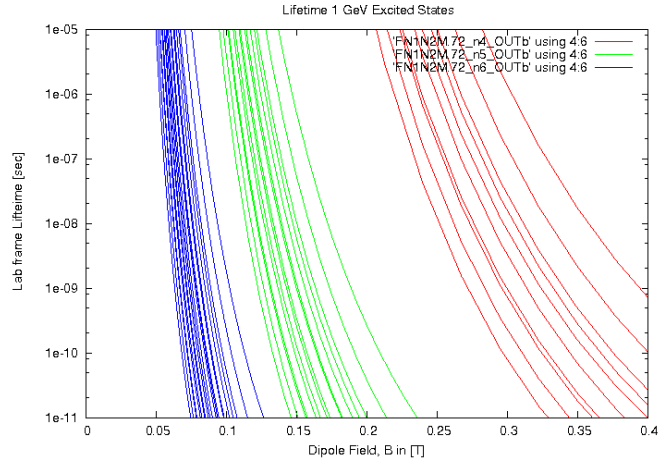

**Figure 11**: Lifetime of excited states with n=4, 5, and 6 of hydrogen at 1 GeV.

*Transverse phase space painting*

The expected 95% normalized emittance of the 1 GeV linac is on the order of 1.7 $\pi$-mm-mr. The design final emittance of the painted beam in the Booster is 20 $\pi$-mm-mr or a ratio of 0.085. There are two main options for performing transverse phase space painting into the Booster, painting in both dimensions in the ring and painting in one dimension in the ring and steering from the beam line in the other dimension. Both of these techniques are utilized in existing machines: SNS paints in both dimensions and JPARC paints and steers. In each of these scenarios the painting algorithms may be either correlated or anti-correlated in direction of painting of small to large or large to small amplitude. In addition, the functional form needs to be optimized to minimize over painting of the phase space as well as producing a KV-like distribution and minimizing the number of parasitic interactions between the circulating protons and the foil.

A Twiss parameter mismatch between the injection line and ring at the point of injection can be useful in minimizing the foil size as well as the number of parasitic hits by matching the orientation of the injected beam within the phase space of the circulating emittance. The first condition assures the injected beam is optimally positioned within the desired beam emittance of the ring and is given by:

$$\frac{\alpha_i}{\beta_i} = \frac{\alpha_r}{\beta_r} = \frac{X'_C - X'_o}{X_C - X_o}$$

where $\alpha$ and $\beta$ are the standard lattice functions in the injection line (*i*) and ring (*r*) and $X$ and $X'$ are the beam position and angle for the injected beam centroid (*C*) and the closed orbit (*o*). This effectively demands that the orientation of the injected and circulating be matched as shown in the left hand plots in Figure 12.

The next condition assures the injected beam has the proper aspect ratio.

$$\frac{\beta_i}{\beta_r} \geq \left(\frac{\varepsilon_i}{\varepsilon_r}\right)^{1/3}$$

Given the expected linac 95% normalized emittance of ~ $1.7\pi$ mm-mr, the final painted 95% normalized emittance in the Booster has been specified to be $20\pi$ mm-mr. This is a ratio of 0.085 and allows the definition of the transport lattice functions at the position of the foil to be $\beta_x = 2.1$ m and $\beta_y = 8.0$ m that are optimized for painting in both dimensions in the ring. For the paint/steer option the horizontal injected beam is matched as in the ring painting, but the vertical $\beta$ for the injected beam needs to be increased to roughly 26m to get a better match. Note that in the paint/steer option that the vertical position of the injected beam remains constant while the angle is varied from large to small during the injection process, hence the vertical angle offset. The two cases are shown in Figure 12.

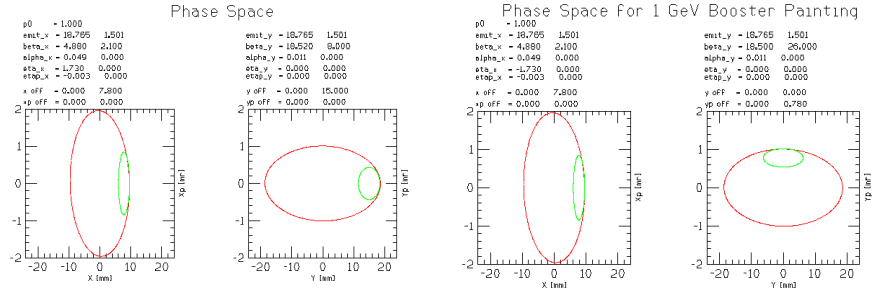

**Figure 12**: Two phase space painting options. Painting in both planes in the ring (left) and horizontal painting and vertical steering (right).

The usual injection orientation is to inject in the horizontal direction. However, in the Booster the horizontal beam size is about a factor of three smaller than the vertical. This implies that the rate the beam is removed from the foil is a factor three smaller for the horizontal direction than for the vertical, which leads to greater parasitic beam hits on the foil. The details of various painting options remain to be simulated.

Investigation of injection in the vertical plane using both a three bump chicane and four bump chicane indicate that vertical separation of the neutrals off the foil and the circulating protons is too small before the downstream gradient magnet to allow the insertion of an injection absorber. However, if the geometry of the injected beam can be arranged to miss the upstream ORBUMP magnet and gradient magnet, then this geometry would all an injection absorber to be installed just upstream of the last chicane dipole. Here, we move the first and last chicane dipoles as far out as possible and center the central chicane dipoles with the foil between them.

### Foil Issues

The thickness, geometry, mounting and orientation of the injection foil will be dependent on the particular painting scenario adopted.

The objective in selecting the foil thickness is to maximize the stripping probability while minimizing the losses due to secondary hits from the circulating protons. Thicker foils maximize the probability of stripping thus reducing the amount of unstripped $H^0$ that the injection absorber must dispose but increase losses from circulating beam hits. So the strategy will be to minimize losses due to each mechanism.

There are two classes of beam power related to foil heating. The first is where the linac injects a high current beam for a small number of turns and the second is where the linac current is low and requires many more turns to accumulate the required charge in the synchrotron. Stage 1 of the Project X will inject a low current (1 mA average) from a CW linac for 600 turns. In this scenario, most of the foil heating will be due to parasitic hits from the already circulating beam. The single pass injection will be only a small component. The parasitic hit density is determined by the specifics of the foil size and geometry and on the particular painting algorithm used. We can get an estimate of the minimum number of hits per proton by

$$h_{min} = \frac{1}{4} N_t \left(\frac{\varepsilon}{A}\right)^{\frac{4}{3}}$$

where $N_t$ is the number of injection turns, $\varepsilon$ is the injection beam emittance and $A$ is the final painting emittance. For the given values of the injection and Booster emittance and assuming 600 turns, the minimum number of hits is 5.6/injected proton.

The injection beam power is 17.5 kW. If 1% of this beam remains un-stripped at injection, an average power of 175 W would be deposited into the injection absorber.

### Booster Injection Energy

In order to accumulate the required intensity in Booster, the injection period needs to be on the order of 600 turns. At 1 GeV the revolution period is approximately 1.78 μs which yields an injection time of about 1.1 ms. Several scenarios have been discussed on how to handle this long injection time in a ring operated as a resonant circuit at 15 Hz. The current in the dipole bus, thus energy, goes as

$$I(t) = I_{inj} + dI \sin(2\pi f t)$$

where $I_{inj}$ is the current at injection, $dI$ is the differential current between injection and extraction, and $f$ is the Booster repetition rate of 15 Hz. One option discussed is to start injection 500 μs before the minimum field and end injection 500 μs after, on the Booster ramp. Here the linac energy and Booster RF frequency would track the change in momentum due to the Booster ramp to keep the central orbit centered. The other option would create an injection front porch such that the energy remains constant during injection. An initial model of a single Booster magnet cell indicates that it is possible to flatten the injection front porch to on the order of a millisecond. This has not been applied to the model of the full ring and no design exists of such. This will require dedicated time for such activity.

### *Chopping patterns*

At 1 GeV, the RF frequency in Booster for a harmonic $h = 84$ is 46.46 MHz. The ratio of the linac bunch frequency (162.5 MHz) to the Booster is 3.497. Figure 13 shows the Booster RF voltage as a function of time in red. The impulses (blue) represent the 162.5 MHz linac bunch spacing. The negative impulses represent empty linac beam buckets where the bunches have been removed by the chopper. Due to the harmonic ratio being close to half-integer (3.5), the filling pattern is constant during a single turn Since we are attempting synchronous transfer we need to restrict the phase at which the bunches land within the Booster bucket to less than ±90 degrees.

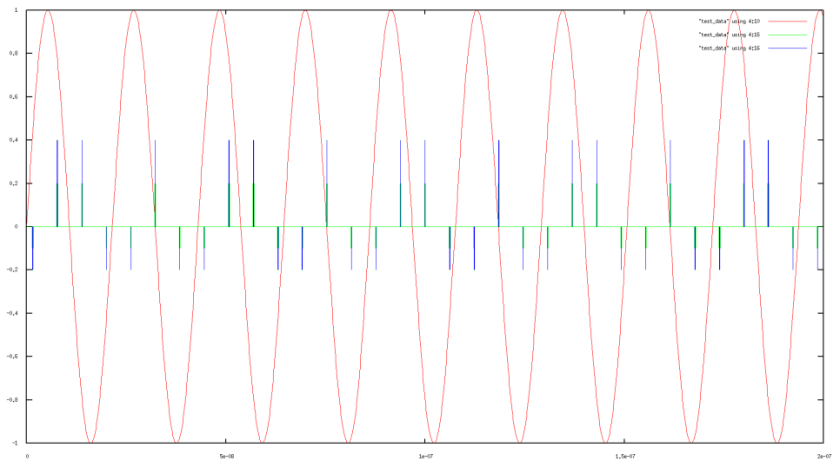

**Figure 13**: Booster RF and linac bunches injection scheme.

With this chopping pattern and an even harmonic number, alternate Booster buckets will be filled unevenly over the full injection time. To remedy this, the harmonic number of Booster, and hence the Recycler and Main Injector may be changed by ±1 such that Booster buckets will be filled alternately by one or two linac bunches on alternate turns which means that two turns would be needed to inject 3 bunches into each bucket (see Table 3). For a fixed number of injection turns the peak linac intensity would have to be increased by 3/2 above the 1 mA average. Longitudinal painting in phase would be required, on a turn by turn basis, in order to spread out the bunch distribution inside the Booster bucket.

Preliminary ESME simulations have been performed utilizing a stationary bucket for three RF configurations, fundamental frequency only, fundamental+$2^{nd}$ harmonic, and fundamental+$3^{rd}$ harmonic while restricting the phase at which linac bunches can be injected. The injected distributions are then accelerated. One finds that restricting the injection phase to ±60° allows 97% of the beam to survive to transition for the case of the fundamental + $2^{nd}$ harmonic.

|   | Booster | | | Recycler | |
|---|---|---|---|---|---|
| h | $E_{inj}$ [GeV] | $f_{inj}$ [MHz] | $f_{ext}$ [MHz] | $f_{inj}$ [MHz] | $f_{ext}$ [MHz] |
| 83 | 1.075 | 46.421148 | 52.1775 | 52.182963 | 52.47129 |
| 84 | 0.995 | 46.426612 | 52.8061 | 52.81167 | 53.10348 |
| 85 | 0.925 | 46.421980 | 53.43479 | 53.44039 | 53.73566 |

**Table 3**: Impact of harmonic number choice on RF frequency

**Harmonic RF Systems**

The addition of a $2^{nd}$ or $3^{rd}$ harmonic RF cavity system will promote more efficient capture, acceleration, and transition-crossing in the Booster. Without increased efficiency of those processes the losses will be too great within the Booster enclosure. This section explores the application of these harmonic systems.

*RF for Injection and ramping*

Injection at 1 GeV in Project X stage 1 will take about 1 ms (613 turns) The Linac bunch frequency at injection is 162.5 MHz while the Booster RF frequency at 1 GeV is 46.47 MHz. Therefore, there must be three Linac bunches injected into one single Booster RF bucket at each turn in order to meet the intensity requirements. However, complications arise because the Linac RF and the Booster RF are not integer multiples and so phase locking is not possible. Each injection will be at a random Booster RF phase and so after 1 ms of continuous injections, the bucket will be completely full. To maintain the high injection efficiency and minimize losses, injection into a stationary Booster RF bucket is needed. Since the Booster has a continuous sinusoidal ramp during injection, a stationary bucket can be created if the following relation is satisfied:

$$\frac{\dot{f}}{f} = \alpha \frac{\dot{B}}{B}$$

where $f$ is the RF frequency and $B$ is the magnetic field and the differentials $\dot{f}$ and $\dot{B}$ are with respect to the time variable $t$. For the Booster, the momentum compaction factor $\alpha = 0.072$. If the injection starts 500 $\mu s$ before the minimum $B$ field and finishes 500 $\mu s$, then the RF frequency change is approximately:

$$\Delta f = \left[\left(\frac{B}{B_0}\right)^\alpha - 1\right] f_0$$

If $B/B_0 \cong 1.004$, then $\Delta f \approx 13.36$ kHz. And during this time, the synchronous radius changes by the following amount:

$$\Delta R \cong R_0 \, \alpha \frac{\dot{B}}{B_0} t$$

With $R_0 = 75.47$ m, $\dot{B}/B_0 \cong 15.54$, $\Delta R \approx 42$ mm, which means the average aperture needed is about 42 mm at injection. For stationary injection capture, the minimum RF voltage required for a given beam energy spread is shown in Figure 14. If the maximum injection energy error is ± 1 MeV, then the minimum voltage is about 40 KV.

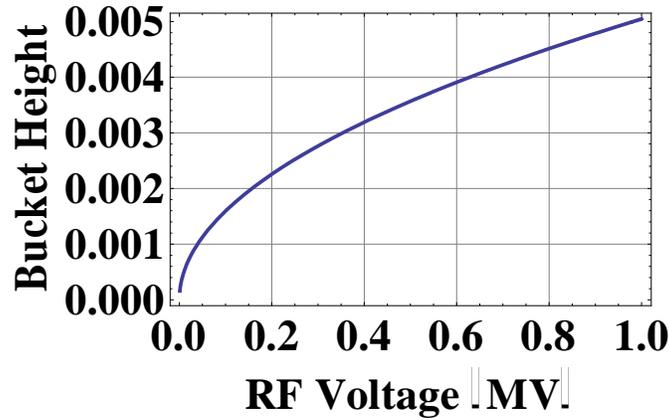

**Figure 14**: The minimum RF voltage required for given beam moment spread.

If the Linac can deliver over 3.6 mA peak current, chopping can be applied to the Linac beam so that two bunches are injected per bucket. However, bunch patterns will need to be programmed per turn so that each injected single or pair of bunches are within $\pm\theta$ degrees of the bucket. Figure 15 shows the injection distribution when one or two bunches are injected and confined to ± 90° of the bucket. There is a double hump in the injected beam distribution because two bunch injections can only be supported in certain ranges of longitudinal phase. The beam distribution at the end of injection is shown in Figures 16-18.

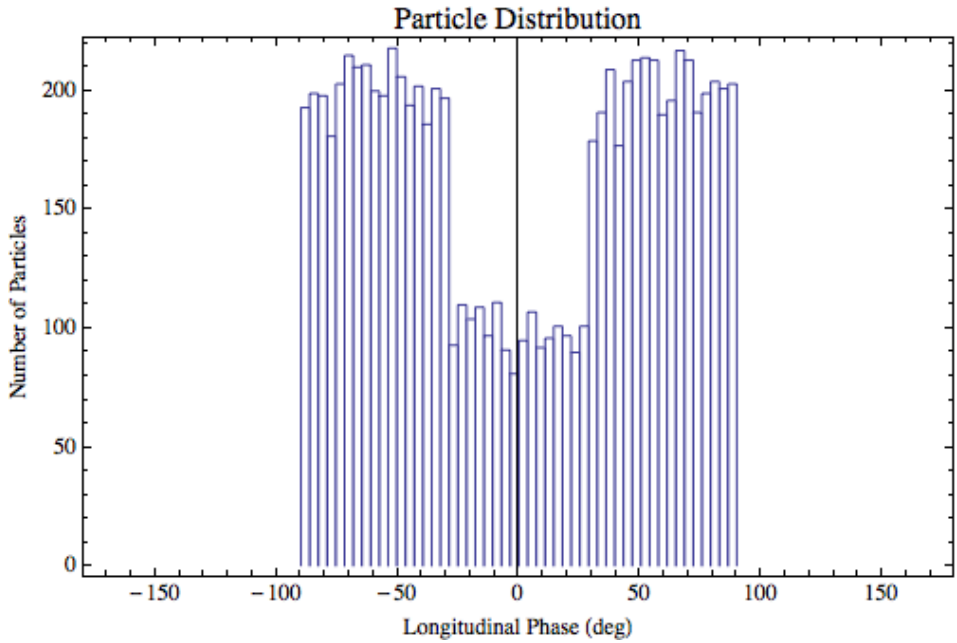

**Figure 15**: The longitudinal phase distribution that is injected into Booster. (Note: this is not the phase space projection onto the phase axis but is how the bunches will be injected). The chopped beam has been confined to ±90° of the bucket and each injection consists of either 1 or 2 bunches.

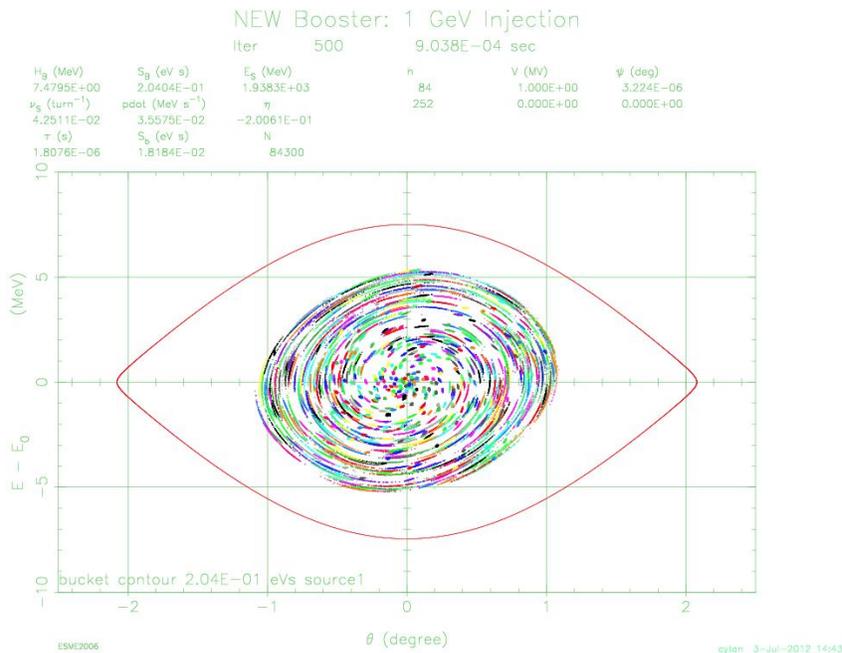

**Figure 16**: The phase space at the end of injection. The projections in longitudinal phase and energy are shown in Figures 17 and 18.

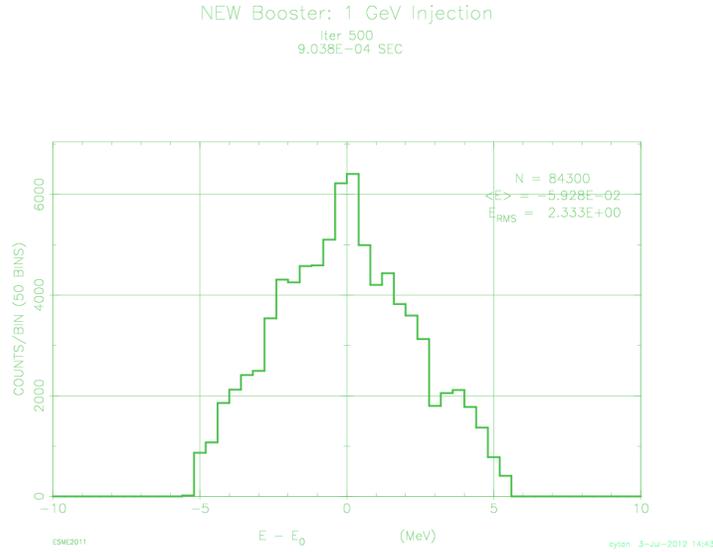

**Figure 17**: The particle distribution projected onto the energy axis.

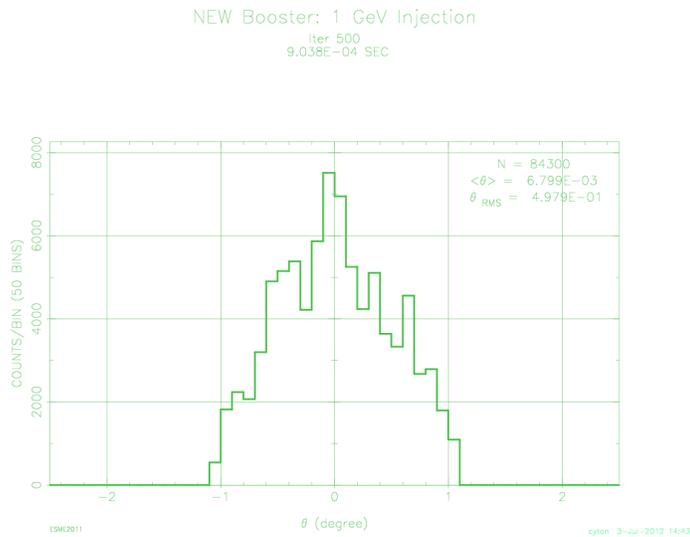

**Figure 18**: The particle distribution projected onto the phase axis.

## 2$^{nd}$ harmonic RF at injection

The addition of a 2$^{nd}$ harmonic RF system can be used to flatten the bucket so that the energy spread is reduced. Figure 19 shows the result of the flattened bucket with the second harmonic at 50% voltage and phase shifted by 180° with respect to the fundamental. The beam distribution is for ±90° chopped beam. Table 4 summarizes the transmission efficiency from the start of ramp to just before transition for different size chopped beam. The largest beam loss comes right at the start of ramp because the stationary bucket is reduced by a factor of 6 when it transitions to an accelerating bucket.

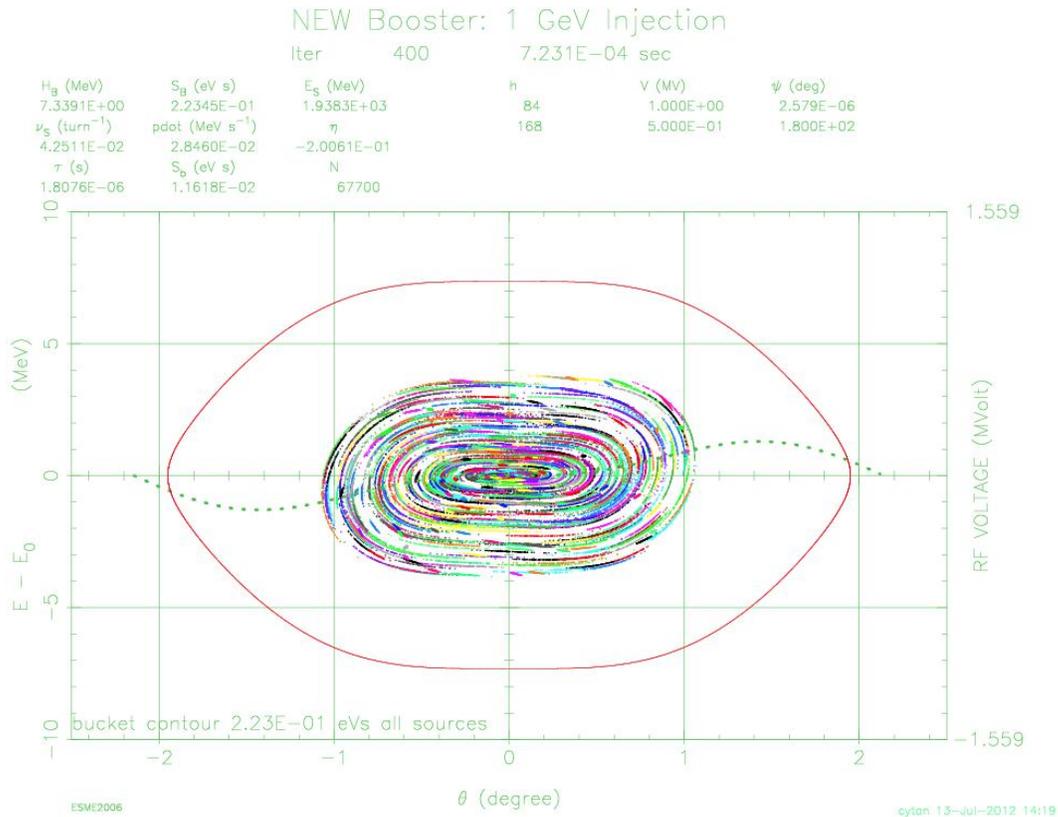

**Figure 19**: The addition of a 2<sup>nd</sup> harmonic at 50% of the voltage and 180° out of phase with respect to the fundamental flattens the bucket and compresses the energy spread of the beam as it evolves in the bucket during injection. The picture shown here is for ±90° chopped beam.

### 3<sup>rd</sup> harmonic RF at injection

A third harmonic RF system can be used instead of the 2<sup>nd</sup> harmonic to flatten the bucket. Compared to the 2<sup>nd</sup> harmonic, the transmission efficiency from the start of ramp to just before transition is worse. However, for ±60° chopped beam, the hit in efficiency is small – see Table 4. There are two advantages in using a 3<sup>rd</sup> harmonic system rather than a 2<sup>nd</sup> harmonic system

- The voltage required to flatten the bucket is 35% of the voltage of the fundamental compared to 50% of the voltage of the fundamental for the 2<sup>nd</sup> harmonic.
- If a 3<sup>rd</sup> harmonic system is required at transition then the cost of using it at injection is minimal.

Figure 20 shows the flattened bucket with a 3<sup>rd</sup> harmonic system at 35% and 180° out of the phase with respect to the fundamental. The beam distribution is for ±90° chopped beam.

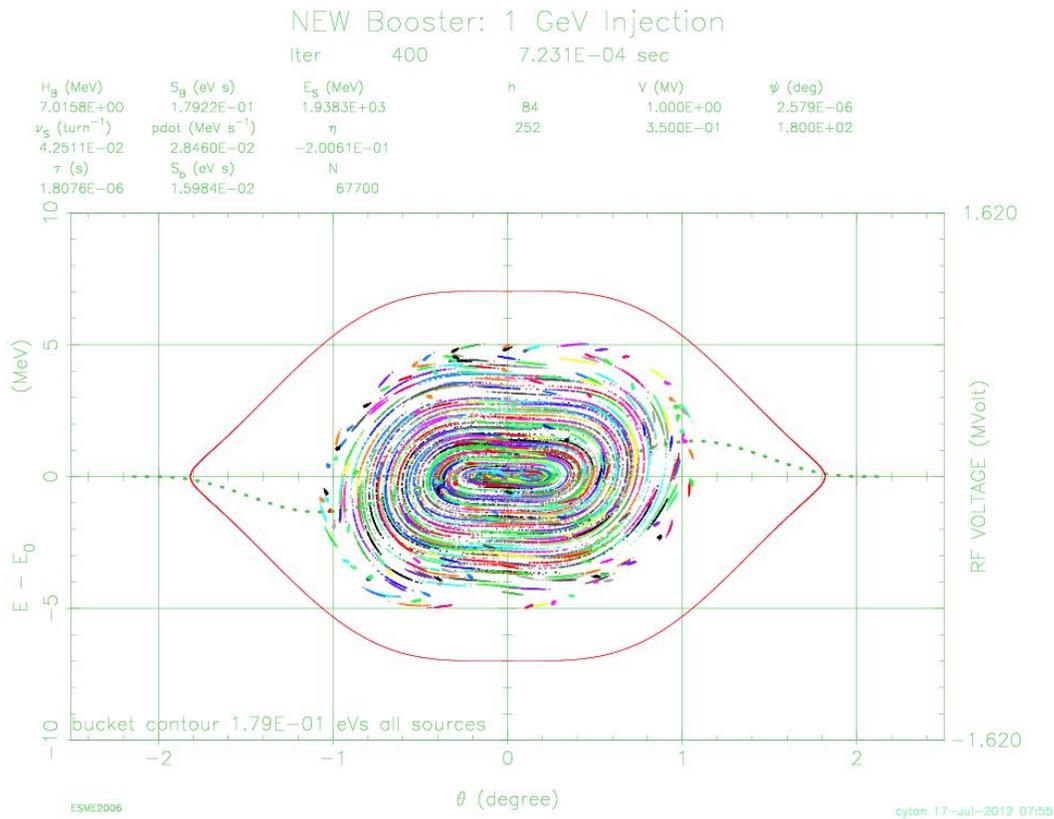

**Figure 20**: The addition of a 3<sup>rd</sup> harmonic at 35% of the voltage and 180° out of phase with respect to the fundamental. The bunch distribution is flattened but not to the extent as the 2<sup>nd</sup> harmonic system. The picture shown here is for ±90° chopped beam.

*Transport efficiency from the start of ramp to the beginning of transition*

The ramp used here is the biased sinusoid defined as KURVEB=3 in ESME. During the ramp, higher harmonics are turned off. In the simulation, beam falls out of the bucket at the start of ramp because the stationary bucket is much larger than the accelerating bucket. In this simulation, with the higher harmonics off, the bucket area just before the start of ramp is 0.2 eV-s, while at the start of ramp, the bucket area is 0.03 eV-s, i.e. more than a factor of 6 smaller.

Table 4 summarizes the results for the cases when only the fundamental RF, 2$^{nd}$ and 3$^{rd}$ harmonics are used at injection. In the case of the 2$^{nd}$ harmonic, it is set to 50% of the voltage and 180° out of phase with respect to the fundamental. In the 3$^{rd}$ harmonic case, it is set to 35% of the voltage and 180° out of phase with respect to the fundamental.

It is clear from this table that the beam must be chopped and that the 2$^{nd}$ harmonic has a better performance than the 3$^{rd}$ harmonic in terms of transmission efficiency. However, there are other considerations for using the 3$^{rd}$ harmonic, particularly transition crossing.

| Size of chopped beam | Transmission Efficiency |
|---|---|
| **Fundamental RF only** | |
| ±180° | 30% |
| ±90° | 52% |
| ±60° | 84% |
| **Fundamental + 2$^{nd}$ harmonic RF** | |
| ±180° | 39% |
| ±90° | 71% |
| ±60° | 97% |
| **Fundamental + 3$^{rd}$ harmonic RF** | |
| ±180° | 35% |
| ±90° | 60% |
| ±60° | 94% |

**Table 4**: Contrasting the ramp efficiencies for the various possible injection scenarios with and without higher order harmonics.

### 3$^{rd}$ Harmonic RF at Transition

In the Booster, transition crossing is strongly dominated by space charge forces. In this regime, the space charge force is defocusing before transition and focusing afterwards. This causes a mismatch to the bucket and as a consequence, there will be longitudinal emittance growth and beam loss.

Near transition energy in the Booster, partial loss of the focusing force in synchrotron motion can be alleviated by either temporally increase RF voltage just before the transition [54] or by flattening the RF wave [55] via the third harmonic RF system. In the flattened RF wave, all beam particles gain an equal amount of energy at each turn. This is called the focus free transition crossing method [56] and has been demonstrated in the Main Ring for preserving the beam intensity through transition crossing.

Figure 21 shows an ESME simulation carried out by using the FFTC scheme for the Booster transition crossing. In this simulation, the proton beam intensity is at the design value of 8.15E10/bunch with an initial longitudinal emittance of 0.02eV-s. In the figure, the beam emittance is larger for the FFTC scheme (0.059 eV-s) as in (c) than the normal transition without 3$^{rd}$ harmonic RF system (0.045eV-s) as in (a). However, more beam survived in the FFTC transition (d) than the normal transition (beam survival rate of 96% in (d) compared to 80.5% in (b)).

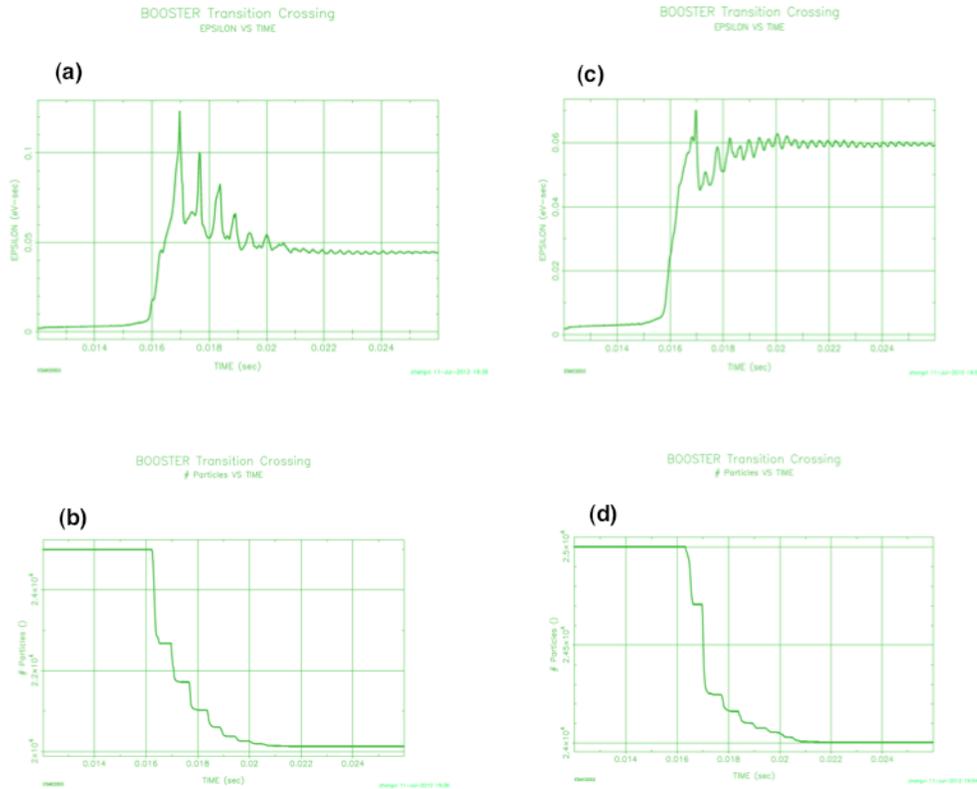

**Figure 21**: Comparison of the normal transition crossing shown in (a) and (b) with the FFTC scheme shown in (c) and (d).

The paper study was also done using RF voltage jumps for Booster operations with a lower beam intensity of 6E10 proton/bunch in which the corrections of the beam to bucket mismatch were simulated. It shows that the mismatch can be reduced by two methods:

- 120 kV of $3^{rd}$ harmonic voltage when applied before and during transition reduces the mismatch. See red traces of Figure 22 and Figure 23.

- At 60 kV, the voltage pattern shown in Figure 22 also reduces the mismatch. See green trace in Figures 22 and 23.

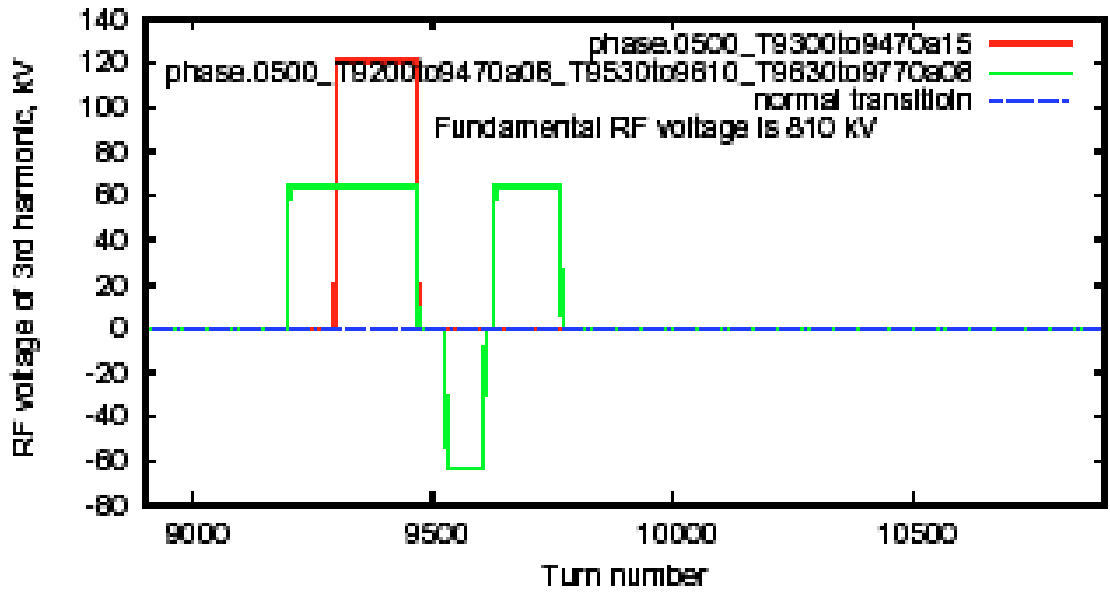

**Figure 22**: Voltage jumps used in the simulations. The red trace is for 120 kV of 3$^{rd}$ harmonic voltage. It is only applied before and during transition. The green trace is for 60 kV of 3$^{rd}$ harmonic voltage. It is applied before and after transition.

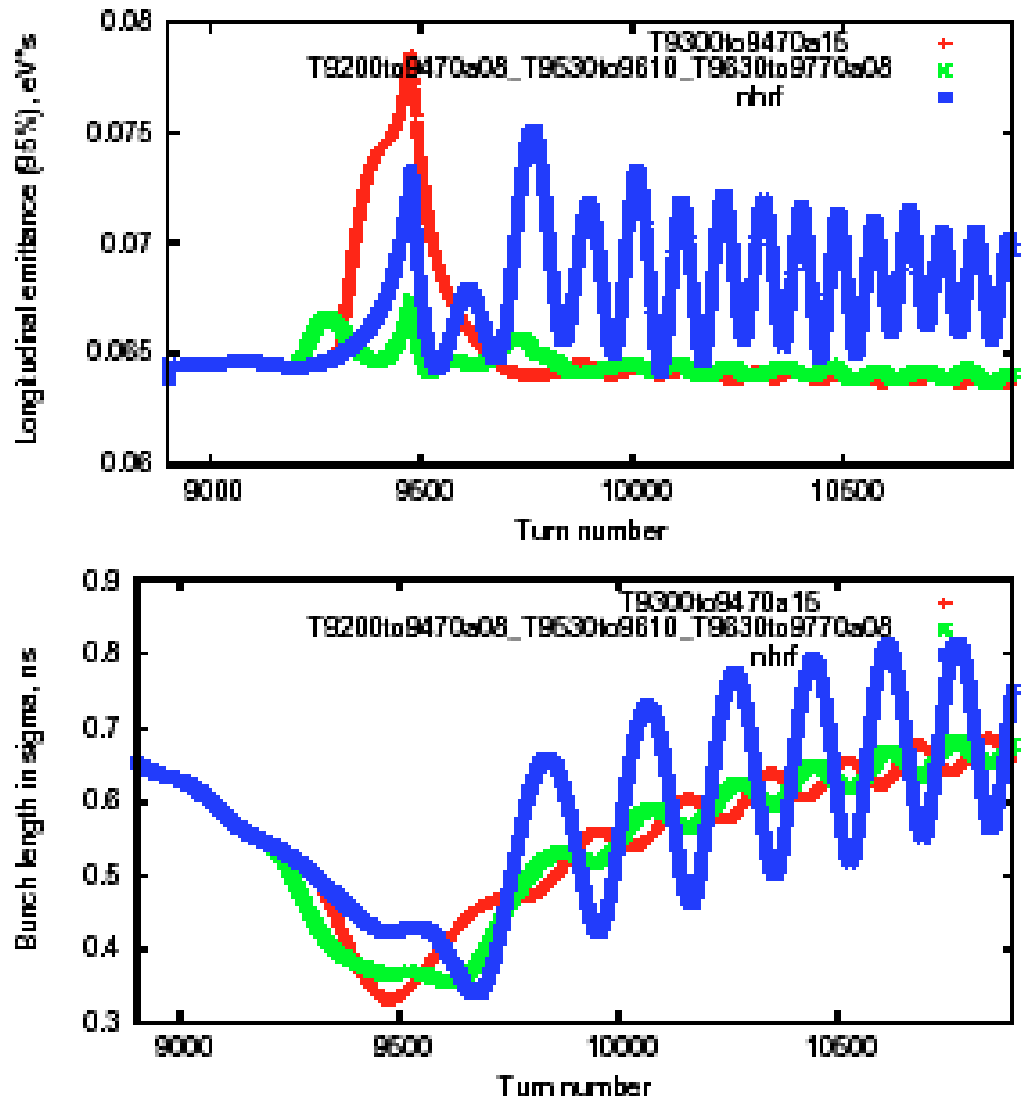

**Figure 23**: 3$^{rd}$ harmonic at 120 kV (red) and 60 kV (green) compared to no 3$^{rd}$ harmonic (blue). The top graph shows the effect on longitudinal emittance and the bottom graph shows the effect on bunch length**.**

# Appendix II: Possible Performance/Scope Enhancements

Among the four Project X mission elements, the third is "provide a path toward a muon source for a possible future Neutrino Factory and/or a Muon Collider." To produce the quantities of muons needed for these possible future facilities, 4 MW of proton beam power is required. Fortuitously, the number of usable muons (collected by a muon capture system, cooled and accelerated) per unit beam power peaks near 8 GeV. Thus, it is natural to look at increasing the Project X pulsed linac output beam power to provide the proton beam for a possible future Neutrino Factory and/or a Muon Collider. The International Design Study for Neutrino Factory (IDS-NF) has provides a list of requirements [57]. Table 1 shows the IDS-NF proton beam requirements compared to Project X pulsed linac capabilities. The Muon Collider's requirement [58] is nearly the same as IDS-NF and is also listed in Table 1.

|  | Project X Capabilities | IDS-NF (Muon Collider) Criteria |
| --- | --- | --- |
| Particle | H minus | proton |
| Beam Energy (GeV) | 8 | 5-15 |
| Average Beam Power (kW) | 320 | 4000 |
| Repetition Rate (Hz) | 10 | 50-60   (12-15) |
| Bunch Rate | 162.5 MHz | -- |

**Table 1**: Project X pulsed linac capabilities compared to the IDS-NF and Muon Collider criteria for proton beam.

The final proton beam requirement for the Neutrino Factory and Muon Collider is for a single proton bunch with an rms length of less than or equal to 3 ns to hit the production target at the repetition rate.

## II.1 High power at 8 GeV

The order of magnitude increase required in Project X beam power at 8 GeV can be addressed by
- increasing the particles per linac bunch;
- increasing the average beam current in the CW linac; and
- increasing the pulsed linac duty factor and repetition rate.

Table 2 shows how Project X parameters can be enhanced to achieve the greater beam power.

|  | Initial | Enhancement |
|---|---|---|
| # Particles/Linac Bunch ($10^8$) | 1.9 | 3.8 |
| Ion Source Current (mA) | 5 | 10 |
| Average Beam Current (mA) | 1 | 5 |
| Pulsed Linac Repetition Rate (Hz) | 10 | 15 |
| Pulsed Linac RF Period (ms) | 100 | 66.7 |
| Pulsed Linac Duty Cycle (%) | 4.3 | 10 |
| Pulsed Linac Beam Pulse Length (ms) | 4.3 | 6.67 |
| 8 GeV Beam Power (kW) | 320 | 4000 |

**Table 2**: Project X operational parameters: initial and enhancement.

The doubling of the number of particles per bunch is achieved by simply doubling the ion source current. The Project X ion source will be capable of delivering 10 mA. The rest of the front end is designed to transport 10 mA as well as the chopper system beam absorber is designed for 25 kW (see Section IV.1.1).

The increase of the ion source beam current is not enough to increase the average linac beam current by a factor of five. The medium energy beam transport section chopper system (see Section IV.1.1) will allow up to half of the beam to exit the front end. The Project X CW linac RF couplers (see Section IV.1.2) will be capable of handling the larger beam current; the couplers may be tuned in the future to optimize the amount of reflected beam power dependent upon the operational physics program that runs in conjunction with a high powered muon facility.

The increased repetition rate and duty factor of the pulsed linac will require more RF power. It is expected that providing the pulsed RF power for ~7.5 ms at 15Hz will require 125 kW per cavity. A straight forward solution is to provide each cavity its own Klystron. The pulsed linac RF couplers (see Section IV.3.1), located inside the cryogenic-modules, will have to be replaced with a 15 kW coupler.

The increased pulsed RF power along with increasing the number of RF power sources will have an impact upon the infrastructure concerning the Project X pulsed linac. The power sources will require more floor space; service buildings can be designed to fit the future power sources or so that the buildings can be expanded. More importantly, the RF power conductors will also increase, requiring more penetrations than will initially required; ideally, Project X will be designed with enough penetrations for the larger number of power conductors. In addition, the increase number of pulsed RF power sources will need additional electrical power and water cooling; both utilities should initially be designed with the possibility of future expansion.

The increased RF pulsed power will also require more cryogenic cooling of the SRF. There will need to be an increase of 65% cryogenic cooling power. The cryogenic plant (see Section IV.5) and associated infrastructure should be designed with possible expansion as a criterion.
The linacs collimation systems and beam aborts will have to be designed (see Section IV.2.1), or replaced in the future, to handle the increased beam power.

## II.2 Accumulator rings

If the enhancements to Project X outlined in the previous section are done, then a majority of the Neutrino Factory and Muon Collider criteria in Table 1 will be met. The output of the enhanced Project X pulsed linac will be 4 MW of H minus beam at 162.5 MHz. The ions need to be converted to protons and the linac bunches need to be combined to form a few intense bunches. An accumulation ring will be required.

At a point along the beam transport line between the pulsed linac and the Main Injector enclosure, a new switching magnet will direct the H minus beam into a new transport line. This 8 GeV H minus transport line will have the same physics considerations as the transport line exiting the Project X pulsed linac. The destination of this new transport line will be a new Accumulator Ring. Figure 1 shows a possible implementation of a new Accumulator Ring.

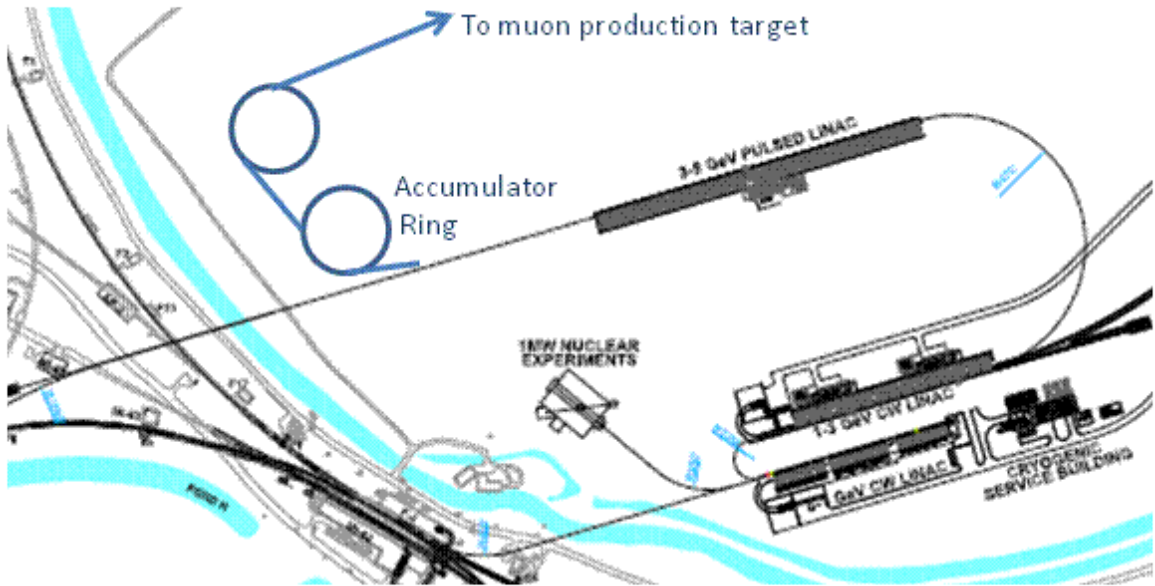

**Figure 1**: Possible layout of proton rings needed for future Neutrino Factory and/or Muon Collider

The Accumulator Ring will have ion stripping injection and will accumulate linac pulses into predefined RF bunches in the same manner as the Recycler. The main difference is that the ring will be much smaller with only four bunches. The front end chopper system will be programmed to allow half of the linac pulses through Project X and be synchronized to fill the RF buckets (similar to the scheme outlined earlier in II.2). An early design has the Accumulator Ring with a circumference of ~300 m [5911]. Initial calculations show the resulting intense proton bunches will be stable.

A challenge will be converting 4 MW of H minus beam into protons. Rotating foil and laser stripping systems will be investigated.

Even with the small circumference of the ring and small number of bunches, bunch rotation will need to be performed to achieve the very short (less than or equal to 3 nsec) bunch length desired for muon production and capture. A second accelerator ring is thought to be appropriate to perform this task as well as final proton beam collimation. The second ring would be located with the muon production and capture system; a proton beam transport system between the two rings would become necessary.

The Neutrino Factory and Muon Collider have different requirements for the repetition rate (see Table 2). The above enhancements result in the Accumulator Ring results in four proton bunches being formed at 15 Hz. Extraction from the Accumulator Ring and the delivery transport lines from the second ring are different for the Neutrino Factory and Muon Collider.

The Neutrino Factory requires proton bunches on target at a rate of 50-60 Hz. After the Accumulator Ring is filled, a single bunch is extracted to the second ring. The single bunch is then bunch rotated and then extracted to the target. At intervals of 16.7 ms, each of the remaining bunches are extracted from the Accumulator Ring, bunch rotated and sent to the muon production target. The resulting proton beam on target is at a rate of 60Hz, appropriate for a Neutrino Factory.

The 15 Hz repletion rate from the pulsed linac is appropriate for a Muon Collider. All four bunches will be extracted from the Accumulator Ring, bunch rotated in the second ring and then all four bunches will be extracted. Each bunch would be directed into a separate beam line. Each transport beam line would be a different length such that the four proton bunches would arrive simultaneously at the target (Figure 2).

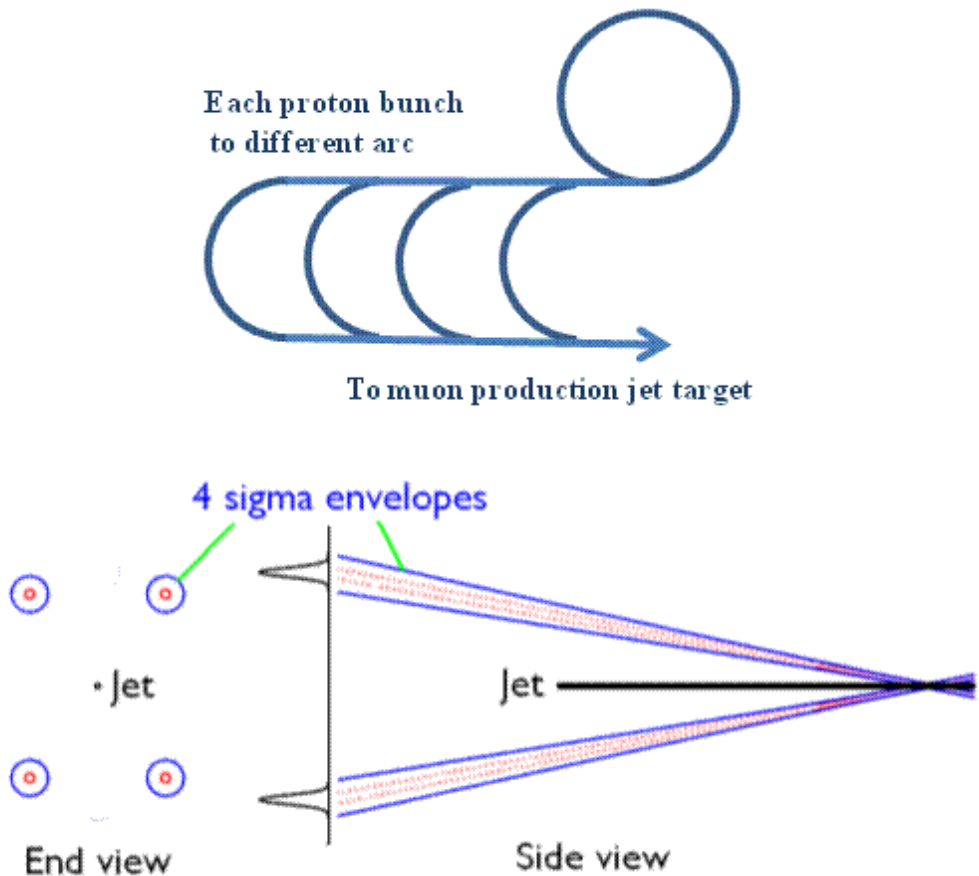

**Figure 2**: Top schematic shows a possible layout of four transport lines. The bottom shows four different beam trajectories to jet target for muon production.